\documentclass[manyauthors,nocleardouble,COMPASS]{cernphprep}

\usepackage{bm}
\usepackage{color}

\usepackage{enumerate}
\usepackage{amssymb}
\usepackage{amsmath}
\usepackage{amsbsy}
\usepackage{times}
\usepackage{epsfig}
\usepackage{colordvi}
\usepackage{graphicx}
\usepackage{wrapfig}
\usepackage[numbers, square, comma, sort&compress]{natbib}
\usepackage{multicol}
\usepackage{booktabs}
\usepackage{multirow}
\usepackage{lineno}
\usepackage{wasysym}
\usepackage{graphicx}
\usepackage{mathrsfs}
\usepackage[caption=false]{subfig}
\usepackage{tabularx}
\usepackage{upgreek}
\usepackage[colorlinks=false,linktocpage=true,breaklinks=true]{hyperref} 
\RequirePackage[T1]{fontenc}

\hypersetup{
     bookmarks=true,         
     unicode=false,          
     pdftoolbar=true,        
     pdfmenubar=true,        
     pdffitwindow=false,     
     pdfstartview={FitH},    
     pdfauthor={The COMPASS Collaboration},     
     pdfsubject={PWA},   
     pdfnewwindow=true,      
     colorlinks=false,       
     linkcolor=red,          
     citecolor=green,        
     filecolor=magenta,      
     urlcolor=cyan           
}

\makeatletter

\DeclareSymbolFont{letters}     {OML}{cmm}{m}{it}
\DeclareSymbolFont{symbols}     {OMS}{cmsy}{m}{n}
\DeclareSymbolFont{largesymbols}{OMX}{cmex}{m}{n}

\input myunits.sty

\newcommand{\figref}[1]{Fig.~\ref{#1}}
\newcommand{\Figref}[1]{Figure~\ref{#1}}
\newcommand{\secref}[1]{Section~\ref{#1}}
\newcommand{\Secref}[1]{Section~\ref{#1}}
\begin{document}

\begin{titlepage}

\PHnumber{2014--247}
\PHdate{05 October 2014}

\title{The COMPASS Setup for Physics with Hadron Beams}

\Collaboration{The COMPASS Collaboration}
\ShortAuthor{The COMPASS Collaboration}

\begin{abstract} 
The main characteristics of the COMPASS experimental setup for physics with
hadron beams are described.  This setup was designed to perform exclusive
measurements of processes with several charged and/or neutral particles in the
final state. Making use of a large part of the apparatus that was previously
built for spin structure studies with a muon beam, it also features a new target
system as well as new or upgraded detectors. The hadron setup is able to operate
at the high incident hadron flux available at CERN.  It is characterised by
large angular and momentum coverages, large and nearly flat acceptances, and
good two and three-particle mass resolutions.  In 2008 and 2009 it was
successfully used with positive and negative hadron beams and with liquid
hydrogen and solid nuclear targets. This article describes the new and upgraded
detectors and auxiliary equipment, outlines the reconstruction procedures used,
and summarises the general performance of the setup.

\vspace{3em}
{\it key words:}
fixed target experiment,  
hadron spectroscopy, 
silicon microstrip detectors , 
GEM detector, 
drift chambers, 
RICH detector, 
calorimetry, 
front-end electronics, 
data acquisition, 
data reconstruction,
Monte-Carlo simulation
\\[3em]

\\
\end{abstract}

\vfill 
\Submitted{(to be submitted to Nucl. Instr. and Meth. A)}

\end{titlepage}

{\pagestyle{empty} }
%
%
\section*{The COMPASS Collaboration}
\label{app:collab}
\renewcommand\labelenumi{\textsuperscript{\theenumi}~}
\renewcommand\theenumi{\arabic{enumi}}
\begin{flushleft}
Ph.~Abbon\Irefn{saclay},
C.~Adolph\Irefn{erlangen},
R.~Akhunzyanov\Irefn{dubna}, 
Yu.~Alexandrov\Irefn{moscowlpi}\Deceased,
M.G.~Alexeev\Irefn{turin_u},
G.D.~Alexeev\Irefn{dubna}, 
A.~Amoroso\Irefnn{turin_u}{turin_i},
V.~Andrieux\Irefn{saclay},
V.~Anosov\Irefn{dubna}, 
A.~Austregesilo\Irefnn{cern}{munichtu}
% ...
\end{flushleft}

%
%

\begin{Authlist}
\item \Idef{bielefeld}{Universit\"at Bielefeld, Fakult\"at f\"ur Physik, 33501 Bielefeld, Germany\Arefs{f}}
\item \Idef{bochum}{Universit\"at Bochum, Institut f\"ur Experimentalphysik, 44780 Bochum, Germany\Arefs{f}\Arefs{ll}}
\item \Idef{bonniskp}{Universit\"at Bonn, Helmholtz-Institut f\"ur  Strahlen- und Kernphysik, 53115 Bonn, Germany\Arefs{f}}
\item \Idef{bonnpi}{Universit\"at Bonn, Physikalisches Institut, 53115 Bonn, Germany\Arefs{f}}
\item \Idef{brno}{Institute of Scientific Instruments, AS CR, 61264 Brno, Czech Republic\Arefs{g}}
\item \Idef{calcutta}{Matrivani Institute of Experimental Research \& Education, Calcutta-700 030, India\Arefs{h}}
\item \Idef{dubna}{Joint Institute for Nuclear Research, 141980 Dubna, Moscow region, Russia\Arefs{i}}
\item \Idef{erlangen}{Universit\"at Erlangen--N\"urnberg, Physikalisches Institut, 91054 Erlangen, Germany\Arefs{f}}
\item \Idef{freiburg}{Universit\"at Freiburg, Physikalisches Institut, 79104 Freiburg, Germany\Arefs{f}\Arefs{ll}}
\item \Idef{cern}{CERN, 1211 Geneva 23, Switzerland}
\item \Idef{liberec}{Technical University in Liberec, 46117 Liberec, Czech Republic\Arefs{g}}
\item \Idef{lisbon}{LIP, 1000-149 Lisbon, Portugal\Arefs{j}}
\item \Idef{mainz}{Universit\"at Mainz, Institut f\"ur Kernphysik, 55099 Mainz, Germany\Arefs{f}}
\item \Idef{miyazaki}{University of Miyazaki, Miyazaki 889-2192, Japan\Arefs{k}}
\item \Idef{moscowlpi}{Lebedev Physical Institute, 119991 Moscow, Russia}
\item \Idef{munichlmu}{Ludwig-Maximilians-Universit\"at M\"unchen, Department f\"ur Physik, 80799 Munich, Germany\Arefs{f}\Arefs{l}}
\item \Idef{munichtu}{Technische Universit\"at M\"unchen, Physik Department, 85748 Garching, Germany\Arefs{f}\Arefs{l}}
\item \Idef{nagoya}{Nagoya University, 464 Nagoya, Japan\Arefs{k}}
\item \Idef{praguecu}{Charles University in Prague, Faculty of Mathematics and Physics, 18000 Prague, Czech Republic\Arefs{g}}
\item \Idef{praguectu}{Czech Technical University in Prague, 16636 Prague, Czech Republic\Arefs{g}}
\item \Idef{protvino}{State Scientific Center Institute for High Energy Physics of National Research Center `Kurchatov Institute', 142281 Protvino, Russia}
\item \Idef{saclay}{CEA IRFU/SPhN Saclay, 91191 Gif-sur-Yvette, France\Arefs{ll}}
\item \Idef{telaviv}{Tel Aviv University, School of Physics and Astronomy, 69978 Tel Aviv, Israel\Arefs{m}}
\item \Idef{triest_u}{University of Trieste, Department of Physics, 34127 Trieste, Italy}
\item \Idef{triest_i}{Trieste Section of INFN, 34127 Trieste, Italy}
\item \Idef{triest_ictp}{Abdus Salam ICTP, 34151 Trieste, Italy}
\item \Idef{turin_u}{University of Turin, Department of Physics, 10125 Turin, Italy}
\item \Idef{turin_p}{University of Eastern Piedmont, 15100 Alessandria, Italy}
\item \Idef{turin_i}{Torino Section of INFN, 10125 Turin, Italy}
\item \Idef{warsaw}{National Centre for Nuclear Research, 00-681 Warsaw, Poland\Arefs{n} }
\item \Idef{warsawu}{University of Warsaw, Faculty of Physics, 00-681 Warsaw, Poland\Arefs{n} }
\item \Idef{warsawtu}{Warsaw University of Technology, Institute of Radioelectronics, 00-665 Warsaw, Poland\Arefs{n} }
\item \Idef{yamagata}{Yamagata University, Yamagata, 992-8510 Japan\Arefs{k} }
\end{Authlist}
%
%
\vspace*{-\baselineskip}\renewcommand\theenumi{\alph{enumi}}
\begin{Authlist}
\item \Adef{a}{Also at Instituto Superior T\'ecnico, Universidade de Lisboa, Lisbon, Portugal}
\item \Adef{aa}{Also at Department of Physics, Pusan National University, Busan 609-735, Republic of Korea and at Physics Department, Brookhaven National Laboratory, Upton, NY 11973, U.S.A. }
\item \Adef{bb}{Supported by the DFG Research Training Group Programme 1102  ``Physics at Hadron Accelerators''}
\item \Adef{b}{Also at Chubu University, Kasugai, Aichi, 487-8501 Japan\Arefs{k}}
\item \Adef{c}{Also at KEK, 1-1 Oho, Tsukuba, Ibaraki, 305-0801 Japan}
\item \Adef{c1c}{Present address: Universit\"at Bonn, Helmholtz-Institut f\"ur Strahlen- und Kernphysik, 53115 Bonn, Germany}
\item \Adef{cc}{Also at Moscow Institute of Physics and Technology, Moscow Region, 141700, Russia}
\item \Adef{x}{present address: RWTH Aachen University, III. Physikalisches Institut, 52056 Aachen, Germany}
\item \Adef{f}{Supported by the German Bundesministerium f\"ur Bildung und Forschung}
\item \Adef{g}{Supported by Czech Republic MEYS Grants ME492 and LA242}
\item \Adef{h}{Supported by SAIL (CSR), Govt.\ of India}
\item \Adef{i}{Supported by CERN-RFBR Grants 08-02-91009 and 12-02-91500}
\item \Adef{j}{\raggedright Supported by the Portuguese FCT - Funda\c{c}\~{a}o para a Ci\^{e}ncia e Tecnologia, COMPETE and QREN, Grants CERN/FP/109323/2009, CERN/FP/116376/2010 and CERN/FP/123600/2011}
\item \Adef{k}{Supported by the MEXT and the JSPS under the Grants No.18002006, No.20540299 and No.18540281; Daiko Foundation and Yamada Foundation}
\item \Adef{l}{Supported by the DFG cluster of excellence `Origin and Structure of the Universe' (www.universe-cluster.de)}
\item \Adef{ll}{Supported by EU FP7 (HadronPhysics3, Grant Agreement number 283286)}
\item \Adef{m}{Supported by the Israel Science Foundation, founded by the Israel Academy of Sciences and Humanities}
\item \Adef{n}{Supported by the Polish NCN Grant DEC-2011/01/M/ST2/02350}
\item [{\makebox[2mm][l]{\textsuperscript{*}}}] Deceased
\end{Authlist}

\newpage

\section{Introduction}
\label{sec:intro}
The goal of the COMPASS experiment at CERN is a better understanding of the
structure and dynamics of hadrons. At the relevant length scales of
$\sim\EE*{-15}\,\m$ the strong coupling constant $\alpha_{\mathrm s}$ approaches
unity, which is the domain of non-perturbative Quantum Chromodynamics (QCD).
Using a $160-200\,\GeV/c$ muon beam, COMPASS studies the nucleon spin structure
by deep inelastic scattering off a polarised $^6$LiD or NH$_3$
target~\cite{Abbon:2007pq}.  Experiments with hadron beams of $190\,\GeV/c$,
which started in 2008, aim at precision spectroscopy of light mesons and baryons
with masses up to $3\,\GeV/c^2$, the identification and systematic study of
possible exotic configurations with gluonic degrees of freedom or multi-quark
systems, as well as the study of processes governed by chiral dynamics and tests
of predictions of chiral perturbation theory.

\begin{figure}[bp]
  \centering
  \includegraphics[width=0.3\columnwidth]{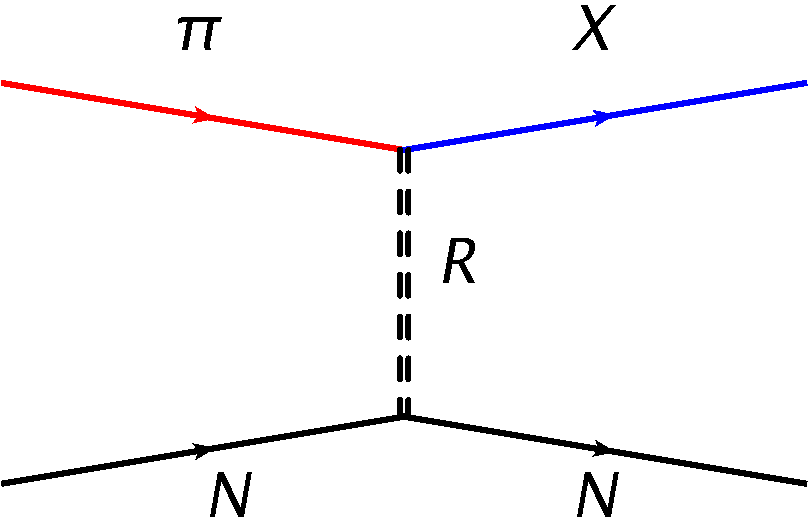}
  \hfill
  \includegraphics[width=0.3\columnwidth]{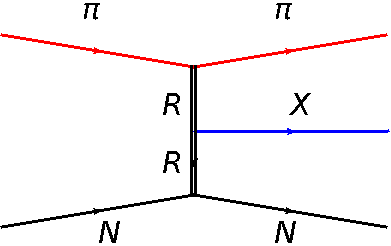}
  \hfill
  \includegraphics[width=0.3\columnwidth]{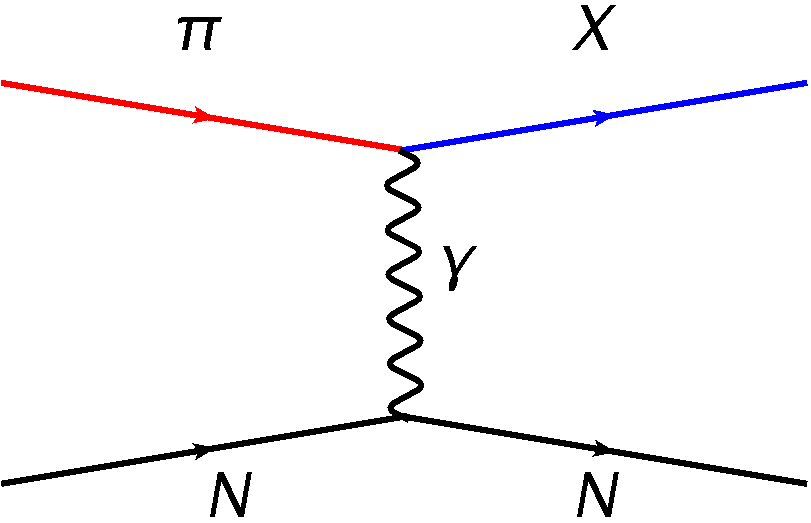}
  \caption{Production mechanisms employed in COMPASS for (left)
    diffractive dissociation, (middle) central production, (right)
    photo-production by quasi-real photons $\gamma$, with $\pi$
    denoting the beam 
    particle (can be also $p$, $K$), and $N$ the target nucleon or nucleus.}  
  \label{fig:intro.production}
\end{figure}

These experiments require a state-of-the-art spectrometer with high acceptance
and high resolution for charged and neutral particles in order to perform
exclusive measurements of multi-particle final states over a wide kinematic
range.  Three different mechanisms contribute to the production of a system $X$,
as shown in Fig.~\ref{fig:intro.production}: diffractive dissociation and
central production, which can be described to proceed via the exchange of one or
two Reggeons $\mathbb{R}$, respectively, between beam hadron and target nucleus
$N$, and photo-production in the Coulomb field of a nucleus at very low values
of momentum transfer.
In all these processes the final-state particles are emitted mostly in forward
direction, which requires an excellent angular resolution of the spectrometer
very close to and even within the beam envelope.  For the interpretation of the
data using partial-wave analysis (PWA) tools, a large and uniform acceptance
over the whole kinematic domain under study is mandatory.

The relative contributions of the above-mentioned processes to a data sample can
be varied by applying different trigger conditions and by adjusting kinematic
selection criteria in the analysis.  At intermediate-to-large values of momentum
transfer, the cross section for reactions mediated by Pomerons,
\mbox{i.e.}\ Reggeons with vacuum quantum numbers, is large, of the order of
$1-2\,\mathrm{mb}$.  The contribution of individual partial waves, however, may
be several orders of magnitude smaller, thus requiring the collection of large
data sets.  Of particular interest are states $X$ that do not fit into the naive
constituent quark model but are allowed by QCD, like glueballs or hybrids which
carry gluonic degrees of freedom, or multi-quark systems.  States with gluonic
degrees of freedom are generally believed to be enhanced in reactions in which
Pomerons are exchanged.  A small but significant contribution of a spin-exotic
partial wave with non-$q\overline{q}'$ quantum numbers $J^{PC}=1^{-+}$,
consistent with the $\pi_1(1600)$, was confirmed by COMPASS using data taken in
2004 \cite{Alekseev:2009xt}.  However, an unambiguous understanding of the
underlying structure of this and many other light-hadron states requires
experiments with higher statistical accuracy, employing different production
mechanisms and observation of the same system $X$ in different decay channels.
At very small values of momentum transfer, the cross section is dominated by
Primakoff reactions, \mbox{i.e.}\ Coulomb scattering of pions or kaons off
quasi-real photons emitted from a nuclear target.  The dynamics of the
scattering of a beam $\pi$ into $\pi\gamma$, $\pi^-\pi^0$, $3\pi$, etc., at
low-energy, i.e.\ from threshold up to a few pion masses, is predicted by chiral
perturbation theory (ChPT).  COMPASS can thus scrutinise ChPT predictions of
chiral dynamics \cite{Adolph:2011it} and of fundamental low-energy parameters
such as the polarisabilities of mesons.

Compared to previous experiments, the main advantages of the COMPASS setup are
the possibility to study reactions with different projectiles in high-intensity
beams with up to $\EE*{7}\,\mathrm{part.}/\s$ and to reconstruct final states
containing both neutral and charged particles.  Different charges and types of
beam particles, e.g.\ $\pi^\pm$, K$^\pm$, and (anti)protons, can be selected by
tuning the COMPASS beam line and by tagging them with differential Cherenkov
counters.  The possibility to switch between pion and muon beams of the same
momentum is a unique asset for the measurement of pion polarisabilities at
COMPASS, where the systematic error of the measurement can be significantly
reduced through regular reference measurements with incident muons,
i.e.\ point-like particles.  As target material either liquid hydrogen or
various solid-state nuclear targets are used.  A recoil proton detector (RPD) is
installed around the target to ensure the exclusivity of the final state.  A set
of double-sided silicon microstrip detectors positioned upstream and downstream
of the target is used to reconstruct the interaction vertex and the angles of
the outgoing particles.  Here, the required angular resolution is dictated by
Primakoff reactions, where pions or muons scattered by angles of a few hundred
$\upmu\mathrm{rad}$ have to be detected.  A high momentum resolution for charged
particles is provided by a two-stage magnetic spectrometer.  For the tracking in
the beam region new pixelised Gas Electron Multiplier (GEM) detectors with a
minimised material budget along the beam were built, in replacement of the
thicker scintillating fibre detectors.  For the tracking at small angles, the
existing Micromegas trackers were adapted to the hadron beam conditions.  A
major upgrade of the Ring-Imaging Cherenkov (RICH) counter was carried out,
which largely improved the performance of particle identification at high rates.
Photons are detected in two electromagnetic calorimeters, which have been
optimised for stability and uniformity in order to achieve good resolution.
Several new trigger elements were built and implemented into the trigger system.

The present paper describes the modifications and upgrades of the experimental
setup required for the hadron programme of COMPASS.  Some of these upgrades were
already realised for the nucleon spin programme after 2005.  After a brief
overview of the layout of the spectrometer in \secref{sec:layout}, the beam line
and associated detectors are described in \secref{sec:beam} and the target
region in \secref{sec:target}. The newly installed tracking detectors are
discussed in \secref{sec:tracking}. \Secref{sec:pid} deals with the systems used
for particle identification, namely the RICH counter and the two electromagnetic
calorimeters. The various trigger systems are explained in \secref{sec:trigger}
and the data acquisition in \secref{sec:daq}. The algorithms for event
reconstruction and the performance of individual detector components are
summarised in \secref{sec:reconstruction}, while the global spectrometer
performance and Monte Carlo simulations of the apparatus are discussed in
\secref{sec:performance}.
 
Throughout this paper, the following kinematic variables will be used: the
squared four-momentum transfer from the incident beam particle to the recoiling
target nucleus $t=(p_\mathrm{beam}-p_X)^2$; the reduced squared 4-momentum
transfer $t' = \vert t \vert - \vert t \vert_\mathrm{min}$ to the recoiling
target nucleon beyond the kinematic minimum $\vert t \vert_\mathrm{min}$; the
Gottfried-Jackson angle $\theta_\mathrm{GJ}$\label{gj_angle_def}, defined as the
polar angle of the three-momentum of the isobar (i.e.\ di-pion) from the decay
of $X$, and the corresponding azimuthal Treiman-Young angle
$\phi_\mathrm{TY}$. These two angles are calculated in the centre-of-momentum
frame of $X$ with the $z$-axis along the beam direction and the $y$-axis
perpendicular to the production plane, formed by the momentum vectors of target
and recoil particles.

\section{Layout of the Spectrometer}
\label{sec:layout}

\begin{figure*}[tbp]
\centering
\includegraphics[width=\textwidth]{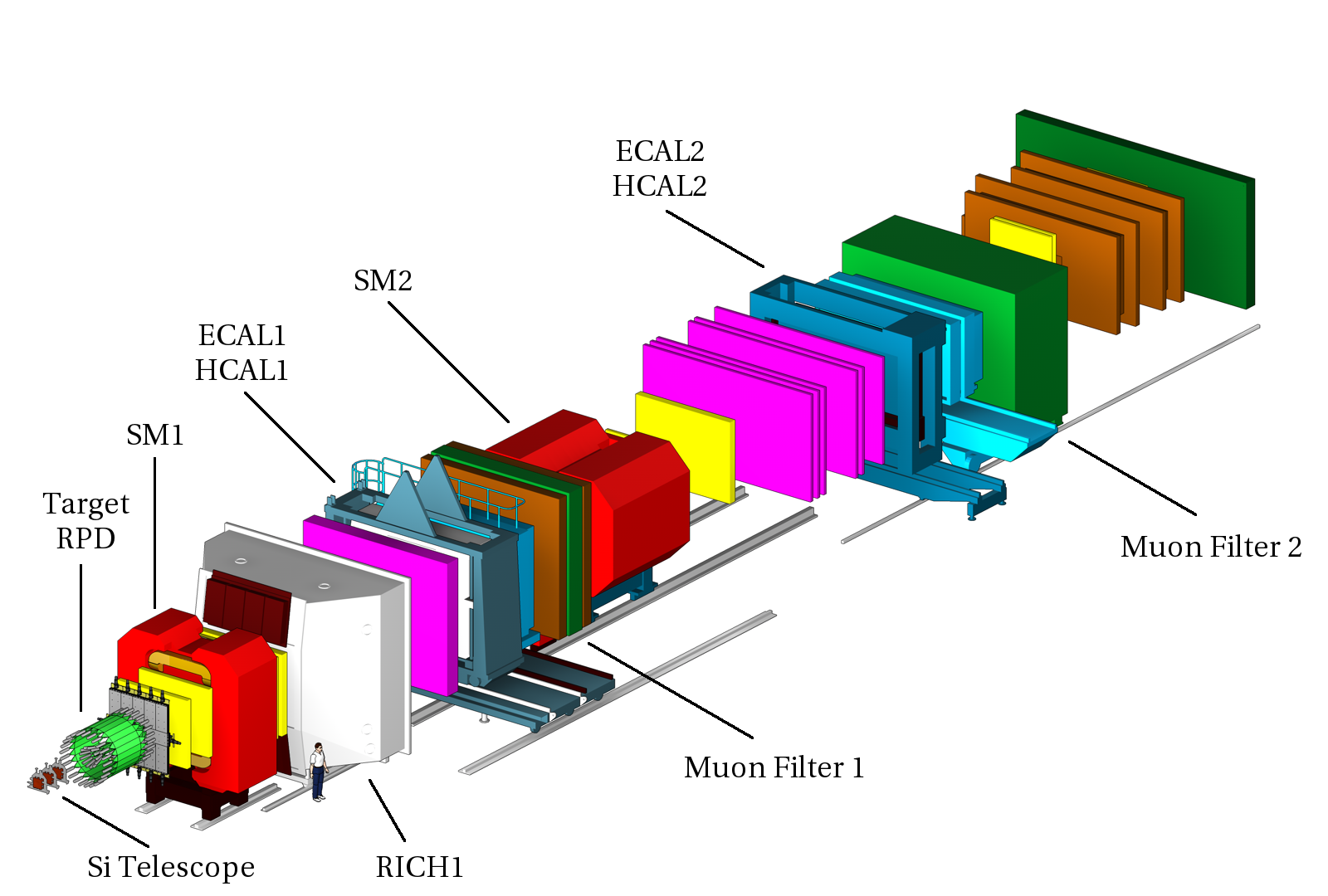}
\caption{Three-dimensional view of the COMPASS setup for measurements with
  hadron beams. The beam comes from the left side. The upstream part of the
  setup (beam line) is not shown here. The different colours indicate different
  detector types.}
\label{fig:layout.3d} 
\end{figure*}

The main features of the COMPASS experimental setup and most of the detectors as
used until 2004 are described in Ref.~\cite{Abbon:2007pq}. In this section a
short overview of the apparatus is given, with particular emphasis on detectors
that are either specific to the data taking with hadron beams in 2008 and 2009,
or were added to the setup after 2005 to be used in both muon and hadron
programmes.
 
\begin{figure*}[tbp]
\centering \includegraphics[width=0.95\textwidth]{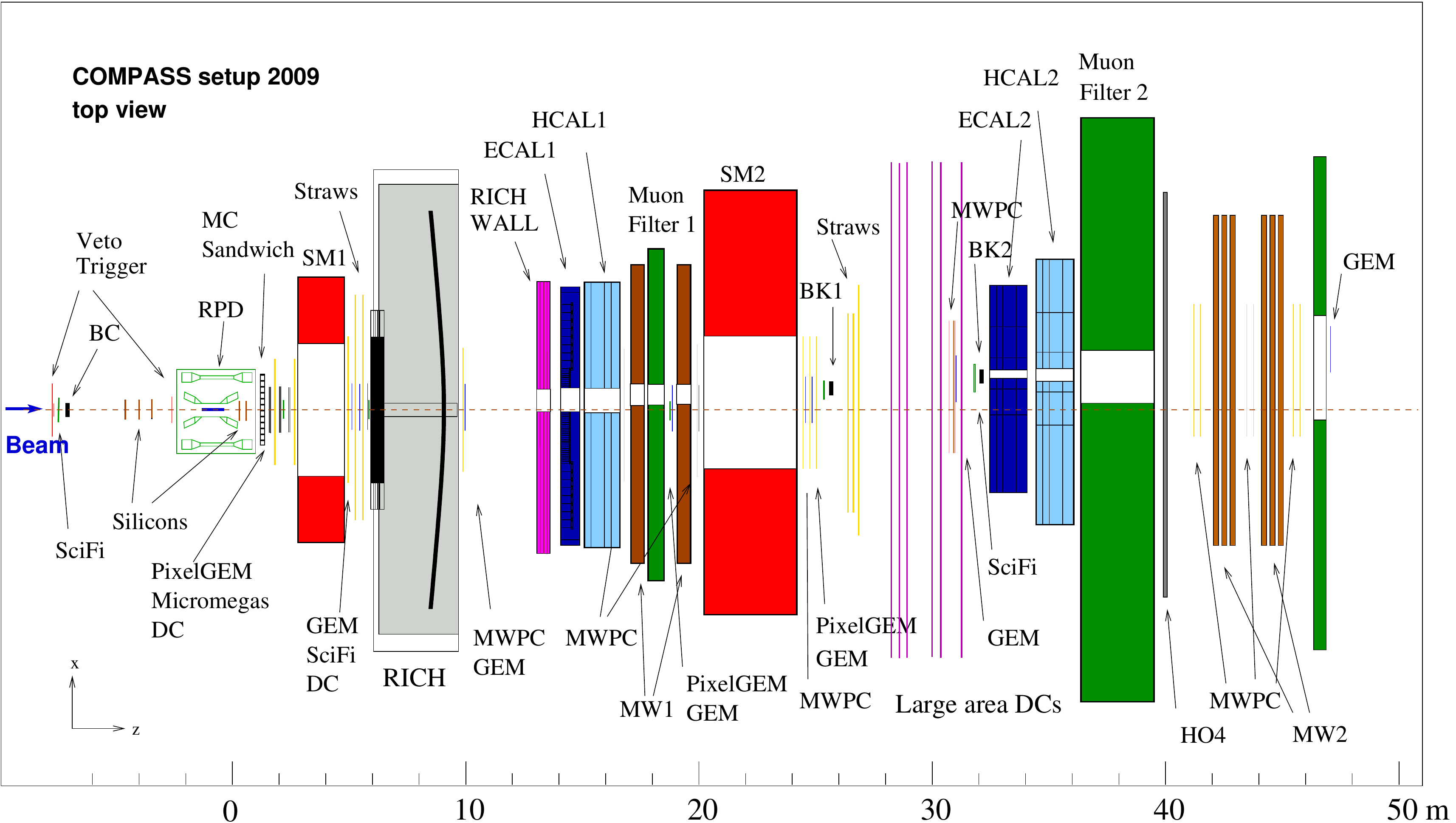}
\caption{Top view of the COMPASS setup for data taking with hadron beams.  The
  labels indicate the various detectors, as referenced throughout this paper.
  The vertical scale is only indicative of the relative detector sizes.  The
  colour code follows that of Fig.~\ref{fig:layout.3d}.}
\label{fig:layout_setup} 
\end{figure*}

The COMPASS setup can be divided into four parts along the beam, starting with
the beam line section and the detectors that identify the incoming beam
particles.  It is followed by the target region, which is specific for each of
the COMPASS physics programmes. It comprises the target and the detectors
located in its near vicinity.  The third part, called Large Angle Spectrometer
(LAS) includes the first dipole magnet, SM1, the tracking detectors around it,
and the RICH-1 counter. The fourth part, called Small Angle Spectrometer (SAS),
occupies the downstream part of the setup. It is built around the SM2 dipole
magnet and includes several tracking detectors.  Both LAS and SAS comprise a
pair of electromagnetic and hadron calorimeters, and a muon filter.  Figures
\ref{fig:layout.3d} and \ref{fig:layout_setup} show the three-dimensional and
top views of the COMPASS setup, respectively.

\subsection{Beam line} 

The COMPASS setup is located at the end of the M2 beam line of the CERN SPS
accelerator.  The M2 beam line can be tuned for beams of different particles,
including secondary hadron beams and tertiary muon or electron beams. Hadron and
muon beams can be either of negative or positive charge. Switching between beams
takes typically thirty minutes.

During data taking with hadron beams only the trajectory of the incident beam
particle is measured. The Beam Momentum Station (BMS), which is used for the
determination of the incident momentum during measurements with a muon beam, is
moved out of the beam in order to minimise the material budget along the beam
path. However, the muon beam is also used during Primakoff measurements in order
to complement the data taken with pions. The BMS is then moved back into the
beam line. Downstream of the BMS location, two differential Cherenkov counters
identify the hadrons (pions, kaons, or protons) that are present in the hadron
beam.

\subsection{Target region}

\begin{figure*}[tbp]
\centering \includegraphics[width=\textwidth]{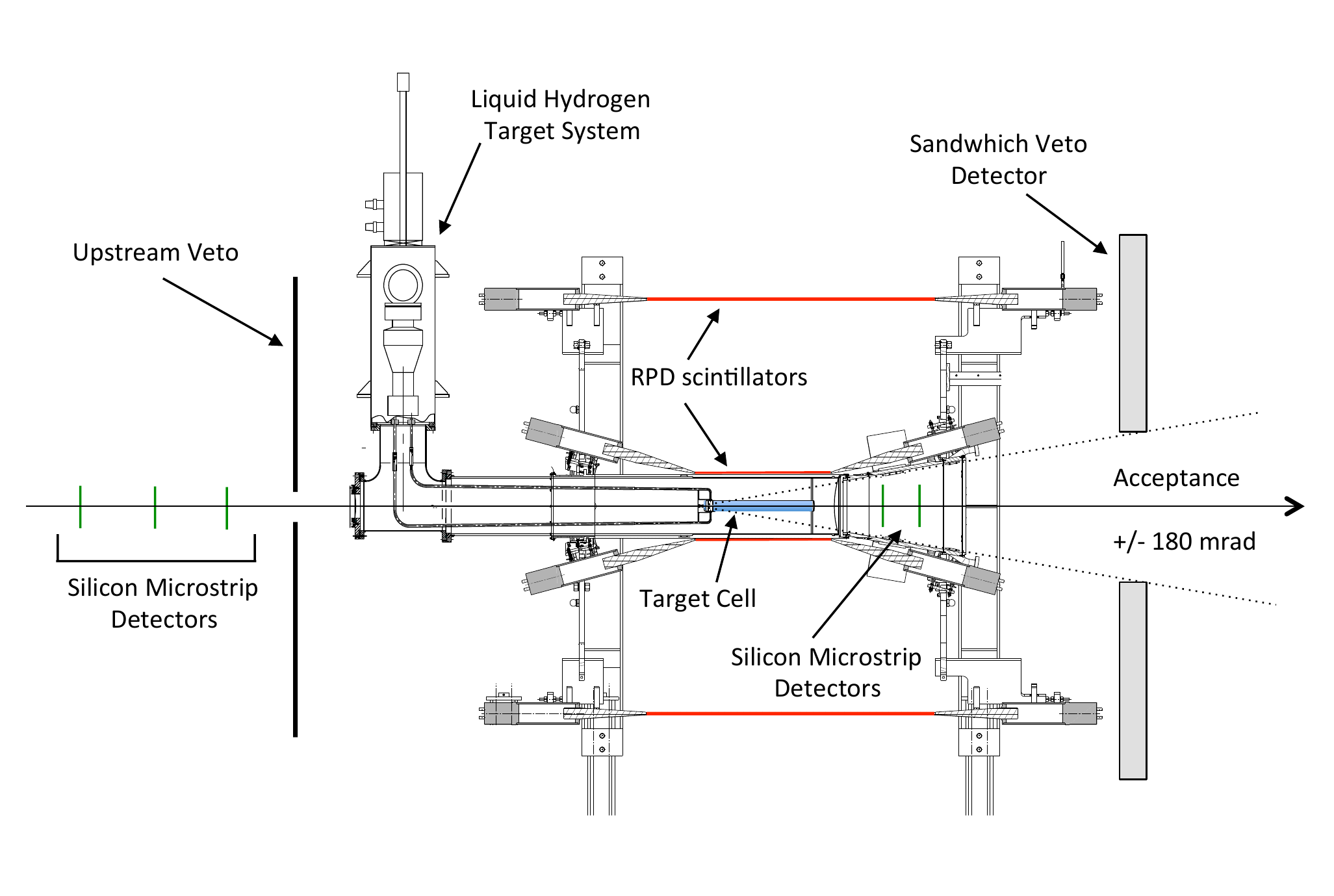}
\caption{Side view of the target region with the liquid hydrogen target system.}
\label{fig:side-target} 
\end{figure*}

Most of the data with hadron beams were collected using a liquid hydrogen
target. The target region comprises the target itself and the detectors around
it (Fig.~\ref{fig:side-target}).  The target is surrounded by a time-of-flight
detector that is called Recoil Proton Detector (RPD). Measuring the recoil
protons from the target, this detector ensures the exclusivity of the processes
under investigation.  The RPD covers the momentum transfer range down to $\vert
t \vert =0.07\,\GeV^2/c^2$.  Three silicon stations operating at a temperature
of 200~K are mounted upstream of the target.  Together with a scintillating
fibre counter, these detectors determine the trajectory of the beam particle
before it enters the target.  Two other silicon stations are located immediately
downstream of the target, inside the RPD.  A scintillator/iron sampling
detector, called Sandwich Veto, is installed downstream of the RPD.  Used as
part of the trigger, this detector vetoes particles detected outside of the LAS
acceptance.  A dedicated Multiplicity Counter (MC) is positioned downstream of
the RPD, behind the Sandwich Veto.  This counter, which measures the number of
charged particles in the final state, extends the momentum transfer range
towards values smaller than $\vert t \vert =0.07\,\GeV^2/c^2$.

The liquid hydrogen target can be easily removed and replaced with a specially
designed solid-target holder. Up to 16 solid targets with different atomic
numbers and different thicknesses can be mounted on the holder and used
simultaneously during data taking.

\subsection{Large angle spectrometer}

The large angle spectrometer includes the detectors located both upstream and
downstream of the SM1 magnet. The LAS tracking detectors measure scattered
particles with polar angles of up to 180 mrad. In the region near the beam, a
PixelGEM detector with low material budget was installed in replacement of the
thicker scintillating fibre counter (SciFi) previously used with the muon beam.
The design of the Micromegas detectors that are located upstream of SM1 was
modified in order to minimise their discharge rate in the hadron beam.  A new
large-size drift chamber, DC4, is installed downstream of the SM1 magnet, in
order to improve the resolution of the tracking at large angles. A major upgrade
of the RICH-1 counter was accomplished~\cite{Abbon:05a}, which considerably
improves its performance.  The central region of RICH-1 was instrumented with
multi-anode photomultipliers, in replacement of the previously used CsI
photodetectors.  A new analog readout with a reduced dead time was implemented
in its peripheral region.  The tracking downstream of RICH-1 was supplemented
with an additional drift-tube detector, called Rich Wall (RW). A new
electromagnetic calorimeter, ECAL1, was added to the LAS setup. ECAL1 extends
the coverage of ECAL2 for detection of photons and electrons to larger
angles. Its position was chosen with the aim of achieving a continuous angular
coverage for both ECAL1 and ECAL2.  Installed since 2006, DC4, ECAL1, RICH-1 and
RW are part of the apparatus that is common to both hadron and muon physics
programmes.

\subsection{Small angle spectrometer}
The SAS detectors are essentially identical to the detectors used during the
data taking with muon beam~\cite{Abbon:2007pq}. In order to minimize the
material budget along the beam path, two new PixelGEM detectors replace two
SciFi counters.  In ECAL2, the inner-most lead glass blocks were replaced with
radiation-hard Shashlik-type lead/scin\-tillator modules of the same transverse
size. In order to maximise the photon detection acceptance near the beam, the
size of the ECAL2 central hole was reduced.  The central hole of the hadron
calorimeter HCAL2, located immediately behind ECAL2, was reduced accordingly.

\subsection{Trigger}
The trigger system for hadron beam was designed to select the processes listed
in \Secref{sec:intro}.  Several new trigger counters were built and combined
with those already available~\cite{Abbon:2007pq}. A beam counter (BC) was
installed upstream of the target, as a part of the beam-definition trigger. Both
RPD and ECAL2 detectors were included in the trigger. The information from the
RPD is used to identify diffractive scattering events.  High energy photons,
particularly important for the Primakoff reaction, are selected by the central
part of ECAL2. Triggering in the region of the lowest momentum transfer values
is provided by a new Multiplicity Counter (MC). The existing veto system was
extended to veto non-interacting beam particles by adding two beam-killer (BK1
and BK2) scintillator counters along the beam path.

With the nominal hadron beam intensity the trigger rate reaches values of up to
$30\,\kHz$.  The corresponding data flow is as high as $350\,$MB/s.  The COMPASS
data acquisition system was upgraded to meet these conditions.

\section{Beam Line}
\label{sec:beam}
The CERN SPS M2 beam was originally built as a high-energy, high-intensity muon
beam.  For the COMPASS experiment, the beam line was partly rebuilt to include a
high-intensity hadron beam option as well as the possibility to use
low-intensity electron beams.  For beam particle identification, two CEDARs were
added just before the COMPASS spectrometer.  Modifications relevant to the muon
mode of operation were described in \cite{Abbon:2007pq}. In this section, the
hadron and electron beam modes of operation are summarised. In addition, the
detectors used for the identification of the particles in the hadron beam are
described.

\subsection{Hadron beams}
\label{sec:beam.hadron}
In order to produce a secondary hadron or tertiary muon beam,
$400\,\GeV/c$~protons from the CERN SPS are slowly extracted onto a primary
production target (T6).  These protons arrive during a time period of $9.6\,\s$
once every $30\,\s$ to $48\,\s$, depending on other users of the SPS.  The
primary target station allows the selection of five positions, an empty target
or one out of four Be targets, $80\,\mm$ wide, 2 to $3\,\mm$ high and 40, 100 or
$500\,\mm$ thick. Measurements with hadron beams used a nominal intensity of $5
\cdot 10^6\,\s^{-1}$, which e.g. for a negative beam of $190\,\GeV/c$ central
momentum is achieved by using $9 \cdot 10^{12}$ protons/cycle on T6 and the
$500\,\mm$ target.

\begin{figure*}[htpb]
\centering
\includegraphics[width=0.9\textheight,angle=90]{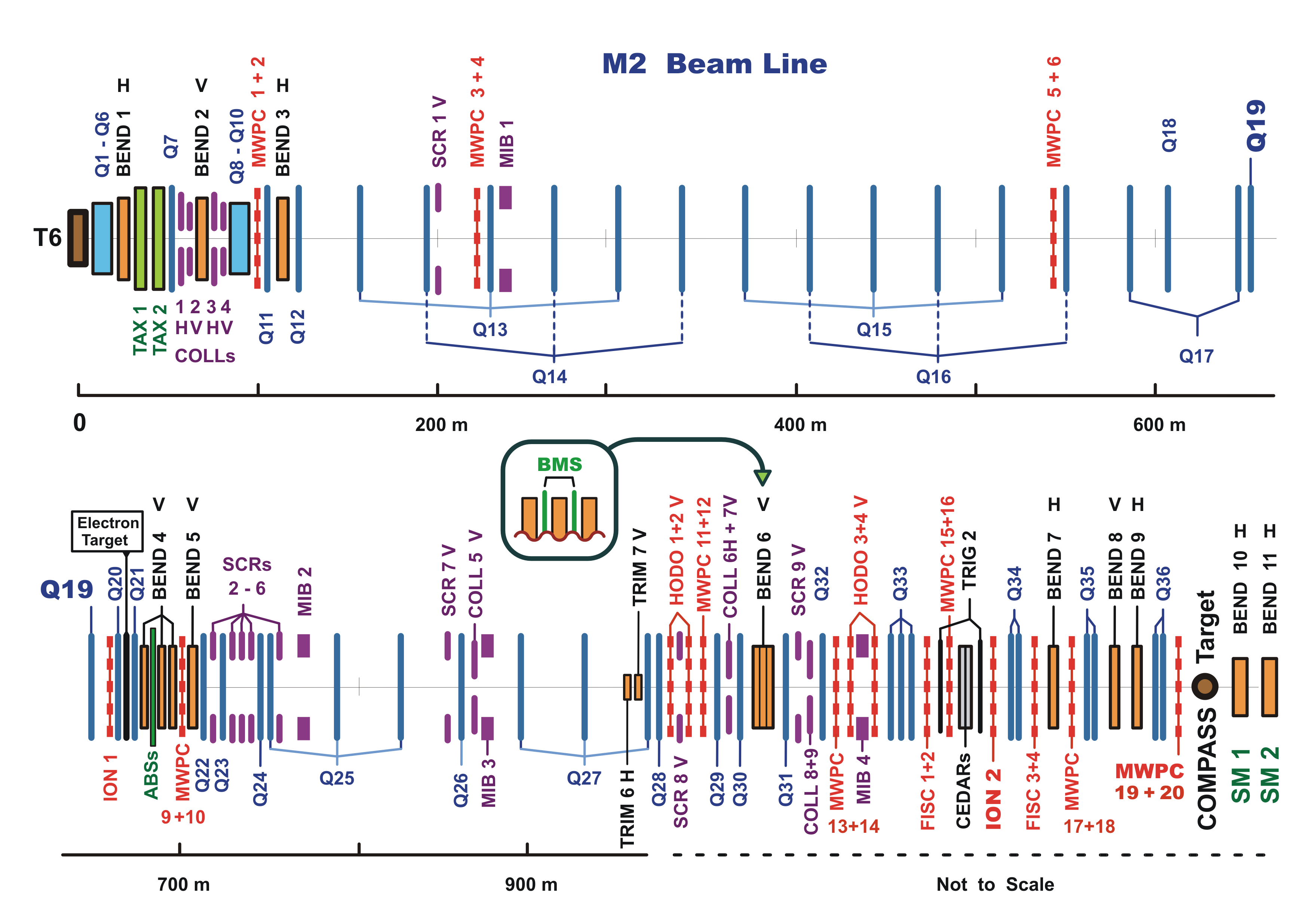}
\caption{The CERN M2 beam line.}
\label{fig:beam.optics} 
\end{figure*}

The M2 beam line (Fig.~\ref{fig:beam.optics}) starts with a series of six
high-gradient quadrupoles (Q1--Q6) for maximum acceptance and a set of three
dipoles (BEND 1).  The highest-acceptance optical mode works only up to
225~GeV/$c$ due to limited quadrupole gradients.\footnote{ A different optical
  mode is available for higher momenta, but the angular acceptance of that mode
  is about $40$\% lower.} The beam optics is optimised to achieve highest
possible momentum resolution.

A pair of massive collimators, which allows a first momentum selection of the
beam particles or is also used to dump the beam in case of access to the
experiment (TAX 1,2), is located downstream of the first dipoles (BEND 1). The
particles passing through theses collimators are transported to an about
$430\,\m$ long array of alternately focusing and defocusing (FODO) quadrupoles
(Q12--Q18). Before entering the FODO array, two pairs of horizontal (H1, H3) and
vertical (V2, V4) collimators define the angular and momentum acceptances of the
beam. At the end of the FODO section, the beam is focused into a set of four
vertical dipoles (BEND 4,5).  The Be absorbers (ABS), which are used in the muon
beam operation, are moved out in the hadron mode.  After these dipoles, the beam
is transported into a second $250\,\m$ long FODO channel (Q22--Q30) that
contains a second momentum-defining collimator. The beam is then bent
horizontally (BEND 7) and parallelised at the location of the two CEDARs (for
more details, see \Secref{sec:beam.cedar}), while restricting the momentum
spread to below 1\%.  Behind the CEDARs, the beam is focused (Q34--Q36) onto the
entrance of the electromagnetic calorimeter in the second spectrometer stage,
which is located $33\,\m$ downstream of the target.  The main beam
characteristics are listed in Table~\ref{tab:beam.parameters}.
\begin{table*}[tbp]
  \centering
  \caption{The main parameters of the M2 hadron beam. }
  \begin{tabular*}{\textwidth}{@{\extracolsep{\fill}}lc} \hline \hline
    {\bf Parameter} & {\bf Value} \\ \hline
    Length of beam line from primary target to COMPASS target & $1131.8\,\m$ \\ 
    Maximum beam momentum (high-energy mode)& 
$280\,\GeV/c$ \\ 
    Maximum beam momentum (normal mode)& 
$225\,\GeV/c$ \\ 
    Angular acceptance: Horizontal & 
$\pm1.0\,\mrad$ \\ 
    Angular acceptance: Vertical &
$\pm2.5\,\mrad$ \\ 
    Angular acceptance: Solid angle &
$7.8\,\musr$ \\ 
    Momentum acceptance & 
$\pm8\%$ \\ 
    Momentum resolution & 
$1\%$ \\ 
    Spot size at COMPASS target ($\sigma_x \times \sigma_y$) & 
$7\times 8\,\mm^{2}$\\ 
    Divergence at COMPASS target ($\sigma_x \times \sigma_y$) &
$80\,\murad \times 200\,\murad$\\
\hline \hline
\end{tabular*}
\label{tab:beam.parameters}
\end{table*}

Negative beams are mainly composed of pions, while for momenta larger than
150~$\GeV/c$ the positive beam have a dominant proton component.  In both cases,
kaons and electrons may be present at a level of a few percent, depending on the
energy chosen. The particle composition of the hadron component of the beam is
given in Table~\ref{tab:beam.composition} for a few typical beam momenta.

\begin{table*}[tbp]
  \centering
  \caption{The relative composition of the hadron beam at the COMPASS target for
    some typical momenta. It does not include the $e^{\pm}$ component, which is
    still present at 100~$\GeV/c$ but rapidly decreasing at higher momenta due
    to synchrotron radiation. The composition values are calculated from
    measured values \cite{Atherton:1980vj} and their relative uncertainties
    amount to 1\% for pions and proton, and 2--3\% for kaons and antiprotons.}
  \label{tab:beam.composition}
  
  \small
  \begin{tabular*}{\textwidth}{@{\extracolsep{\fill}}ccccccc} \hline \hline
    Momentum &\multicolumn{3}{c}{Positive beams} &\multicolumn{3}{c}{Negative beams} \\ 
    ($\GeV/c$) & $\pi^{+}$ & $K^{+}$ & $p$ & $\pi^{-}$ & $K^{-}$ & $\bar p$ \\ \hline
   100 & 0.618 & 0.015 & 0.367 & 0.958 & 0.018 & 0.024 \\ 
   160 & 0.360 & 0.017 & 0.623 & 0.966 & 0.023 & 0.011 \\ 
   190 & 0.240 & 0.014 & 0.746 & 0.968 & 0.024 & 0.008 \\ 
   200 & 0.205 & 0.012 & 0.783 & 0.969 & 0.024 & 0.007 \\ 
\hline \hline
  \end{tabular*}
  \normalsize
\end{table*}

\subsection{Electron beam}
\label{sec:beam.electron}
A tertiary electron beam can be produced on demand. For this purpose, a $5\,\mm$
thick lead plate (also called ``electron target") equivalent to 90\% of a
radiation length is moved into the beam line at the end of the first FODO
section, about $680\,\m$ downstream of the primary production target (see
Fig.~\ref{fig:beam.optics}).  A high-intensity negative hadron beam of
$100\,\GeV/c$, which contains electrons at the level of 10\%, is directed to the
electron target.  The hadrons mostly traverse the lead target, as its thickness
is equivalent to only about 3\% of an interaction length. In contrast, most of
the electrons of the beam lose energy by bremsstrahlung. The outgoing electrons
have a momentum spectrum that extends up to the momentum of the parent beam but
with very low intensity yielding useful electron energies of up to
$50\,\GeV$. The required electron momentum is selected with the beam line
magnets located downstream of the lead target. For the nominal energies of 15,
20 and $40\,\GeV$, which are used for the calibration of the electromagnetic
calorimeters, intensities up to a few $10^4$ electrons per spill are routinely
reached.

\subsection{Beam particle identification}
\label{sec:beam.cedar}

Two CEDAR detectors are installed $30\,\m$ before the COMPASS target region.
They were designed to provide fast beam particle identification at high rates
for particle momenta up to $300\,\GeV/c$ \cite{Bovet:82}.

\subsubsection{The principle of operation and the mechanical design}
\begin{figure}
  \centering \resizebox{0.7\columnwidth}{!}{
    \includegraphics{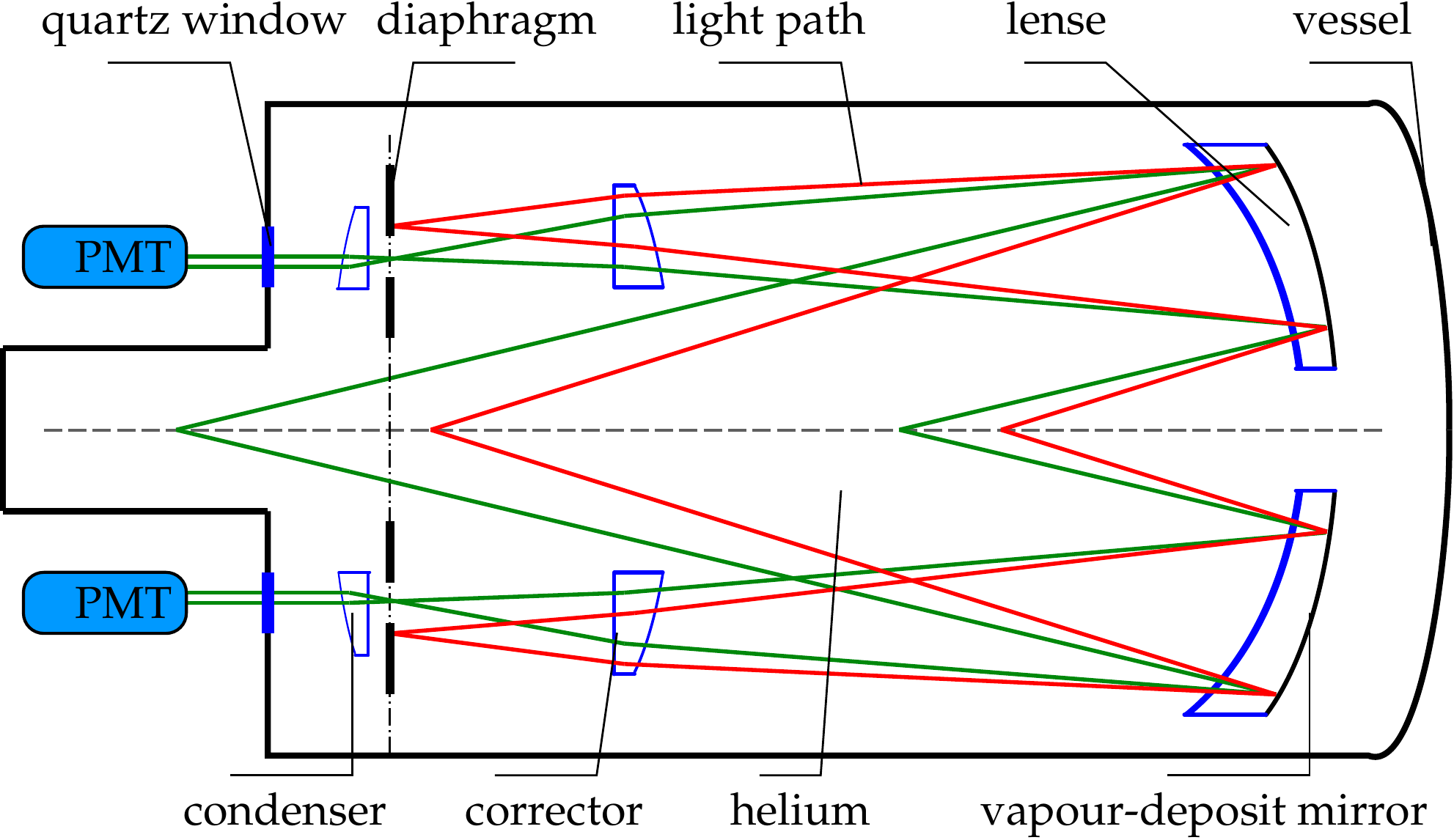}}
  \caption{The basic principle of a CEDAR counter. Two particles with the same
    momentum but with different masses (here red and green lines) radiate
    Cherenkov photons at different angles, resulting in rings with different
    radii. A diaphragm selects the rings from the required particle type.}
  \label{fig:CEDAR_rays} 
\end{figure}
The principle of operation of a CEDAR detector is illustrated in
\figref{fig:CEDAR_rays}. For a beam containing particles of different types but
the same momentum, the angles of the emitted Cherenkov photons differ due to the
different masses.  The Cherenkov photons are focused onto the photon detectors
using a a mirror and a system of lenses (lens, corrector, condenser). This
results in rings of photons at the focal plane whereby compensating for the
chromatic aberration in the gas, which is mandatory for a proper separation of
the rings.  A ring shaped diaphragm, which is located in the focal plane
perpendicular to the beam direction, selects photon rings with a fixed radius.
The radius of the photon ring is matched to the radius of the diaphragm by
adjusting the pressure of the helium gas in the vessel.

\begin{figure*}
  \centering \resizebox{0.75\columnwidth}{!}{ \includegraphics{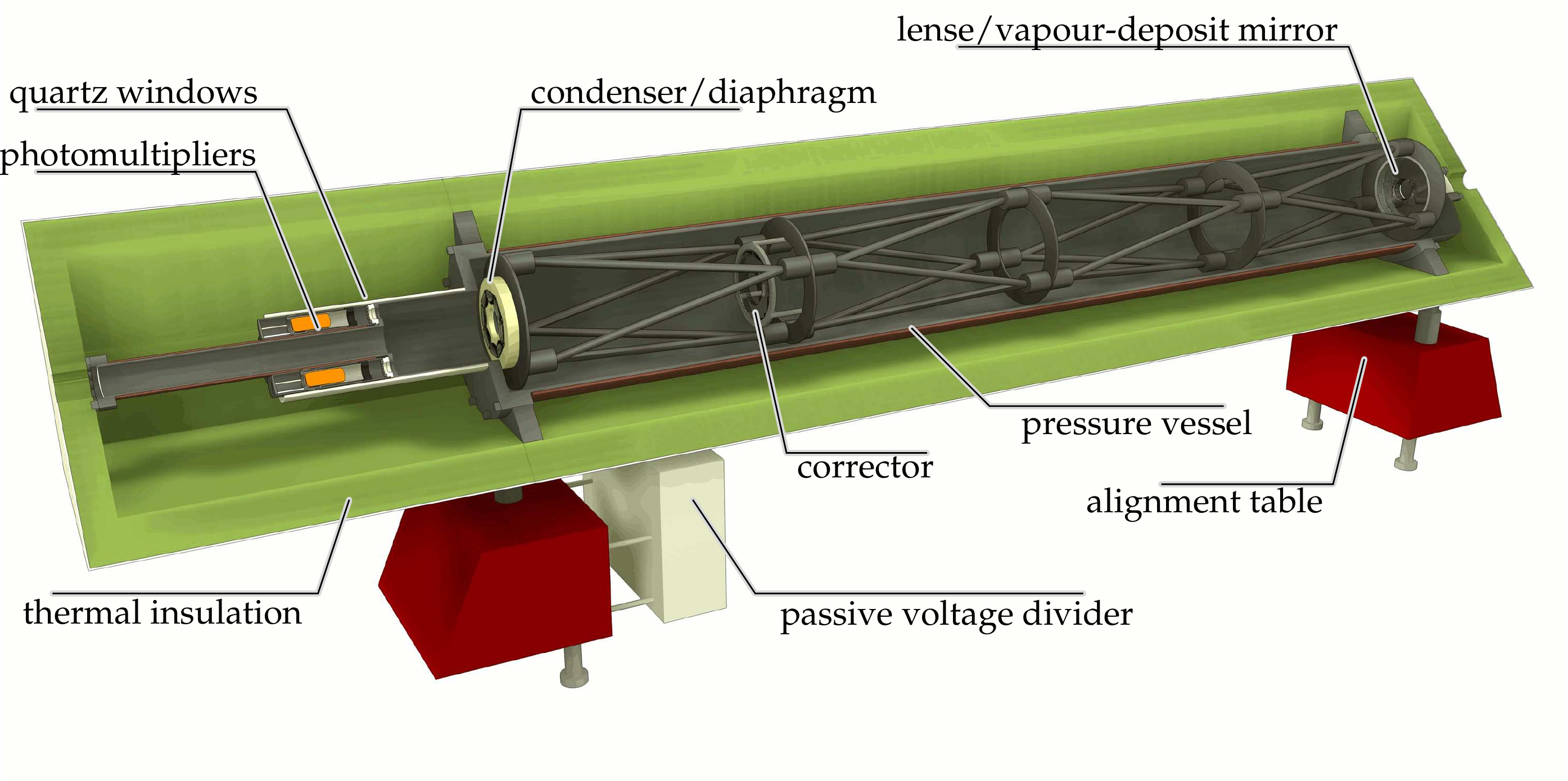} }
  \caption{ A cut through one of the CEDAR detectors.  }
  \label{fig:CEDAR}
\end{figure*}

COMPASS operates two CEDAR detectors. Each consists of a $6\,\m$ long vessel
containing pressurised He gas, a mirror, a lens system and a diaphragm
(\figref{fig:CEDAR}).  The nominal pressure at 190\,GeV/$c$ beam momentum is
10.5~bar.  The photons are detected with eight PMTs (Thorn-EMI-9820) equipped
with passive voltage dividers.

The photon rings are smeared by several effects, e.g. temperature changes,
beam divergence and limited precision of alignment. In order to keep the density
constant along the $6\,\m$ long vessels and thus the refractive index, good
thermal insulation and conduction is mandatory. The vessel is covered with
copper filaments for thermal conduction and surrounded by a $10\,\Cm$ thick
polyethylene foam layer for insulation. In addition, the PMT voltage dividers
are mounted outside the vessel.  Particles travelling not parallel to the
optical axis will produce shifted photon rings that do not match the diaphragm.
A tilt of the beam with respect to the principal axis of the optical system can
be corrected by adjusting the detector position with the help of a motorised
base. The beam divergence could only be compensated by opening the diaphragm at
the expense of a lower purity of the particle identification.

As the parallelism of beam tracks is of great importance for an efficient
operation of the CEDARs, the beam divergence is monitored using pairs of single
scintillating fibre detectors (one horizontal, one vertical) that were installed
upstream (FISC1,2) and downstream (FISC3,4) of the CEDARs. Their position in the
beam can be adjusted to measure the track angles by a coincidence between an
upstream and a downstream fibre hit.  Furthermore, two scintillating discs (TRIG
2) are installed as beam counters.  They are used to normalise the CEDAR count
rates during so-called pressure scans.  While taking physics data, the discs and
single-fibre detectors are moved out of the beam in order to reduce the material
budget in the beam line.

As the ratio of pressure over temperature, $p/T$, is proportional to the
refractive index, the working point of the CEDAR detectors is determined by
performing pressure scans. In a pressure scan, the count rate normalised to the
rate in the FISC counters is determined as a function of the pressure in the
vessel and the multiplicity of PMT signals.  Using the known beam composition,
this yields also an online estimate for the particle identification
efficiency. A more refined offline method will be discussed in
\secref{sec:reconstruction.cedar}.  During data taking the He pressure in the
CEDARs is regularly adjusted to compensate for He leakage and to keep $p/T$
constant.

\begin{figure}
  \centering \resizebox{0.9\columnwidth}{!}{
    \includegraphics{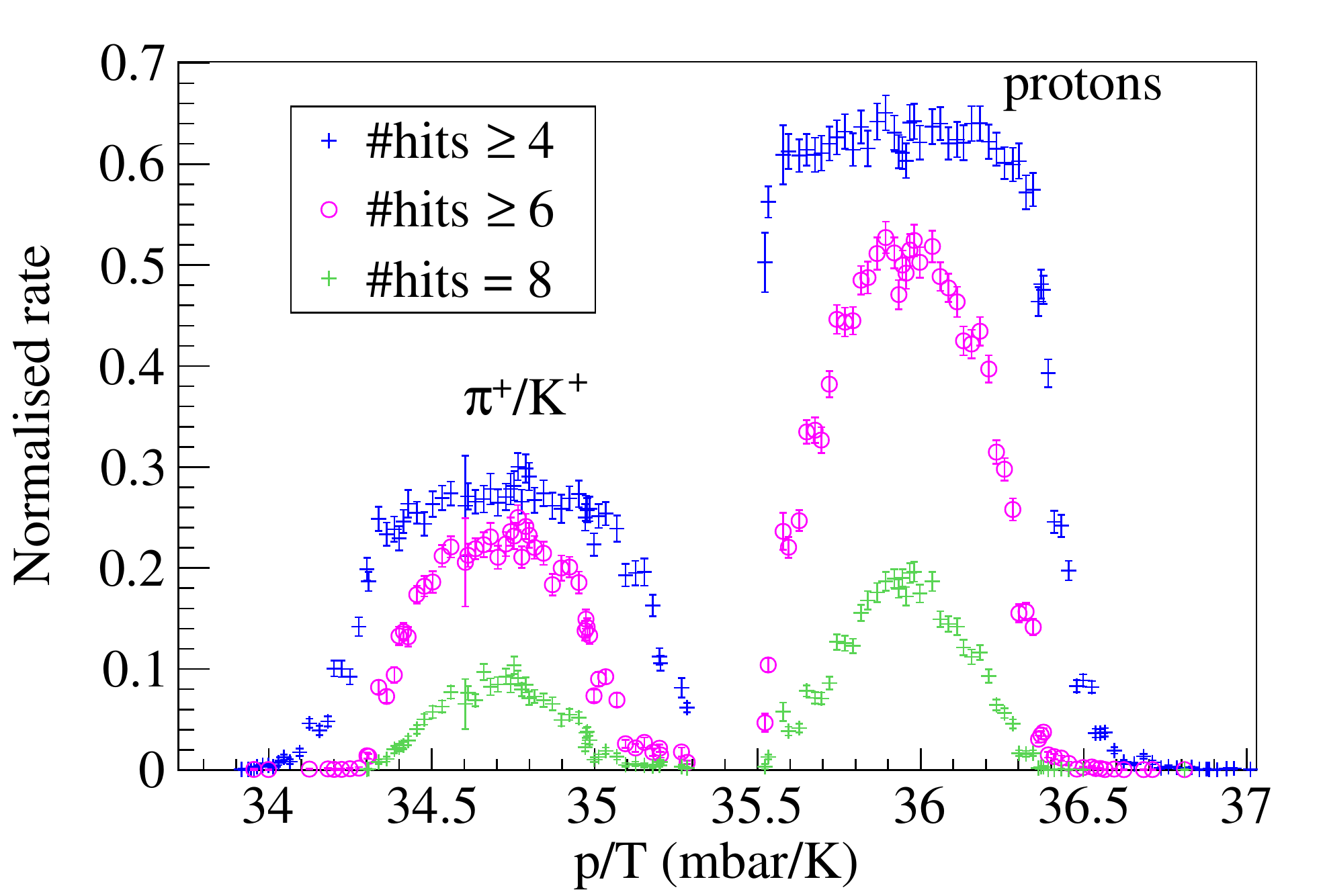} }
  \caption{Pressure scan with CEDAR1 for a positive hadron beam with at least 4,
    6 or 8 PMTs in coincidence. The kaon peak cannot be distinguished from the
    pion peak.  }
  \label{fig:CE1proton_pionscan}
\end{figure}

\subsubsection{Separation of protons, pions and kaons (positive beam)}
In the high-energy positive hadron beam, the proton component is dominant.  For
the CEDARs, a difference of $1\,\mm$ is expected between the ring radii of
protons and kaons at $190\,\GeV/c$. The plateau of the efficiency is reached
with a slit width of $1.2\,\mm$.  \Figref{fig:CE1proton_pionscan} shows an
example of a pressure scan for three different requirements on the minimum
number of PMT hits (multiplicity). The clear separation of pions and proton is
obvious, while the small kaon component is hidden under the pion signal and
cannot be distinguished here. From the measured rates and the known beam
composition, a particle identification efficiency of almost 90\% for protons is
estimated using a multiplicity of $\ge 4$ with a high purity of larger than 95\%
for the chosen working point of the CEDAR.  During the data taking one CEDAR was
set to identify protons, the other to identify pions.
\begin{figure}
  \centering \resizebox{0.9\columnwidth}{!}{ \includegraphics{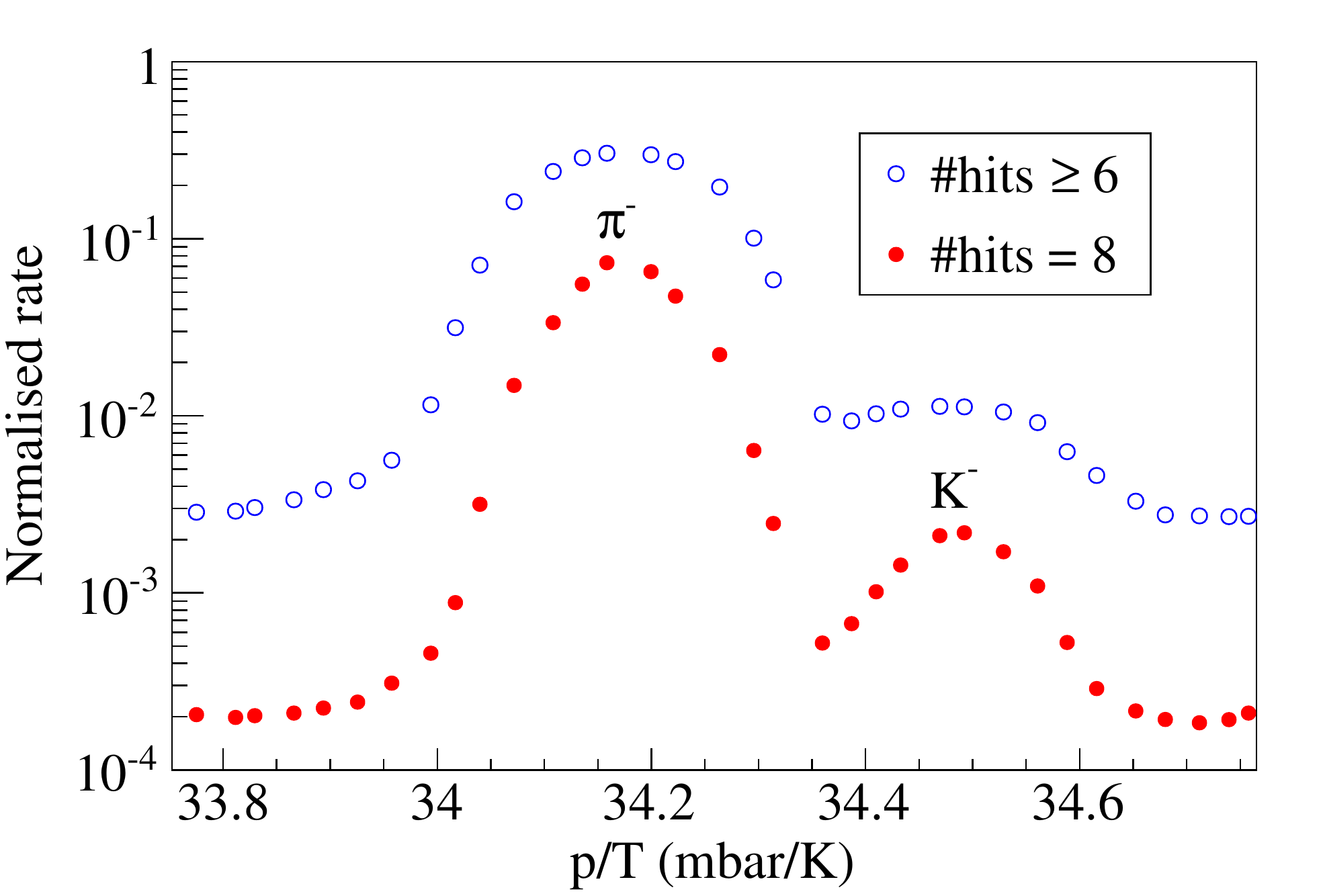}
  }
  \caption{Pressure scan with CEDAR1 for a negative hadron beam with at least 6,
    or with 8 PMTs in coincidence.}
  \label{fig:pressscan}
\end{figure}

\subsubsection{Separation of kaons and pions (negative beam)}
\begin{figure*}
  \centering \resizebox{1.0\columnwidth}{!}{
    \includegraphics{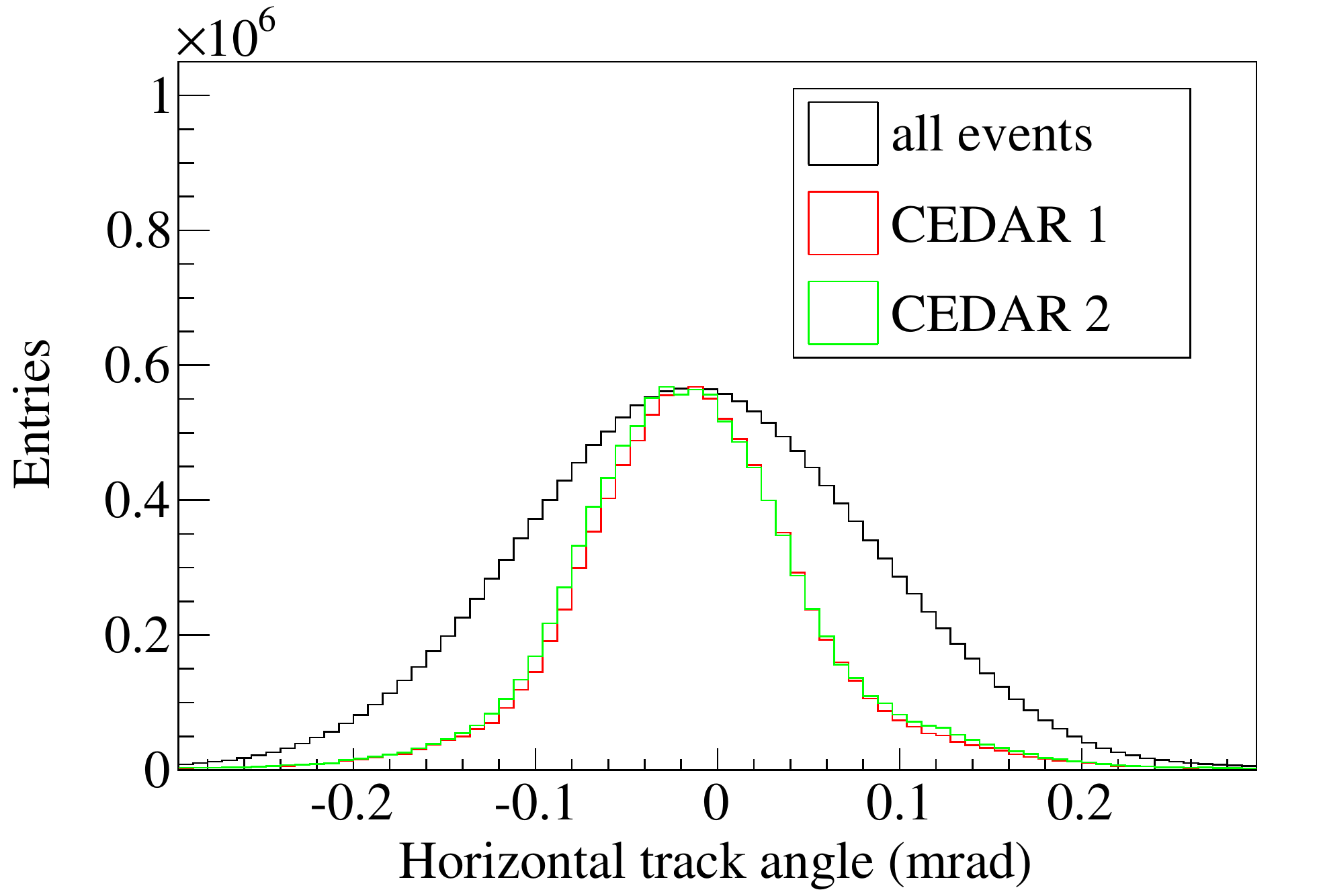} \includegraphics{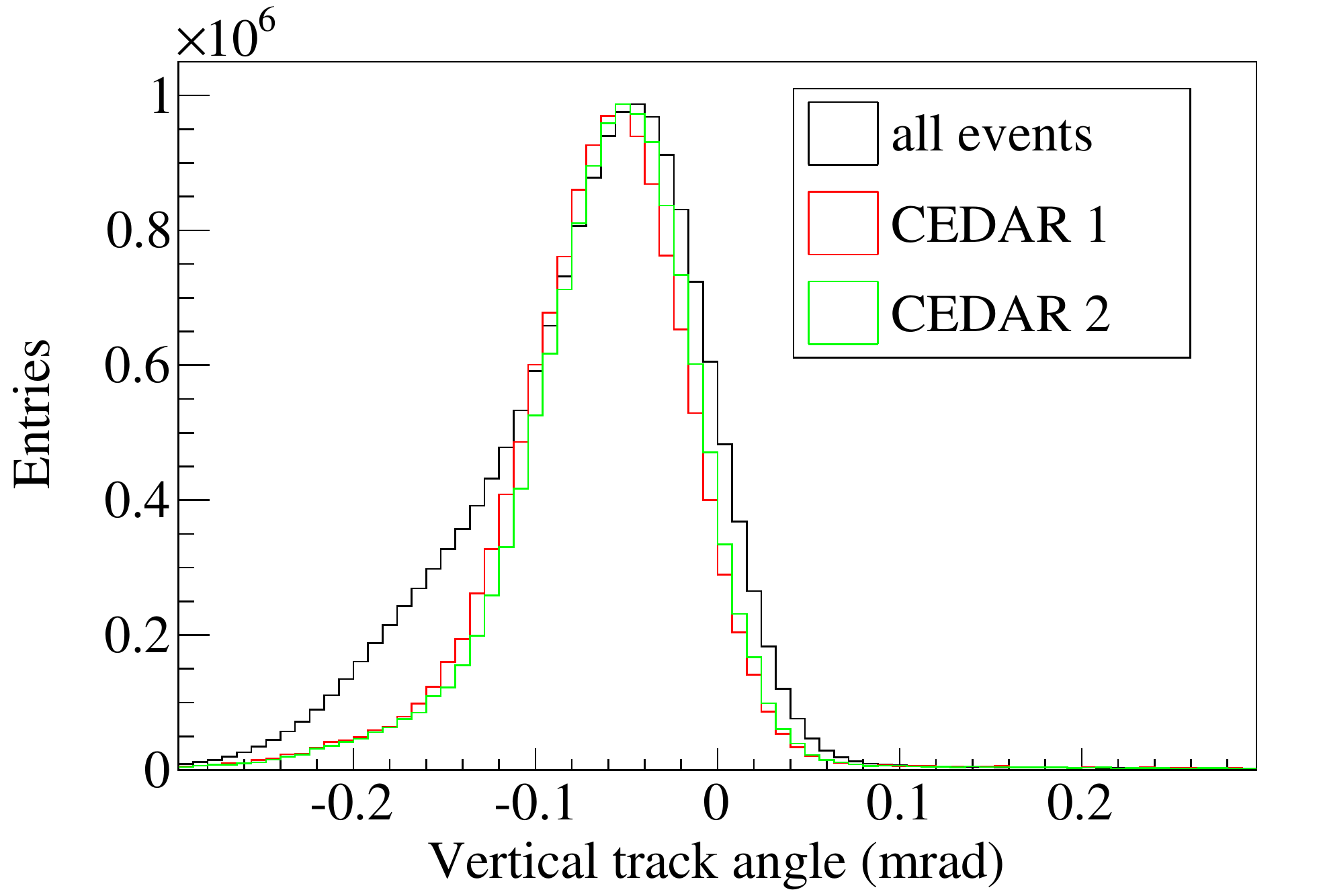} }
  \caption{Horizontal (left) and vertical (right) track angles at the
    CEDARs. The angles for all tracks measured by the Silicon beam telescope and
    propagated back to the CEDAR positions are compared to the angles of the
    tracks accepted by CEDAR1 or CEDAR2.  The acceptance of the CEDARs is
    reduced significantly for very divergent beam tracks.}
  \label{fig:divergence}
\end{figure*}

Negative hadron beams contain mainly pions with a small admixture of kaons and
antiprotons. In this case, the CEDARs are used to identify kaons.  Although the
difference between the mean radii of the photons rings of kaons and pions at
$190\,\GeV/c$ is less than $0.5\,\mm$, the diaphragm was set to $0.5\,\mm$ as a
compromise between efficiency and purity.  The pressure scan in
\figref{fig:pressscan} shows the obtained separation for multiplicities of
$\ge6$ and $8$.  An online efficiency of about 35\% is estimated for pions with
a multiplicity of $\ge 6$.  Such a low efficiency is due to the high beam
divergence of the very long beam line, in combination with the narrow slit width
of the diaphragm.  The loss due to the beam divergence is illustrated in
\figref{fig:divergence} which shows the distributions of beam track angles as
measured by the Silicon detectors upstream of the target and propagated back to
the CEDAR position. The distribution for all beam particles are compared to
those for beam particles identified online by CEDAR1 or CEDAR2.

\begin{figure}
  \centering \resizebox{0.7\columnwidth}{!}{ \includegraphics{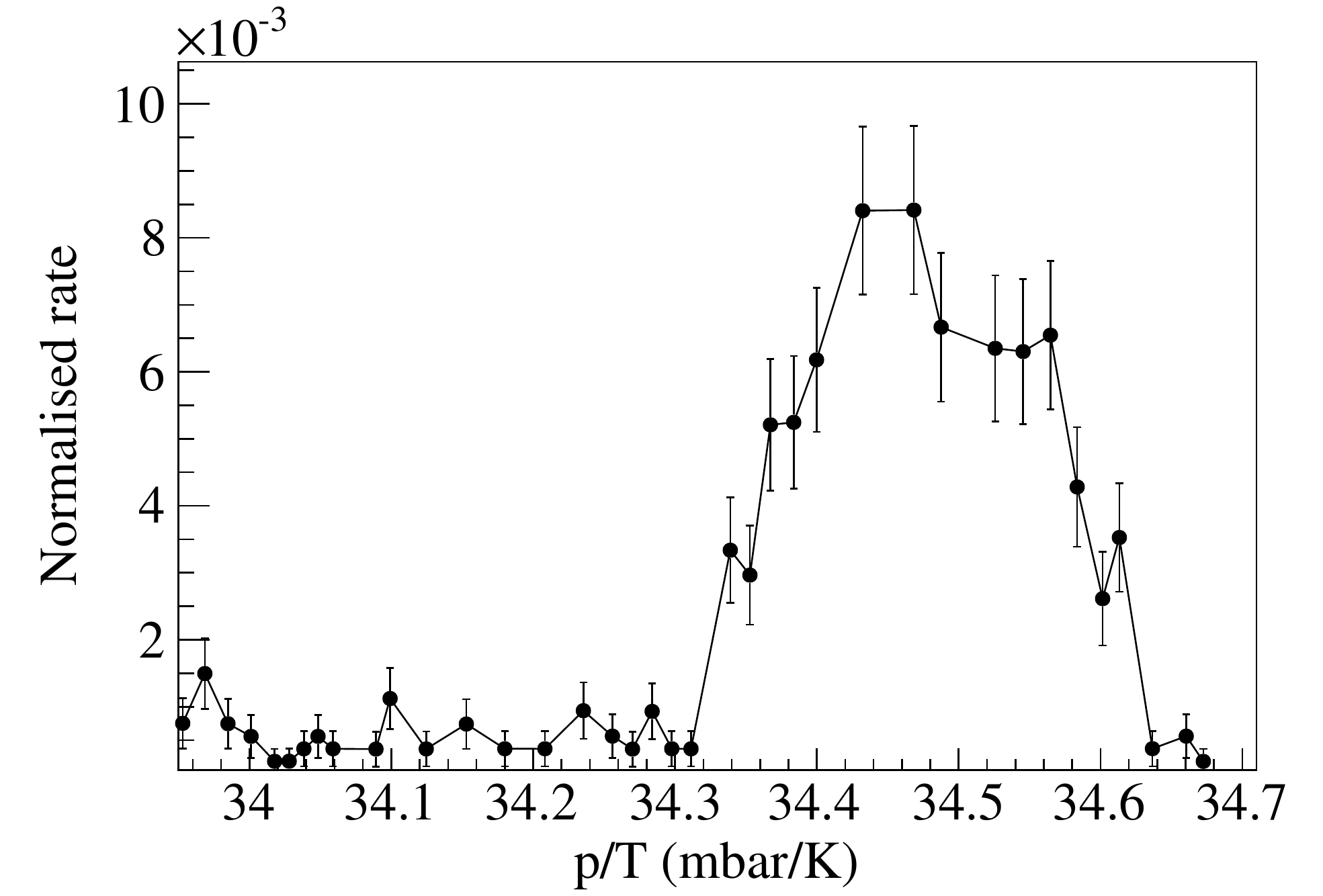} }
  \caption{Count rate of coincident events recorded with CEDAR1 and CEDAR2. The
    pressure of CEDAR2 was scanned while CEDAR1 was set to detect kaons.}
  \label{fig:CE12scan}
\end{figure}

In order to reduce and to separate the background that mainly originates from
pions, both CEDARs are set to detect kaons. The background can be measured by
setting one CEDAR to detect kaons and performing a pressure scan with the other
CEDAR. As illustrated in Fig.~\ref{fig:CE12scan}, the pion background is below
7\%.

\section{Target Region}
\label{sec:target}

The target region comprises the target systems, the Recoil proton detector, the
Sandwich veto detector and the Silicon detectors (\figref{fig:side-target}).
Either liquid hydrogen or solid targets can be used during the measurements
performed with hadron beams.  The hydrogen and lead targets are used for
diffractive dissociation and central production measurements.  A nickel target
is used for the study of the Primakoff reaction.  The Silicon detectors, which
are also located in the target region, will be described along with other
tracking detectors in \secref{sec:tracking}.

\subsection{Liquid hydrogen target}
\label{sec:target.hydrogen}
\begin{figure}[tbp]
    \centering
    \includegraphics[width=\textwidth]{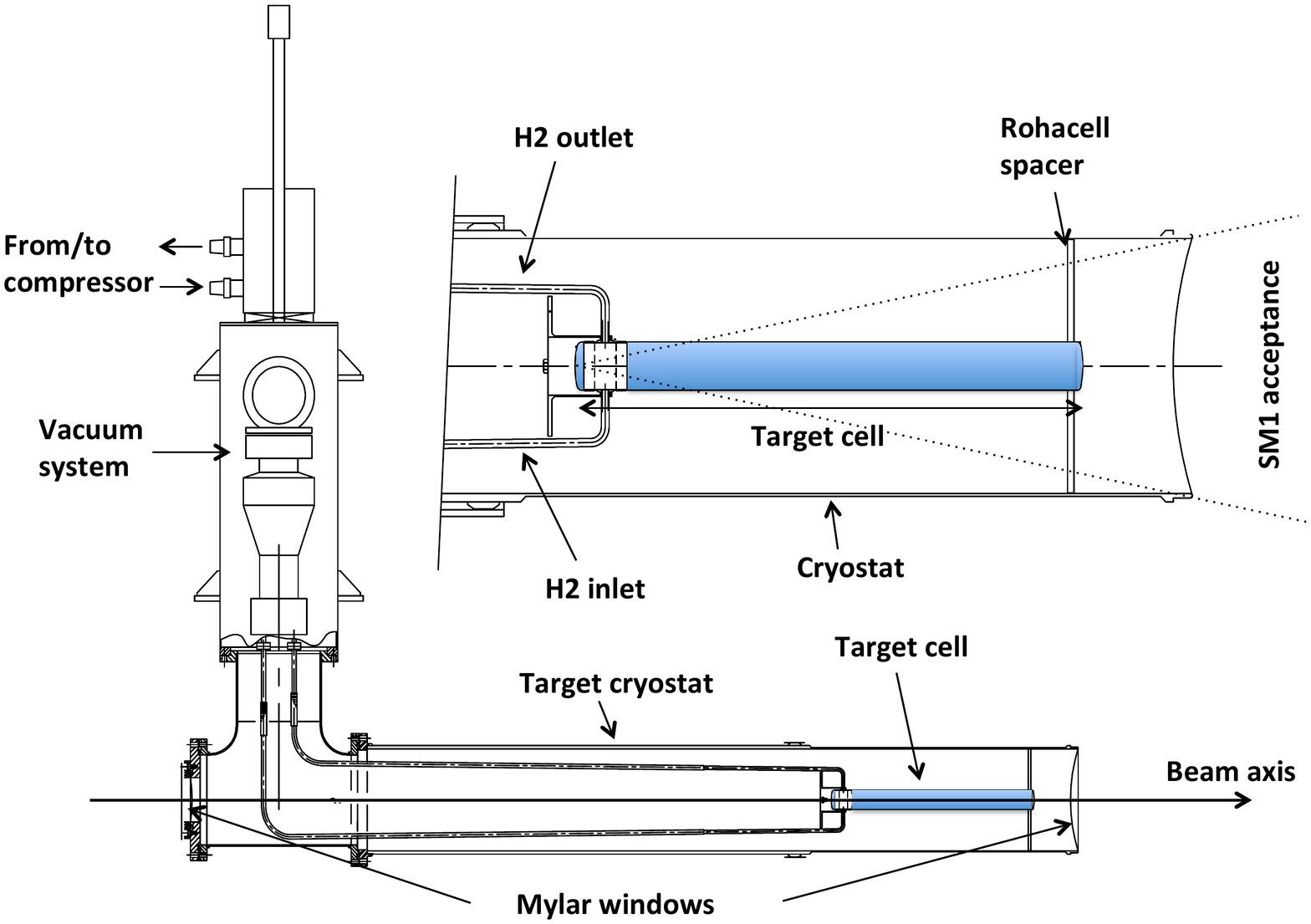}
    \caption{Side view of the liquid hydrogen target system.  A closer view of
    the cylindrical Mylar cell and hydrogen piping is shown in the inset.}
    \label{fig:targets.lh2.pic2}   
\end{figure}

For scattering on protons, a liquid hydrogen target is used. The target cell has
a cylindrical shape with a length of $400\,\mm$ along the beam and a diameter of
$35\,\mm$, which corresponds to a volume of $0.4\,$l
(Fig.~\ref{fig:targets.lh2.pic2}).  The thickness of hydrogen along the beam
axis is equivalent to $4.5\%$ of a radiation length ($X_0$) and $5.5\%$ of a
nuclear interaction length ($\lambda_I$). The diameter of the target is matched
to the dimensions of the beam spot ($\sigma \approx 8\,\mm$) and kept small in
order to reduce the amount of material traversed by recoil protons. The hydrogen
is enclosed by a Mylar cylinder of $125\,\mum$ thickness ($ 5 \cdot 10^{-4}X_0 ,
2 \cdot 10^{-4}\lambda_I$). The liquid hydrogen inlet and gas outlets are
constructed from stainless steel pipes, which are connected to a stainless steel
ring surrounding the target Mylar cell. The hydrogen cell and the stainless
steel pipes are wrapped with 10 layers of heat superinsulation foils (with
thickness of $\le 1\,\mum$/foil).

The target system with target cell, cryostat, and refrigerator is shown in
Fig.~\ref{fig:targets.lh2.pic2}.  The target cell is surrounded by a cryostat
tube made from aluminium.  The cryostat has a diameter of $185\,\mm$ and is
terminated towards the spectrometer by a $250\,\mum$ thin Mylar window. Its
diameter was chosen large enough so that forward going particles detected by the
COMPASS spectrometer pass through the window and not through the aluminium
cryostat itself.  In order to reduce the amount of material traversed by the
recoil particles, the thickness of the aluminium tube surrounding the target
cell is $1.8\,\mm$ (see Fig.~\ref{fig:side-target}). The upstream $250\,\mum$
Mylar window of the cryostat has a diameter of $80\,\mm$.

During operation, the target cell is filled with about $0.4\,$l of liquid
hydrogen.  The necessary hydrogen is taken from a large buffer volume of about
$850\,$l, which is kept at room temperature.  In warm mode (only hydrogen gas),
the pressure of the hydrogen buffer is set to about $1800\,\mbar$. During the
filling process, the hydrogen gas is cooled down inside a refrigerator and
liquefied until the pressure in the buffer volume reaches about
$1200\,\mbar$. This corresponds to about $0.6\,$l of liquid hydrogen. During
operation, the pressure in the buffer volume is kept constant by a simple
control unit regulating the effective cooling power of the refrigerator (5 W).
The biggest fraction of the liquid drops into the target cell by gravity, while
about 1/3 of it is kept in the $2.3\,$l condenser as a reserve. The heat
exchange between the target cell and the condenser is driven by evaporation and
condensation. The liquid inside the cell is just at the phase transition point
and slightly bubbling.  New liquid is continuously flowing from the condenser
into the cell via the lower pipe, while the evaporated gas streams back into the
condenser via the upper pipe where it is liquefied again.  The target was
designed such that bubbling should not affect the effective density by more than
$3\%$.  Evaporation of gas from the surface of the liquid occurs in the upper
$2-3\,\mm$ of the target cell. The cell is tilted by $1\,\mm$ to insure that the
gas flows back to the outlet. The resolution on the reconstructed vertex
position is sufficient to discard events originating from the top of the cell.

Starting from room temperature, the filling of the liquid hydrogen target takes
about $7$ hours.  The time for the evaporation of the liquid from the system is
about $3$ hours when the refrigerator is switched off. In order to allow for
background measurements that require an empty target cell, the target system
provides a fast emptying mode at low temperatures. For this purpose, the return
pipe for the gas vapour is closed remotely by a pneumatic valve. Then, the
pressure builds up rapidly by evaporation and pushes the liquid back from the
cell into the condenser volume within a few minutes. The target is refilled by
opening the valve.

The leakage rate of hydrogen was measured to be around $40\,\mg$ per day and
allows for an operation of the target without any refill during the six months
of a typical yearly data taking period. It has been observed that the vacuum in
the cryostat degrades due to the diffusion of gases and water vapour through the
thin Mylar vacuum windows ($0.1\,\g$ water per day).  No indication for freezing
of residual gases inside the cryostat was found in the offline analysis of
physics data.

\subsection{Nuclear targets}
\label{sec:target.nuclear}
\begin{figure}[tbp]
    \centering
    \includegraphics[trim=0 4cm 0 4cm, clip=true,width=\textwidth]{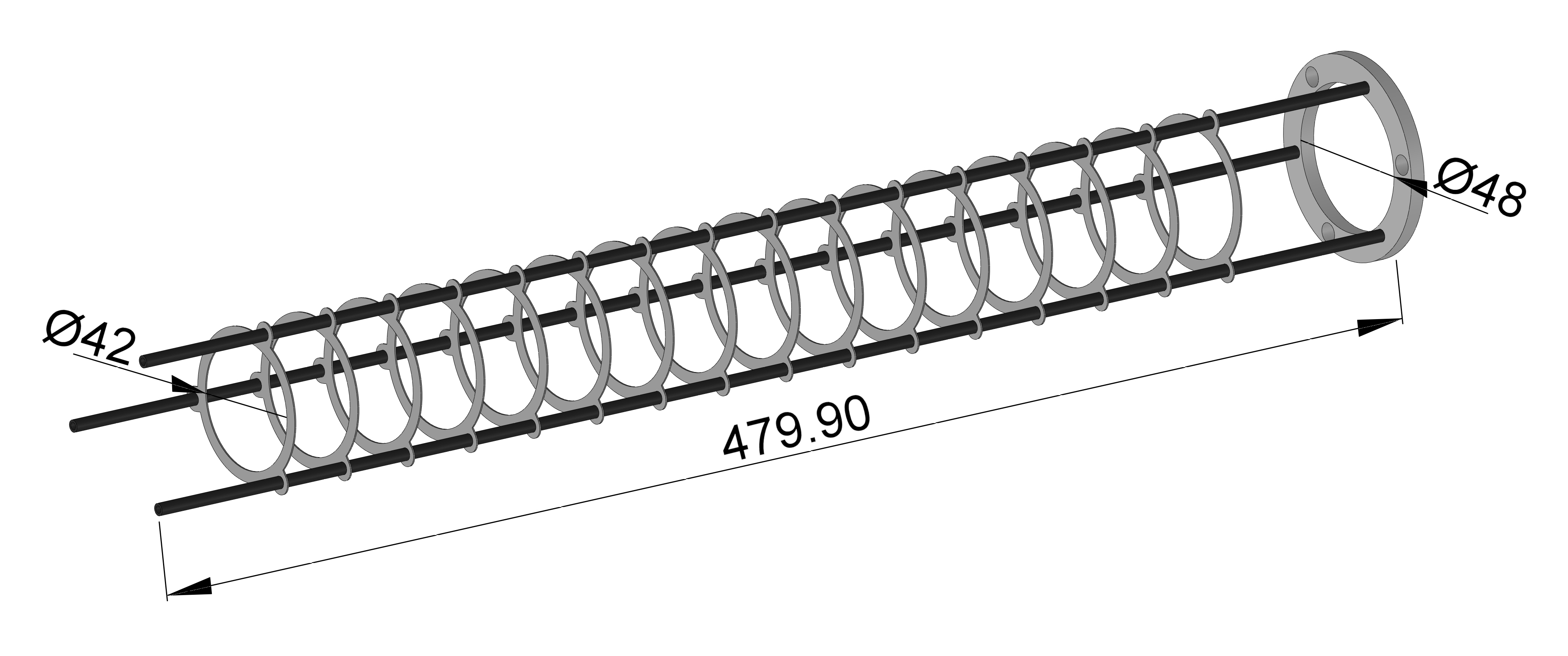}
    \caption{Schematic view of the target holder used for measurements
      with nuclear targets.}\label{fig:targets.pb.scheme}
\end{figure}

For measurements with nuclear targets, a light-weight target holder made of
carbon fibre rods and thin frames of fibreglass reinforced epoxy (FR4) was
used. Housing up to 16 target disks, the target holder is inserted into the RPD
instead of the liquid hydrogen target.  Figure~\ref{fig:targets.pb.scheme} shows
a schematic view of the target holder and the frames onto which the foils were
glued.
 
\begin{table}[tbp]
  \centering
  \caption{Overview of target materials used during the
    measurements with hadron beams in 2008/2009. }
  \begin{tabular}{lcrrrr} \hline\hline
    Material & Number   & \multicolumn{4}{c}{Thickness $x$} \\
             &  (elements)        & $(\mm)$ & $(\g/\Cm^2)$ & $(\lambda_\mathrm{I})$ & $(X_0)$
             \\ \hline
             Liquid H$_2$ & 1 & $400$ & 2.84 & $5.5\%$ & $4.5\%$ \\ \hline 
             Pb           & 6 & $0.250$ & 0.284 & $0.14\%$ & $4.45\%$ \\
             Pb           & 2 & $0.125$ & 0.142 & $0.07\%$ & $2.23\%$ \\
             Pb           & 2 & $0.025$ & 0.028 & $0.01\%$ & $0.45\%$ \\
             Pb           & 2 & $0.050$ & 0.057 & $0.03\%$ & $0.89\%$ \\
             W            & 2 & $0.050$ & 0.097 & $0.05\%$ & $1.43\%$ \\
             W            & 2 & $0.025$ & 0.048 & $0.03\%$ & $0.71\%$ \\ \hline
             Ni           & 1 & $4.2$   & 3.74 & $2.8\%$  & $29.5\%$ \\
             W            & 1 & $0.025$ & 0.048 & $0.03\%$ & $0.71\%$ \\ 
             W            & 1 & $0.050$ & 0.097 & $0.05\%$ & $1.43\%$ \\ \hline\hline        
  \end{tabular}
  \label{tab:target.overview}
\end{table}

The specifications of all targets used are listed in
Table~\ref{tab:target.overview}. Two different sets of nuclear targets were
mounted on the target holder. The first set consisted of 16 thin disks made of
Pb and W of natural isotopic composition.  The thicknesses of the disks and the
distance between them was chosen such that recoil protons from each individual
disk with momenta above $200\,\MeV/c$ could be detected over the full acceptance
of the RPD. The more downstream disks were made thinner in order to minimise the
effect of multiple scattering and conversion for events originating from the
more upstream targets.  The 12 Pb targets were used for diffractive dissociation
measurements, while the four W targets were used for feasibility studies of a
measurement of the $\pi^0$ lifetime.

The second set of target disks consisted of one $4.2\,\mm$ thick Ni disk and two
thin disks of W. The W disks were separated by $30\,\Cm$ from the Ni disk. This
target set was used for Primakoff measurements and as an additional
configuration for diffractive studies.

\subsection{Recoil proton detector}
\label{sec:target.rpd}
The Recoil Proton Detector (RPD) measures the velocity and energy loss of the
recoiling particles emitted at large angles. For particles produced in the
middle of the target, the region of full acceptance covers polar angles from
$50^\circ$ to $90^\circ$ (see \figref{fig:side-target}).  The energy loss of
protons in the target walls and in the inner ring limits the lowest detectable
momentum to $270\,\MeV/c$ in the case of the liquid hydrogen target. The RPD is
also used in the trigger system (see \secref{sec:trigger}) for the
identification of protons.
 
The design of the RPD closely follows the design of the detector used for the
GAMS NA12/2 experiment at CERN~\cite{Alde:1993nh}. It is made of two concentric
cylindrical barrels of plastic scintillators that surround the target and are
referred to as ``rings" in the following.  The inner ring is segmented in 12
slabs of BC404\textsuperscript{\textregistered} scintillator of dimensions
$50\times6.6\times0.5\,\Cm^{3}$, which are positioned at a radius of
$12\,\Cm$. Light guides for the inner ring are made of Plexiglas and have a
fish-tail geometry. They are tilted at an angle of $15^\circ$ with respect to
the longitudinal axis in order to stay outside of the acceptance in the forward
region.

The outer ring is segmented in 24 slabs of plastic scintillators with dimensions
of $115\times20\times1\,\Cm^{3}$, produced at IHEP
Protvino~\cite{azimova:1986}. Each slab is made of a single piece of material
that also includes a $29\,\Cm$ long light guide on each side. The ends are cut,
twisted and molded to fit into a $3.9\,\Cm$ diameter cylinder, to which the PMTs
are attached. The radius of the outer ring is $75\,\Cm$. Each element covers an
azimuthal angle of $15^\circ$. In order to optimise the azimuthal angle
resolution, the outer ring is positioned such that each inner ring counter faces
three outer ring slabs as viewed from the target centre (see
\figref{fig:trigger.pointing}).

Each scintillator is read-out at both sides using EMI 9813B photomultiplier
tubes. The PMTs are equipped with active voltage dividers to cope with the high
rate and high light output. The PMT signals are split using 8-fold active
splitters~\cite{bernhard:2007a} and sent to ADCs (2 dynamic ranges) and TDCs (2
threshold levels). The remaining outputs are used for the trigger system.  Two
outputs are connected to leading edge discriminators with two different
thresholds.  Furthermore, the signals from the inner ring downstream PMTs have
the smallest time jitter with respect to the incoming track since light in the
scintillator propagates in the same direction as the scattered particle. The
signals from these are sent to Constant Fraction Discriminators to preserve
their good timing properties.  All logic signals are then fed into a FPGA-based
system for triggering (see \secref{sec:trigger}).

The properties of each individual counter were measured during earlier tests
using muons from the beam halo with the RPD positioned transversely to the beam.
The resolutions obtained are $\sigma(t)=200\,\ps$ and $\sigma(z)=2.7\,\Cm$ for
the inner ring elements and $\sigma(t)=400\,\ps$ and $\sigma(z)=5.0\,\Cm$ for
the outer ring elements. The attenuation lengths of both types of counters were
measured and found to be of the order of $60\,\Cm$.

\begin{figure}[tbp]
  \begin{center}
    \includegraphics[width=0.8\columnwidth]{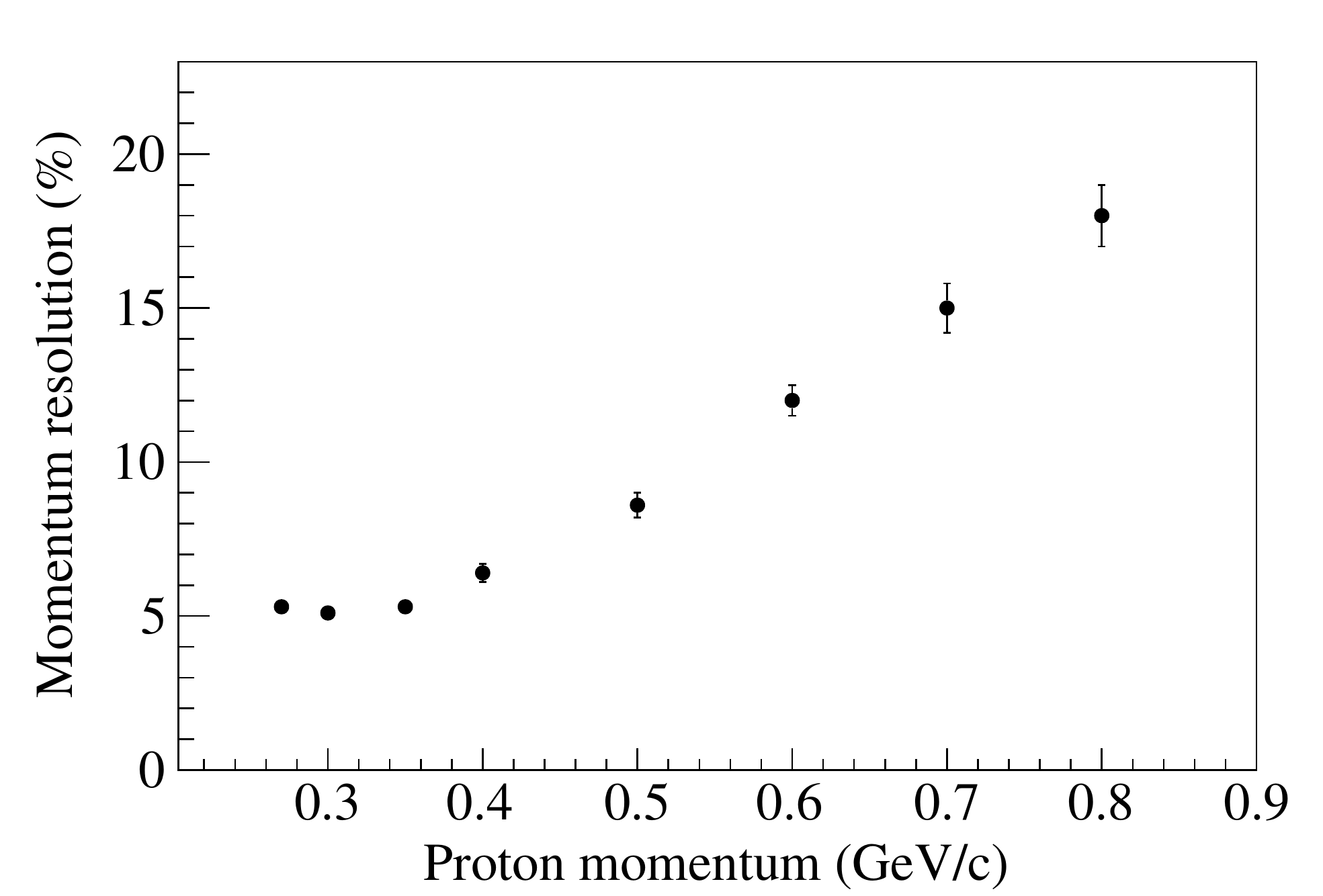}
  \end{center}
  \caption{\small Momentum resolution of the RPD for protons detected at an
    angle of $70^\circ$ relative to the beam axis.  }
  \label{fig:target.rpd.rpd-mom-resol}
\end{figure}

The momentum resolution as a function of the proton momentum is measured using
pion-proton elastic scattering events. The proton transverse momentum is
measured in the RPD and then compared to the value calculated from the
kinematics of the pion that is detected in the spectrometer.  The width of the
resulting momentum distribution is shown as a function of the proton momentum in
Fig.~\ref{fig:target.rpd.rpd-mom-resol}.  It is in agreement with an estimation
performed using the time and position resolutions of the RPD, as determined from
test beam data.

\begin{figure}[bp]
  \begin{center}
    \includegraphics[width=0.7\columnwidth]{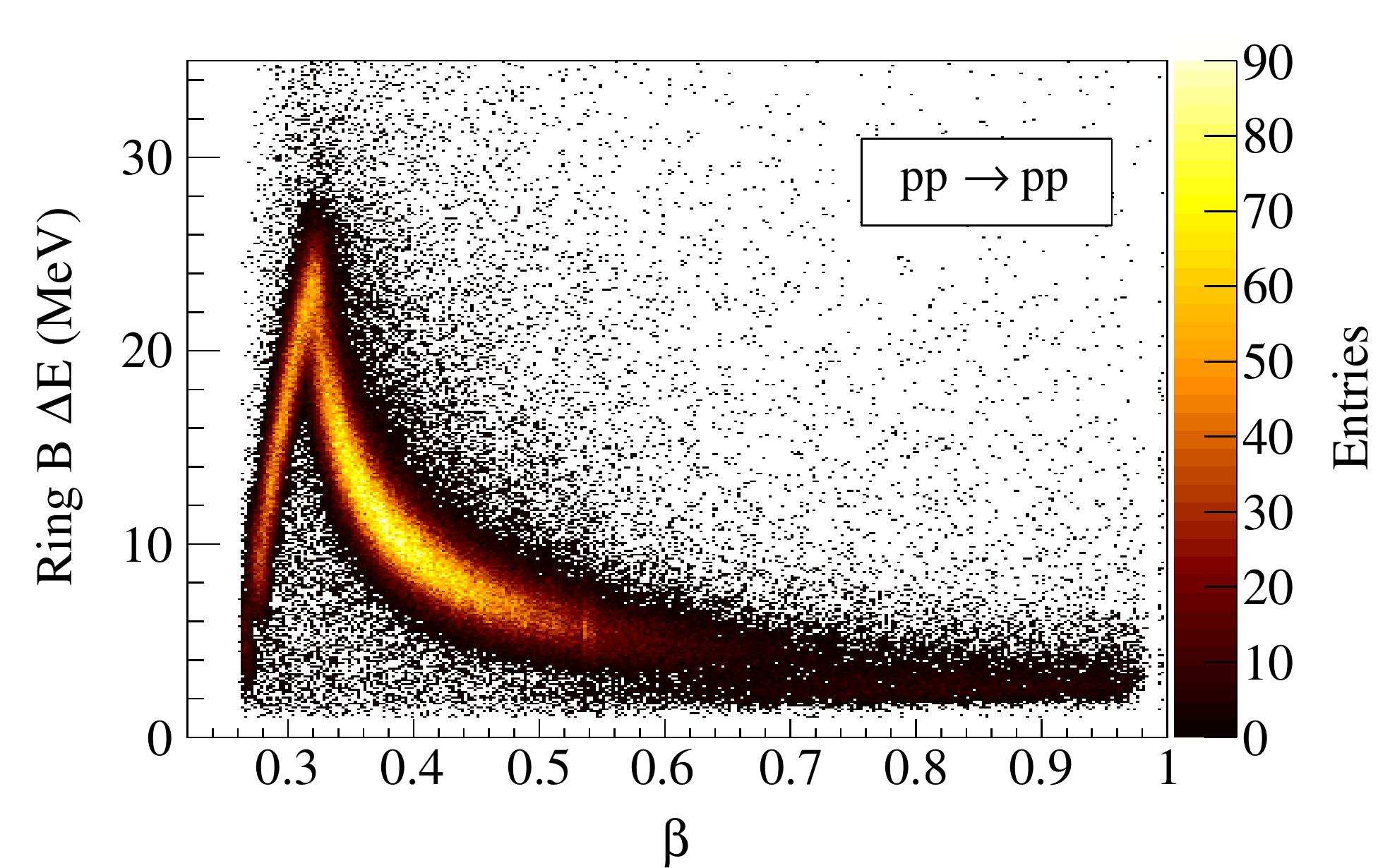}
  \end{center}
  \caption{\small Energy loss $\Delta E$ in the outer ring of the RPD as a
    function of the velocity of the particle in elastic pp scattering.}
  \label{fig:target.rpd.rpd-deb-beta}
\end{figure}

Figure~\ref{fig:target.rpd.rpd-deb-beta} shows the energy loss in the outer ring
as a function of the velocity measured with the RPD for elastic pp scattering.
For velocities of up to $\beta=0.34$ the protons are stopped in the outer
ring. Above this value the protons escape the scintillator and deposit in it
only part of their energy.  The figure for pions would be similar and the energy
loss for stopping pions would reach a maximum value of $10\,\MeV$ for
$\beta=0.4$. Therefore, proton particle identification is ensured only for
$\beta<0.4$. In Fig.~\ref{fig:target.rpd.rpd-deb-beta} there is no indication
for presence of pions, as expected in pp elastic scattering.

\subsection{Sandwich veto detector}
\label{sec:target.sandwich}
\begin{figure}[tbp]
    \centering \includegraphics[width=0.45\textwidth]{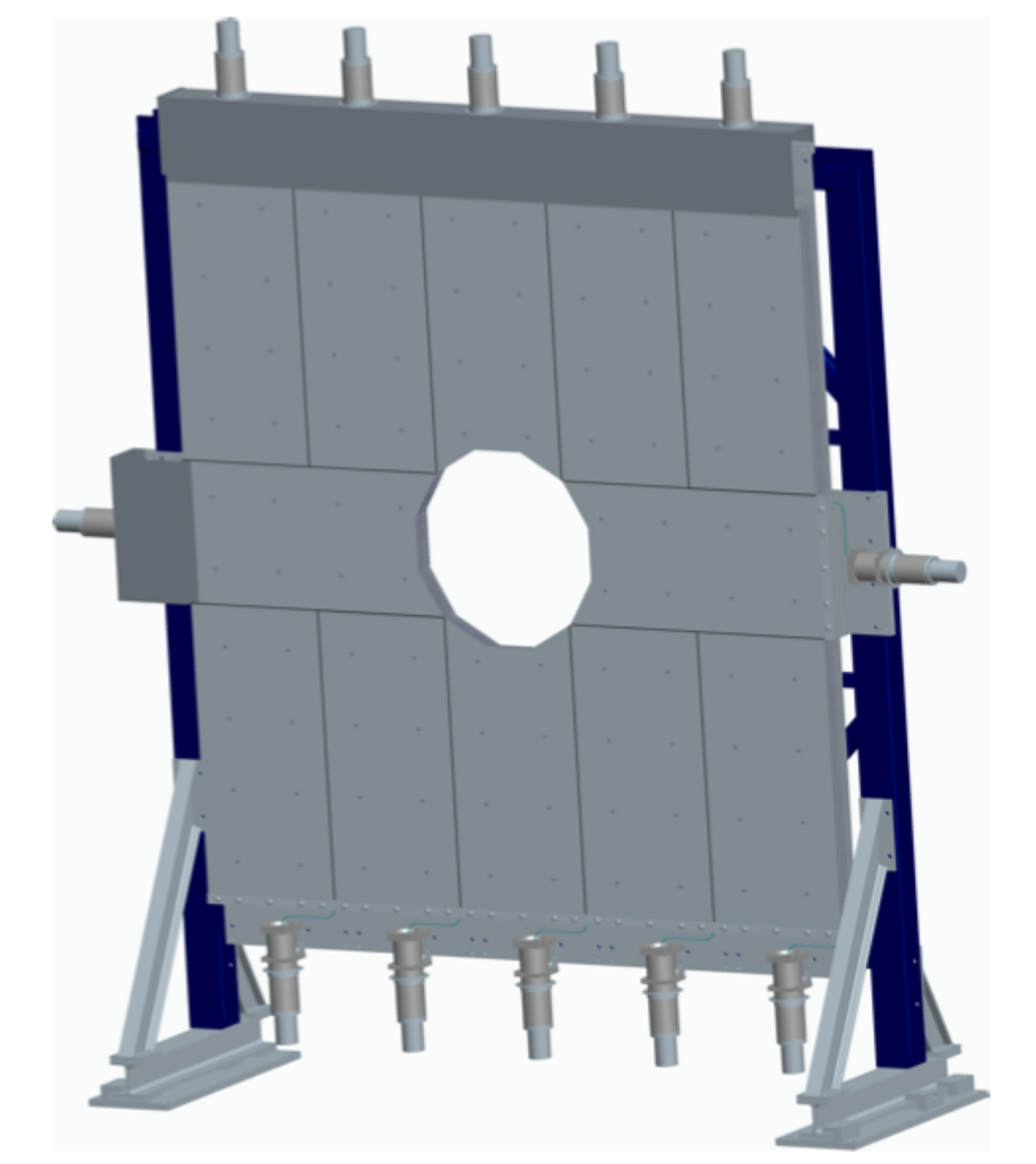}
    \caption{Sketch of the Sandwich veto detector. The active area of the
      detector (depicted in grey) has dimensions of 200$\times 200\,\Cm^2$.}
    \label{fig:targets.target_sandwich}
\end{figure}

The role of the Sandwich veto detector~\cite{Schluter:11} lies in vetoing events
in which photons or charged particles reach the acceptance gap between RPD and
LAS (see Fig.~\ref{fig:side-target}).  This detector is a $2\,\textrm{m}\times
2\,\textrm{m}$ stack of five layers of steel-covered lead plates and
scintillators with a total thickness of 5.1 radiation lengths. Segmented in 12
elements (\figref{fig:targets.target_sandwich}), the detector has a central hole
that matches the acceptance of the spectrometer.  Each lead layer consists of
$5\,\mm$ Pb plates, with $1\,\mm$ steel plates on each side to insure the
stiffness of the assembly.  Each scintillator layer is formed of a pair of
$80\times\,20\,\Cm^2$ scintillator bars lying side-by-side. The first three
layers are $1\,\Cm$ thick, the last two $0.5\,\Cm$. The light is extracted from
the scintillators using wavelength-shifting fibres connected to photomultiplier
tubes.  The Sandwich veto efficiency was measured to be 98\% for minimum
ionising particles~\cite{Schluter:11}. Using MC simulation, the efficiency was
also determined to be higher than 95\% for pions with energies larger than
$50$\,MeV and higher than 90\% or 80\% for photons with energies larger than
$100$\,MeV or $50$\,MeV, respectively.  The effect of the Sandwich veto detector
on the trigger rates is described in \secref{sec:trigger.veto}.

\section{Tracking Detectors}
\label{sec:tracking}
The tracking system of COMPASS is composed of many tracking stations, each
consisting of a set of planar tracking detectors of a given type located at
approximately the same $z$-coordinate along the beam.  Many different detector
technologies with different sizes, granularities and resolutions are in use.
Far from the beam in the outer region, large areas of several square meters have
to be covered in order to detect low-momentum particles scattered at large
angles.  Close to the beam in the inner region, the particle rates quickly
increase with decreasing distance to the beam, requiring fast detectors with
good resolution.  The large-area tracking is provided by several variants of
wire-based gas detectors such as Multiwire Proportional Chambers (MWPC), Drift
Chambers (DC), and Straw Tube Chambers. The region closer to the beam, where the
particle rates are too high for wire-based detectors, is covered by two types of
Micropattern Gaseous Detectors with strip readout, namely the Micromegas and Gas
Electron Multiplier (GEM) detectors. The beam region itself, where rates above
$10^{5}\,\mm^{-2}\s^{-1}$ are observed, is equipped with Scintillating Fibre
Detectors and novel GEM detectors with pixel readout, the PixelGEMs. Tracking
immediately upstream and downstream of the target is performed by silicon
microstrip detectors.

This section focuses on the upgrades of the tracking system for the hadron
program as compared to the setup used for muon beams, detailed in
\cite{Abbon:2007pq}.  For some detectors, like the straw tube chambers, the
multiwire proportional chambers, and the large area drift chambers, no
mentionable changes were introduced, and therefore they are not described here.
These detectors are however discussed in detail in \cite{Abbon:2007pq}.  The
wire and strip detectors measure different projections of a particle penetration
point. They are are called $X$ and $Y$-planes when measuring horizontal and
vertical coordinates, respectively. Detector planes measuring coordinates that
are rotated clock or and anti-clockwise by a given angle with respect to the
$x$-axis, are called $U$ and $V$-planes, respectively.

\subsection{Silicon Microstrip Detectors}
\label{sec:tracking.silicon}

The COMPASS silicon microstrip tracking system consists of three stations
upstream of the target, which are used as a beam telescope, and two stations
downstream of the target, which are used for vertex reconstruction.  As these
detectors are traversed by the beam particles and by the forward-boosted
reaction products, they are prone to radiation damage. The damage affects the
bulk material in terms of change of doping, and the surface in terms of decrease
of insulation, resulting in an increase of the depletion voltage and of the
leakage current, respectively.  In order to minimise these effects, the
detectors are cooled with liquid nitrogen.  Since the leakage current decreases
with temperature, noise caused by radiation damage is suppressed.  In addition
to this, the cooling leads to a significant improvement of the spatial and time
resolution compared to room-temperature operation, as discussed below.  While
the system was designed to cool the detectors down to $130\,\K$, the desired
performance is already achieved at $200\,\K$, which reduces the thermal stress
on the modules.

\begin{figure}[tbp]
	\begin{center}
	\includegraphics[width=0.475\textwidth]{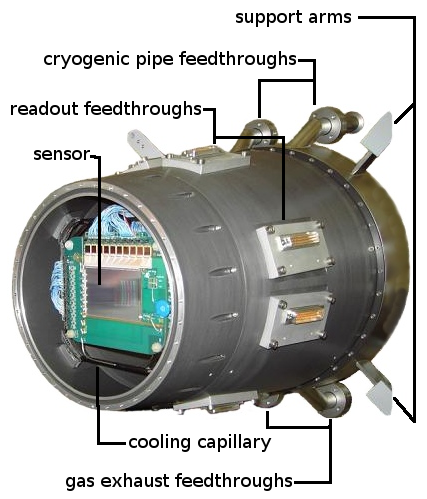}
	\end{center}
	\caption{The conical cryostat with the upstream beam window
          dismounted. The height of the (green) PCB frame that holds the
          detector (sensor) is about $100\,\mm$, the length of the full cryostat
          about $400\,\mm$. The bent cooling capillary is fixed to the PCB close
          to the sensitive area of the detector.  Inside the cryostat, the
          readout cables are directly soldered to the detector module and
          plugged to vacuum-sealed feedthrough connectors also visible on the
          outer surface of the cryostat.}
	\label{fig:tracking.SICC}
\end{figure}

One station comprises two Silicon detectors with a stereo angle of $5^\circ$
between their respective strip orientations to resolve multi-track hit
ambiguities.  Each detectors consist of a $300\,\mum$ thick silicon sensor with
an active area of $50\times70\,\mm^2$.  The signals are picked up on both sides,
by 1280 strips on the $n$-side and 1024 perpendicularly-oriented strips on the
$p$-side.  The sensors are glued onto two L-shaped FR4 printed circuit boards
(L-boards) that hold the APV25-S1~\cite{French:01} based readout electronics.
There are three cryostats for the beam stations upstream of the target and one
conically shaped cryostat housing the two stations downstream of the target (see
Fig.~\ref{fig:side-target}).

\begin{figure}[tbp]
  \begin{center}
    \includegraphics[width=0.8\textwidth]{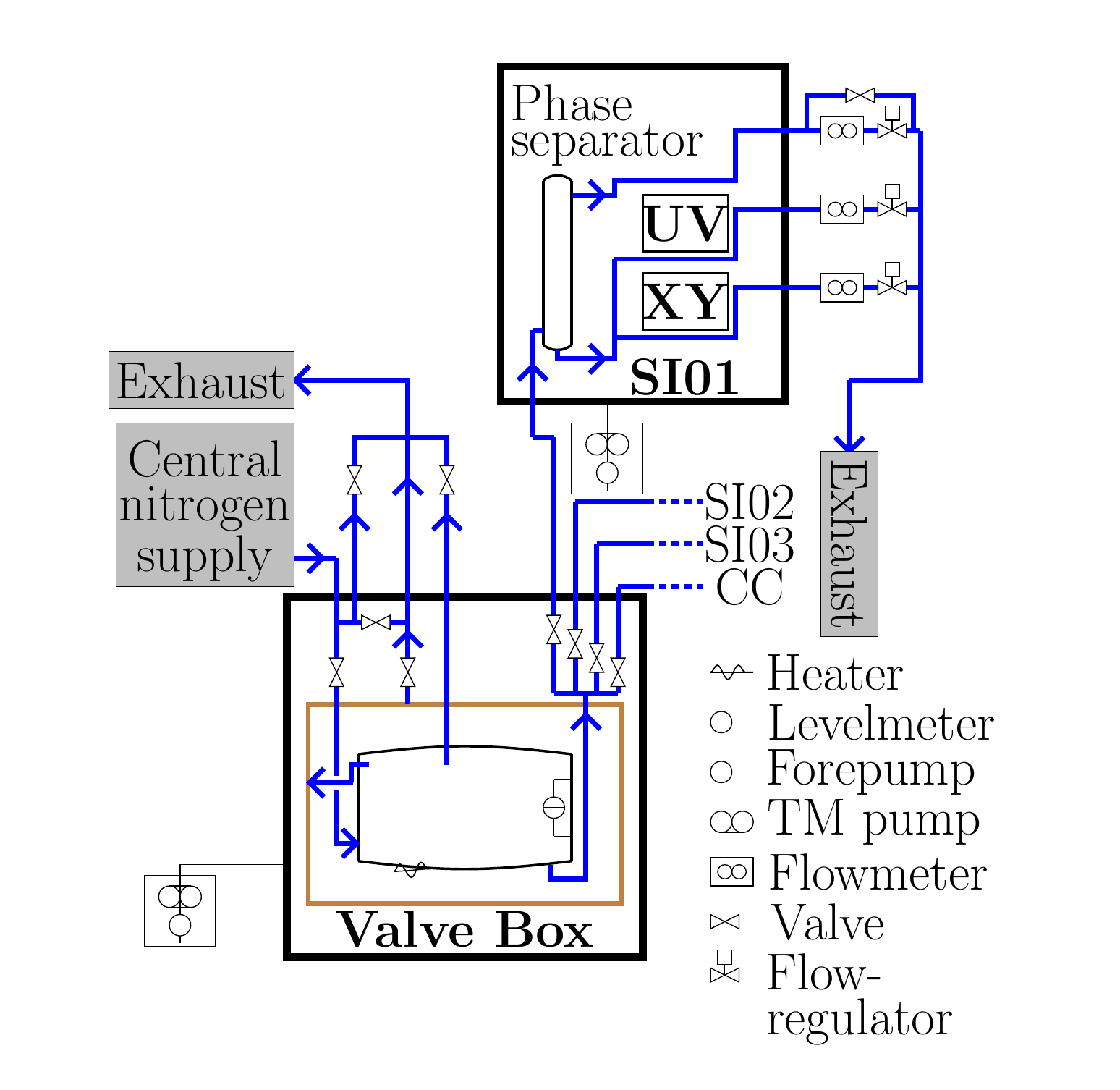}
  \end{center}
  \caption{Block diagram of the valve box and the first upstream cryostat
    labelled SI01. The other two upstream cryostats SI02 and SI03 are equipped
    analogously. The downstream conical cryostat (CC) is shown in
    Fig.~\ref{fig:tracking.SI_CC_schematics}.  The phase separators are
    integrated in the cryostats near the detectors, but outside the acceptance.}
  \label{fig:tracking.SI_valvebox_schematics}
\end{figure}

\begin{figure}[tbp]
  \begin{center}

    \includegraphics[width=0.8\textwidth]{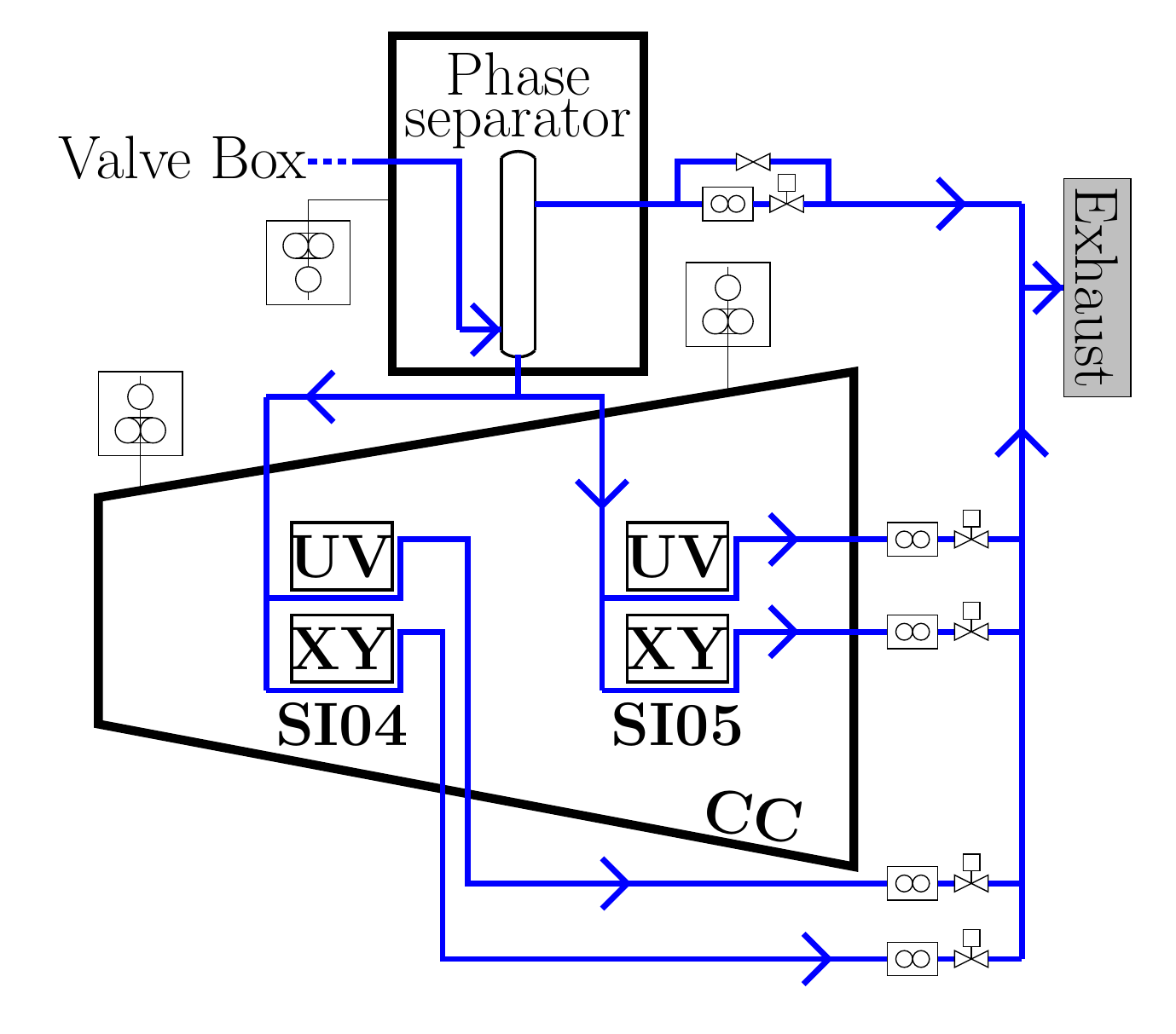}
  \end{center}
  \caption{Block diagram of the conical cryostat (CC), symbols as in
    Fig.~\ref{fig:tracking.SI_valvebox_schematics}.  The phase separator is
    mounted in an extra housing outside the spectrometer acceptance with a
    vacuum connection to the cryostat.}
  \label{fig:tracking.SI_CC_schematics}
\end{figure}

The cooling system of the Silicon detectors has to fulfil the requirement of a
minimal amount of material within the acceptance of the spectrometer.  This
prevents a solution, in which the detectors are connected to a massive cold head
to dissipate the electronic heat.  The technology developed for these detectors
is based on the evaporation of liquid nitrogen in thin capillaries on the PCBs.
The schematic layout of the Silicon cooling system is shown in
Figs.~\ref{fig:tracking.SI_valvebox_schematics} and
\ref{fig:tracking.SI_CC_schematics}.  In order to dissipate about $8\,\W$ from
each detector, purely liquid nitrogen must be provided to the capillary.  For
this purpose, a dedicated phase separator that removes the gaseous nitrogen is
incorporated in each cryostat.  The whole cooling infrastructure increases the
material thickness of the PCB on average by $0.1\%$ of a radiation length.

The nitrogen arrives from a central liquid nitrogen dewar located in the
vicinity of the experimental hall. It is transferred by a $100\,\m$ long
vacuum-isolated transfer line to a valve box near the Silicon stations.  The
valve box (Fig.~\ref{fig:tracking.SI_valvebox_schematics}) also acts as a buffer
for the liquid nitrogen that is kept at $1.8\,\textrm{bar}$ absolute pressure.
The liquid nitrogen is brought to the Silicon stations using $2$--$3\,\m$ long
transfer lines.  When the nitrogen reaches the phase separator, the gas that
evaporated in the transfer line is allowed to escape.  At the bottom of the
phase separator, where there is only liquid nitrogen, two capillaries are
connected, one for each detector.  The thermal contact to the L-boards is made
by soldering dots.  The temperature of the detector is regulated through the
gaseous exhaust flow with a feedback time in the order of one second.  All
components are operated by a Programmable Logic Controller (PLC, SIMATIC S7
300), utilising a Proportional-Integral-Derivative algorithm for the temperature
regulation.  The software used is a Java\textsuperscript{\texttrademark}-based
application called Muscade\cite{Anibus} which provides real-time monitoring,
remote control, data storage, and an alarm system.

In 2009, all Silicon stations were cooled to $200\,\K$.  The temperature of the
system was stabilised to within $\pm 1$K for all detectors of the upstream
stations.  Slightly larger variations were observed for the detectors in the
conical cryostat, where a partly blocked capillary prevented good cooling for
one of the detectors, limiting the temperature to $\sim220\,\K$ only.  This
detector also exhibited slow drifts following the daily temperature variations.

\begin{figure}[tbp]
	\begin{minipage}[t]{0.49\textwidth}
		\centering
                \includegraphics[width=\textwidth]{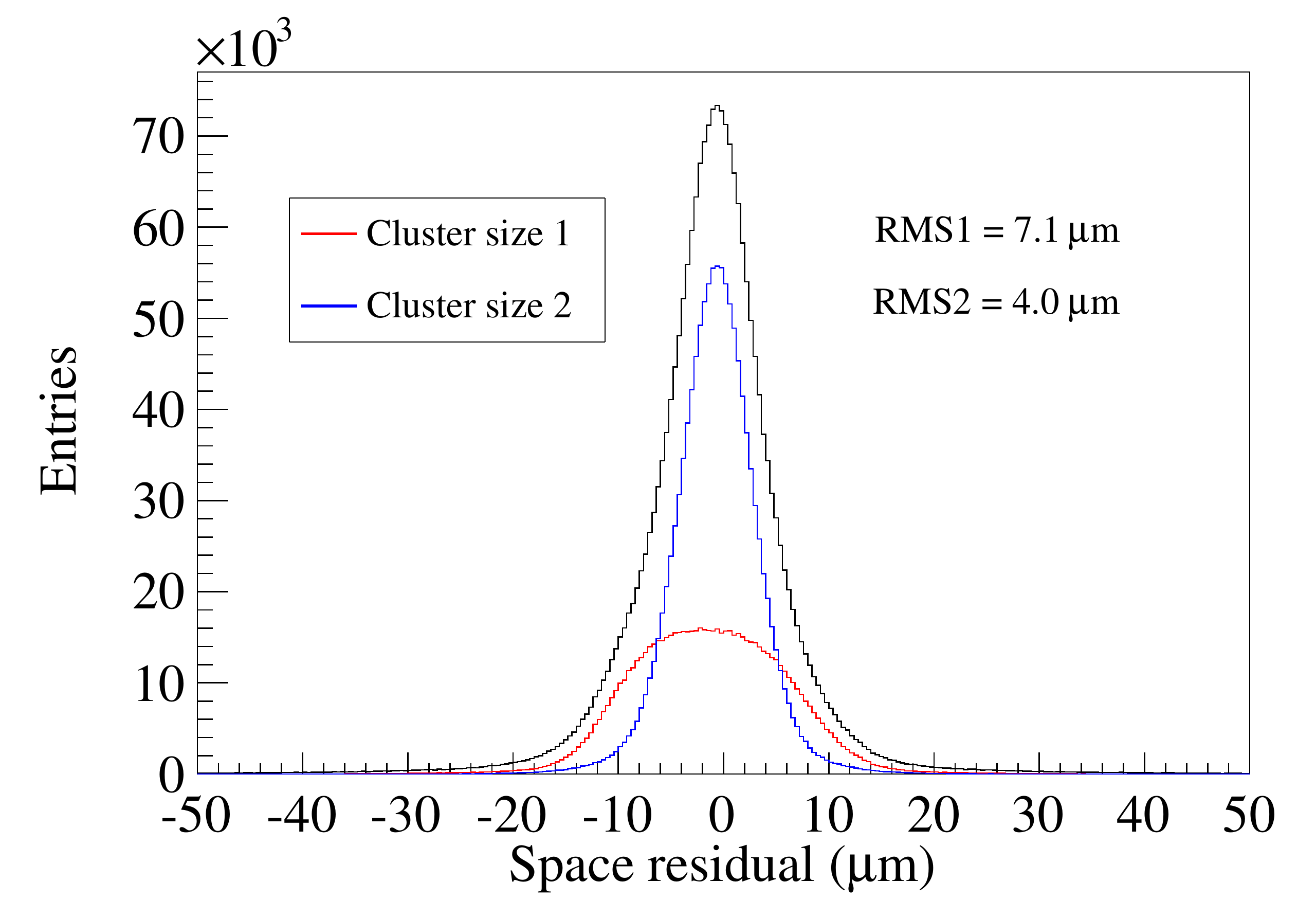}
		\caption{Spatial resolution as determined for a single Silicon
                  detector plane. ``RMS1'' and ``RMS2'' refer to the cases of
                  clusters with one and two hit strips, respectively.}
		\label{fig:tracking.SI05residual}
	\end{minipage}
	\hfill
	\begin{minipage}[t]{0.49\textwidth}
		\centering
                \includegraphics[width=\textwidth]{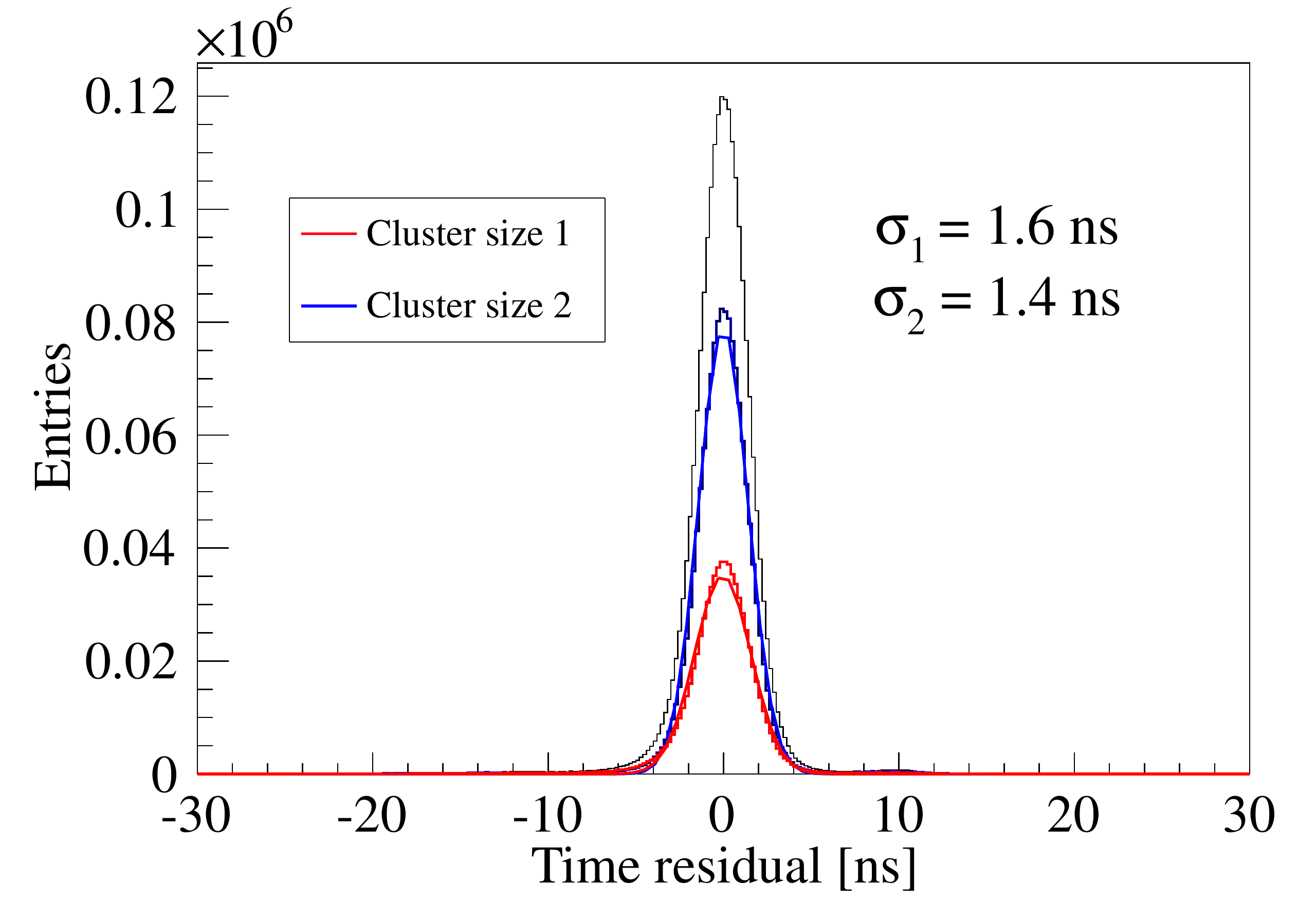}
		\caption{Time resolution of a single Silicon detector
                  projection.}
		\label{fig:tracking.SI01Xtime}
	\end{minipage}
\end{figure}

The spatial resolution of the cold Silicon detectors is in the range $4-6\,\mum$
for clusters when two strips are hit and amplitude weighting can be employed to
determine the track position~\cite{Grabmueller:12}.  When only one strip is hit,
the resolution is in the range $7-11\,\mum$.  This spatial resolution is
illustrated in Fig.~\ref{fig:tracking.SI05residual} for one of the detectors. It
represents an improvement of 15-20\% compared to room-temperature
operation~\cite{Abbon:2007pq}.  The reduction of the leakage current and the
increase of the signal each contribute of about $10\%$ to this improvement.  The
time resolution, displayed in Fig.~\ref{fig:tracking.SI01Xtime}, is improved for
the same reason and is in the range $1.4-1.8\,\ns$.

\begin{figure}[tbp]
  \centering
  \includegraphics*[width=0.6\textwidth]{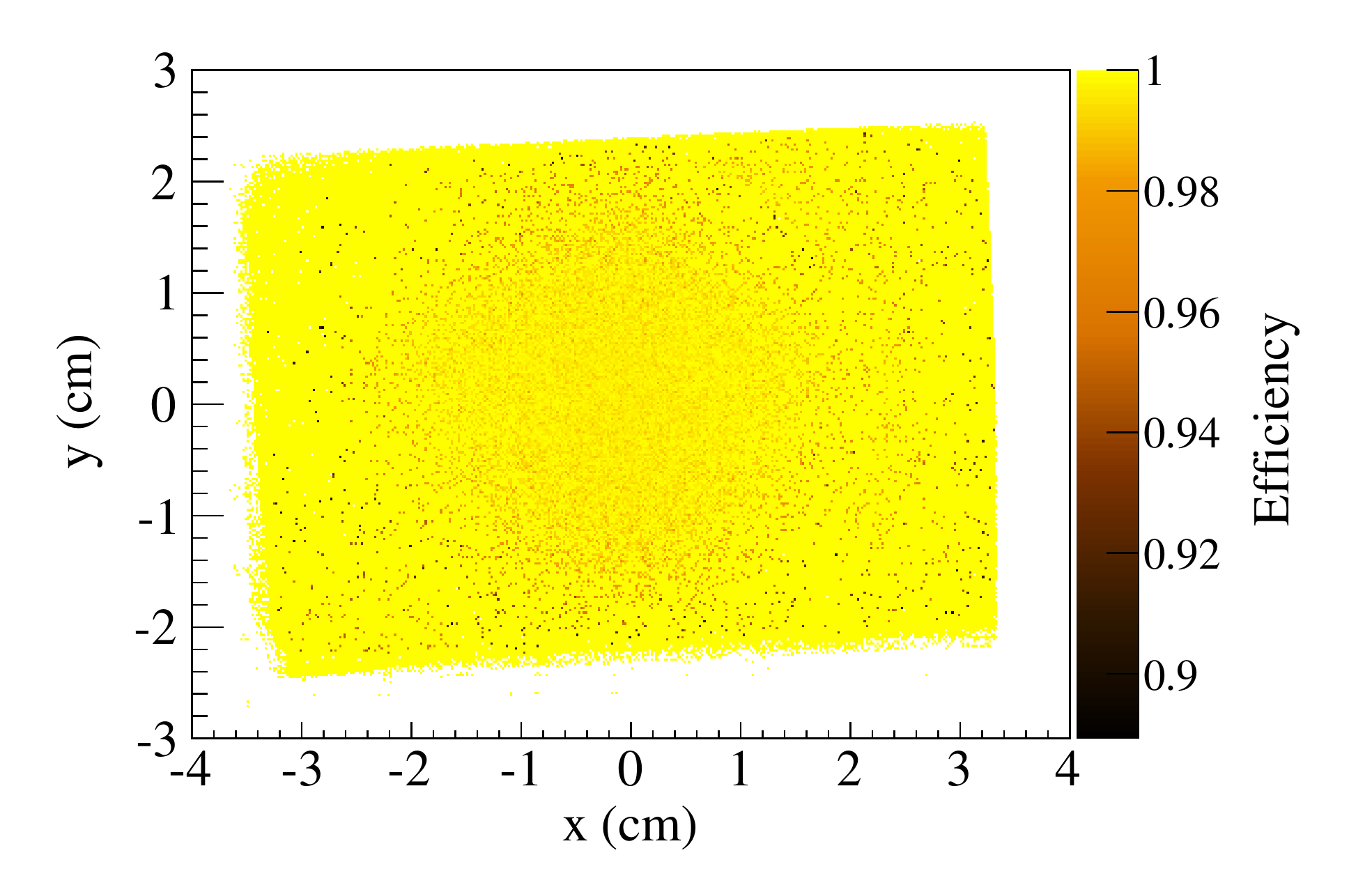}
  \caption{Two-dimensional efficiency distribution for a plane in the beam
    telescope. The stereo-angle tilt of the sensitive area is visible.  }
  \label{fig:tracking.SI03Xeff}
\end{figure}

The efficiency was studied in detail with a high-intensity muon beam.  Tracks
reconstructed without taking into account hits of the detector under study are
used to measure the efficiency by providing the expected hit position on the
detector. The presence of a hit is then checked within a $\pm 3\sigma$ window
around the expected position. The measured efficiency is above $99\%$ as shown
in Fig.~\ref{fig:tracking.SI03Xeff} for one of the planes. Similar results were
also obtained for operation with hadron beams.

\subsection{PixelGEM Detectors}
\label{sec:tracking.pixelgem}

In order to minimise the material from detectors directly exposed to the hadron
beam, some of the scintillating fibre detectors that were used with the muon
beam were replaced by thinner detectors based on Gas Electron Multiplier (GEM)
foils \cite{Sauli:97}. Starting with the first hadron run in 2008, five GEM
detectors with a novel kind of readout and a thickness in the beam region of
$0.26\,\%$ of a radiation length ($X_0$) and $0.1\,\%$ of a nuclear interaction
length each ($\lambda_I$) were installed, thereby reducing the material budget
of the whole system by a factor of $5$--$10$ compared to the scintillating fibre
detectors.

\begin{figure}
 \begin{minipage}[t]{.48\textwidth}
    \centering \includegraphics[trim=0 0.8cm 0 0,clip=true,width=0.95\textwidth]{fig_s05_023.pdf}
    \caption{Schematic view of the pixel and strip region of the readout
      circuit. Note that the pixel region consists of $32\times 32$ pixels of
      $1\,\mm^2$ size each, while only $4\times 4$ are shown for clarity. Figure
      not to scale.}
    \label{fig:tracking.pixelgem.schematic}
  \end{minipage}
  \hfill
  \begin{minipage}[t]{.48\textwidth}
    \centering
    \includegraphics[width=.95\textwidth]{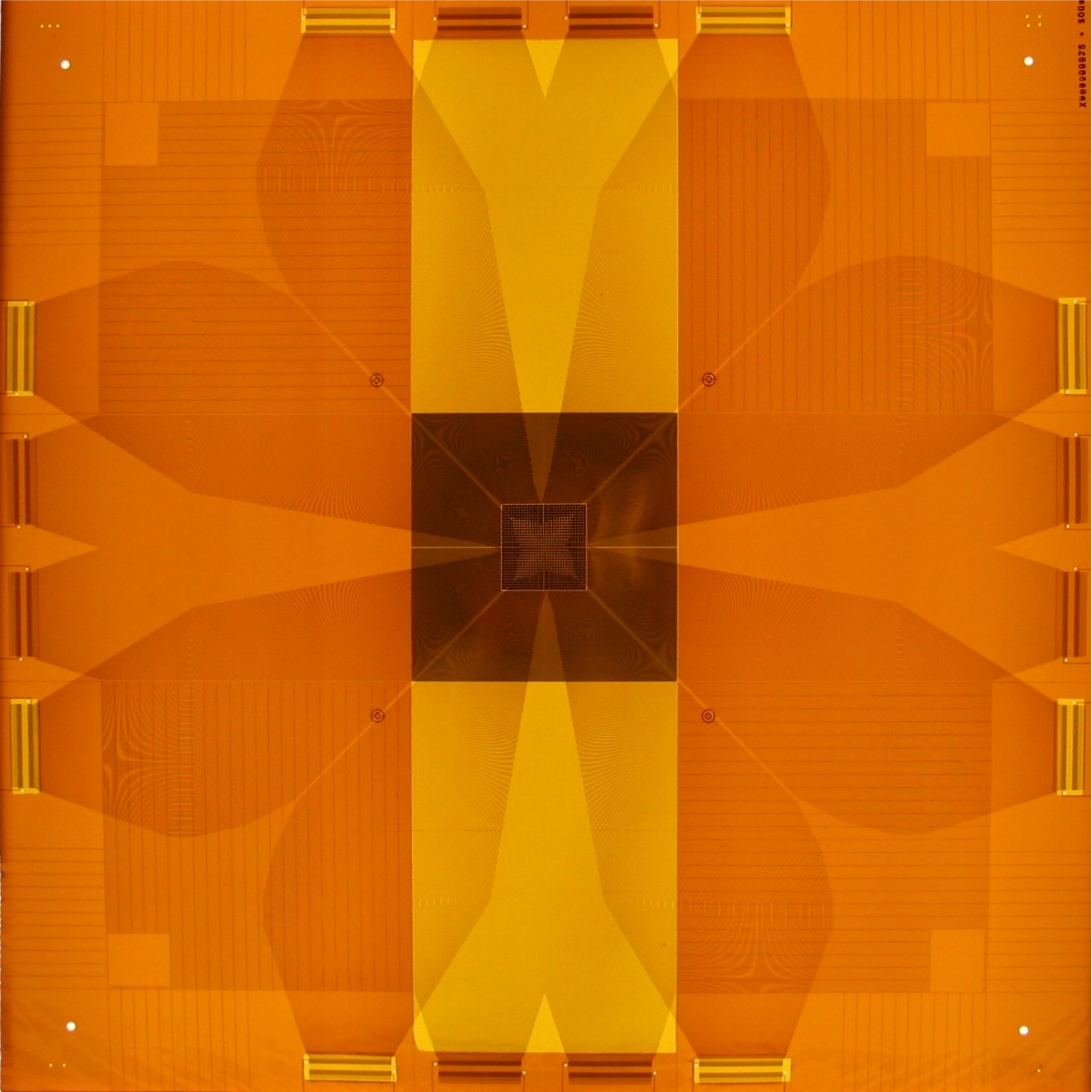}
    \caption{The PixelGEM read-out foil. The inner $10\times 10\,\Cm^2$ darkest
      part is the active area. The symmetric wires connecting the pads and the
      strips to the read-out electronics surround this part.}
    \label{fig:tracking.pixelgem.readout}
  \end{minipage}
\end{figure}

GEM detectors with a two-dimensional strip readout have been used in COMPASS
since its start-up \cite{Ketzer:04a}. These gaseous detectors have proved to be
able to cope with the high particle fluxes in the beam centre, but the strip
readout makes it impossible to separate individual hits close to the beam due to
a too high occupancy.  In order to overcome this limitation, a novel read-out
structure has been realised on a polyimide basis using the GEM patterning and
wet-etching printed-circuit board (PCB) technologies \cite{Ketzer:07a}. The
central are of $32\times 32\,\mm^2$ with 1024 pixels of $1\times 1\,\mm^2$ size
each are patterned on one side of $50\,\upmu\m$ thick polyimide foil. The signal
traces from the pixels to the readout electronics are routed on the other side
of the foil, with an extremely small width of only $50\,\upmu\m$ and a pitch of
$100\,\upmu\m$. The rest of the total active area of $100\times 100\,\mm^2$,
where the occupancy is sufficiently low, is covered by two orthogonal sets of
512 strips with a pitch of $400\,\upmu\m$, realised on a second $50\,\upmu\m$
thick polyimide foil.  The strip foil is then glued onto the one with the
pixels, with the central area completely removed from the strip foil in order to
open the pixels for charge collection.  The strips are split in the middle and
read out on both sides in order to equalise their capacitances, also for the
ones not ending at the pixel region.  In
Fig.~\ref{fig:tracking.pixelgem.schematic}, the pixel and strip regions are
displayed schematically, while Fig.~\ref{fig:tracking.pixelgem.readout} shows a
photograph of the complete readout foil.
The readout foil is glued onto a light honeycomb sandwich panel of $610\times
610\,\mm^2$ size, which serves as support plate and also carries the front-end
electronics, the high voltage distribution circuit, and the GEM stack.

The GEM stack consists of three GEM foils of $10\times 10\,\Cm^2$ active area,
stretched and glued onto larger frames of fibre glass material with $316\times
316\,\Cm^2$ inner dimensions. These frames with a thickness of $2\,\mm$ are
piled up and glued on top of each other.  The active part of a GEM foil is
sectorised on one side into four parallel sectors of equal size, and a fifth
sector in the centre matching the pixel area of $32\times 32\,\mm^2$. The foils
are mounted such that the segmented sides face the drift cathode. The potentials
on the foils are defined through an external resistive divider. They are
adjusted such that the largest gain is provided by the first foil and it is
stepwise decreasing for the second and third foil.  The segmented sides of a
foil are supplied through individual $10\,\MOhm$ loading resistors, while there
are no loading resistors for the non-segmented side. This configuration allows
for an operation of the detector even with a potential permanent short circuit
in one of the sectors, and avoids a high electric field between the last foil
and the readout circuit in case of a discharge. The central sector of the third
GEM foil is powered by a separate supply through a $1\,\MOhm$ serial resistor,
which allows an independent adjustment of the gain for the central region. This
takes into account the fact that a smaller effective gain is necessary for the
central region because the signal is induced on pads instead of two sets of
strips for the peripheral region. Efficiency scans performed with prototype
detectors showed that an effective gain of $8000$ is required for the strip
region for fully efficient detection of minimum ionising particles, while a gain
of $6000$ is sufficient for the pixel region.  The triple amplification together
with the non-uniform gain distribution and the segmented GEM foils, which were
already used for the large-size COMPASS GEM detectors \cite{Altunbas:02a},
ensures operation of the PixelGEM detectors without electrical discharges even
in a high-intensity hadron beam.

In order to minimise the material in the region near the beam, the gas-filled
volume extends to cover a total area of $316\times 316\,\mm^2$.  It is enclosed
by a frame defining the conversion volume, and a smaller honeycomb panel of
$330\times 330\,\mm^2$ size, which carries the cathode foil made of Cu-coated
polyimide.  The material exposed to the beam is minimised by central holes of
$30\,\mm$ diameter in both honeycomb panels and by reducing the thickness of
each of the Cu layers on the drift cathode and the GEM foils from originally
$5\,\upmu\m$ to about $1\,\upmu\m$.

\begin{figure}
  \begin{minipage}[t]{.63\textwidth}
    \centering
    \includegraphics[width=.99\textwidth]{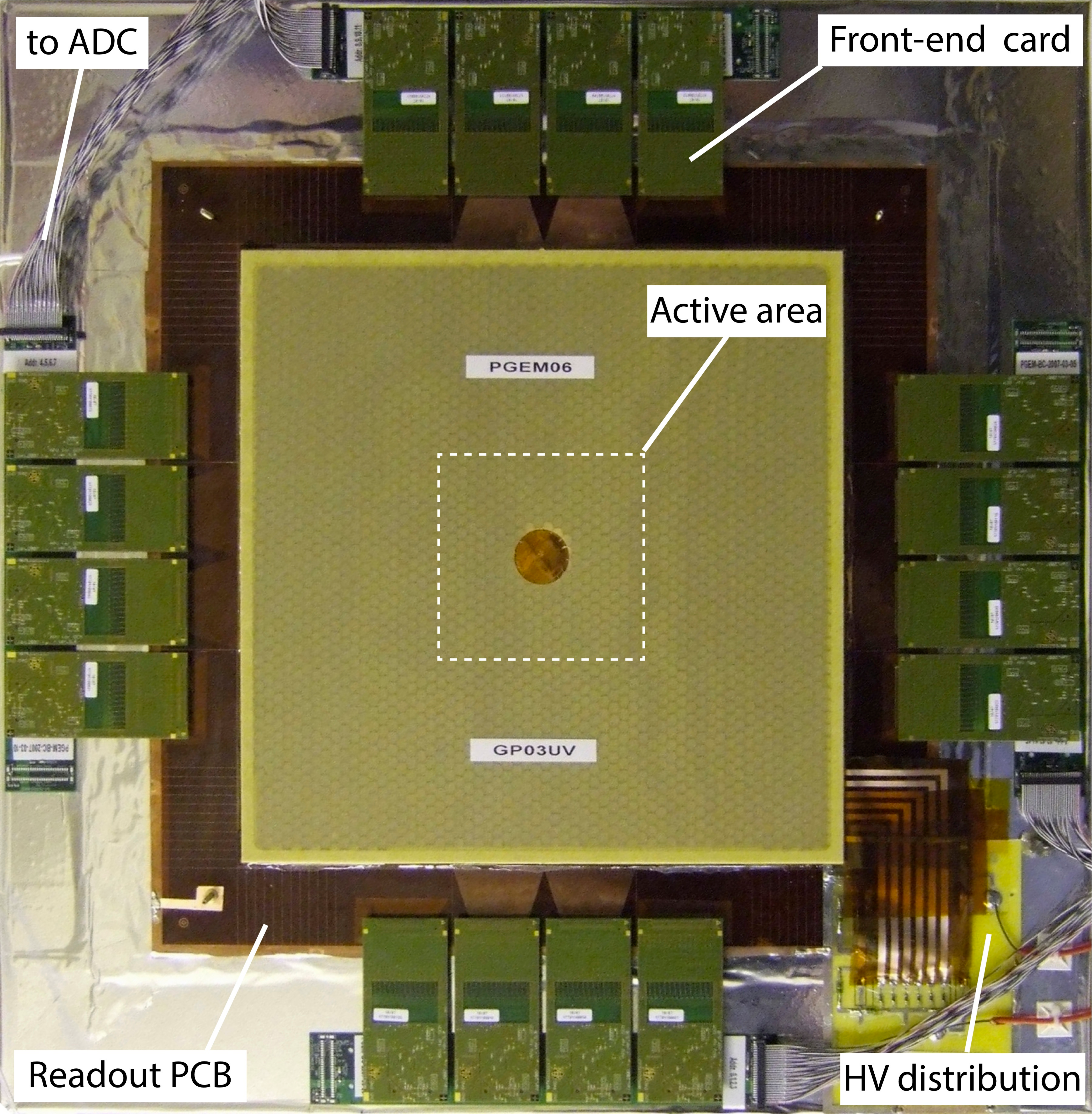}
    \caption{A fully assembled PixelGEM detector, equipped with 16 APV front-end
      cards. The digitisation of the analog signals from the APVs is done at an
      external ADC card, which is connected via the grey cables.}
    \label{fig:tracking.pixelgem.detector}
  \end{minipage}
  \hfill
  \begin{minipage}[t]{.33\textwidth}
    \centering
    \includegraphics[width=0.94\textwidth]{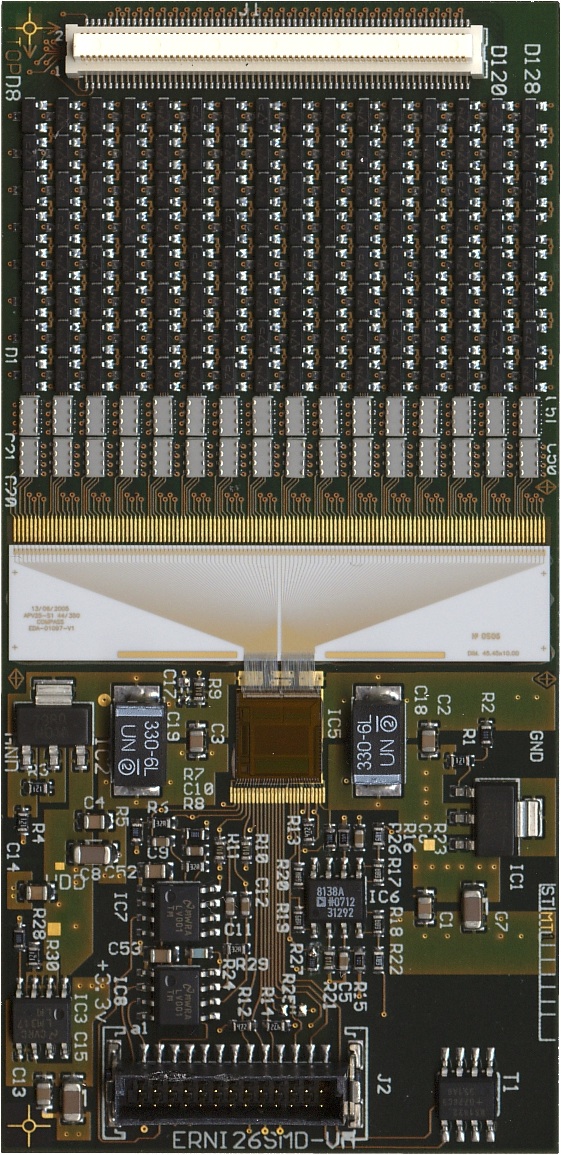}
    \caption{Front-end card carrying (from top to bottom) the 130-pin connector,
      the protection network, a ceramic pitch adaptor, and the APV25-S1 ASIC for
      analog sampling of the signals induced on the readout electrodes.}
    \label{fig:tracking.pixelgem.frontend}
  \end{minipage}
\end{figure}

Figure~\ref{fig:tracking.pixelgem.detector} shows a top view of an assembled
detector, with the high voltage distribution board (lower right corner) and the
16 front-end electronics cards mounted upside-down. During operation the whole
detector is shielded from external electronic noise by a thin aluminium-coated
Mylar foil. As the large-area GEM detectors, the PixelGEM detectors are operated
in a gas mixture of Ar/CO$_2$ (70\%/30\%).

A total of 2048 channels per detector are read out using the APV25-S1
preamplifier/shaper ASIC \cite{French:01}, which samples the input signal at a
frequency of $38.88\,\MHz$ into an analog pipeline with a depth of 160 samples.
Each chip is mounted onto a separate front-end card connected to the readout
circuit using high-density 130-pin connectors, of which two pins are used to
connect the ground level of the chip to the detector ground. In contrast to the
large-area detectors, where the front-end cards were directly wire-bonded to the
readout circuit, the connector solution provides much more reliability and
allows for a simple replacement of faulty cards. The front-end cards also
contain an external protection network consisting of a pair of high-speed
switching diodes (BAV99) and an AC coupling using a $220\,\pF$ capacitor for
each channel, and a ceramics pitch
adaptor. Figure~\ref{fig:tracking.pixelgem.frontend} shows a photograph of the
front-end card.  Three signal amplitudes per channel are multiplexed onto a
single differential line for each APV25-S1 chip and digitised by a pipelined
12-bit differential ADC at a sampling rate of $40\,\MHz$. The signals from
sixteen APV25-S1 chips are digitised on a custom-made ADC card. This card also
includes a Virtex FPGA~\cite{Xilinx}, which performs pedestal subtraction with
individual values for each channel, common mode noise correction and zero
suppression by applying individual thresholds for each channel.

After a successfully operated prototype, which was tested in a muon beam with a
flux up to $1.2\cdot10^5\,\mu^+/\left(\mm^2\s\right)$
\cite{Kraemer:08a,Austregesilo:2009zz}, five PixelGEM detectors were installed
in the spectrometer in 2008.  One detector was placed about $2.5\,\m$ downstream
of the target. Two pairs of detectors mounted back to back, with the second
rotated by $45^\circ$ with respect to the first, were installed around SM2, at
$19\,\m$ and $24\,\m$ downstream of the target, respectively (see
\figref{fig:layout_setup}).

In the offline analysis, a pulse-shape analysis technique is used to extract the
signal amplitude and time for each channel, a feature of great importance in a
high-intensity environment.  Signals from neighbouring pixel channels on the
detector are then weighted by their amplitudes and grouped into clusters.
Corrections for a non-linear charge sharing between pixels are applied.  These
corrections have been determined in a dedicated test beam experiment, using
high-resolution silicon microstrip detectors \cite{Kraemer:08a}.  At this stage
also a small (percent-level) cross-talk between channels, arising from the
narrow and long PCB traces between the pixels and the front-end cards, is
removed.  For the strip signals a simpler clustering algorithm based on a
centre-of-gravity method is applied.
 
\begin{figure}
  \centering
  \includegraphics[width=.45\textwidth]{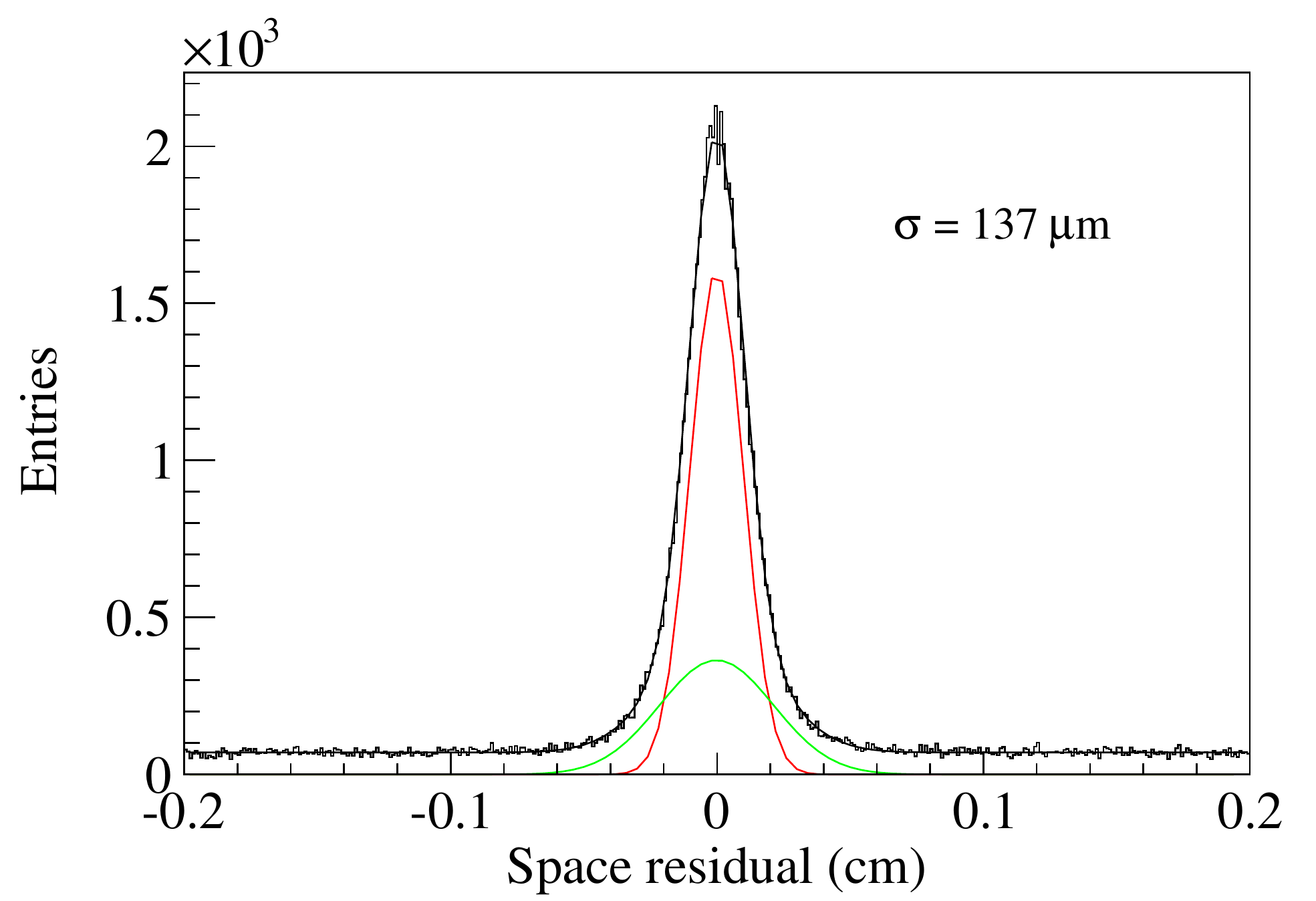}
  \includegraphics[width=0.45\textwidth]{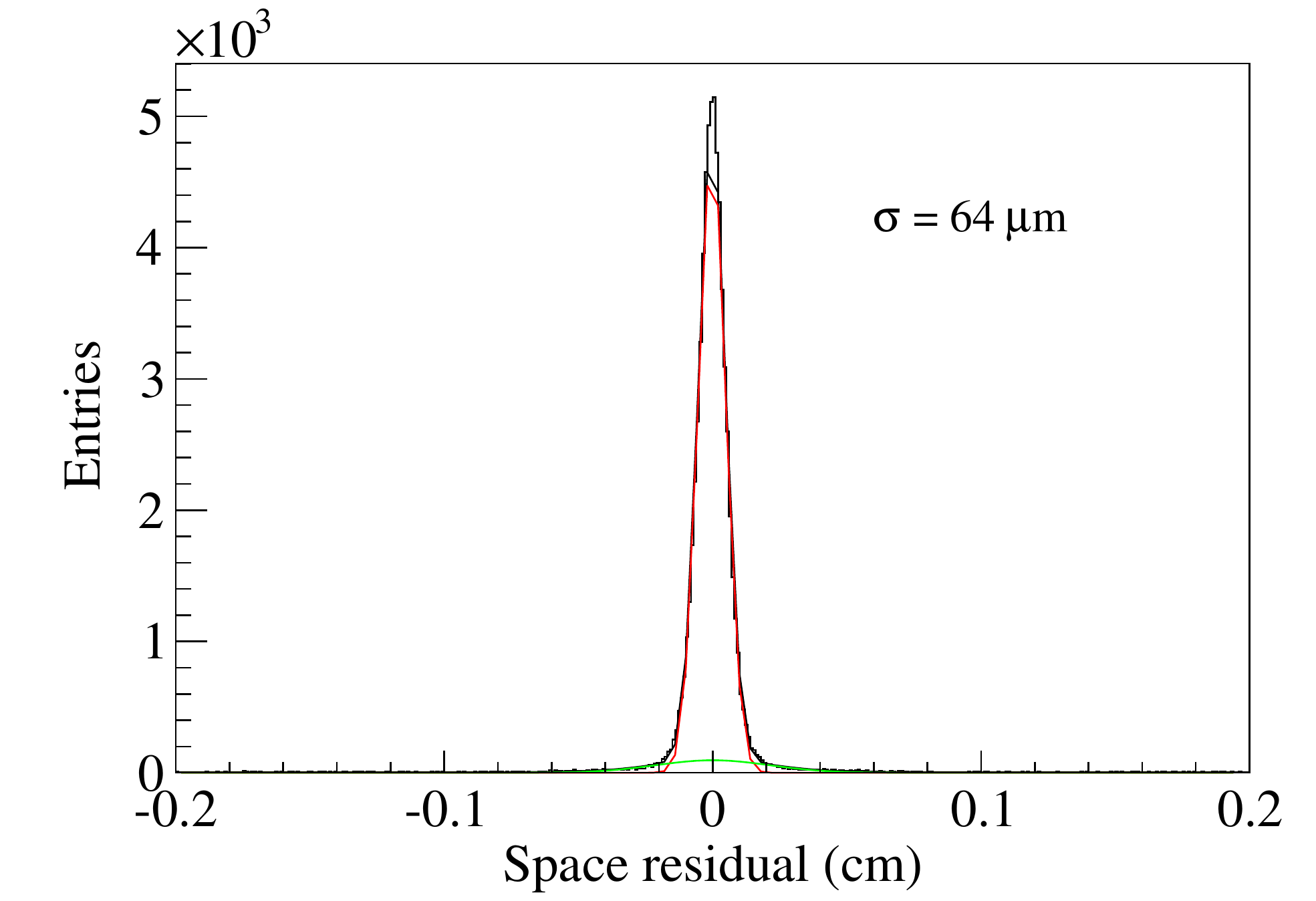}
  \caption{Residual distribution (difference between measured cluster position
    and track penetration point) in $x$-direction for (left) the pixel region
    and (right) the strip region of a PixelGEM detector. The quoted residual
    widths are obtained from fits of a sum of two Gaussians. When corrected for
    the track uncertainties, spatial resolutions of $106\,\mum$ (pixels) and
    $54\,\mum$ (strips) are obtained for this particular detector.  }
  \label{fig:tracking.pixelgem.residual}
\end{figure}

\begin{figure}
  \centering
  \includegraphics[width=.45\textwidth]{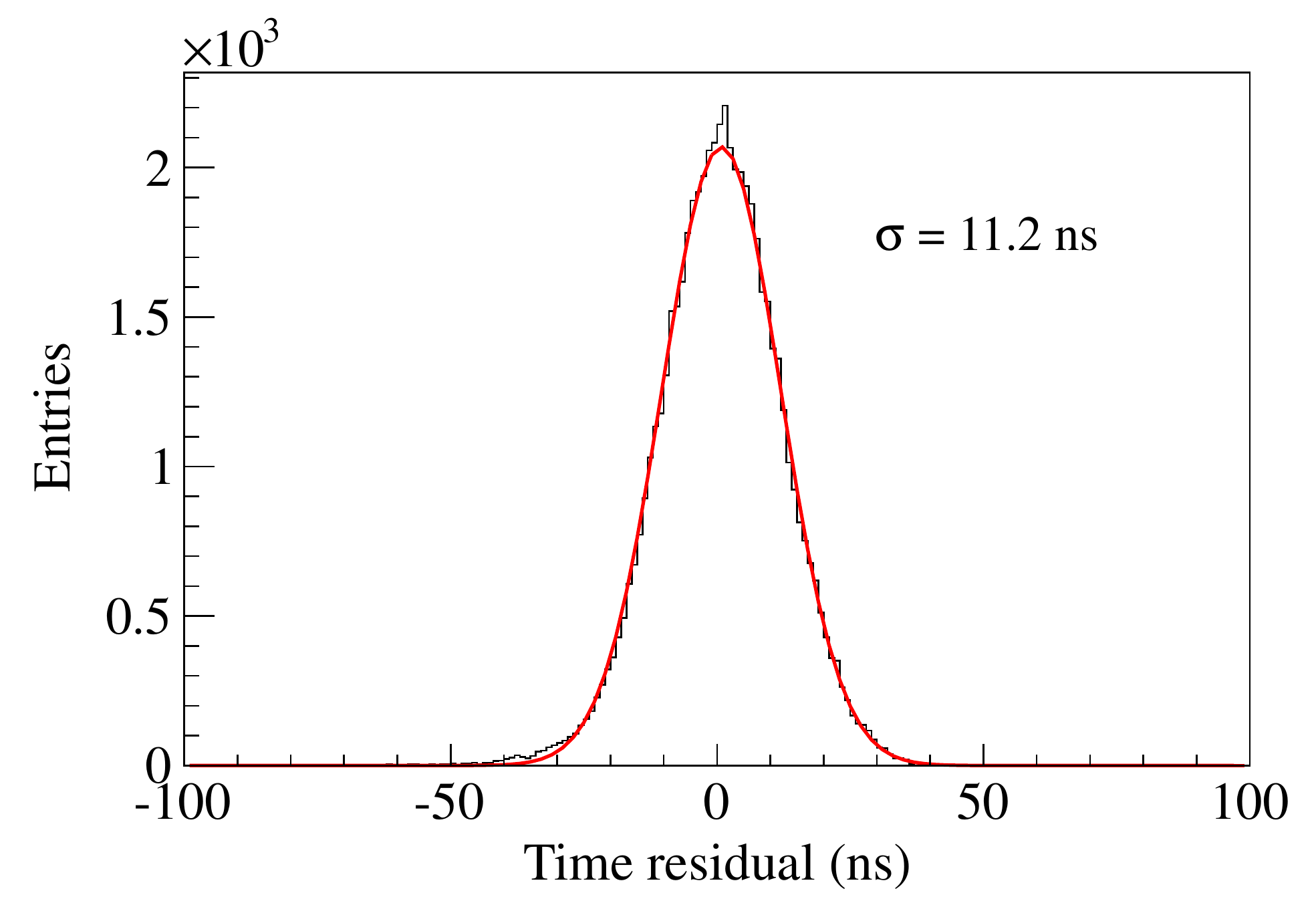}
  \includegraphics[width=.45\textwidth]{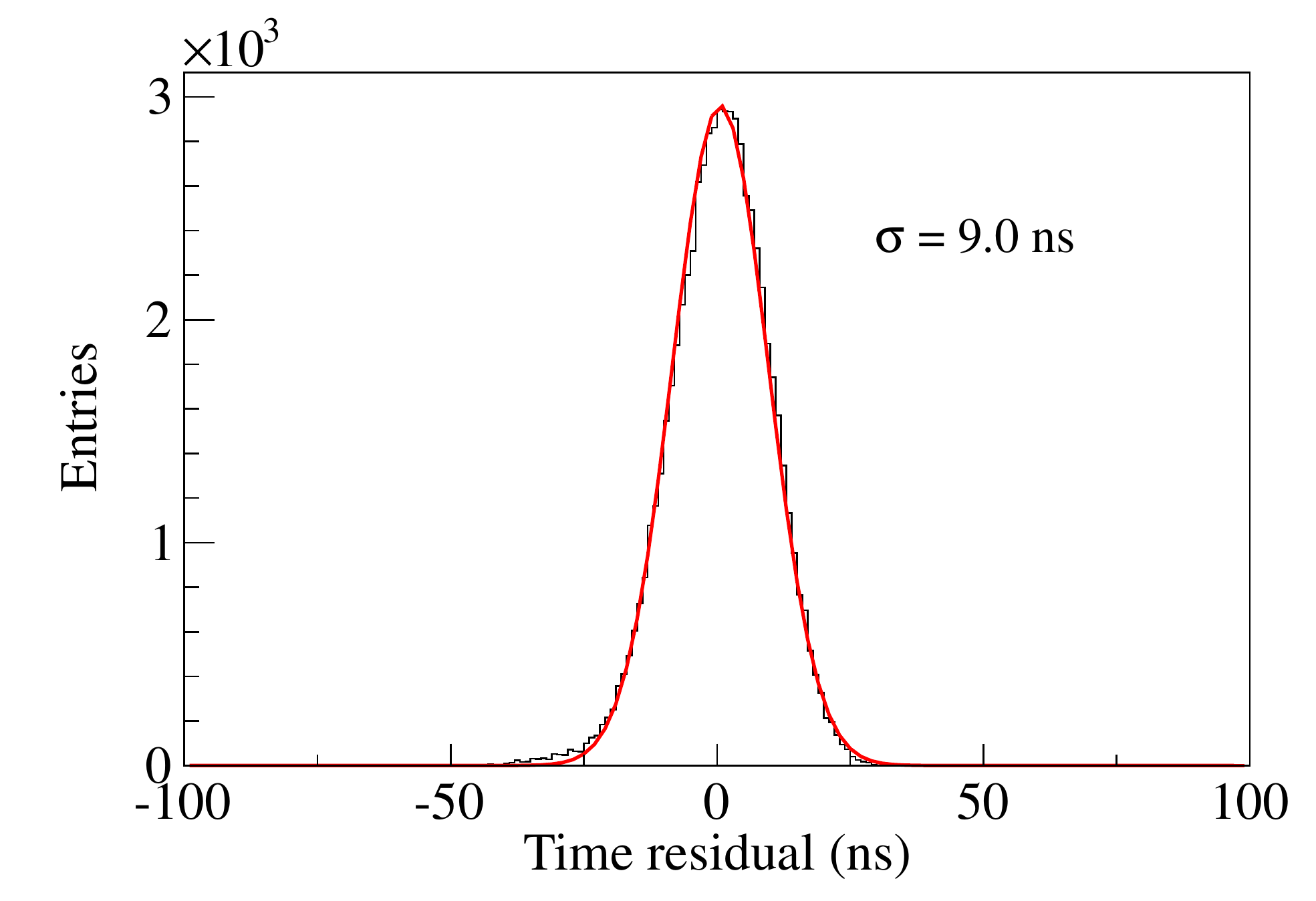}
  \caption{Time residual distribution (difference between measured cluster time
    and track time) for (left) the pixel region and (right) the strip region
    ($x$-direction) of a PixelGEM detector.  }
  \label{fig:tracking.pixelgem.timing}
\end{figure}

In the following, the performance of the PixelGEM detectors at a hadron beam
flux density of $2.4\cdot 10^4\,\pi^-/\left(\mm^2\s\right)$ (total flux of
$6.3\EE{5}\,\pi^-/\s$), used for data taking, is shown.
Figure~\ref{fig:tracking.pixelgem.residual} shows the residual distribution,
i.e.\ the difference between the measured cluster position and the extrapolated
penetration point of a reference track, for the pixel region (left) and the
strip region (right).
As for the Silicon detectors, the detector under investigation has been excluded
from the track reconstruction, such that unbiased residuals are obtained. After
deconvolving the uncertainty on the reconstructed tracks, one obtains, for all
five PixelGEM detectors, spatial resolutions distributed around an average value
of $125\,\mum$ with a standard deviation of $13\,\mum$ for the pixel regions,
and an average value of $65\,\mum$ with a standard deviation of $12\,\mum$ for
the strip regions.  From the pulse-shape analysis of the three samples read out
per channel per event, one can extract the time of the signal and thus
efficiently remove background hits due to pile-up. The time resolution is then
determined by comparing the time extracted that way with the one measured by
scintillation detectors for a given track, as shown in
\Figref{fig:tracking.pixelgem.timing} for one particular detector.
The time resolutions obtained for the five PixelGEM detectors are distributed
around an average value of $11.2\,\ns$ with a standard deviation of $1.7\,\ns$
for the pixel regions and an average value of $9.4\,\ns$ with a standard
deviation of $1.1\,\ns$ for the strip regions.

\begin{figure}
  \centering
  \includegraphics[width=.8\textwidth]{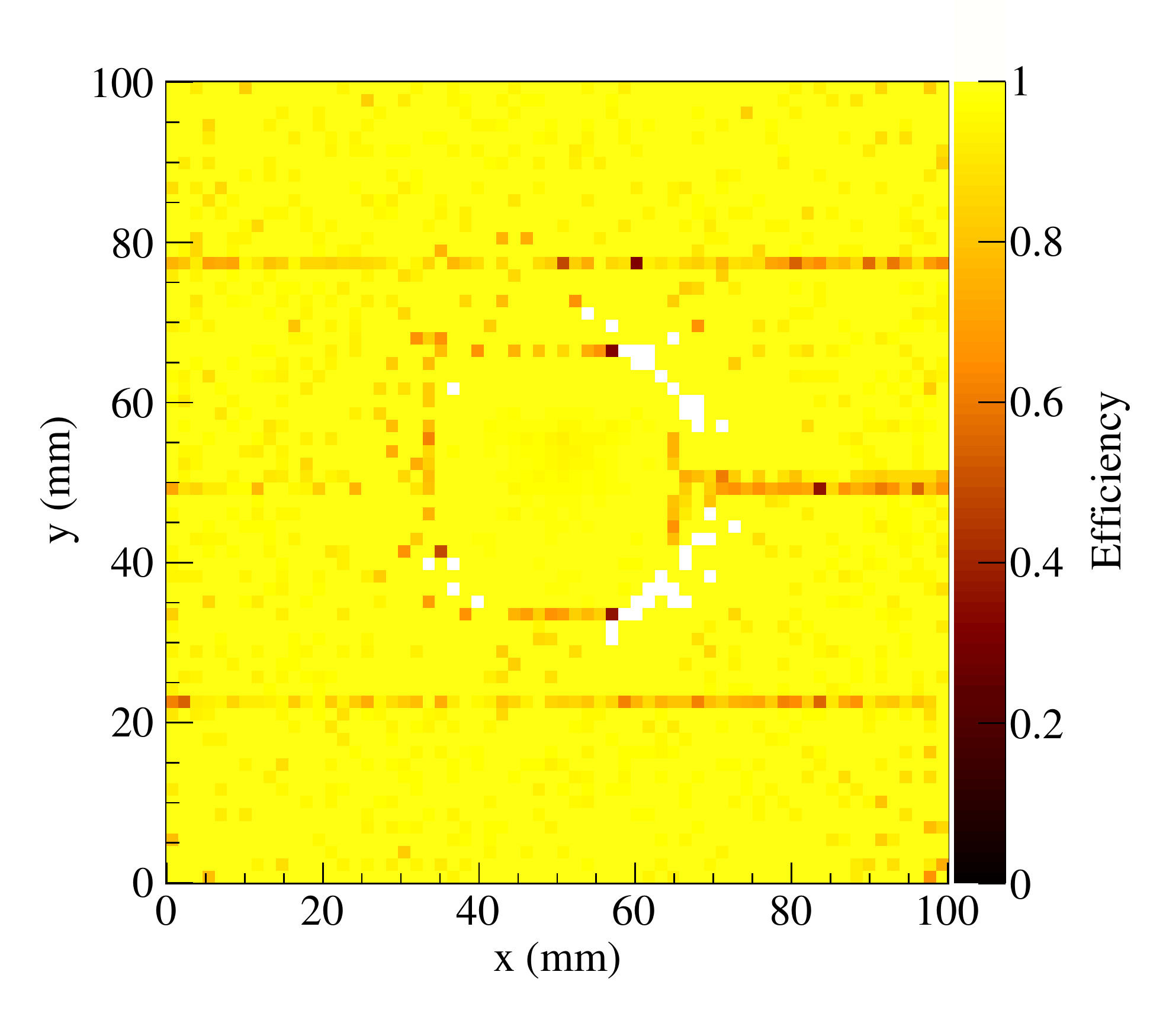}
  \caption{Efficiency of one of the PixelGEM detectors, measured in a
    high-intensity hadron beam. The horizontal lines with reduced efficiency
    correspond to boundaries between GEM sectors. In the white region not enough
    tracks are reconstructed when this particular detector is excluded from the
    tracking.}
  \label{fig:tracking.pixelgem.efficiency}
\end{figure}

In order to determine the efficiency $\epsilon$ of a detector in a
high-background environment, one has to take into account the presence of
uncorrelated background hits that may fall within the road width around a track
with a probability $b$ and thus artificially increase the apparent efficiency
$\epsilon_\mathrm{app}= \epsilon + b (1-\epsilon)$.  Here, the background
probability $b$ at a given position on the detector is determined from hits that
fall outside the road width around a given track used for the efficiency
calculation.  Figure~\ref{fig:tracking.pixelgem.efficiency} shows the
background-corrected efficiency for a complete detector plane. Here, the pixel
region is merged into the strip region, hence the complete active area of
$10\times 10\,\Cm^2$ is shown. The lines of lower efficiency parallel to the $x$
and $y$ axes correspond to the boundaries between the HV sectors on the GEM
foils. Few, or no tracks are reconstructed in the ring-shaped region when this
particular detector is excluded from the tracking to obtain an unbiased
efficiency determination.  Background-corrected efficiencies for the PixelGEM
detectors were found to be above $97\%$ for all detectors during data taking in
2008.
The PixelGEM detectors are also used for data taking with muon beams of
intensities around $10^7\,\mu^+/\s$.

\subsection{Micromegas Detectors}
\label{sec:tracking.micromegas}

Twelve Micromegas (MicroMesh Gaseous Structure, or MM) detectors are used for
tracking particles emitted at small angles.  Assembled in 3 stations of four
detectors each, they are installed in the region between the target and the
first dipole magnet SM1.  Each MM detector covers an active area of $40\times
40\,\Cm^{2}$, except a central dead zone with a diameter of $5\,\Cm$, and
measures a single projection of a particle track crossing the detector.  To this
end, the anode plane is divided in three zones, a central zone with 512 strips
and pitch of $360\,\upmu\m$, and two outer zones, each with 256 strips and pitch
of $420\,\upmu\m$.  The detectors, of 1024 strips each, have a parallel plate
electrode structure, with a volume separated into two regions: a $5\,\mm$
conversion gap with a moderate electric field (less than $1\,\kV/\Cm$), where
the ionising particle produces primary electrons, and an amplification gap of
$100\,\upmu\m$ with a much stronger field (typically $40\,\kV/\Cm$), where the
primary electrons generate an avalanche.  A $5\,\upmu\m$ thin metallic
micro-mesh (grid), which captures most of the ions produced during the
avalanche, separates the two regions. Another grid, which is used as a drift
electrode, defines the conversion gap region.

From 2006 onwards, the original MM detectors \cite{Abbon:2007pq} were modified
in order to satisfy two additional requirements: operate in a strong magnetic
field, and withstand an increased flux of highly-ionising particles during data
taking with hadron beam.  The first requirement comes from the use of a
superconducting magnet with a $2.5\,\T$ solenoid field during data taking
periods with the muon beam and the polarised target. This magnet, designed to
match the large angular acceptance of the COMPASS setup, produces a strong
fringe field that reaches $1\,\T$ at the level of the nearest MM detector. In
order to minimise the effect of the solenoid field on the detector, the nickel
grids used in the first generation MMs were replaced with new $5\,\upmu\m$ thin,
non ferromagnetic copper grid foils. The new mesh used for the amplification gap
has $65\,\upmu\m$ diameter holes and a pitch of $90\,\upmu\m$. The corresponding
values of the drift electrode are $300\,\upmu\m$ and $600\,\upmu\m$.

The second requirement comes from the use of hadron beams, which produce a large
number of highly ionising secondary particles and generate nearly three orders
of magnitude more discharges per incident particle. Since the discharge rate is
proportional to the gain of the detector, the size of the conversion gap was
enlarged from $3.2\,\mm$ to $5\,\mm$ in order to increase the number of primary
electrons.  The detectors could then be operated at a lower gain while
preserving a good efficiency. The gas mixture used is Ne/C$_2$H$_6$/CF$_4$ with
corresponding volume fractions of 85\%/10\%/5\%.  In comparison to data taking
with a muon beam, the CF$_4$ component was decreased from 10\% to 5\%, thereby
further reducing the discharge rate at the expense of a slight decrease of the
electron drift velocity.  At the nominal hadron beam intensity of $5\times 10^6$
particles per second impinging on a $40\,\Cm$ long liquid hydrogen target, each
MM detector sees an integrated flux of up to $30\,\MHz$, reaching $100\,\kHz$
per strip near the central dead zone. The mean discharge rate in such conditions
varies between $0.026\,\Hz$ and $0.050\,\Hz$, depending on the specific plane
and HV settings. During a discharge the micromesh voltage decreases and thereby
reduces the efficiency of the detector. The decrease is recovered several
microseconds after the discharge.

For all MM detectors, a digital readout of the signal using the SFE16 chip is
used. When recording the leading and the trailing edges of a signal, both the
mean time and the amplitude of a hit can be calculated, the latter by using a
time-over-threshold technique. Adjacent hits are then combined to clusters.  The
average cluster size is $2.8$ and $3.2$ in the region with $360\,\upmu\m$ and
$420\,\upmu\m$ pitch, respectively.  The mean value of the time resolution is
$14\,\ns$, compared to $8.5\,\ns$ measured for the first generation of MM
detectors. This loss in performance is due to the combined effect of the
increased conversion gap size, the decrease of the drift-gap electric field, and
the use of a smaller fraction of CF$_4$ in the gas mixture.

\begin{figure}[tp]
  \centering
  \includegraphics[width=0.8\textwidth]{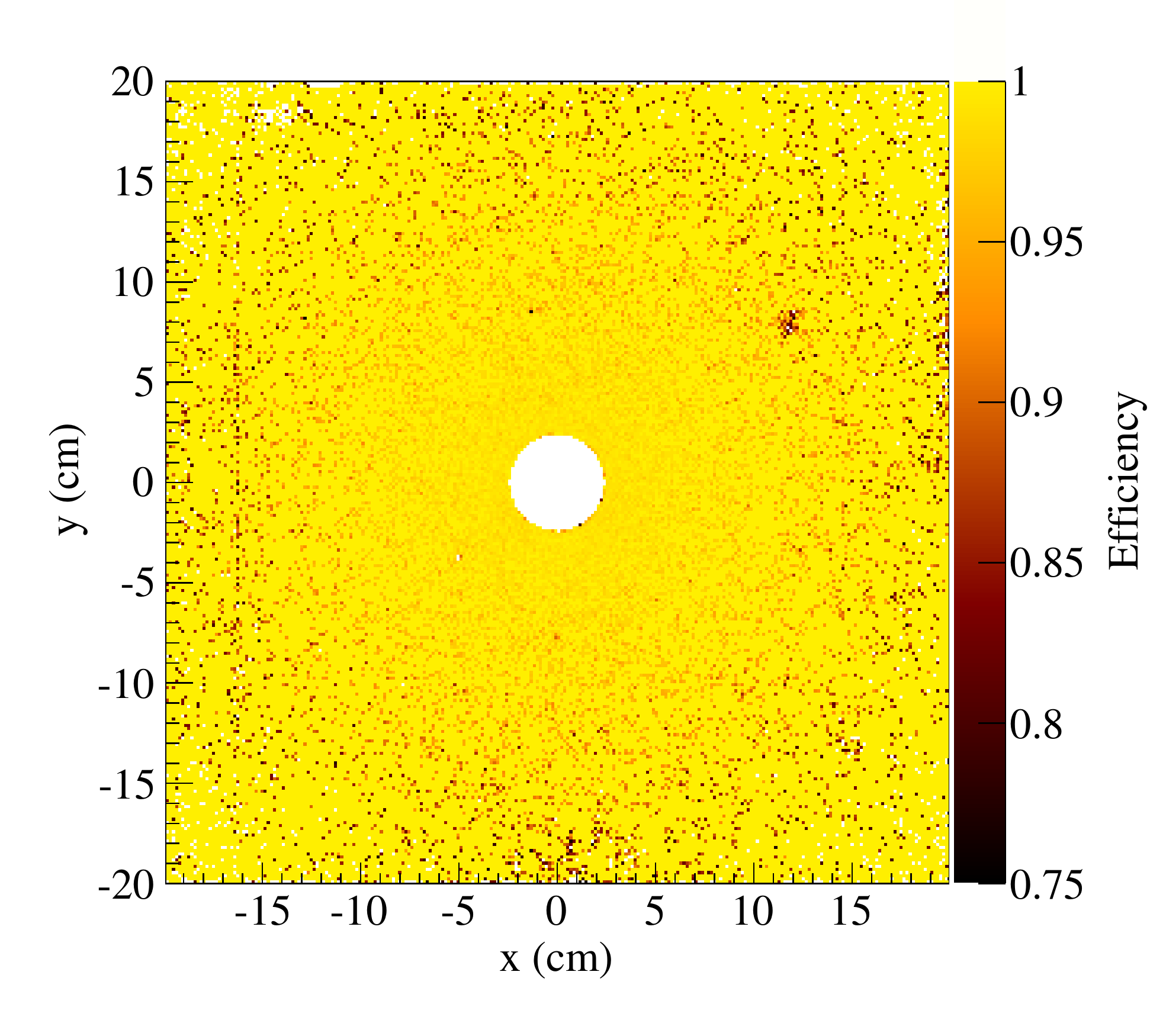}
  \caption{Two-dimensional efficiency of a Micromegas detector. The empty region
    in the middle is the $5\,\Cm$ central dead zone.}
  \label{fig:MM-efficiency}
\end{figure}

The efficiencies of the MM detectors are determined using the same method as
described in \secref{sec:tracking.pixelgem} for the PixelGEM detectors.  A
two-dimensional representation of the efficiency of one of the detectors is
shown in Fig.~\ref{fig:MM-efficiency}. The mean efficiency values obtained for
the other 11 MM detectors are all in the range 97.5\% - 98.5\%.  Note that the
use of copper grids with a less favourable geometric transparency than that of
the nickel grids has negligible effect on the final efficiency values.

\begin{figure}[bp]
  \centering
  \includegraphics[width=0.6\textwidth]{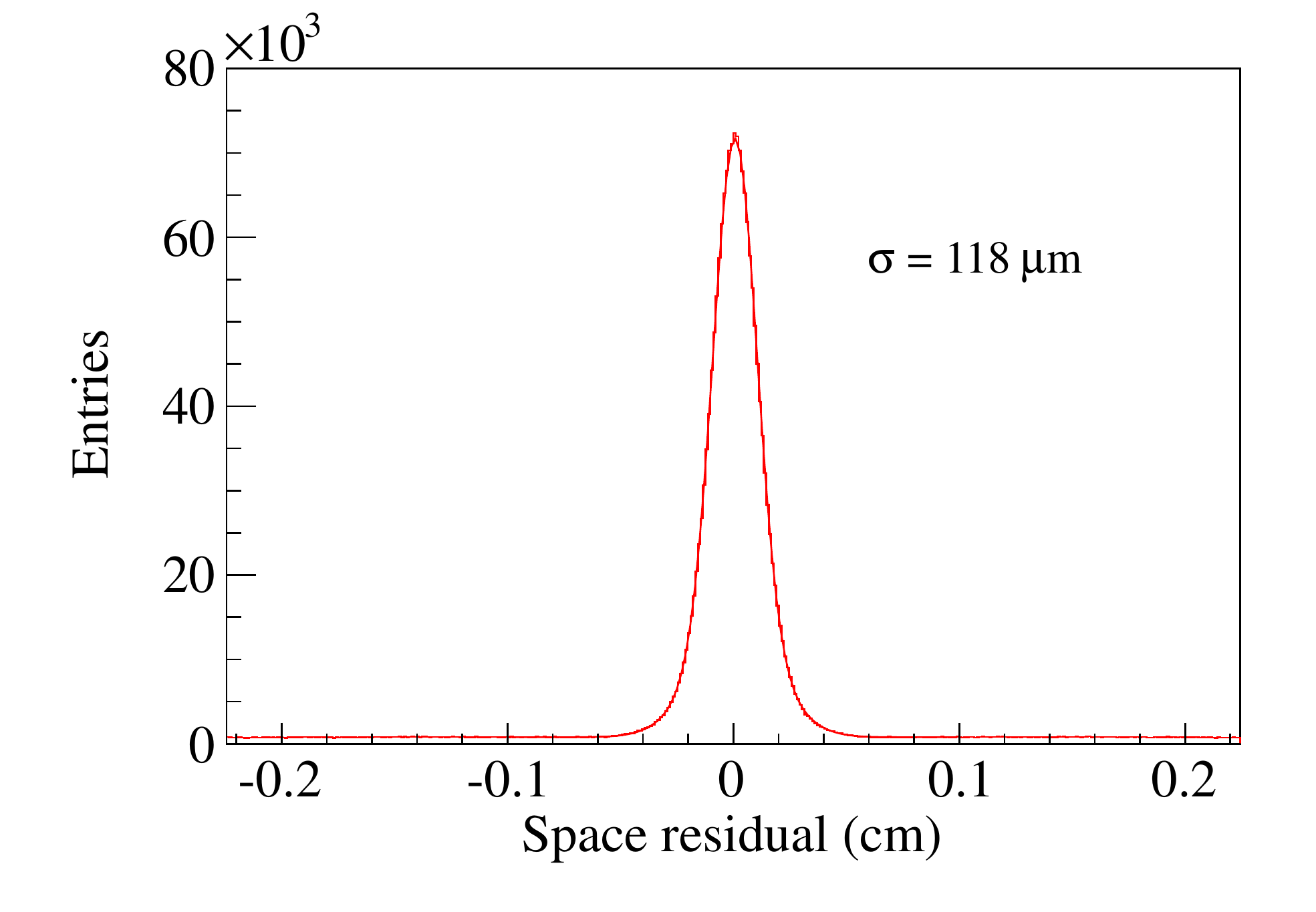}
  \caption {Space residual distribution of a Micromegas detector. The quoted
    residual width is obtained from a fit of a sum of two Gaussians.}
  \label{fig:MM-spaceres}
\end{figure}

\Figref{fig:MM-spaceres} shows the space residual distribution of a MM detector
for nominal beam intensity and running conditions.  After subtracting the
contribution of the track uncertainty, the intrinsic spatial resolution obtained
is $105\,\upmu\m$. The resolution value is a weighted average of the resolution
in the central zone ($360\,\upmu\m$ pitch), and that of the two outer zones
($420\,\upmu\m$ pitch). The four detectors of the third MM station operate in
the fringe field of the SM1 dipole magnet, which exerts a Lorentz force on the
drifting electrons. The resolution of these detectors varies from 110 to
$145\,\upmu\m$, depending on the orientation of the strips relative to the SM1
field lines.

\subsection{Drift Chambers}
\label{sec:tracking.driftchamber}
 
A new large-size multiwire drift chamber (DC4) was installed already in 2006 in
the LAS part of the setup. The design of DC4 closely follows that of the
medium-size DC trackers \cite{Abbon:2007pq} already operational in the COMPASS
set-up, while the overall dimensions were enlarged to match the angular
acceptance downstream of the SM1 magnet. The distance between active wires was
increased by $1\,\mm$ and the angle of the inclined wires was decreased. The DC4
chamber also features a water-cooling system, which ensures a good temperature
stability of its frame.
 
The external dimensions of the DC4 detector are $294\times 254\times
17\,\Cm^{3}$ with an active gas area of $248\times 208\,\Cm^{2}$. The detector
has eight layers of wires and four wire orientations: two vertical layers ($X$-
and $X'$-plane), two horizontal layers ($Y$- and $Y'$-plane), two layers with
wires inclined with respect to the vertical axis by $+10^\circ$($U$- and
$U'$-plane) and two others by $-10^\circ$($V$- and $V'$-plane).  The
configuration of the detector along the beam is $UU'$, $VV'$, $XX'$, $YY'$.
Every second layer is staggered by $4\,\mm$ (half of the cell dimension) in
order to minimise track ambiguities.  Each layer consists of 256 active wires
made of gold plated tungsten and 257 alternating potential wires made of
beryllium, with diameters of $20\,\upmu\m$ and $100\,\upmu\m$, respectively. To
avoid sagging, two nylon wires per plane are fixed perpendicularly to the active
and potential wires.  The distance between every two active wires is
$8\,\mm$. Each wire layer is enclosed between two $25\,\upmu\m$ thick Mylar
cathode foils, at a distance of $\pm 4\,\mm$ from the wire. This configuration
defines drift cells of $8\times 8\,\mm^2$, which are small enough to cope with
counting rates as high as 250~kHz per wire. During operation of the detector,
the active wires are kept at $0\,\V$, whereas both Mylar foils and potential
wires are set at values close to -1700~V.
 
Central zones with a diameter of $28.6\,\Cm$ and independent HV power supplies
are segmented in all cathode foils. During normal operation their HV values are
kept small enough (usually 900~V), which makes them blind to the high particle
flux in the vicinity of the beam.  The central zones are activated only for
detector alignment purposes, when the beam intensity is low.
 
The gas used is a mixture of Ar, C$_2$H$_6$ and CF$_4$ with volume fractions of
45\%, 45\% and 10\% respectively.  This gas ensures a fast charge collection
(drift velocity is $77\,\upmu\m/\ns$) while preserving a good spatial
resolution. Full efficiency is reached for gain values close to $10^4$,
corresponding to HV settings of approximately $1750\,\V$.
 
The read-out electronics of the detector is identical to the electronics used
for the already installed DC trackers. A single front-end card with 64 channels
consists of eight pre-amplifier/amplifier/discriminator chips\cite{Newcomer:93},
called ASD8.  Each ASD8 card is connected to a 64-channel F1-TDC board.  The
thresholds on the ASD8 card are remotely controlled.  The nominal threshold is
set at values between $0.8\,\V$ and $1.1\,\V$, or between 25000 and 34000
electrons.

\begin{figure}[tbp]
  \begin{center}
    \includegraphics[width=0.8\textwidth]{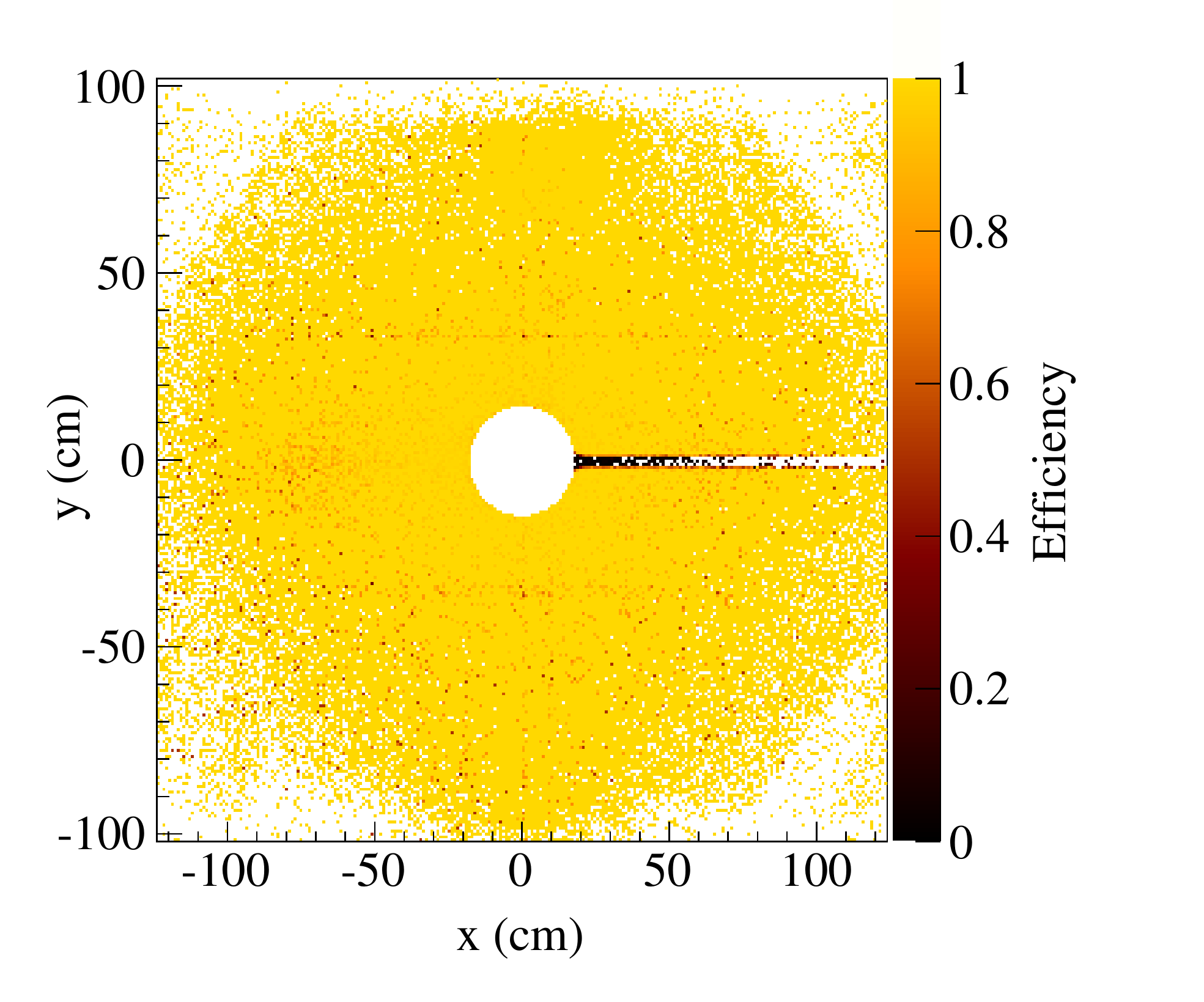}
  \end{center}
  \caption{Two-dimensional representation of the efficiency for one of DC4
    layers.  The half horizontal lines with reduced efficiency indicate the
    position of the power supply lines of the beam killer.}
  \label{fig:tracking.dc4.2Deff}
\end{figure}
 
Due to its large size, the components of the DC4 tracker are sensitive to the
temperature variations in the hall. The thermal expansion or contraction of the
DC4 aluminium frame may alter the nominal wire tension and degrade the detector
performances.  In order to minimise such effects, a specific cooling system was
installed. The system consists of copper pipes running on both sides of the
detector frame. The copper pipes are part of a closed secondary circuit filled
with demineralised water and maintained at constant temperature.  Several
probes, which are installed at various locations, continuously measure the
actual detector temperature.  The cooling system limits the temperature
variation of the frame to within 2~K.
 
\begin{figure}[tbp]
  \begin{center}
    \includegraphics[width=0.6\textwidth]{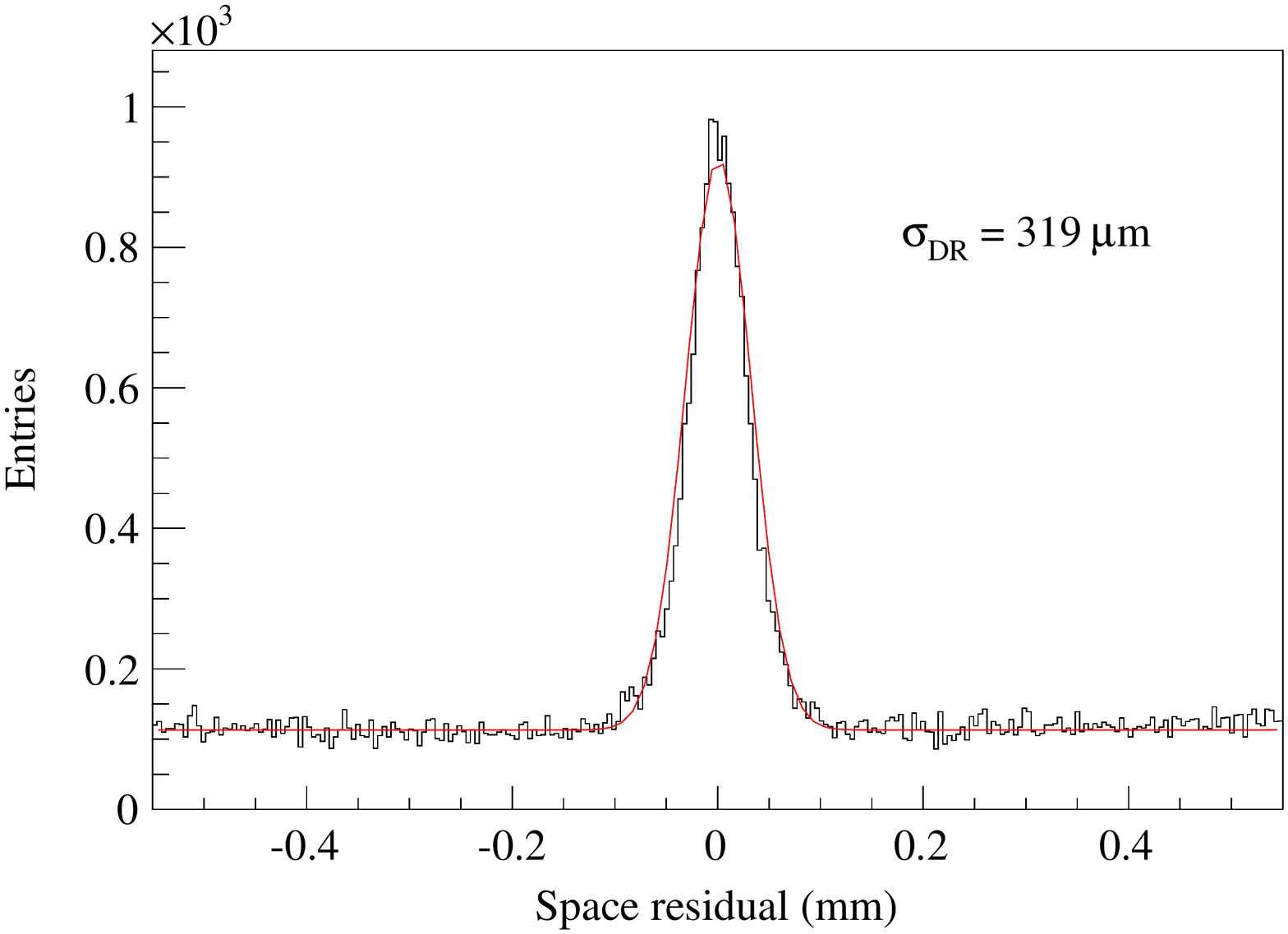}
  \end{center}
  \caption{Double residual (see text) distribution of the DC4 chamber for one of
    its doublets. The quoted width is from the fit of a simple Gaussian.}
  \label{fig:tracking.dc4.resol}
\end{figure}
 
The efficiencies of the eight DC4 planes were measured at nominal running
conditions (hadron beam intensity of $5\times 10^6/\s$). They were found to be
in the range of 95\% to 97\%.  \Figref{fig:tracking.dc4.2Deff} shows the
efficiency of the first vertical plane (Y1) of the DC4 chamber.
 
The spatial resolution of the DC4 drift chamber planes is determined by taking
advantage of the staggered layers with the same orientation (doublet). The
difference between the positions of the hits in the two planes, $x_{1}$ and
$x_{2}$ of a doublet (double residual, or $D\!R$), is independent of the track
uncertainty associated with the other planes or detectors. It includes a
correction $\Delta x_{12}(\theta_{track})$, which accounts for a non
perpendicular incidence of the track, i.e. a position shift when going from one
plane to another. The distance between two planes being $8\,\mm$, this
correction is small. The double residual is therefore $D\!R = x_{1} - x_{2} -
\Delta x_{12}(\theta_{track})$.  Figure~\ref{fig:tracking.dc4.resol} shows the
double residual distribution for the $X$-doublet of DC4, measured under nominal
beam conditions.  Except for a shift of half a drift-cell length, the two layers
have identical characteristics; therefore the resulting $D\!R$ resolution is
$\sigma_{D\!R}^2 = \sigma_{x_1}^2 + \sigma_{x_2}^2 = 2 \sigma_x^2$. The position
resolution for a single DC4 plane is consequently $\sigma_x = 226\upmu\m$. This
result is obtained in the central region of the detector, which corresponds to
about one tenth of the total detector area.

\subsection{Rich Wall}
\label{sec:tracking.richwall}

The Rich Wall detector is a large-area tracker that is positioned between RICH-1
and ECAL1.  The detector was built to improve the tracking accuracy at large
angles ($150 < \theta < 300\,\mrad$) downstream of RICH-1.  The additional track
points measured by the detector provide a better determination of large-angle
particle trajectories through RICH-1 and, as a consequence, improve the accuracy
of Cherenkov ring reconstruction.

\begin{figure}[tbp]
  \centering
  \includegraphics[width=0.7\textwidth]{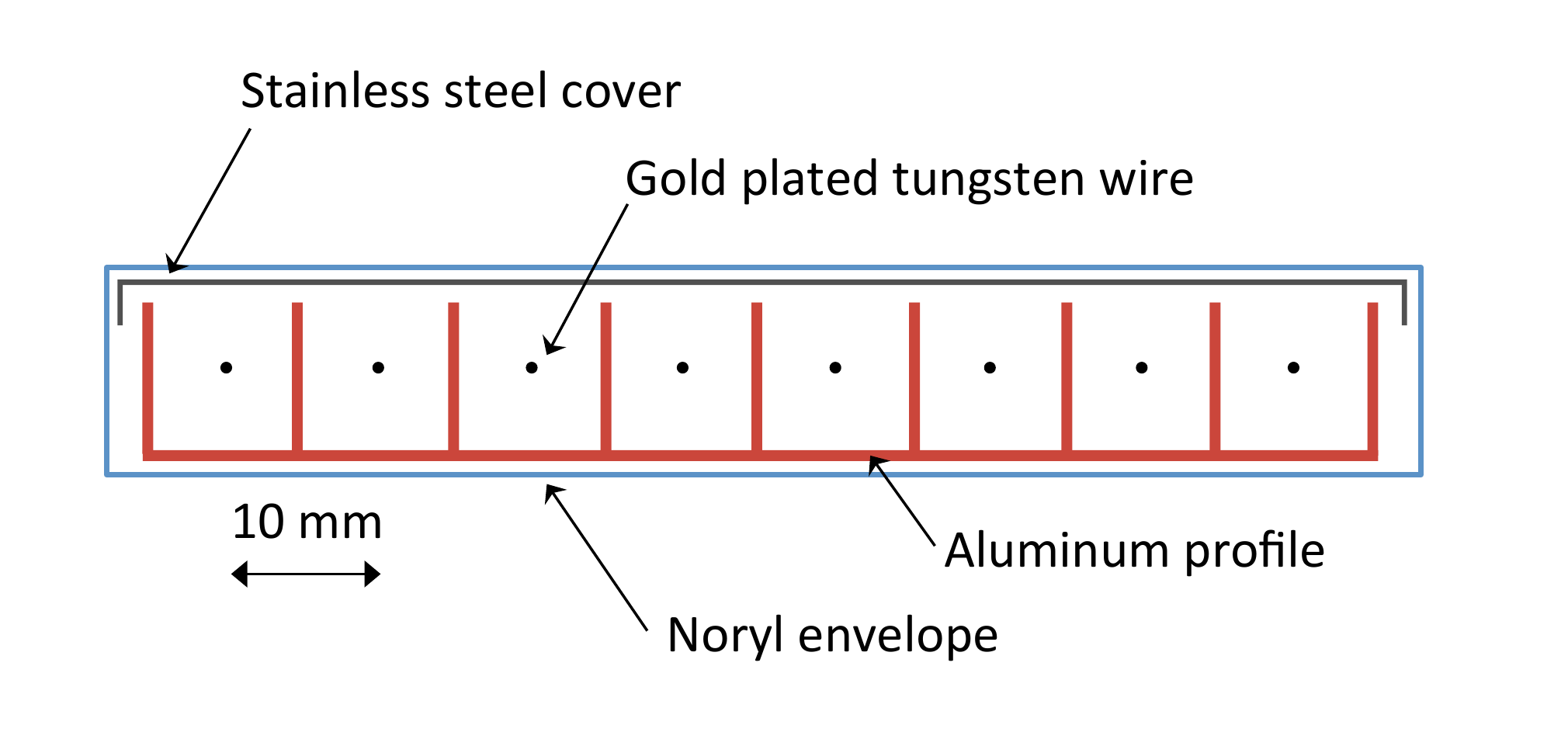}
  \caption{Sketch of a Mini Drift Tube module.}
  \label{rw-MDT}
\end{figure}

The detector has dimensions of $5.27 \times 3.91\,\m^2$ with a central hole of
$1.02 \times 0.51\,\m^2$. It consists of eight planes of Mini Drift Tubes (MDT)
made up of MDT modules. An MDT module consists of an eight-cell aluminium comb
extrusion with a wall thickness of $0.44\,\mm$, which is covered on the top by a
$0.15\,\mm$ thick stainless steel foil. Gold-plated tungsten wires of $50\,\mum$
diameter are strung in the centre of the cells. The wire pitch is $10\,\mm$.  A
Noryl\textsuperscript{\textregistered} plastic envelope with a thickness of
around $1\,\mm$ encapsulates the module.  The wires are thermally glued to
polyethylene plastic spacers (not shown in Fig.~\ref{rw-MDT}) at equal distances
of $1\,\m$ along the length of the MDT to provide electrostatic stability.  A
sketch of one MDT module is shown in Fig.~\ref{rw-MDT}.

\begin{figure}[tbp]
  \centering
  \includegraphics[width=0.8\textwidth]{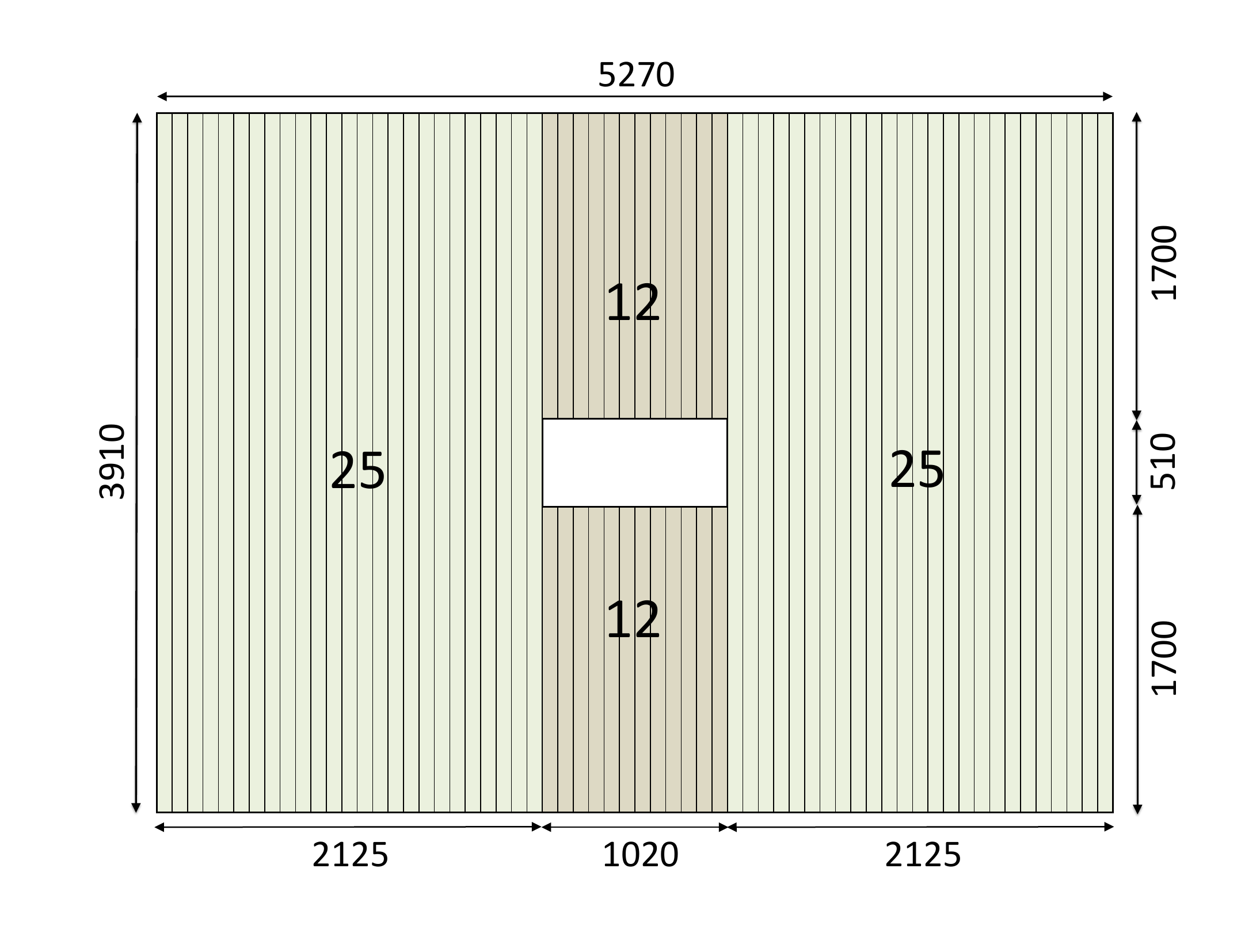}
  \caption{Front view of an $X$-plane of the Rich Wall detector. The large-size
    numbers correspond to the number of MDT modules in each sector, the small
    numbers indicate the dimensions in units of $\mm$.}
  \label{rw-front-view}
\end{figure}

Figure~\ref{rw-front-view} shows a front view of an $X$-plane. It consists of
$2\times 25$ long MDT modules (length $3910\,\mm$), and $2\times 12$ short
modules (length $1700\,\mm$) above and below the central hole. Similarly, a
$Y$-plane comprises $2\times 20$ long MDT modules (length $5270\,\mm$), and
$2\times 6$ short modules (length $2125\,\mm$).  The eight MDT planes are
arranged in four groups of two planes each, i.e.\ two $X$- or two
$Y$-planes. The two planes within one group are staggered by $2.5\,\mm$. Each
group is mounted onto an aluminium frame.

\begin{figure}[tbp]
  \centering
 
  \includegraphics[width=0.7\textwidth,height=.15\textheight]{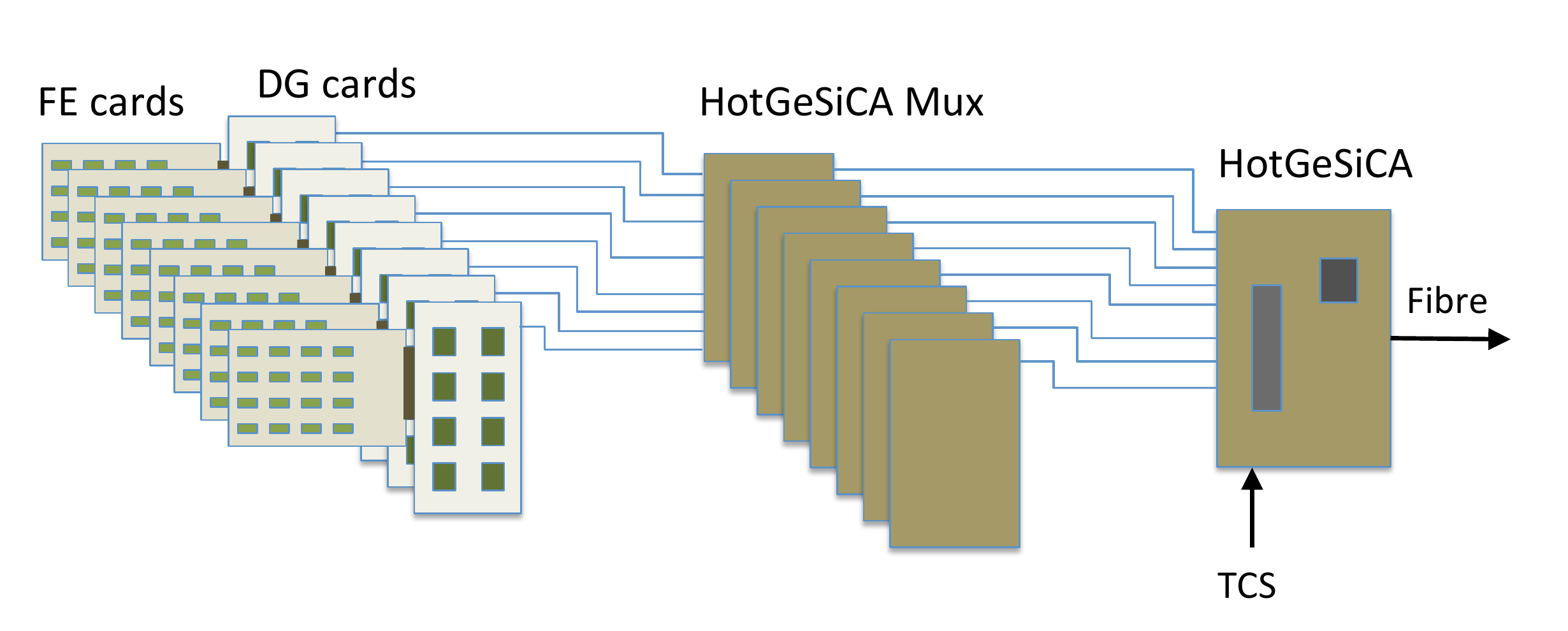}
  \caption{Schematic view of the Rich Wall readout chain.}
  \label{rw_readout}
\end{figure}

The readout electronics consists of front-end (FE) cards fixed on the detector
frame and digital (DG) cards plugged into the FE cards.  The FE cards are
connected to the MDT signal wires via short shielded cables. Each FE card houses
16 MAD4 chips~\cite{Gonella:01}, a threshold digital-to-analog converter (DAC),
a test-pulse generation circuit, power-supply filters and regulators. The FE
card is connected to the DG cards through a high-speed card edge connector. The
DG card houses eight F1 TDC~\cite{Fischer:00a} chips, a high-speed ($40\,\MHz$)
HOTLink chip, and initialisation circuits.  The card reads out 64 TDC channels
in parallel. Two 8-bit Analog Devices DAC8841 chips per DG card are used to
independently set the threshold of each MAD4 chip (common threshold for four
channels) and a third one is used to generate a variable-charge test pulse.  The
readout chain shown in Fig.~\ref{rw_readout} is completed with FPGA-based
HotGeSiCA cards (see \secref{sec:daq.general}) programmed in two different
ways. In the first stage, the data from eight DG cards are multiplexed onto a
single connection. Eight such multiplexers are then connected to the second
multiplexing stage consisting of one HotGeSiCA card equipped with additional
random access memory (RAM), which sends the data to the readout buffers.

\begin{figure}[tbp]
  \centering \includegraphics[width=0.8\textwidth]{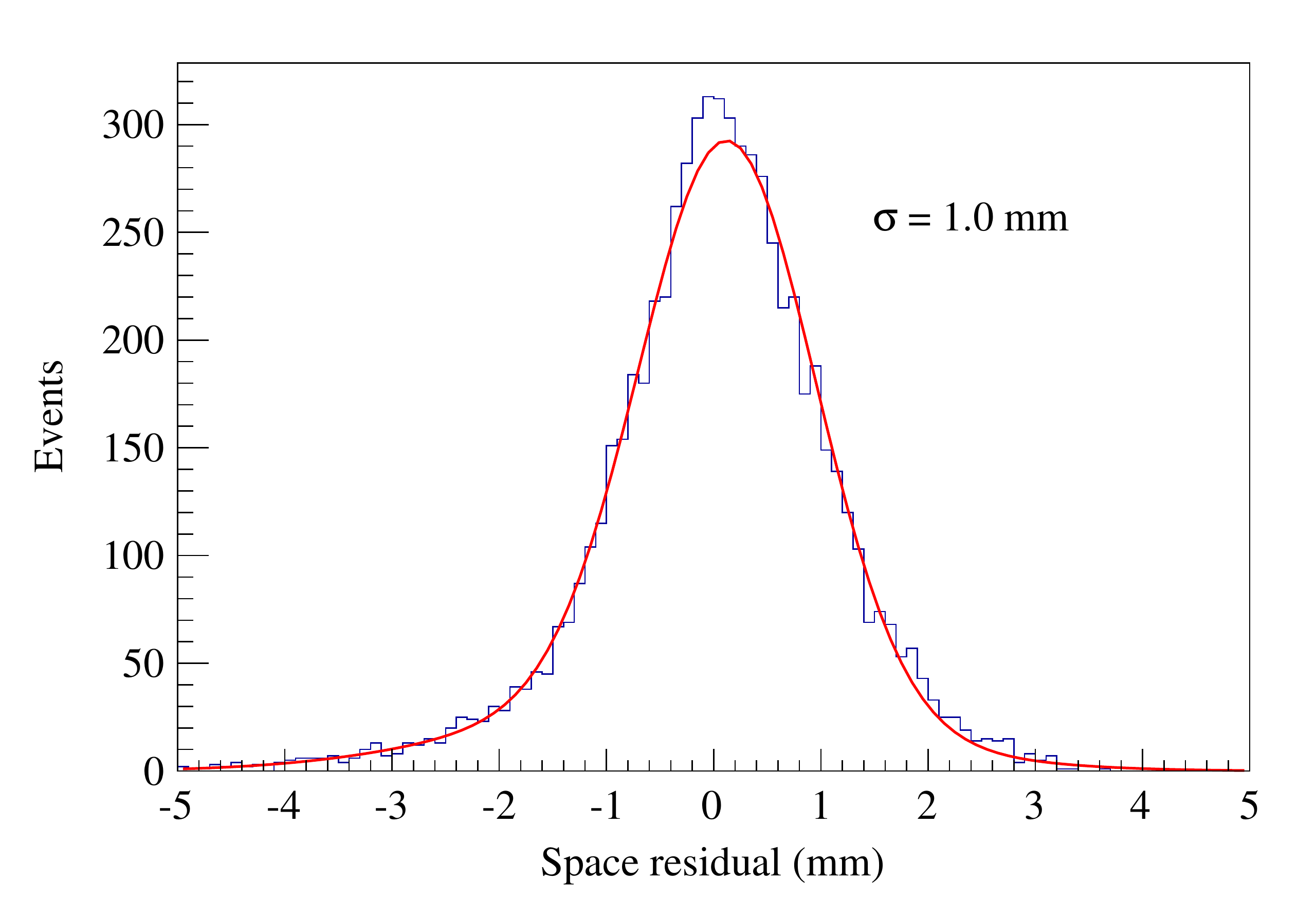}
  \caption{Rich Wall residual distribution, showing the difference between
    reconstructed cluster position and extrapolated track position along the
    axis perpendicular to the wire layer. The quoted sigma is extracted by
    fitting a sum of two Gaussians.  }
  \label{rw_residuals}
\end{figure}

The gas mixture used in the Rich Wall detector is Ar/CO$_2$ (70/30). For this
gas mixture an operating HV of $2050\,\V$ was chosen. Ageing tests performed
with this gas mixture have shown no degradation effects for incident charges of
up to $1\,\C$ per $\Cm$ of anode wire length. The beam-induced MDT charge,
integrated over the lifetime of the COMPASS experiment, is comparable to this
value.  The Rich Wall detector is operated in the drift mode where the
coordinate of a crossing track is calculated from the drift time in the MDT
cell, using the measured RT relation.  Operating the detector in the drift mode
allows us to obtain a single-plane coordinate resolution of the order of
$1.0\,\mm$, determined by fitting a sum of two Gaussians to the residual
distribution (Fig.~\ref{rw_residuals}). During operation the detector has shown
a stable single-plane efficiency of $90$--$91\%$.

\begin{figure}[tbp]
  \centering \includegraphics[width=0.8\textwidth]{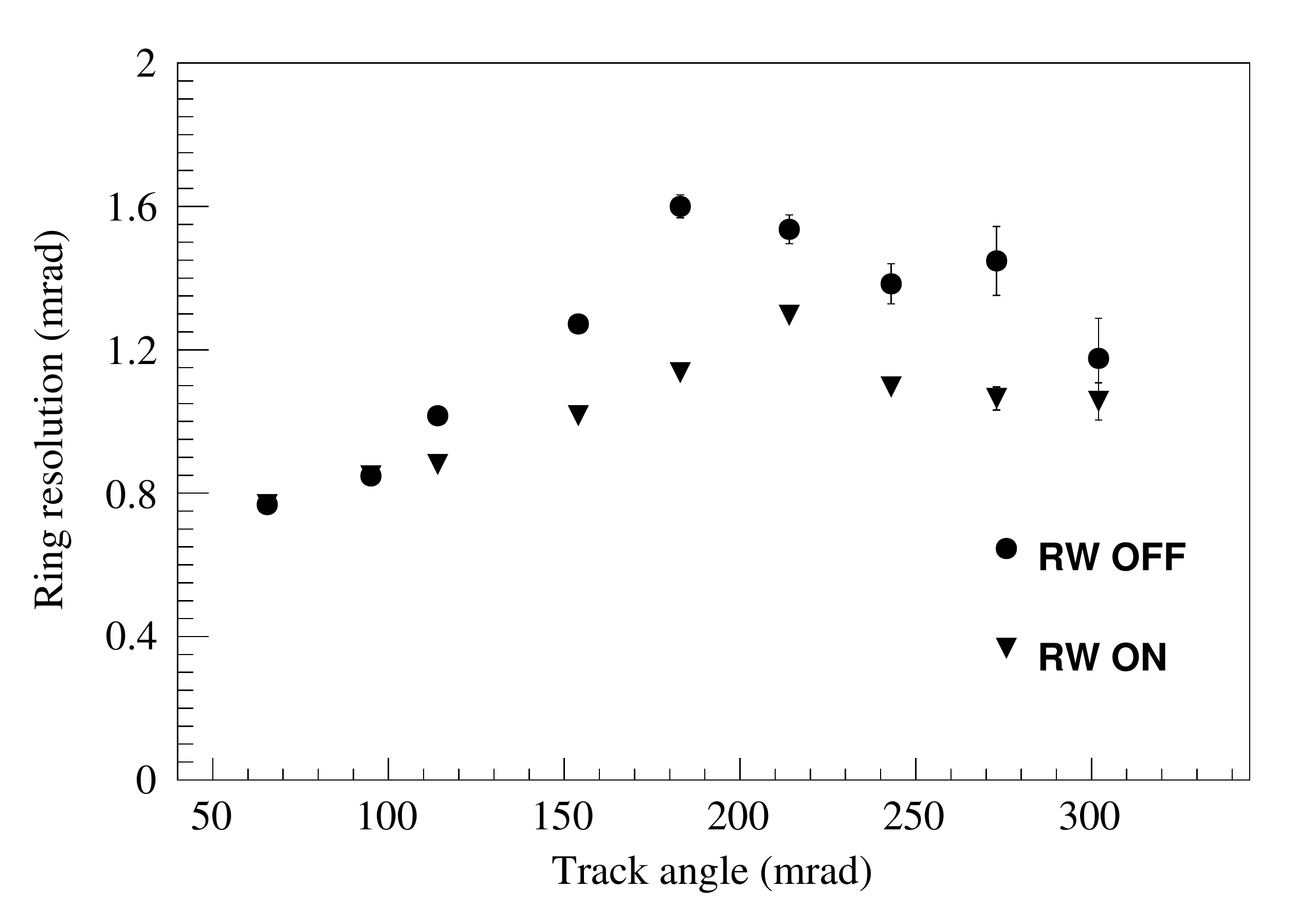}
  \caption{Resolution of the reconstructed Cherenkov ring for pions as a
    function of the track angle.  The two different trends in the curve below
    and above $\sim 175\,\mrad$ are due to the different RICH-1 photon detector
    types (see \secref{sec:pid.rich1}).}
  \label{rw_Rich_Richwall}
\end{figure}

The beneficial effect of the Rich Wall detector on the Cherenkov ring
reconstruction in RICH-1 is shown in Fig.~\ref{rw_Rich_Richwall}. In the polar
angular range $150$--$300\,\mrad$ the Cherenkov ring resolution is improved by
$20\%$ on average.

\section{Particle Identification Detectors}
\label{sec:pid}

The hadron physics programme at COMPASS requires the reconstruction of final
states with charged and/or neutral particles in a large angular range.  Several
types of particle identification detectors (PID) are used to achieve this goal
(see Fig.~\ref{fig:layout_setup}).  Charged pions and kaons, as well as protons,
with momenta of up to $50\,\GeV/c$ are identified in the RICH-1 detector, while
their energy is measured in the two hadron calorimeters, HCAL1 and
HCAL2. Photons emitted during the interaction and decay photons are detected in
two electromagnetic calorimeters, ECAL1 and ECAL2.  Scattered muons are
identified in the two muon identification systems, consisting of drift tubes
detectors (MW1 and MW2) and absorber walls made of iron (Muon Filter 1) or
concrete (Muon Filter 2).

Since the publication of Ref.~\cite{Abbon:2007pq}, the PID part of the setup was
significantly upgraded.  New photon detectors were installed in the central
region of \mbox{RICH-1}, and a new readout system was implemented in its
peripheral region.  The new ECAL1 calorimeter was added, which extends the
acceptance for photon detection to large angles.  The ECAL2 calorimeter was
upgraded with radiation-hard Shashlik modules in its central region and with
fully pipelined electronics. For both calorimeters, the calibration procedure
and the monitoring of the individual modules were significantly improved.  The
hadron calorimeters and the muon identification systems remained unchanged since
their description in Ref.~\cite{Abbon:2007pq} and are hence not discussed here.

\subsection{The RICH-1 detector}
\label{sec:pid.rich1}

The RICH-1 detector~\cite{Albrecht:05} covers the horizontal and vertical
angular acceptances downstream of the SM1 magnet (250 mrad$\times$180 mrad). Its
$3\,\m$ long vessel is filled with C$_4$F$_{10}$ gas as a
radiator~\cite{Albrecht:03c}.  The refractive index of the radiator material
corresponds to Cherenkov thresholds of about $2.5$, $9$, and $17$$\,\GeV/c$ for
pions, kaons, and protons, respectively.  A steel pipe with a radius of $5\,\Cm$
and thickness of $0.15\,\mm$ separates the vessel from the beam path.  Cherenkov
photons produced along the path of a hadron are reflected by a $21\,\m^2$
surface that consists of 116 spherical UV mirror elements that are grouped into
two spherical surfaces~\cite{Albrecht:03b}.  The mirrors are designed such that
the photons are focused onto two arrays of photon detectors (see
\figref{fig:pid.rich.evdisplay}), located outside of the spectrometer
acceptance.

\subsubsection{Upgrade}

Until 2004, Cherenkov photons were detected in Multiwire Proportional Chambers
(MWPC) equipped with solid-state CsI photocathodes that limit the MWPC operation
to gains below $5\times10^4$.  The first stage of the electronics readout
system~\cite{Santiard:94} was characterised by a long integration time; this was
a limiting factor in the COMPASS environment, where a high-rate uncorrelated
background is present due to the large muon beam halo.  High rates and large
correlated background are also typical for measurements with a hadron beam.  In
addition, the long base-line restoration time (about $3.5\,\mus$) generated a
non-negligible dead time.

In order to overcome these limitations, a major upgrade of the RICH-1 detector
was undertaken.  Details can be found in Refs~\cite{Abbon:06,Abbon:08,Abbon:10}.
Two different technologies were chosen in order to minimise the overall cost of
the project.  In the peripheral regions that cover 75\% of the photo-detection
surface, where the level of the uncorrelated background is small, the MWPC/CsI
photon detectors were kept. However, their front-end electronics was replaced by
a new system~\cite{Abbon:06} that is based on the 128 channel APV25
chip~\cite{French:01}. The new system provides two major improvements. First, it
reduces the effective time window from $3\,\mus$ to $400\,\ns$ and decreases the
dead-time losses of the readout system to values close to 5\%. Second, the APV25
chip performs a triple sampling of the MWPC signal, which results in a much
improved time resolution and in an increase in the signal-to-background
ratio~\cite{Abbon:06} from 0.35 with the old system to 2.13 with the new one.

The central region of RICH-1, which covers 25\% of the photo-detection surface,
is instrumented with a detection system based on Multi-Anode PhotoMultiplier
Tubes (MAPMTs)~\cite{Abbon:10}.  The MAPMTs are coupled to individual telescopes
of fused silica lenses which consist of a prismatic field lens followed by a
concentrator lens, thereby enlarging the effective active area of the photon
detectors by a factor of seven.  The effective pad size that results from the
MAPMT pixel-size and the lens telescope magnification is about
$12\times12\,\mm^2$. The new system detects about four times more Cherenkov
photons than the old one and reaches values as high as 60 photons per ring.  The
MAPMT detectors are intrinsically fast and have time resolutions better than
$1\,\ns$.  They are coupled to a readout system~\cite{Abbon:08} based on the
MAD4 high sensitivity amplifier/discriminators and the standard COMPASS F1 TDCs.
 
\subsubsection{Characterisation}
\label{sec:pid.rich1.char}

\begin{figure}[!htb]
\begin{center}
\includegraphics[width=0.4\textwidth]{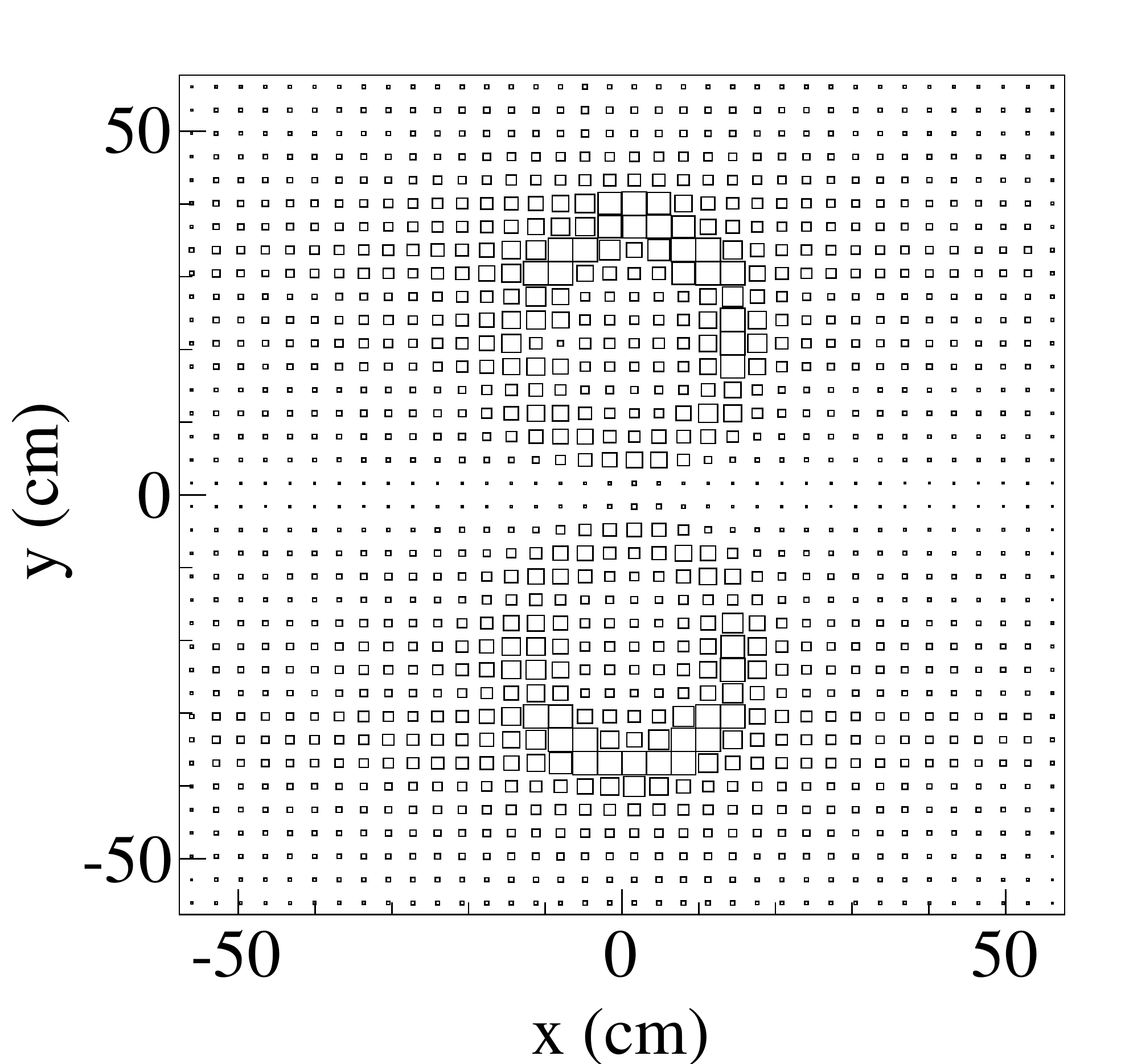}
\includegraphics[width=0.4\textwidth]{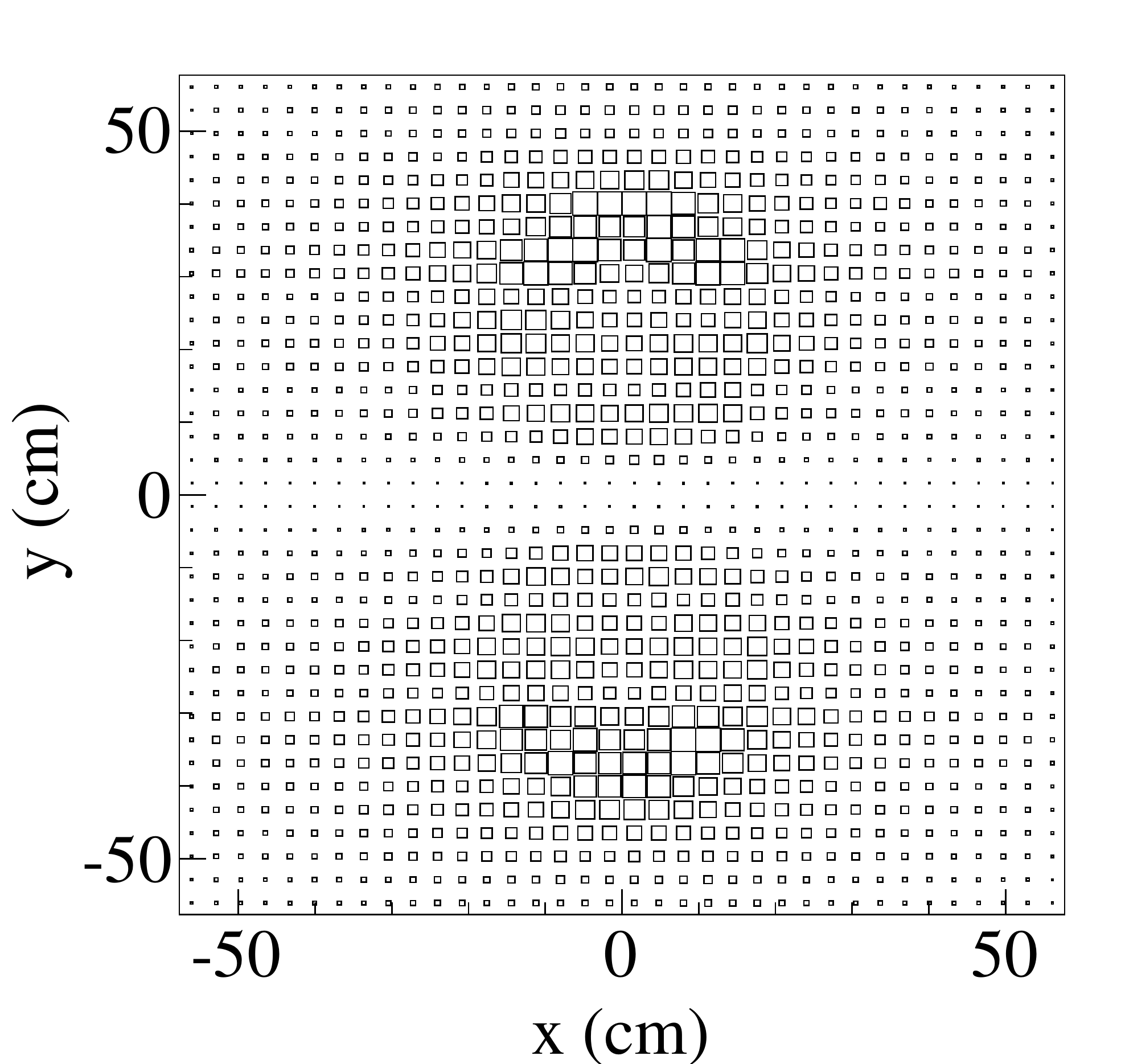}
\caption{Two-dimensional hit distributions in the central part of the RICH-1
  photon detectors for (left) data taken with a muon beam and (right) data taken
  with a positive hadron beam.
\label{fig:pid.rich.map}}
\end{center}
\end{figure}

\begin{figure}[!htb]
\begin{center}
\includegraphics[width=0.5\textwidth]{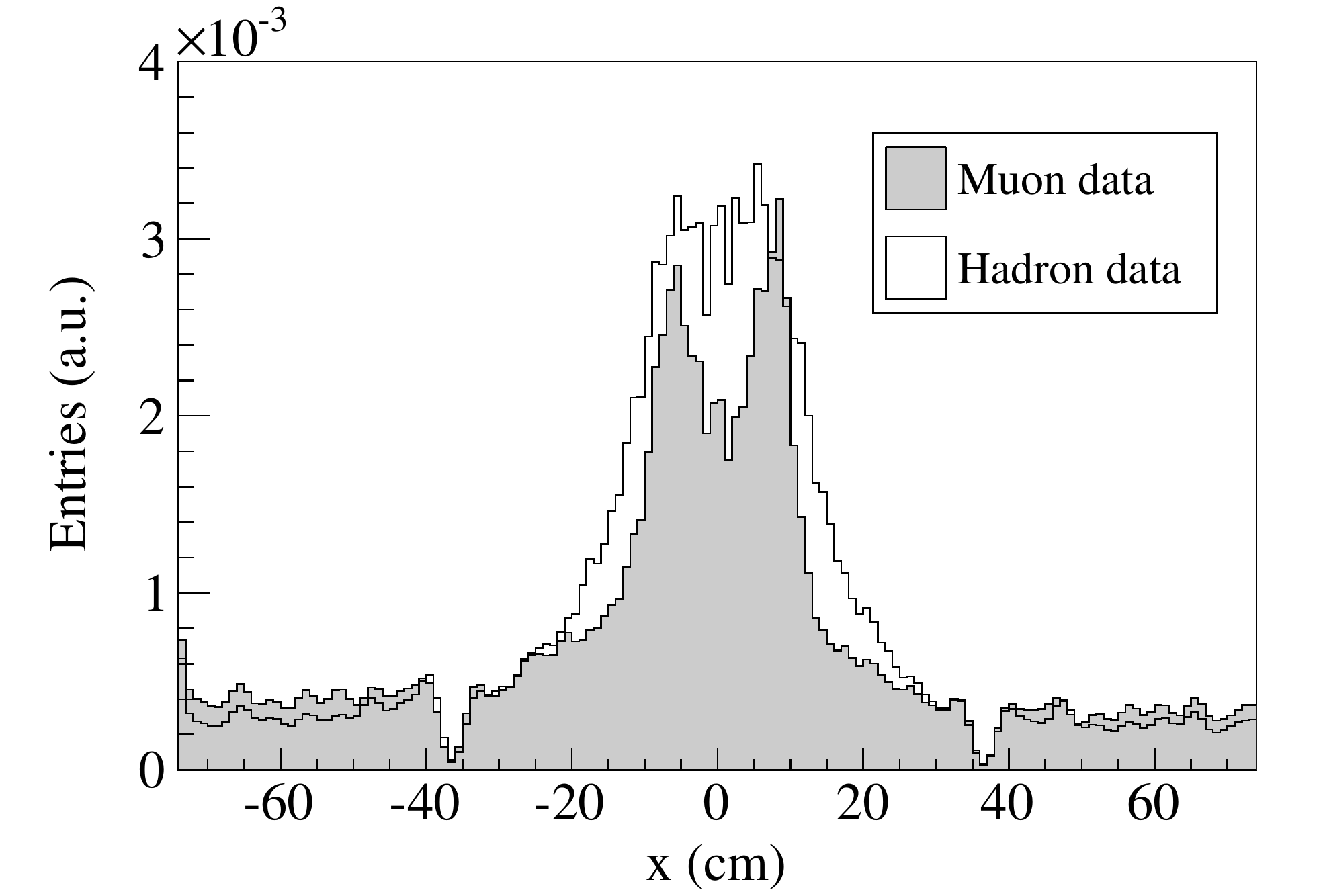}
\caption{Horizontal axis projection of the integrated hit distributions for the
  lower photon detectors.  Both central and peripheral parts of RICH-1 are
  included.  The shaded histogram refers to the muon environment, the open to
  the hadron one.  The small dips in the hit distributions correspond to the
  dead zones between the detector parts equipped with MAPMTs and with MWPCs.
\label{fig:pid.rich.map-proj}}
\end{center}
\end{figure}

\begin{figure}[!htb]
\begin{center}
\includegraphics[width=0.6\textwidth]{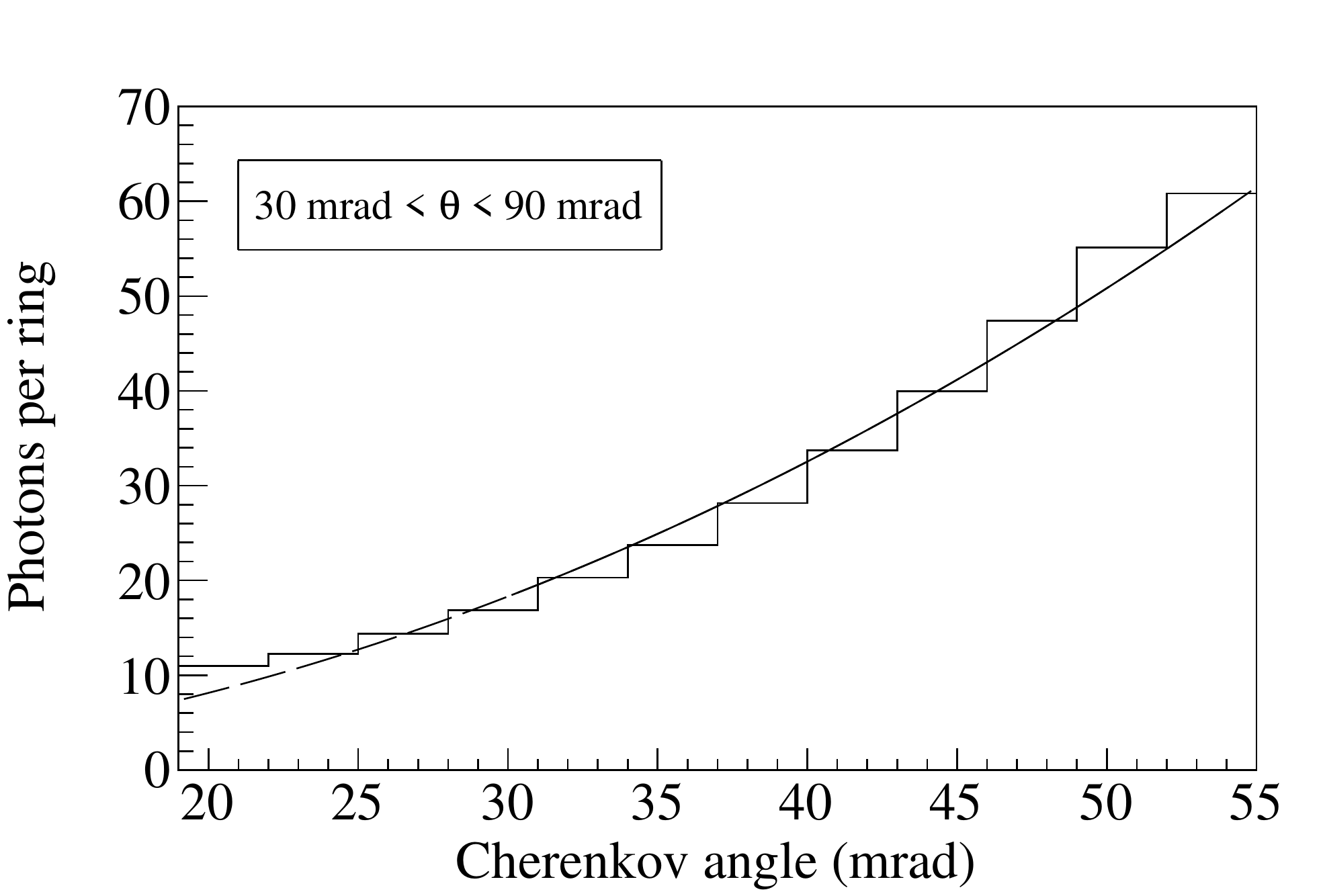}
\caption{Mean number of detected photons per reconstructed ring as a function of
  the corresponding Cherenkov angle $\theta_\mathrm{Ch}$ in the central region
  of the RICH-1 detector for track angles $\theta$ between $30\,\mrad$ and
  $90\,\mrad$.  The line is a fit with the functional form
  $N=N_0$sin$^2(\theta_\mathrm{Ch})$.
\label{fig:pid.rich.npho}}
\end{center}
\end{figure}

A dedicated software package, called RICHONE~\cite{Abbon:11}, was developed for
the RICH-1 data reduction. It performs pattern recognition and particle
identification, and characterises the detector response.
\begin{figure}
\begin{center}
\includegraphics[width=0.8\textwidth]{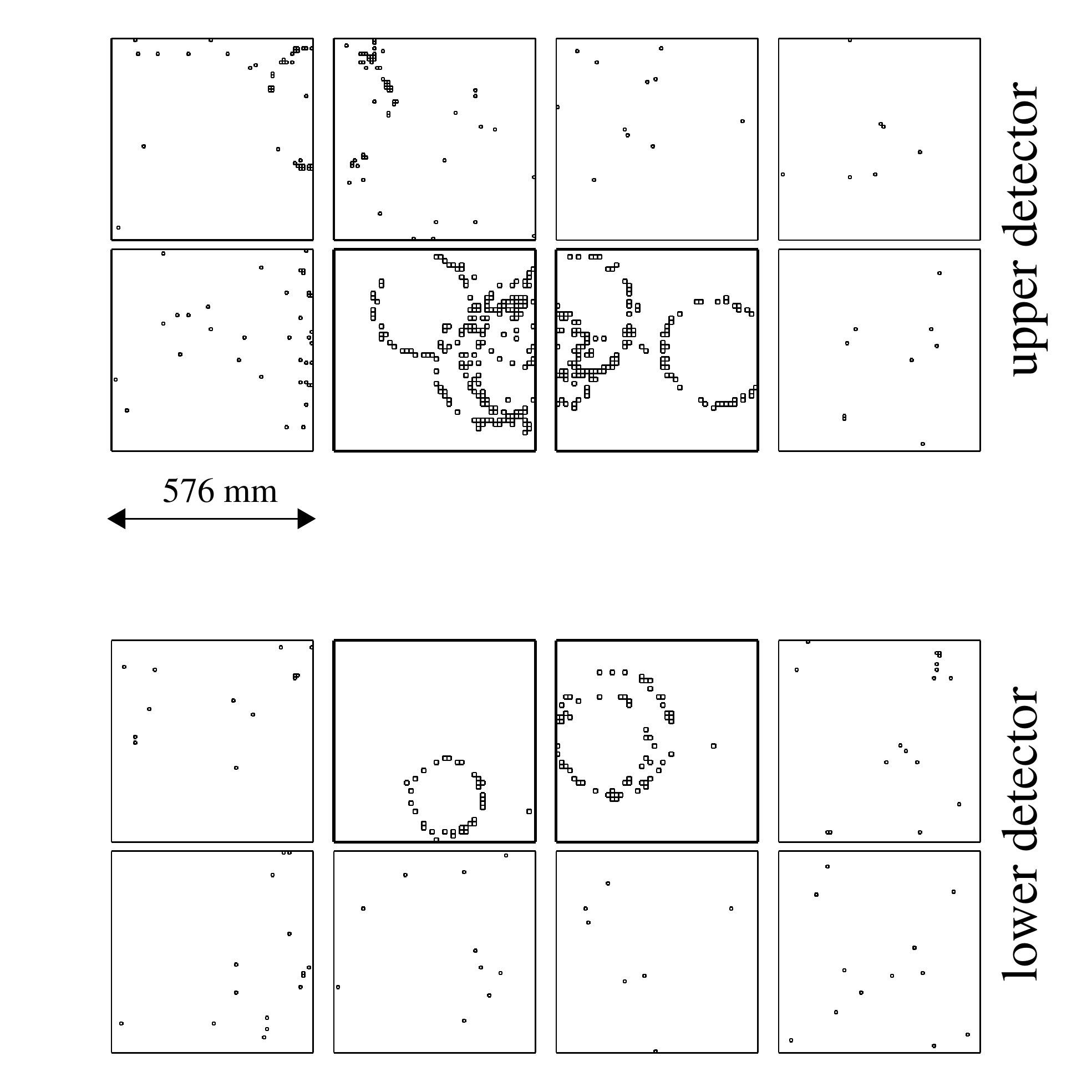}
\caption{A typical event display during hadron data taking. The 16 squares
  represent the detector areas; the four central ones are equipped with MAPMTs.
  The small squares represent the hits detected in the photon detectors.
\label{fig:pid.rich.evdisplay}}
\end{center}
\end{figure}
Figure~\ref{fig:pid.rich.evdisplay} shows an example of a RICH-1 event in the
hadron beam environment showing many rings in the central detectors.  The time
windows applied are the same as used in data reconstruction, namely $10\,\ns$
for the MAPMT part and $250\,\ns$ in the MWPC part.  Each visible ring belongs
to a detected particle.  A major difference between the use of RICH-1 with muon
and with hadron beam is the different particle population in the events, which
is due to the different event multiplicity and particle phase space.  The muon
beam is characterised by a wide halo, which extends over all photon detectors
and has a flux comparable to that of the focused beam.  The Cherenkov photons
that are emitted by the halo particles travelling parallel to the beam are
focused into the central zone of the RICH-1 photon detectors, which results in a
large background.  The hadron beam can be better focused and has hence much less
halo.  Nevertheless, due to the higher interaction rate in hadron scattering, a
large number of particles is emitted at small polar angles, i.e. in the very
forward region.  These particles also populate the central region of the RICH-1
photon detectors.  A map of the integrated hit distribution in the central part
of the RICH-1 photon detectors is shown in \figref{fig:pid.rich.map} for data
taken with muon and positive hadron beam. The distributions are normalised to
the number of entries and the same scale is used for the comparison. Both
distributions show large occupancies for photons emitted from particles
traversing RICH-1 under small polar angles.  The ring images in the muon
environment have more overlap since they are mostly produced by the parallel
halo particles, while in the hadron case the particles have a slightly broader
polar angle distribution.  Figure~\ref{fig:pid.rich.map-proj} shows the
projection of the hit distribution in the lower photon detectors onto the
horizontal axis, for both the muon and the hadron environment.  Even if the
origin of the background is different in the two environments, the overall
background distributions are similar.  The same was observed when changing to
another hadron beam or target, so that the general properties of the detector
response remain the same as the ones measured with a muon beam~\cite{Abbon:11}.
The uncertainties in the reconstructed angle of the individual Cherenkov photons
is $2\,\mrad$ in the central region and $2.5\,\mrad$ in the periphery, while the
uncertainties in the determination of the mean Cherenkov angle (ring angle) are
$0.3\,\mrad$ and $1.6\,\mrad$, respectively.  These resolutions allow pion-kaon
separation at 95\% confidence level for momenta up to $45\,\GeV/c$.  The average
number of photons per ring at saturation, i.e. for $\beta \rightarrow 1$, is 56
in the central and 14 in the peripheral region.  The dependence of the mean
number of detected photons per ring versus the corresponding Cherenkov angle is
shown in \figref{fig:pid.rich.npho} for the detectors equipped with MAPMTs.


Part of the data taking in 2009 was devoted to a test of the Primakoff
measurement.  For this test, RICH-1 was filled with N$_2$ gas in order to have a
smaller material budget in the acceptance region.  The response is largely
different in this case as the refractive index of N$_2$ is lower than that of
C$_4$F$_{10}$.  In particular, the number of emitted photons at saturation is
expected to be lower by a factor of 4.8 for the N$_2$ radiator. The number of
detected photons is then sufficient to allow for particle identification only in
the central part of RICH-1, which is equipped with MAPMTs.  In this region, the
average number of photons per ring at saturation is 11.7, which has to be
compared with 56 for the C$_4$F$_{10}$ radiator (\figref{fig:pid.rich.npho}).
The lower number of detected photons leads to an uncertainty in the
determination of the ring angle, which is larger by a factor 2.2 with respect to
the operation with the C$_4$F$_{10}$ radiator.  Nevertheless, the upper momentum
limit for pion-kaon separation is very similar for the two radiators as the
poorer resolution of N$_2$ is compensated by a larger difference between the
corresponding Cherenkov angles.  The thresholds of the Cherenkov effect are
$5.6$, $20$, $38\,\GeV/c$ for pions, kaons, and protons respectively.  Thus, in
comparison to the values for C$_4$F$_{10}$ quoted above, the momentum range for
pion-kaon separation is severely reduced.

\subsection{Electromagnetic calorimeter ECAL1}
\label{sec:pid.ecal.ecal1}

The ECAL1 calorimeter is part of the Large Angle Spectrometer. It consists of
1500 lead glass (LG) modules.  For reasons of availability and cost, three types
with different dimensions are used, see Table ~\ref{tab:ecal1-overview}. The
calorimeter ECAL1 has a width of $3.97\,\m$ and a height of $2.86\,\textrm{m}$,
which corresponds to the angular acceptance for photons coming from the centre
of the liquid hydrogen target of $37\,\mrad$ to $136\,\mrad$ in the horizontal
direction and of $21\,\mrad$ to $98\,\mrad$ in the vertical direction.  The
central hole has a size of $1.07\times 0.61\,\textrm{m}^2$. The ECAL1
calorimeter is installed on a motorised platform that allows horizontal and
vertical movements orthogonal to the beam direction, which is used mainly for
calibration purposes.

\begin{figure}[b]
\begin{center}
  \includegraphics[width=\textwidth]{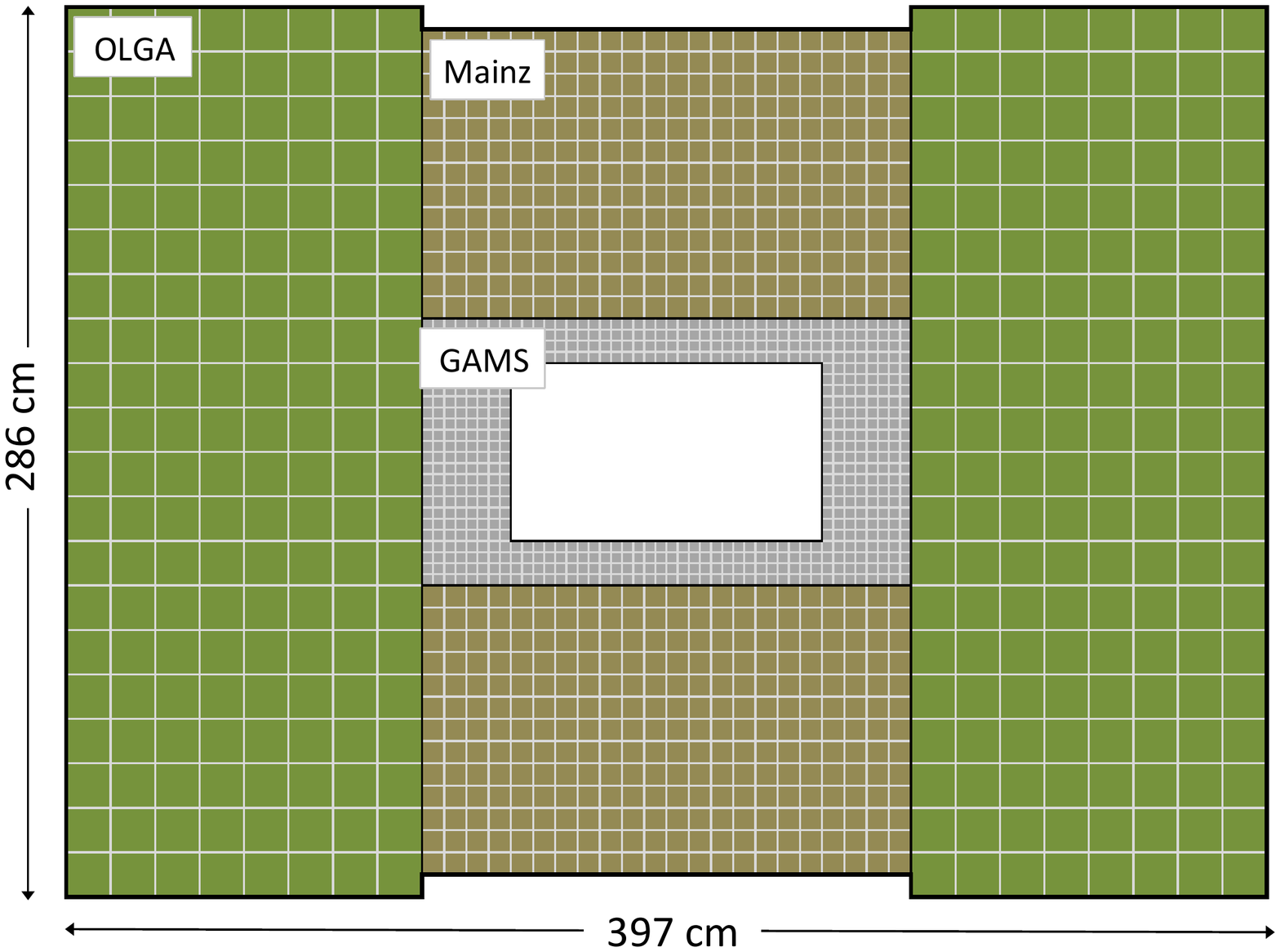}
  \caption{Configuration of ECAL1. The central area is equipped with GAMS
    modules. The MAINZ modules are installed above and below the GAMS area.  The
    OLGA modules cover the outer left and right regions.}
  \label{fig:pid.ecal.ecal1-structure}
\end{center}
\end{figure}

\subsubsection{Design and electronics}

A front view of the ECAL1 calorimeter is shown in
Fig.~\ref{fig:pid.ecal.ecal1-structure}.  The central part of ECAL1 consists of
608 LG modules of transverse dimensions $3.83\times 3.83$~cm$^2$, which are
denoted GAMS modules~\cite{Binon:86}.  They are arranged in a $44\times 24$
matrix with its central $28\times 16$ array left empty. Above and below this
central part, two $22\times 13$ matrices of ``MAINZ''
modules~\cite{Adamovich:1996ih} are installed, which contain in total 572 LG
modules.  One MAINZ module has the size of nearly four GAMS modules.  In order
to compensate for the small difference in size, $1.6\,\textrm{mm}$ vertical gaps
have been left between all columns of MAINZ modules, except the two central
columns. On both sides of the central columns, the nearest two gaps are filled
with iron plates.  The two outermost parts of ECAL1 consist of two matrices of
$8\times 20$ large-size ``OLGA'' modules~\cite{Astbury:1985pa}. Each OLGA module
has the
size of nearly four MAINZ modules. 
\begin{table}[tbp]
\caption{Parameters of the ECAL1 lead glass modules.}
\centering
  \begin{tabular}{lcrrr} 
    \hline
    Parameter          & Units     & GAMS               & MAINZ            &   OLGA               \\
    \hline                                                                                          
    LG type            &           & TF1                & SF57             & SF5                  \\
    Density            & g/cm$^3$  & 3.86               & $5.51$           & $4.08$               \\
    Rad. length ($X_0$)& cm        & 2.74               & $1.55$           & $2.55$               \\
    Total thickness    & $X_0$     & $16.4$             & $23.3$           & $18.5$               \\
    Moliere radius     &  cm       & $4.7$              & $2.61$           & $4.3$                \\
    Refractive index   &           & $1.65$             & $1.89$           & $1.67$               \\
    Length             & cm        & $45$               & $36$             & $47$                 \\
    Surface            & cm$^2$    & $3.83 \times 3.83$ & $7.5 \times 7.5$ & $14.1 \times 14.1$   \\
    PMT type           &           & FEU-84-3           & EMI 9236KB       & XP2050               \\
    \hline         
\end{tabular}
\label{tab:ecal1-overview}  
\end{table}
Table~\ref{tab:ecal1-overview} summarises all relevant parameters of the LG
modules used.  It also contains the type of PMT that detect their Cherenkov
light.  The analogue signals coming from the PMTs pass through shaper
modules. The shaper modules preserve the integral value of a signal and enlarge
its width to $80\,\ns$ FWHM in order to match with the SADC sampling rate of
77.76 million samples per second.

In the offline event reconstruction the SADC information is used to extract the
amplitude and time of a signal relative to the trigger time.  After subtracting
the ADC-baseline that is determined for even and odd samples separately
averaging the first 3 samples for each hit, the separation between even and odd
samples is implemented in order to account for interleaved readout as described
in \secref{sec:pid.ecal2.design}.  The amplitude of a signal is determined as
the difference between its absolute maximum and the calculated baseline.  The
hit time is determined for each calorimeter module by selecting those SADC
samples with amplitudes $A_{n}$ and $A_{n+1}$, which surround the position in
time that represents one-half of the maximum amplitude $A_{max}$.  In order to
improve timing accuracy, the time at which the signal is reaching 50\% of
$A_{max}$ is derived from an interpolation between the samples $n$ and $n+1$:

\begin{equation}
   t = \left( n + \frac{A_{max}/2-A_{n}}{A_{n+1}-A_n}\right) \times
   12.86\,\text{ns}\,,
\end{equation}

where 12.86\,ns is the sampling period. For photon energies larger than
$1\,\GeV$ the resulting time resolution is about 1\,ns.

\subsubsection{Calibration and monitoring}
\label{subsec:pid.ecal.ecal1.calib}

For the calibration of ECAL1 LG modules the field of SM1 is set to zero.  A
$15\,\GeV$ electron beam is used, which is a compromise that accounts for the
different dynamic ranges of the three types of modules.  An automatised
calibration procedure changes the position of the calorimeter between two
consecutive spills, so that every module is exposed during calibration.  Up to
several thousands electrons per module are collected within each spill.  The
total cluster charge deposited, i.e. the sum of the charges of the module being
calibrated and its neighbouring modules, is compared to the incident electron
energy. Several iterations are necessary to determine the HV settings for all
modules.

The calibration coefficients, which relate the charge measured by each SADC to
the energy deposited in the corresponding module, are calculated taking into
consideration the energy range of the photons detected in that module during the
experiment.  Since the energy of the photons decreases as the angle between the
photon direction and the beam axis increases, three different HV settings are
applied. For the incident beam energy of $190\,\GeV$, the settings are chosen
such that the corresponding dynamic ranges for the three types of modules extend
up to 60, 30, and $20\,\GeV$ for GAMS, MAINZ and OLGA modules.  The whole
calibration procedure is applied once or twice per data taking period of several
months.

\begin{figure}[tbp]
\begin{center}  
\includegraphics[width=1\columnwidth]{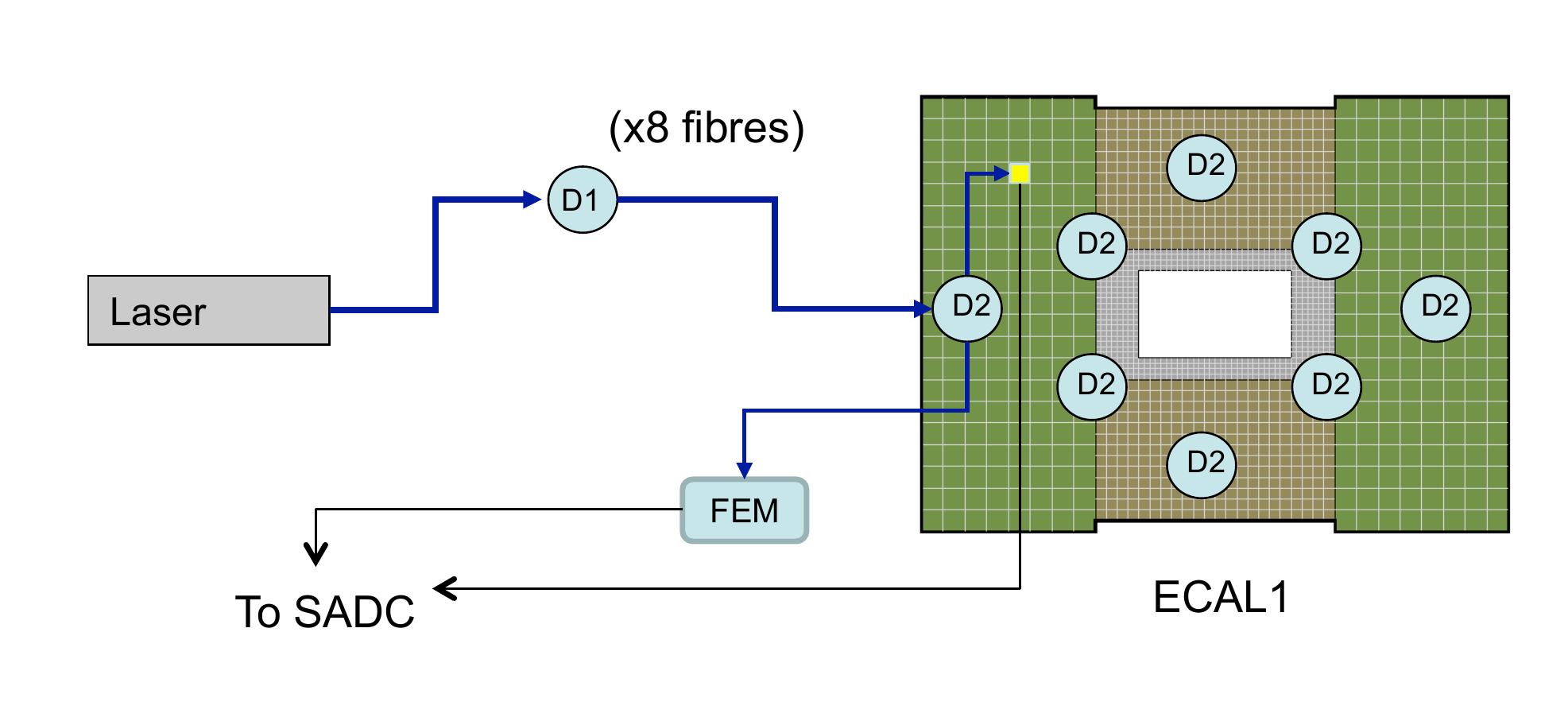}
  \caption{Schematic view of the LASER monitoring system for ECAL1. The laser
    beam is distributed to the ECAL1 modules using one primary (D1) and eight
    secondary (D2) light diffusion spheres. For clarity, only one of the 8
    primary fibres dispatching the light to D2, only one of the secondary 1500
    fibres transmitting it to the LG modules, and only one of the 8
    front-end-monitoring (FEM) modules are explicitly shown.}
  \label{fig:pid.ecal.ecal1_laser}
\end{center}
\end{figure}

In order to control the light collection efficiency and the photomultiplier
gains of all 1500 LG modules, the ECAL1 calorimeter is equipped with a dedicated
monitoring system that is based on the design of Ref.~\cite{CMS:07}. It uses a
single laser source, namely a Minilite-1 model from
Continuum~\cite{Continuum}. The use of a single light source allows the
detection of possible light collection or PMT instabilities individually in each
ECAL1 channel.  The laser light is transmitted to the LG modules through one
primary and eight secondary optical fibre bundles. The fibre bundles are
interconnected using light diffusers that guarantee a uniform distribution of
the light in the fibres.  Each secondary bundle consists of 240 fibres from
which between 160 and 200 fibres are connected to the LG modules.

\begin{figure}[tbp]
  \begin{minipage}[t]{0.48\textwidth}
    \centering
    \includegraphics[width=\textwidth]{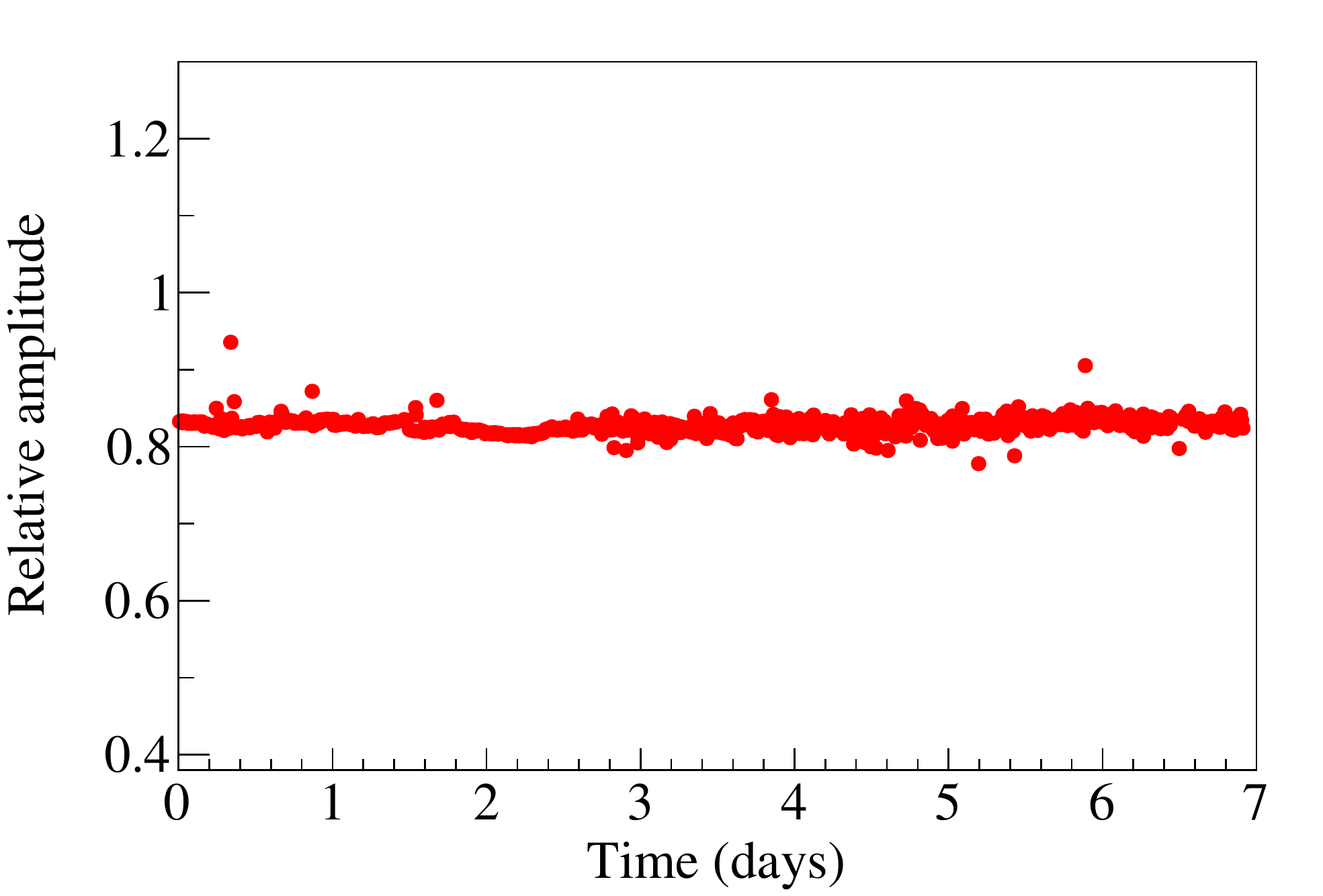}
  \end{minipage}
  \hfill
  \begin{minipage}[t]{0.48\textwidth}
    \centering \includegraphics[width=\textwidth]{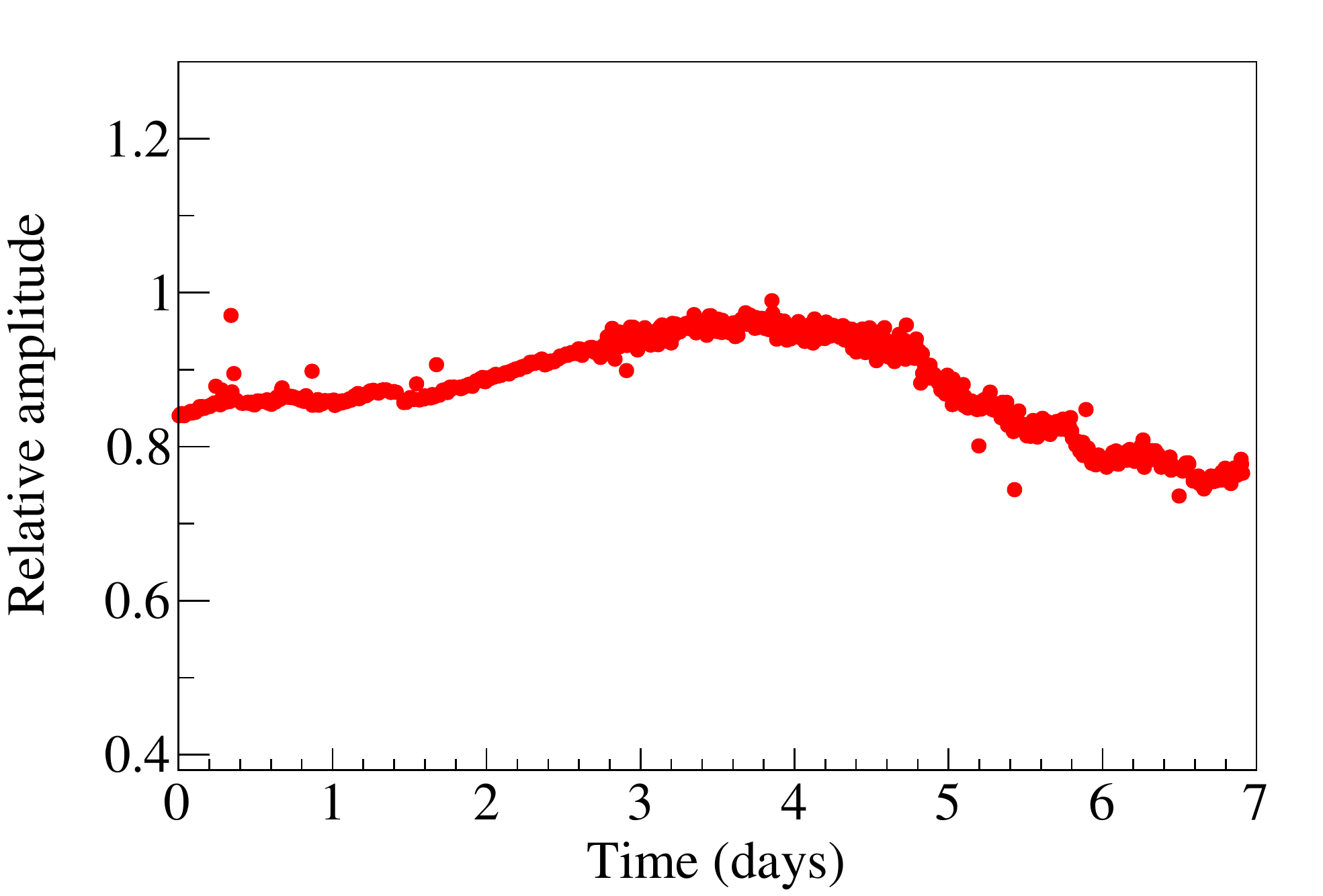}
  \end{minipage}
\caption{ECAL1 module responses as monitored during a period of one week for
  (left) a stable module and (right) an unstable module. The vertical scale is
  normalised to the SADC charge measured in the beginning of the period.}
\label{fig:pid.ecal.ecal1_dcs}
\end{figure}
A simplified drawing of the monitoring system is shown in
Fig.~\ref{fig:pid.ecal.ecal1_laser}. The laser injects $532\,\nm$ light pulses
with $5\,\ns$ FWHM into all calorimeter modules at a frequency of $1\,\Hz$ in
the SPS inter-spill periods. The laser energy per pulse is tuned to an amount
that matches the photomultiplier signal amplitudes.  Since the light output of
the laser source itself may vary between two consecutive pulses, an independent
reference measurement of the pulse amplitudes is required. Nineteen fibres, to
make the available light signal strong enough, from each secondary bundle are
plugged into eight reference photodiodes.  Each photodiode is connected to a
temperature-stabilised electronics circuitry~\cite{CMS:07}, which is enclosed in
a Front-End Monitoring (FEM) module. The eight FEM signals are read out by the
same SADC electronics as the calorimeter modules, thus providing an eight-fold
normalisation of the laser pulse amplitude. The amplitude of the signals from
the LG modules is determined as the peak value of the SADC samples as obtained
after pedestal subtraction.

The recorded laser monitoring amplitudes are used to correct the responses of
all ECAL1 modules on a run-by-run basis. The electron beam calibration provides
the starting values of the calibration coefficients $C^{i}$. These coefficients
convert the photomultiplier signal amplitude from ADC channels to GeV using
\begin{equation}
  A_{GeV}^{i}(t)\ =\ A_{ADC}^{i}(t)\times
  C^{i}\times\frac{L^{i}(t)}{L^{i}(0)}\,.
  \label{eq:ecal_amplitude _conversion}
\end{equation}
Here, $A_{ADC}^{i}(t)$ is the ADC amplitude for the module $i$ measured at a
given time $t$, $L^{i}(t)$ is the ADC amplitude of the monitoring amplitude of
the module $i$ at time $t$, and $L^{i}(0)$ is the monitoring amplitude of the
same module, but recorded during the electron beam calibration run. The time $t$
is taken as the time of the run, for which the correction is to be applied. A
display of two LG modules (a good one and an unstable one) as a function of time
is shown in Fig.~\ref{fig:pid.ecal.ecal1_dcs}.
 
\subsection{Electromagnetic calorimeter ECAL2}
\label{sec:pid.ecal.ecal2}

ECAL2 is a part of the Small Angle Spectrometer. It consists of 3068 calorimeter
modules of three different types, all with the same transverse dimensions
($3.83\times 3.83\,\textrm{cm}^2$).  With its dimensions of $2.44\times
1.83\,\m^2$, ECAL2 covers angular ranges between $1.3\,\mrad$ and $39\,\mrad$ in
the horizontal plane and between $1.3\,\mrad$ and $29\,\mrad$ in the vertical
plane.  In both planes the angular ranges are slightly larger than the
corresponding angular ranges of both ECAL1 (\secref{sec:pid.ecal.ecal1}) and the
hadron calorimeter HCAL1. Accordingly, the peripheral rows and columns of ECAL2
are not used; for the corresponding ECAL2 angles the photons from the target are
either detected in ECAL1 or absorbed in HCAL1.  ECAL2 is installed on a
motorised platform allowing movements in both horizontal and vertical
directions.

\subsubsection{Design and electronics}
\label{sec:pid.ecal2.design}

\begin{figure}[tb]
  \centering
  \includegraphics[width=1.0\textwidth]{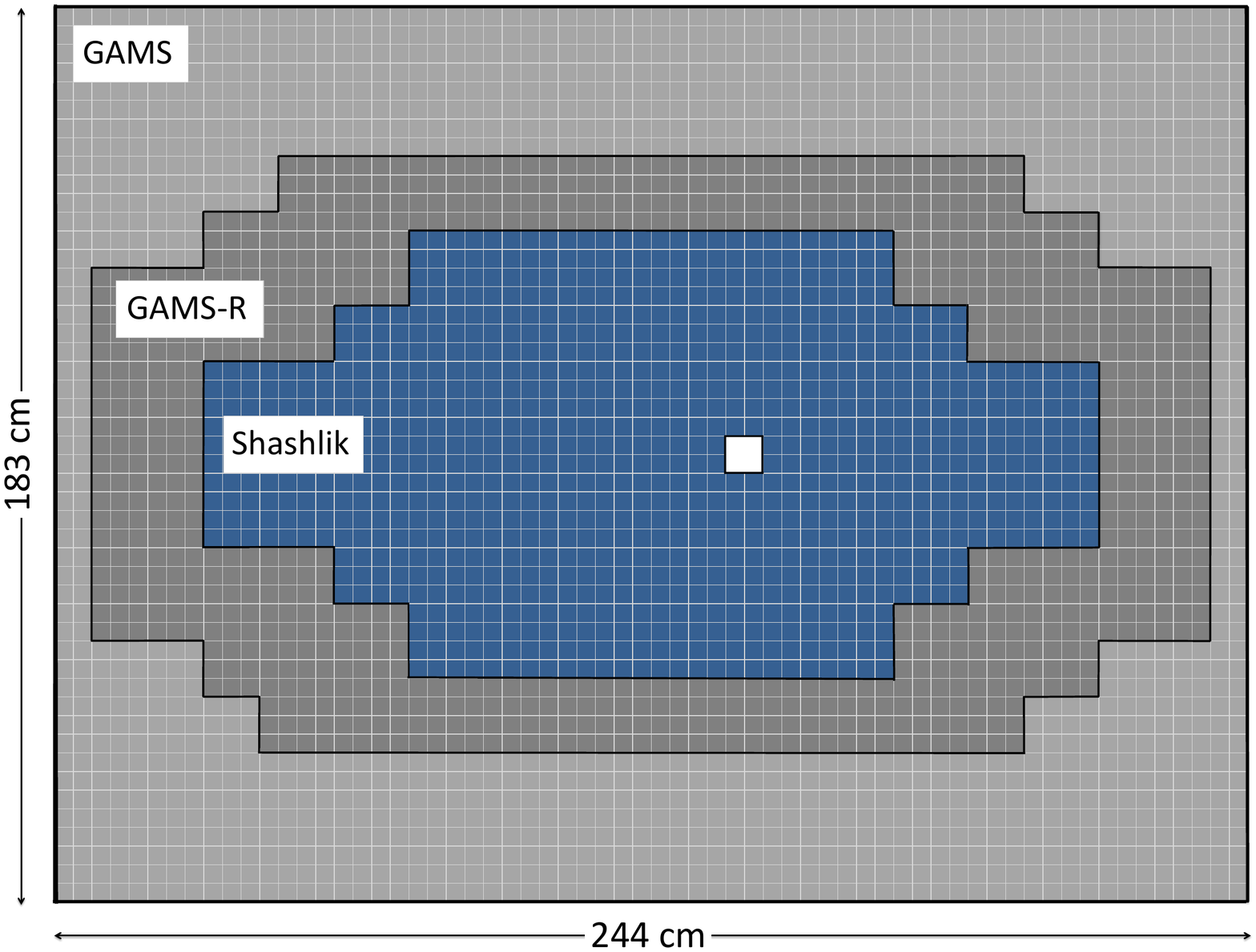}
  \caption{Configuration of ECAL2. The outer and intermediate regions are
    equipped with GAMS and radiation-hardened GAMS modules respectively. The
    inner region is equipped with Shashlik sampling modules. The transverse
    sizes of all three types of modules are identical.  The central hole of
    $2\times 2$ modules can be seen as a white spot.  }
\label{fig:pid.ecal.ecal2-structure}
\end{figure}

\begin{figure}[htbp]
  \centering \includegraphics[width=0.8\textwidth]{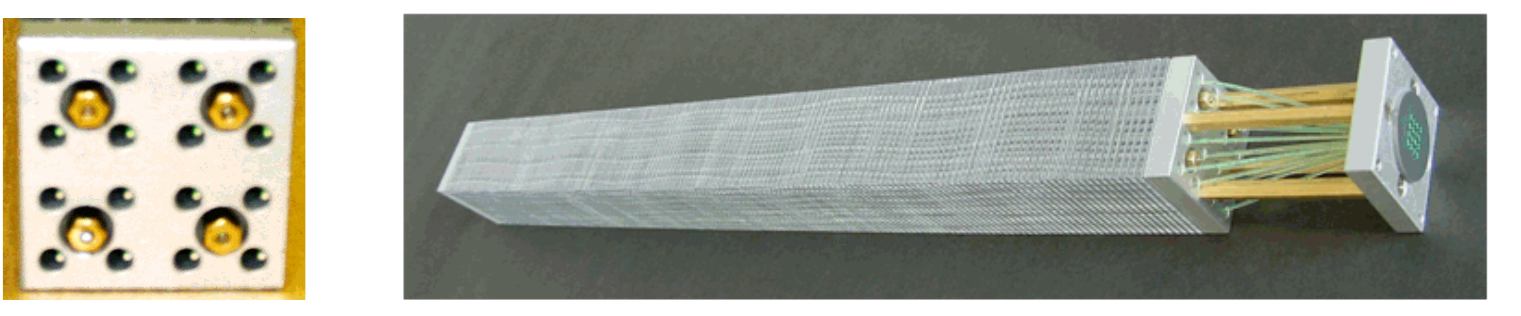}
  \caption{Photographs of a Shashlik-type calorimeter module. Left part: the
    upstream face of the module with its four central rods and 16 light fibres.
    Right part: the module itself with the fibres guide at the downstream face.}
  \label{fig:pid.ecal2.shashlik}
\end{figure}

The ECAL2 modules are arranged in a $64\times 48$ matrix, as shown in
Fig.~\ref{fig:pid.ecal.ecal2-structure}.  For data taking with hadron beams, its
central hole with respect to the nominal beam directions is set to $2\times 2$
modules.  The central hole of the HCAL2 calorimeter is set accordingly to
$10\times 10\,\textrm{cm}^2$.  The outermost part of ECAL2 is equipped with 1332
TF1 lead glass~\cite{Binon:86} modules, which are identical to the GAMS modules
used for ECAL1.  The intermediate part of ECAL2 is filled with 848
radiation-hardened modules (GAMS-R) made out of TF101
material~\cite{Kobayashi:94}.  This material is a standard TF1 type LG, which is
enriched with 0.2\% of cerium. The innermost part is equipped with 888 Shashlik
type modules (see Fig.~\ref{fig:pid.ecal2.shashlik}).  The $39\,\textrm{cm}$
long Shashlik modules are composed of 154 double layers, each consisting of a
$0.8\,\mm$ thick lead plate and of a $1.55\,\mm$ thick scintillator plate.  The
photons from the Shashlik modules are collected by 16 wavelength-shifting light
fibres and guided onto FEU-84-3 photomultipliers.

The different ECAL2 modules have identical transverse dimensions, but different
radiation hardness properties.  Calculations for the present ECAL2 configuration
have shown that with the COMPASS nominal hadron beam intensity and duty cycle
the most exposed modules, i.e., those located closest to the beam, would stand
radiation doses corresponding to several years of data taking for GAMS and
GAMS-R and nearly 20 years for Shashlik, without significant degradation of
their response.

\begin{figure}[h]
  \centering
  \includegraphics[width=0.8\textwidth]{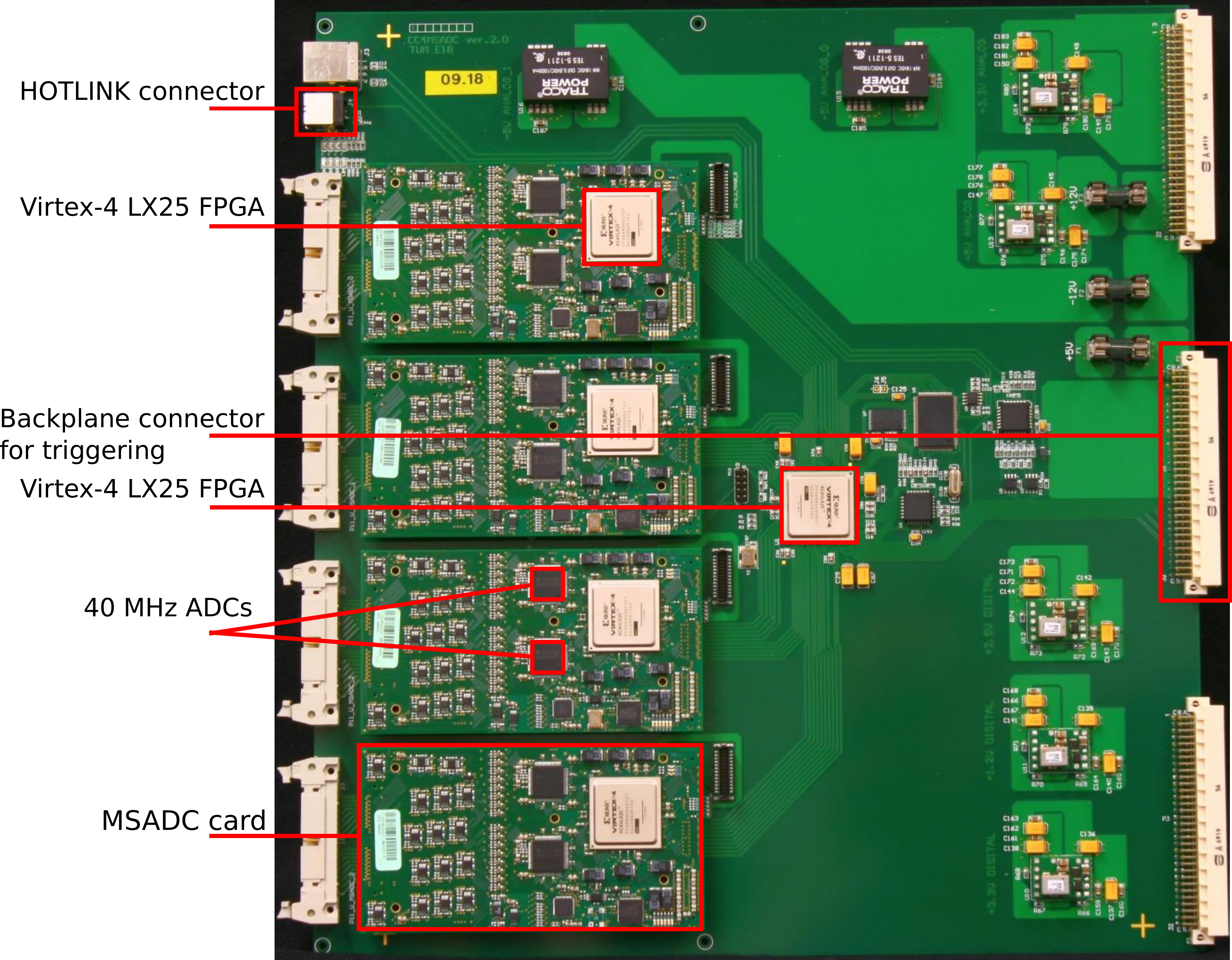}
  \caption{VME carrier card with four mounted MSADC modules.}
  \label{fig:pid.ecal.msadc_cc}
\end{figure}

The photomultiplier signals coming from the 3068 ECAL2 modules are first
transferred to a shaper, which increases the signal width to $120\,\ns$
FWHM. The signals are then digitised by sampling ADCs.  The ECAL2 readout was
upgraded with a new sampling ADC system, which provides a dynamic range of
12~bit and allows more elaborate data processing. The basic building block is a
compact Mezzanine Sampling ADC card (MSADC), which performs a digitisation of
the 16 analogue input channels at 77.76 million cycles per second, with two
interleaved multichannel ADCs~\cite{Mann:07}. Data processing is implemented by
a Xilinx Virtex4 FPGA~\cite{Xilinx}. The MSADC firmware includes a digital ring
buffer to compensate the trigger latency, a pipelined logic for pedestal
correction, zero suppression and data formatting.  In addition, an independent
processing chain is implemented on the FPGA to determine time and amplitude
information for the calorimeter trigger described in \secref{sec:trigger.ecal}.
As shown in Fig.~\ref{fig:pid.ecal.msadc_cc}, four MSADCs are combined on a 9U
VME carrier card, which merges the data from 64 calorimeter channels and
provides a serial 40~Mbit/s HOTLink interface to the HotGeSiCA multiplexer
modules in the next readout stage.  In order to reduce the power dissipation,
all MSADC supply voltages are generated centrally on the carrier card with DC/DC
converters. The resulting noise on the MSADC channels is below 1.5 least
significant bit.

\begin{figure}[b]
    \centering
    \includegraphics[width=0.5\textwidth]{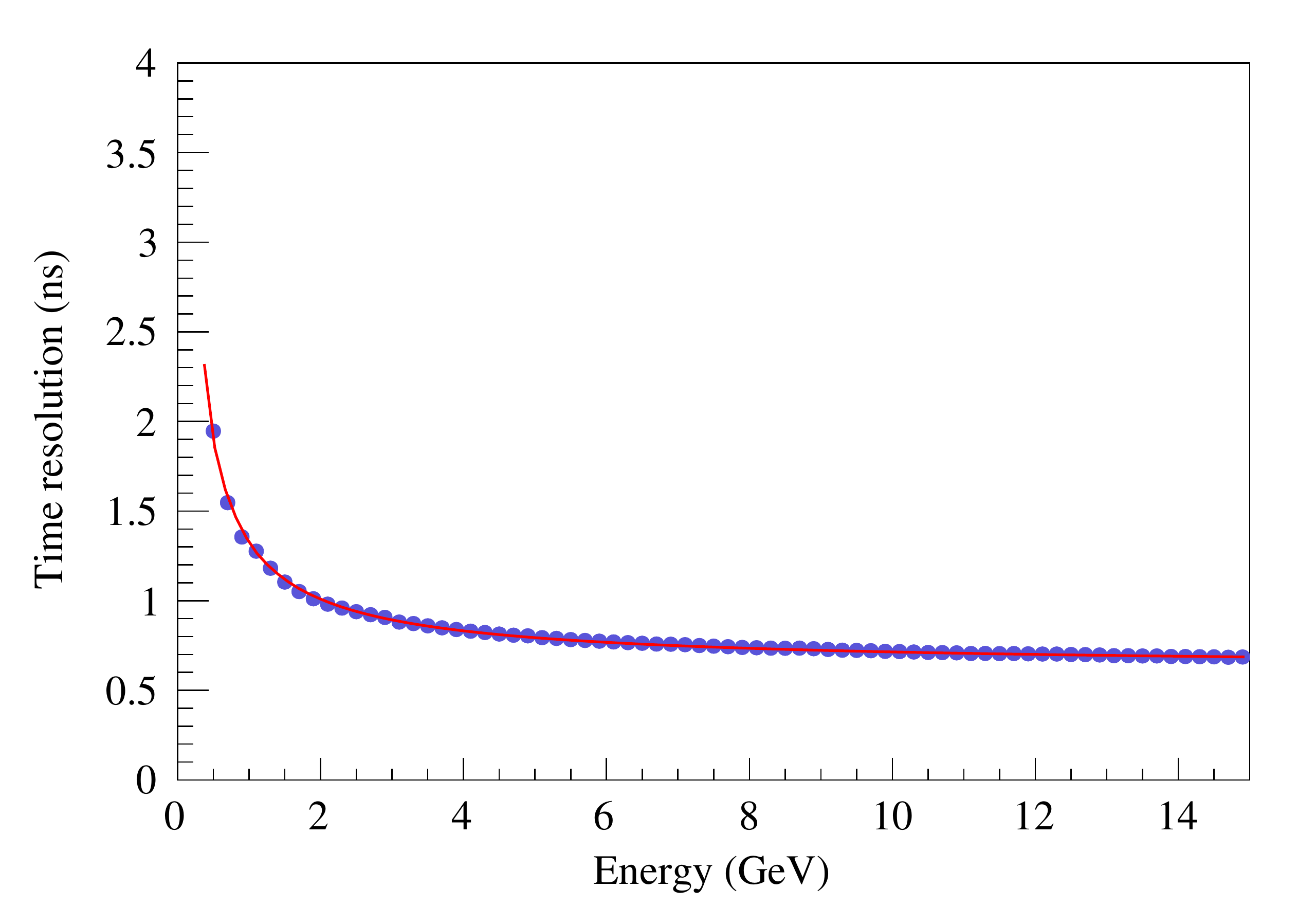}
    \caption{ Standard deviation $\sigma$ for the ECAL2 time resolution as a
      function of the photon energy E. The solid curve is a fit to the data
      points using the expression: $\sigma^{2}(E)=1.13/E+0.22/E^{2}+0.39$.  }
\label{fig:pid.ecal2.time-resolution}
\end{figure}

The information from the MSADCs is also used to calculate the time for each
event.  For each ECAL2 module, the algorithm interpolates between the times of
the two samples around the one-half value of the maximum sampled amplitude (see
\secref{sec:pid.ecal.ecal1}).  The time resolution for ECAL2 is shown in
Fig.~\ref{fig:pid.ecal2.time-resolution}.  For energies higher than $2\,\GeV$,
resolutions of 1\,ns or better are achieved.

\subsubsection{Calibration and monitoring}
\label{sec:pid.ecal.ecal2.calib}
The ECAL2 calorimeter is calibrated by exposing all its modules to a $40\,\GeV$
electron beam.  Apart from different geometry and different number of modules,
the calibration procedure is identical to that used for ECAL1 (see
\secref{subsec:pid.ecal.ecal1.calib}).  The charge deposited in each cluster of
LG modules (as measured in the MSADCs) is calculated and compared to the
incident electron energy.  After the data for all modules are collected, few
iterations are necessary to determine the calibration coefficients for all
Shashlik and LG modules.  The final HV settings are calculated after taking into
account the energy ranges of the photons detected in the different parts of
ECAL2.  The high voltages of the PMTs of the innermost $16\times 16$ modules are
set to measure energies of up to $200\,\GeV$.  In the surrounding part, which
represents a $48\times 48$ matrix, the maximal energy is set at $150\,\GeV$.
Finally, in the two outermost parts with $8\times 48$ modules each, the PMTs are
set to detect energies of up to $60\,\GeV$.

\begin{figure}[tbp]
  \begin{minipage}[t]{0.5\textwidth}
    \centering
    \includegraphics[width=\textwidth]{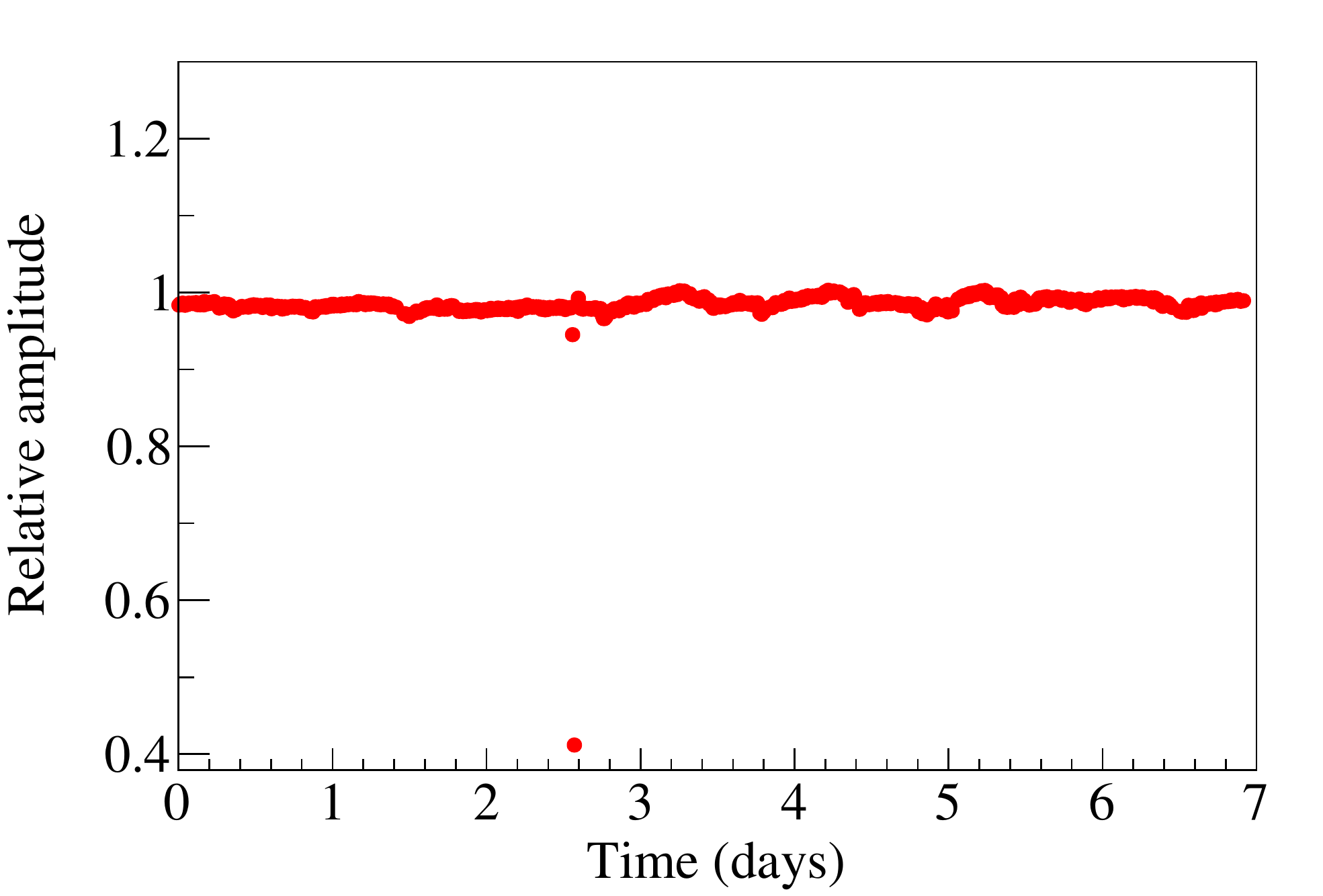}
  \end{minipage}
  \hfill
  \begin{minipage}[t]{0.5\textwidth}
    \centering \includegraphics[width=\textwidth]{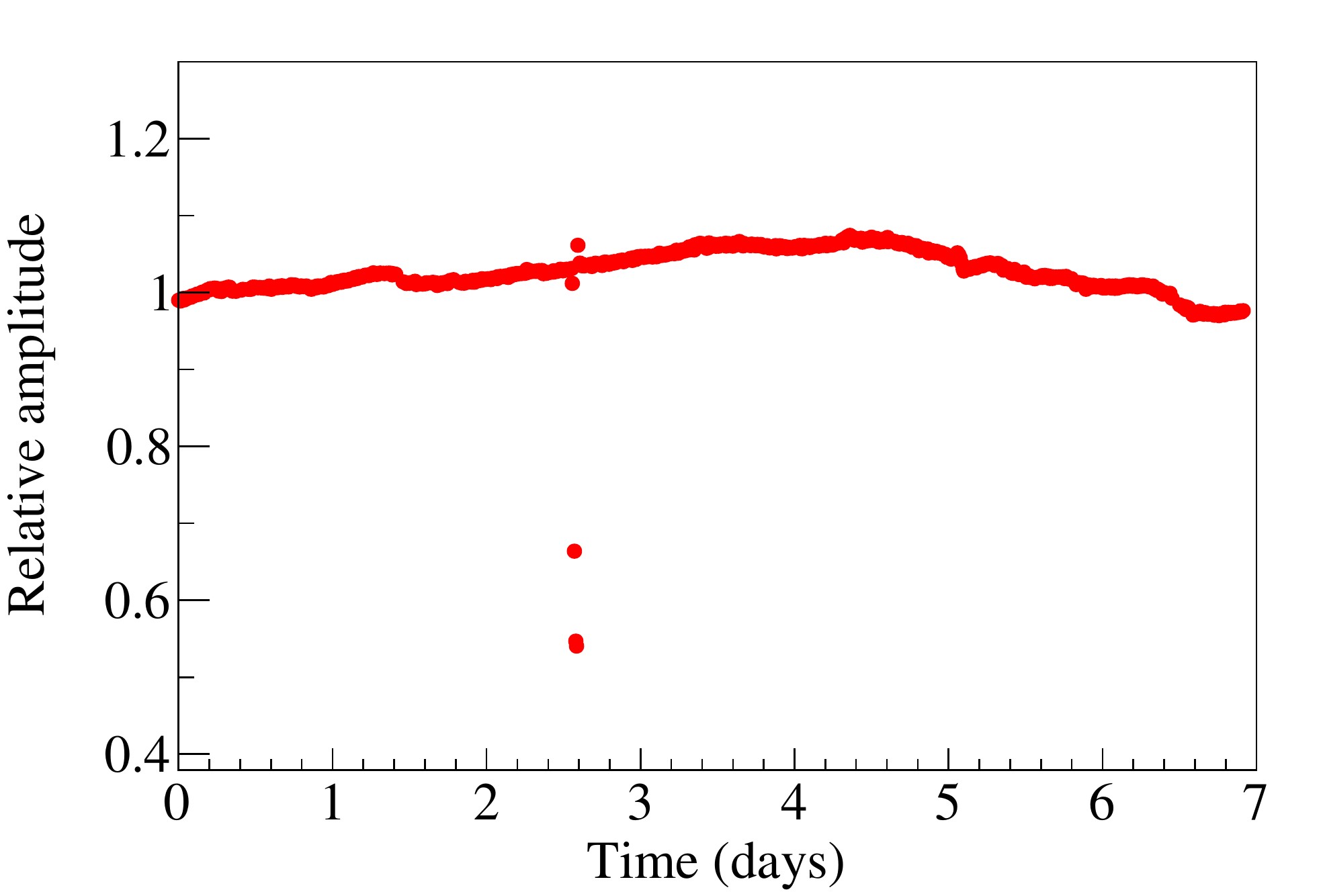}
  \end{minipage}
\caption{ECAL2 module responses as monitored during a period of one week for
  (left) a stable module and (right) an unstable module. The vertical scale is
  normalised to the SADC charge measured in the beginning of the period.}
\label{fig:pid.ecal.ecal2_led}
\end{figure}

The time stability of the ECAL2 calibration is monitored using a LED-based
monitoring system.  This light-distribution system consists of 30 fibre bundles,
where the number of fibres varies between 64 and 140 per bundle.  Each bundle is
illuminated by six simultaneously activated LEDs.  The combined use of six LEDs
increases the available light intensity and minimises possible intensity
fluctuations by averaging out the individual LED instabilities. The system is
activated using a calibration trigger with a frequency of $1\,\Hz$. A display of
two ECAL2 modules with a stable and with an unstable response is shown in
Fig.~\ref{fig:pid.ecal.ecal2_led}.  The information from the monitoring system
is used to correct short and long term drifts of individual cells on a spill by
spill basis.

\section{Trigger}
\label{sec:trigger}
\begin{figure}[htb]
  \centering \includegraphics[width=\textwidth]{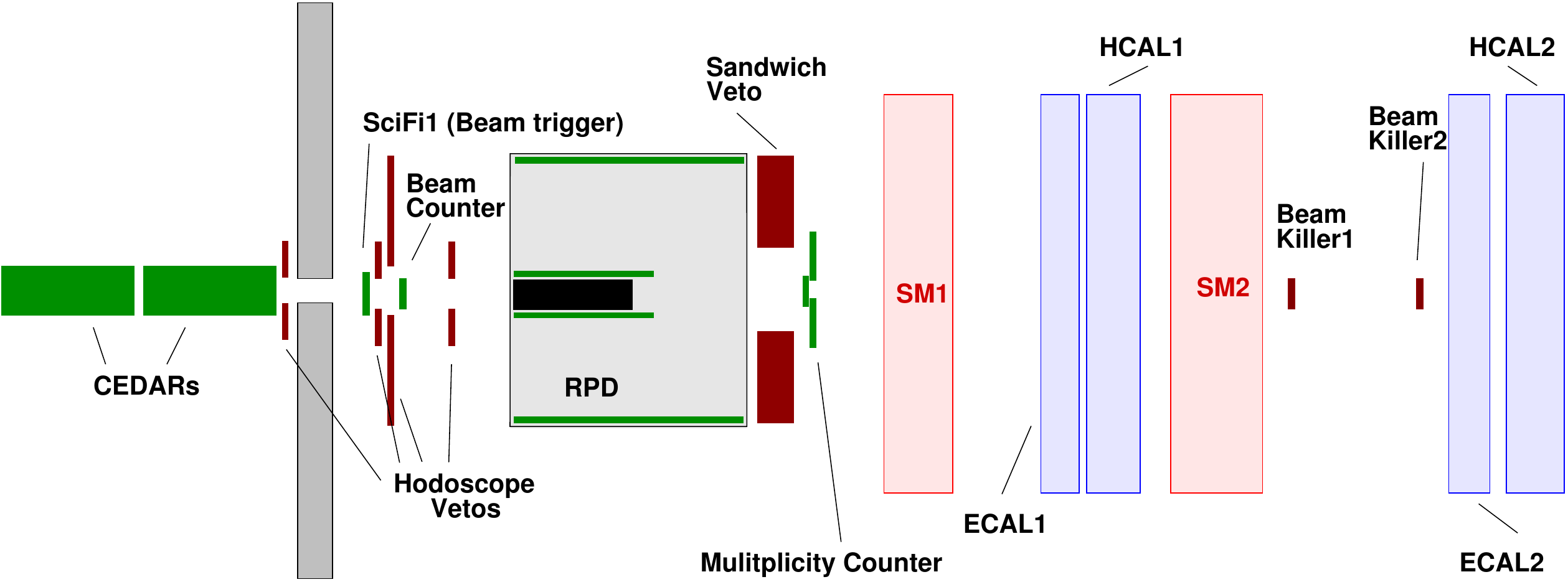}
  \caption{Arrangement of trigger elements in the spectrometer (schematic side
    view, not to scale).}
  \label{fig:trigger.setup}
\end{figure}

The COMPASS trigger system for hadron beams is designed to select events that
carry all the information needed for exclusive measurements.  A fast response is
needed to provide the time reference for the readout of all detectors. A physics
trigger consists of three subsystems: beam-defining elements to select beam
particles crossing the target, veto detectors to reject events containing
particles produced outside of the target or outside of the spectrometer
acceptance, and specific detector systems that account for the particular
physics case. The latter are: i) the proton trigger
(\secref{sec:trigger.recoil}) that is used for measurements of diffractive
scattering and central production processes with momentum transfers $t <
-0.07$\,GeV$^2/c^2$ (see \secref{sec:target.rpd}), ii) the multiplicity trigger
that completes the coverage in $t$ for reactions with higher charged track
multiplicities (\secref{sec:trigger.multiplicity}), and the calorimeter trigger
(\secref{sec:trigger.ecal}) that is used for Primakoff data
taking. Figure\,\ref{fig:trigger.setup} shows schematically the location of the
trigger elements in the spectrometer.

\subsection{Beam Trigger}
\label{sec:trigger.beam}
The beam trigger selects incoming beam particles and is used to define the
reference time of an event.  In addition, it reduces the geometric acceptance of
the beam in the transverse plane to match the target geometry. It consists of a
coincidence of a scintillating fibre detector, SciFi1, with a beam counter.
SciFi1 is located 7\,m upstream of the liquid hydrogen target. It has one
vertical and one horizontal plane.  Each plane is read out by six multi-anode
photomultiplier tubes (PMT) with 16 channels each. In addition, the PMTs are
read out at the last dynode stage, thus providing six analogue sums for each of
two planes of the detector.
    
\begin{figure}[ht]
  \centering \includegraphics[width=0.49\textwidth]{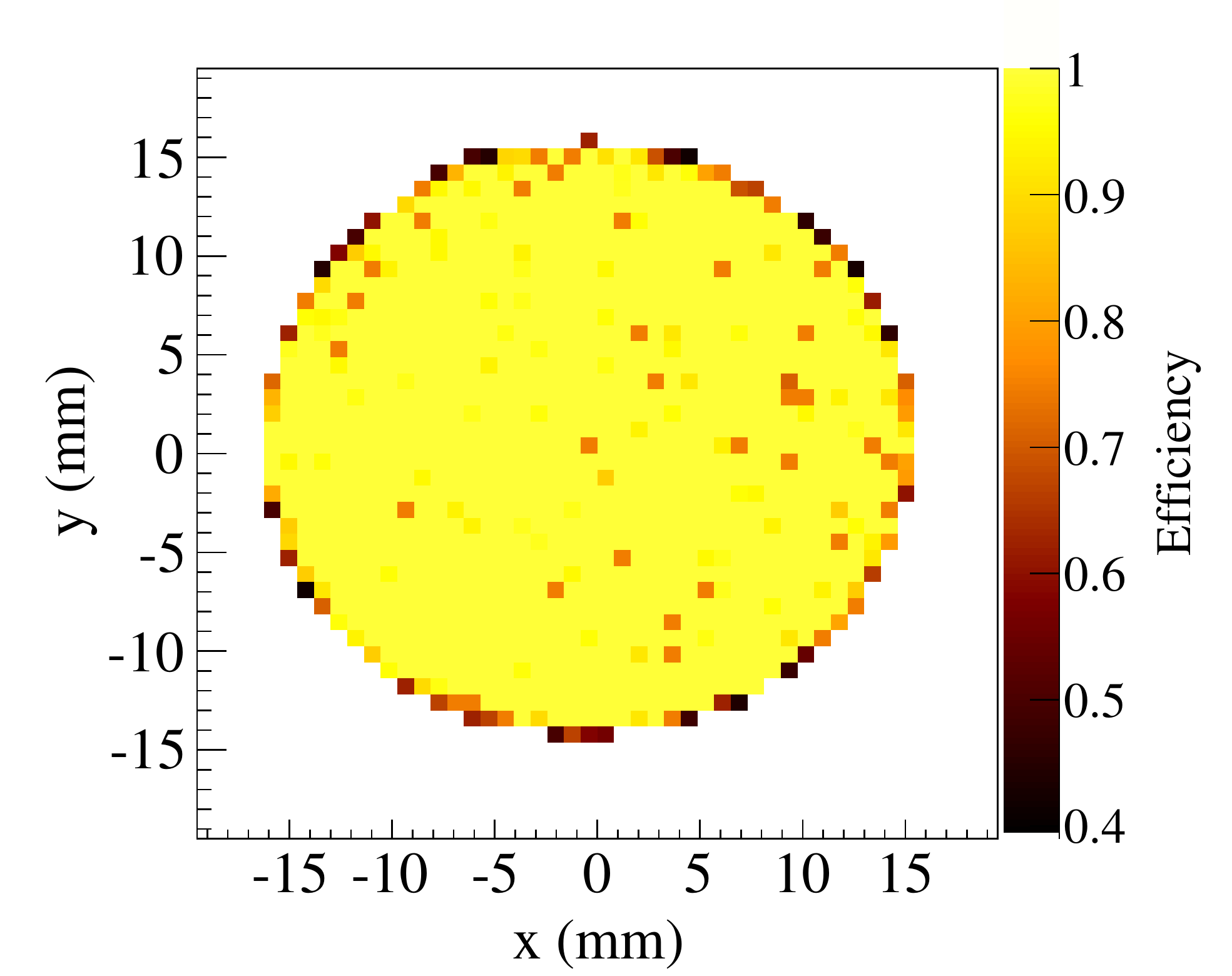}
  \caption{Beam counter efficiency distribution in transverse coordinates.}
  \label{fig:trigger.beam.efficiency}
\end{figure}

The beam counter is a small scintillator disc that is located 50\,cm downstream
of SciFi1. It has a diameter of 3.2\,cm, a thickness of 4\,mm, and is centred at
the beam. It is surrounded by a thin, black PVC tube covered inside by
aluminised Mylar foil with an internal reflection of better than 92\%. A 35\,cm
long tube used as an air light-guide is connected to a single EMI 9813KB
PMT. The PMT is equipped with a voltage divider that stands beam rates of up to
10\,MHz. The efficiency of the beam counter was measured to be 99.5\% all over
the surface of the disk, as shown in Fig.~\ref{fig:trigger.beam.efficiency}.

\begin{figure}[hb]
  \centering
  \includegraphics[width=0.49\textwidth]{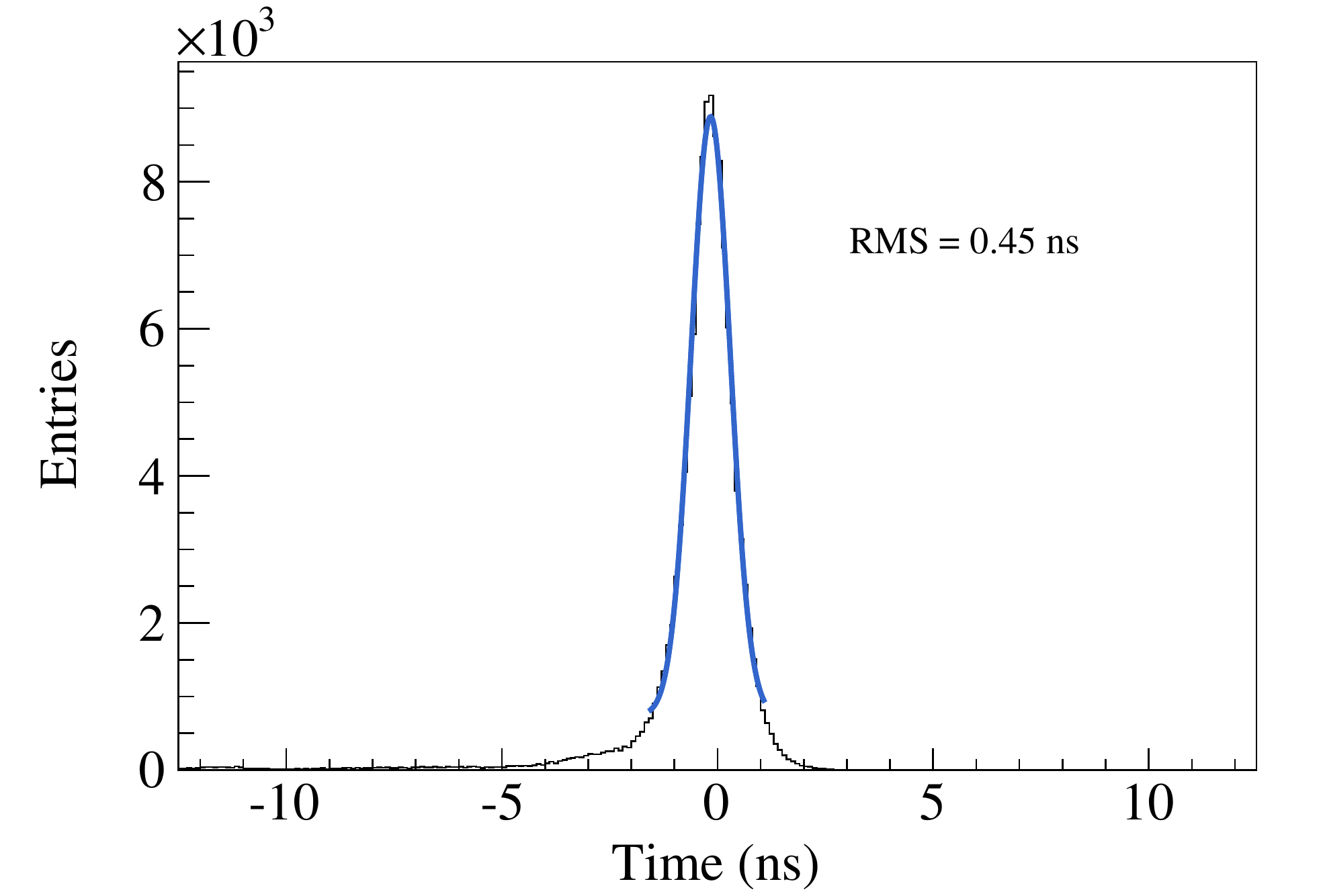}
  \caption{Time residual of the beam trigger.}
  \label{fig:trigger.beam.timing}
\end{figure}

The beam trigger is defined by the coincidence of the beam counter signal and
the logical OR of the 6 analogue sums of the SciFi1 X plane. The time resolution
is measured to be 450\,ps $\pm$ 50\,ps (Fig.~\ref{fig:trigger.beam.timing}).  It
is used as a time reference of the trigger system.

\subsection{Veto Detectors}
\label{sec:trigger.veto}
The veto system consists of two scintillation counters (``beam killers"), a
`sandwich" veto detector (see \Secref{sec:target.sandwich}) and a hodoscope veto
system. It inhibits false physics triggers. The overall dead time of the veto
system was measured to be 13\%-16\% for the nominal beam intensity of $5 \cdot
10^6\,\s^{-1}$. For Primakoff data taking, it is reduced to 8\%-10\% due to the
lower beam intensity.

\subsubsection{Beam killers}
\label{sec:trigger.veto.bk}
Two scintillating counters are positioned along the beam axis of the
spectrometer at $z = $+25\,m (BK1) and $z = $+33\,m (BK2). Both counters have a
diameter of 3.5\,cm and a thickness of 0.5\,cm. Their function is to inhibit a
trigger signal coming from non-interacting beam particles.  The use of the beam
killers introduces an angular cut-off of 0.97\,mrad with respect to the nominal
beam axis.  Including the beam killers in the diffractive trigger (see
Table~\ref{tab:trigger.triggers}), reduces its trigger rate by about a factor of
2.

\subsubsection{Sandwich Veto Detector}
\label{sec:trigger.veto.sandwich}
A Sandwich veto detector (described in \Secref{sec:target.sandwich}) is used to
veto charged and neutral particles that are detected outside of the angular
acceptance of the spectrometer and the RPD. Such particles are dominantly
produced in inelastic, non-diffractive reactions or in reactions in which the
target protons are diffractively excited.  Including the Sandwich veto in the
trigger improves the purity of the physics triggers by a factor of about 3.5.

\subsubsection{Hodoscope Veto System}
\label{sec:trigger.veto.hodo}
The hodoscope veto system is the same as the one used for the muon programme.
It consists of three parts: a beam line hodoscope veto system (V$_\textrm{BL}$)
installed at $z\,=\,-20$\,m, a Veto1 system located at $z\,=\,-7.5$\,m, and a
Veto2 counter at $z\,=\,-1.5$\,m. A detailed description can be found
in~Refs~\cite{Abbon:2007pq} and ~\cite{Bernet:05a}. The hodoscope veto system
removes events with large multiplicities in the RPD.

\subsection{Proton trigger}
\label{sec:trigger.recoil}

\begin{figure}[htpb]
  \centering \includegraphics[width=0.49\textwidth]{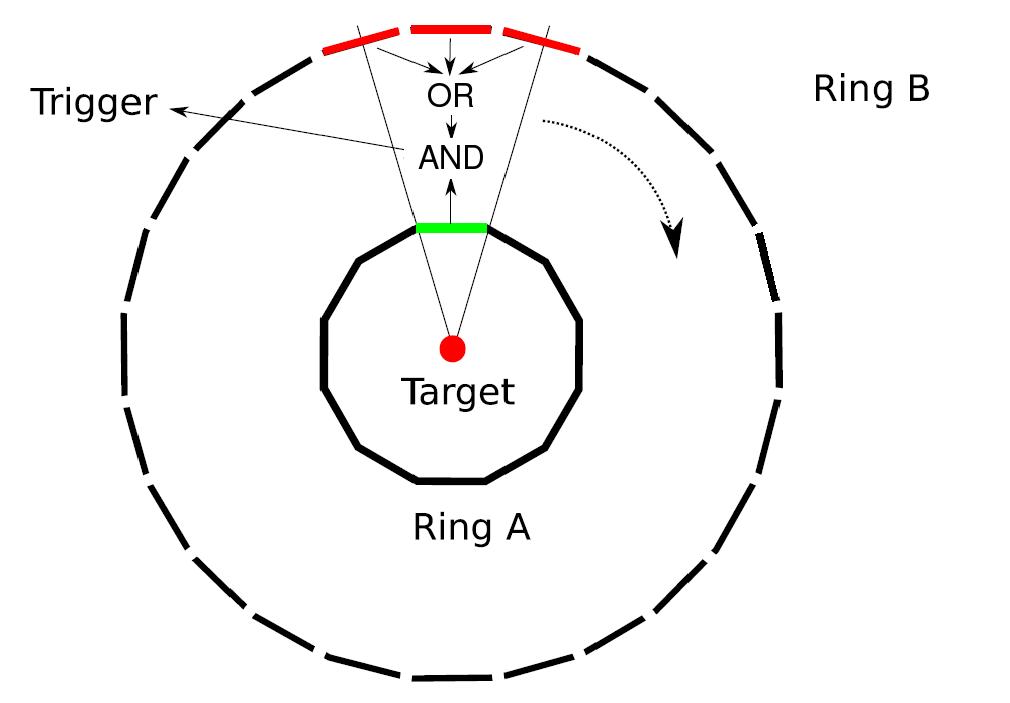}
  \caption{Allowed combinations for target pointing in the RPD part of the
    proton trigger.}
  \label{fig:trigger.pointing}
\end{figure}

The proton trigger selects events with recoiling protons from the target.  The
RPD (described in \secref{sec:target.rpd}) information is used for two purposes:
target pointing and discrimination of protons from pions and delta-electrons by
measuring the energy loss in each ring of the RPD.  Target pointing is
implemented by allowing only for combinations, where hits in one scintillator of
the inner ring are followed by a signal in one of the three corresponding outer
ring scintillators, as shown in Fig.~\ref{fig:trigger.pointing}.

\begin{figure}[hpb]
  \centering
  \includegraphics[width=0.60\textwidth]{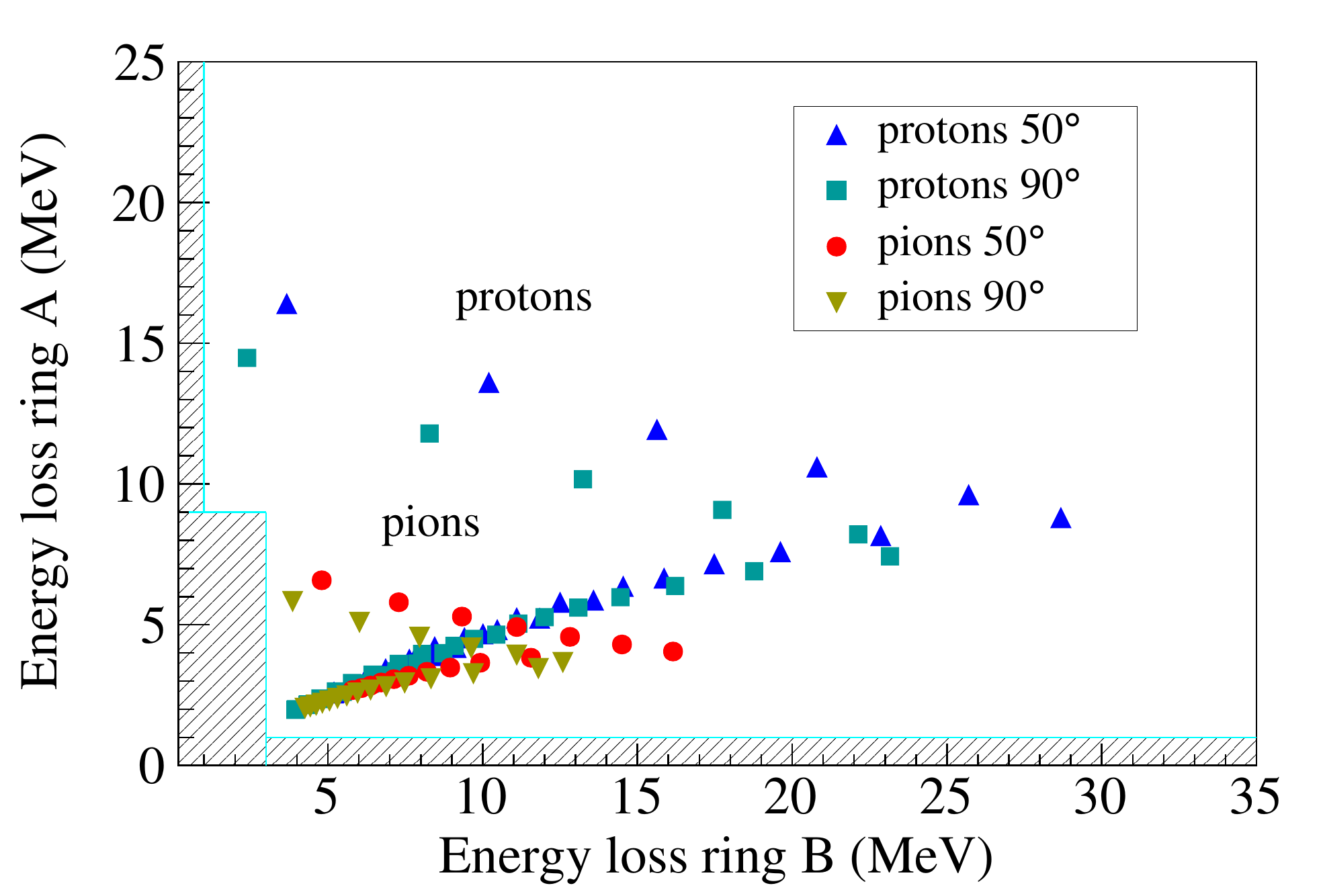}
  \caption{Correlation between the energy losses of protons and pions traversing
    ring A and stopping (or traversing) ring B of the RPD. For each particle
    type the minimum and the maximum polar angles ($50^\circ$ and $90^\circ$)
    are shown.  The shaded area corresponds to the region rejected by the
    trigger logic.}
  \label{fig:trigger.recoil}
\end{figure}

For a particle traversing the RPD, its energy losses in the inner and in the
outer rings are strongly correlated.  This is used to reject electrons coming
from the target as well as part of the low-energy
pions. Fig.~\ref{fig:trigger.recoil} shows the calculated energy losses for both
protons and pions, and for the minimum and maximum polar angles ($50^\circ$ and
$90^\circ$) of the RPD acceptance (see~\secref{sec:target.rpd}).  The area to be
rejected is defined using the two levels of discriminator thresholds in both
rings.  The coincidence of low-threshold signals for upstream and downstream
PMTs of the inner (outer) ring is denoted by $A_{i}^{Low}$ ($B_{j}^{Low}$),
where $i$ and $j$ are the respective scintillator elements.  Similarly, the
coincidence of the two high-threshold signals is denoted by the superscript
``High".  The trigger logic function for recoil protons has the following
expression:
\begin{equation}
RPD=\bigvee_{i=1}^{12} A_{i,down}^{Low}\wedge \bigvee_{j=2i-1}^{2i+1} \left(
A_{i}^{Low}B_{j}^{High} \vee A_{i}^{High}B_{j}^{Low} \right).
\label{eq:proton-trigger}
\end{equation}
Here, the signals from the downstream PMTs of the inner ring,
$A_{i,down}^{Low}$, are used to minimise the time jitter with respect to the
beam trigger.  The trigger logic is set to reject the electrons that cross both
rings as well as pions that cross ring A but leave less than few MeV in ring
B. In \figref{fig:trigger.recoil}, the region rejected by the proton trigger is
indicated by the shaded area.

In order to be able to measure the time-of-flight of all recoil protons,
irrespective of their velocities, a large time window of 50\,ns is required for
the coincidence between any of the inner-ring downstream PMTs with the
geometrically allowed outer-ring PMTs.  The trigger logic function
(\ref{eq:proton-trigger}) is implemented in a single FPGA module that is fed by
the logic signals from all PMTs of the RPD.

\subsection{Multiplicity Trigger}
\label{sec:trigger.multiplicity}
The multiplicity triggers were built to extend the measurements to events with
momentum transfers $-t$ smaller than $-t<0.07\,\GeV^2/c^2$ that are outside the
acceptance of the proton trigger. It uses the multiplicity counter to estimate
the charged-particle multiplicities in the beam region or tag events with at
least one (or two, see Table~\ref{tab:trigger.triggers}) track at large angles.

\begin{figure}[tbp]
  \centering
  \includegraphics[width=0.8\textwidth]{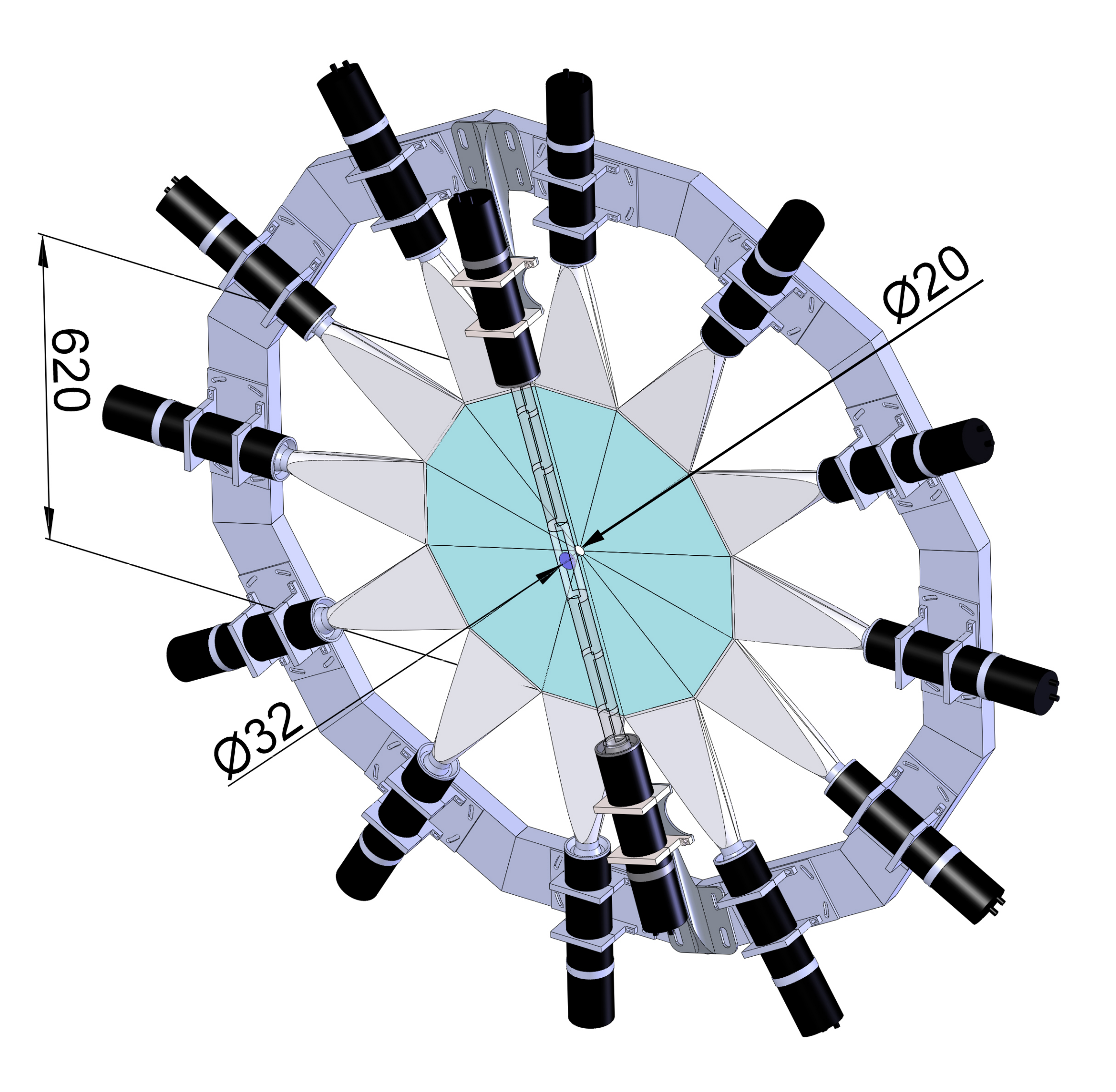}
  \caption{The multiplicity counter. All dimensions are in mm.}
  \label{fig:triggers.multiplicity_counter}
\end{figure}

The multiplicity counter (Fig.~\ref{fig:triggers.multiplicity_counter}) consists
of 12 trapezoidally shaped scintillator slabs with a central hole of $20\,\mm$
diameter.  It covers the charged-particle acceptance of the spectrometer at
1.7\,m, which projects to a disk with a radius of $310\,\mm$. The light is read
out by one photomultiplier per slab. The counter was upgraded in 2009 with a
scintillator disk with a diameter of $32\,\mm$, which is centered at the hole
and is read out by two photomultipliers.  In order to minimise photon conversion
in the active area, all scintillators have a thickness of $3\,\mm$, which
corresponds to $0.71\,\%$ of a radiation length. The photomultipliers for the
inner disk are connected through an $83\,\Cm$ long air light guide made of a
tube skeleton of $15\,\mum$ aluminised Mylar inside a $150\,\mum$ thick plastic
coating.

The two components of the multiplicity detector, the outer and the inner
counters, are used to build two independent triggers, MT1 and MT2.  The MT1
multiplicity trigger requires one hit or more in each element of the outer
multiplicity counter.  For this purpose, the threshold per element is set to
reject noise only, thereby selecting charged particle multiplicities of one or
larger. A logical OR of all elements is then used as the trigger signal.  The
MT2 multiplicity trigger requires an energy deposit corresponding to
1.6~MIPs\footnote{minimum ionising particle} or higher in the inner counter.
Data were also taken in stricter conditions for both MT1 and MT2.  For MT1 this
was achieved by using a multiplicity logic instead of the logical OR, thereby
selecting events where two or more slabs of the outer counter are hit. For MT2 a
higher energy deposit equivalent to 2.5~MIPs was required to select
multiplicities of three or larger. These conditions enrich events that have
final states with higher multiplicities and therefore higher masses.

\subsection{Calorimeter Trigger}
\label{sec:trigger.ecal}
\begin{figure}[htb]
\centering \includegraphics[width=0.8\textwidth]{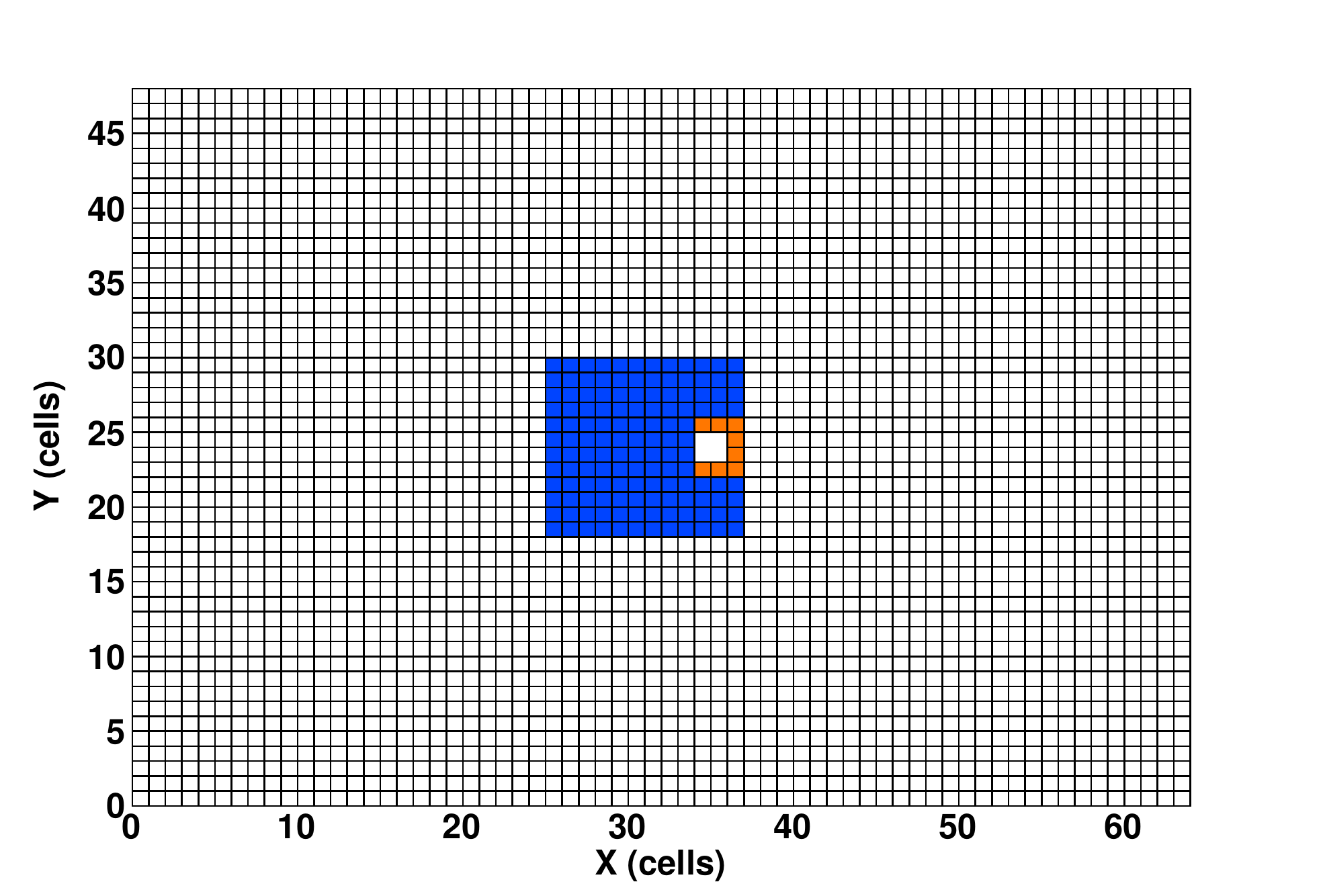}
  \caption{The active area of the ECAL2 trigger (shown in blue).  The cells
    shown in orange are rejected due to high rates.}
\label{fig:trigger.ecal.area}
\end{figure}

\begin{figure}[htb]
\centering \includegraphics[width=0.8\textwidth]{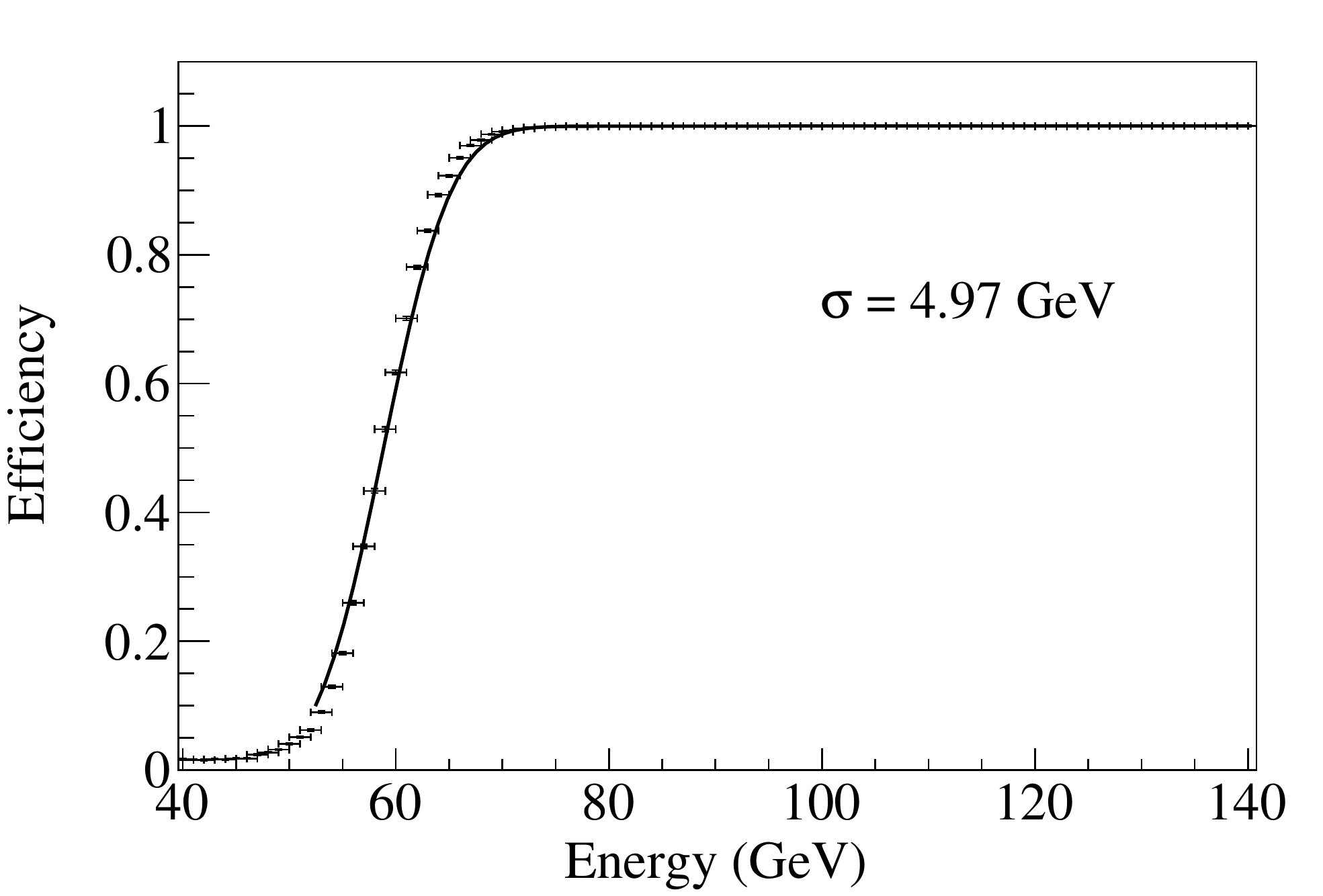}
  \caption{Efficiency of the ECAL2 trigger as a function of the energy.  The
    solid line is a fit to the data with an error function.}
\label{fig:trigger.ecal.efficiency}
\end{figure}

\begin{figure}[htb]
\centering \includegraphics[width=0.8\textwidth]{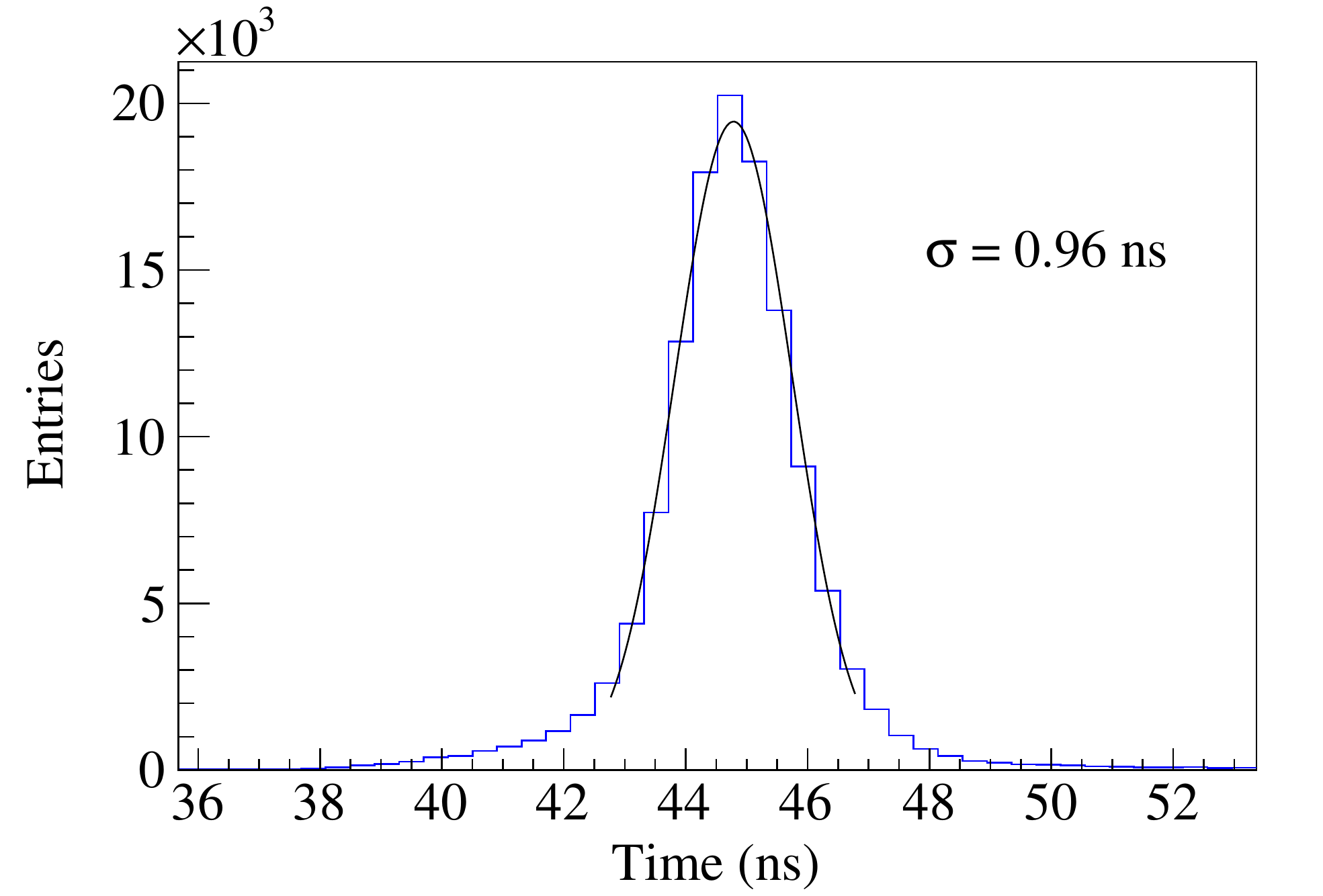}
  \caption{Time resolution of the CFD algorithm for a representative cell in the
    centre and signal amplitudes above 800 MeV.}
\label{fig:trigger.ecal.time_resolution}
\end{figure}

The calorimeter trigger selects high-energy photons detected by ECAL2 within
$12\times 12$ cells, whereby 8 cells surrounding the beam hole are excluded, as
depicted in Fig.~\ref{fig:trigger.ecal.area}. The trigger logic is implemented
in the existing ECAL2 readout module described in
\Secref{sec:pid.ecal2.design}. At the first stage, FPGAs mounted on the MSADC
cards detect a signal and extract, on a cell-by-cell basis, amplitude and time
information.  The time information is obtained using a digitally implemented
constant fraction algorithm.  In order to achieve good time resolution and clean
up noise, signals below $800\,\MeV$ are rejected at the cost of a small
uncertainty of the measurement of the total amplitude.  At a later stage, time
and energy calibrations are applied.  The sum of the time-correlated energies in
a selected region of cells is compared with a programmable threshold.  The
achieved precision is $\sigma=4.97\,\GeV$ for a $60\,\GeV$ threshold, determined
by fitting an error function, as shown in
Fig.~\ref{fig:trigger.ecal.efficiency}.  The main contributions to this
precision are the accuracy of the per-channel thresholds and the preliminary
calibration constants used in the trigger system.  The overall time resolution
could be reduced to about 1\,ns by digital signal processing, as illustrated in
Fig.~\ref{fig:trigger.ecal.time_resolution}.

\subsection{Physics Triggers}
\label{sec:trigger.physics}

The final physics triggers are summarised in Table~\ref{tab:trigger.triggers}
together with their typical rates. A standard physics trigger is generated by a
combination of the beam trigger, the veto system, and one of the specialised
triggers described above.

The diffractive trigger (DT0) is the main physics trigger for spectroscopy data
taking.  Based on the proton trigger, it selects events with recoiling protons
from the target.  Besides the low angle cut-off of the beam killers, DT0
introduces only a minimum bias on the angular acceptance of forward particles.

The "low-$t$" triggers, LT1 and LT2, are especially important for measurements
with solid-state targets. In this case, the recoil proton has to pass dense
material that can lead to large uncertainties due to multiple scattering or
protons stopping in the material. This means that for heavy targets the DT0
trigger is not as efficient as for the hydrogen target.  Therefore, a large part
of the solid-state target data is recorded with prescaled LT1 and LT2.

The Primakoff trigger (Prim1) uses the calorimeter trigger with a $60\,\GeV$
threshold.  A secondary Primakoff trigger (Prim2) is based on a calorimeter
trigger with a threshold of $40\,\GeV$ and a prescaling factor of two. Its
purpose is to monitor the Prim1 trigger threshold.

The kaon trigger (KT) makes use of the CEDAR detectors in the beam line, which
are set to detect beam kaons (see \Secref{sec:beam.cedar}). Signals from both
CEDARs need to be present for the trigger in order to maximise its purity. It is
used as a kaon-enriched beam trigger for luminosity monitoring via $K
\rightarrow 3\,\pi$ decays and for systematic studies.

Further auxiliary triggers are set up for monitoring purposes, systematic
studies and alignment purposes (see \Secref{sec:reconstruction.alignment}). They
include an additional beam trigger with a transverse acceptance of $3.9 \times
3.9$\,cm$^2$, which is required for the alignment procedure. The Veto Inner
trigger and Halo triggers make use of the hodoscope veto system to detect
straight halo tracks for muon data taking, which is utilised in the alignment
procedure, as well.

All inputs to the trigger system and the signals of the individual sub-triggers
and triggers themselves are monitored with TDCs and scalers. In addition, the
individual signals of the multiplicity counters are monitored by sampling ADCs.

\begin{table*}[tbp]
  \centering
  \caption{Overview of trigger subsystems, vetos and physics triggers used for data taking.}
  \scriptsize
  \begin{tabular*}{\textwidth}{@{\extracolsep{\fill}}lll} \hline
{\bf Trigger subsystem} & {\bf Logical composition}                                        & \\ 
\hline
Beam trigger (BT)       & SciFi1 $\wedge$ beam counter                                     & \\
Beam killer veto        & beam killer 1 $\wedge$ beam killer 2                             & \\
Veto                    & Sandwich $\vee$ veto hodoscopes $\vee$ beam killer               & \\
Proton trigger          & see Eq.~\ref{eq:proton-trigger}                                  & \\
Multiplicity trigger MT1& 1 (later 2) el. of outer ring counter                            & \\
Multiplicity trigger MT2& amp. inner disk $>$ 1.6~MIPs (later 2.5~MIPs)                    & \\
Calorimeter trigger     & $\sum_{12\times 12}\mathrm{cell\ amplitude}>\mathrm{threshold}$    & \\
CEDAR trigger           & CEDAR1 multiplicity $\wedge$ CEDAR2 multiplicity                 & \\
\hline  
{\bf Physics trigger}   & {\bf Logical composition}                                        & {\bf Rate / 10\,s spill} \\ 
\hline
Diffractive trigger DT0 & BT $\wedge$ proton trigger $\bar{\wedge}$ veto                   & 180k        \\
Low-$t$ trigger LT1     & BT $\wedge$ MT1 $\bar{\wedge}$ veto                              & 370k (140k) \\
Low-$t$ trigger LT2     & BT $\wedge$ MT2 $\bar{\wedge}$ veto                              & 620K (260K) \\
Primakoff trigger Prim1 & BT $\wedge$ calorimeter trigger ($>60$\,GeV) $\bar{\wedge}$ veto & 260k        \\
Primakoff trigger Prim2 & BT $\wedge$ calorimeter trigger ($>40$\,GeV) $\bar{\wedge}$ veto & 450k        \\
Kaon trigger KT         & BT $\wedge$ CEDAR trigger$\bar{\wedge}$ veto                     & 30k
\\
    \hline
  \end{tabular*}
  \label{tab:trigger.triggers}
\end{table*}

\section{Data Acquisition}
\label{sec:daq}
The COMPASS data acquisition system (DAQ) has been designed to cope with high
trigger rates and large data flow. For data taking with hadron beams both
interaction rate and particle multiplicity per interaction are higher than for a
muon beam, making these requirements even more important. The DAQ is based on a
pipelined architecture, which was fully implemented for the 2008/2009 data
taking. It is complemented with a Detector Control System (DCS), which
permanently monitors all parameters relevant for the operation of the setup. The
general structure of the COMPASS DAQ and DCS systems were described in
Ref.~\cite{Abbon:2007pq}.  In this section their main characteristics are
shortly reviewed; only the most important improvements and modifications are
discussed.

\subsection{General structure of the COMPASS data acquisition system}
\label{sec:daq.general}
\begin{figure}[tbp]
  \begin{center}
    \includegraphics[width=\columnwidth]{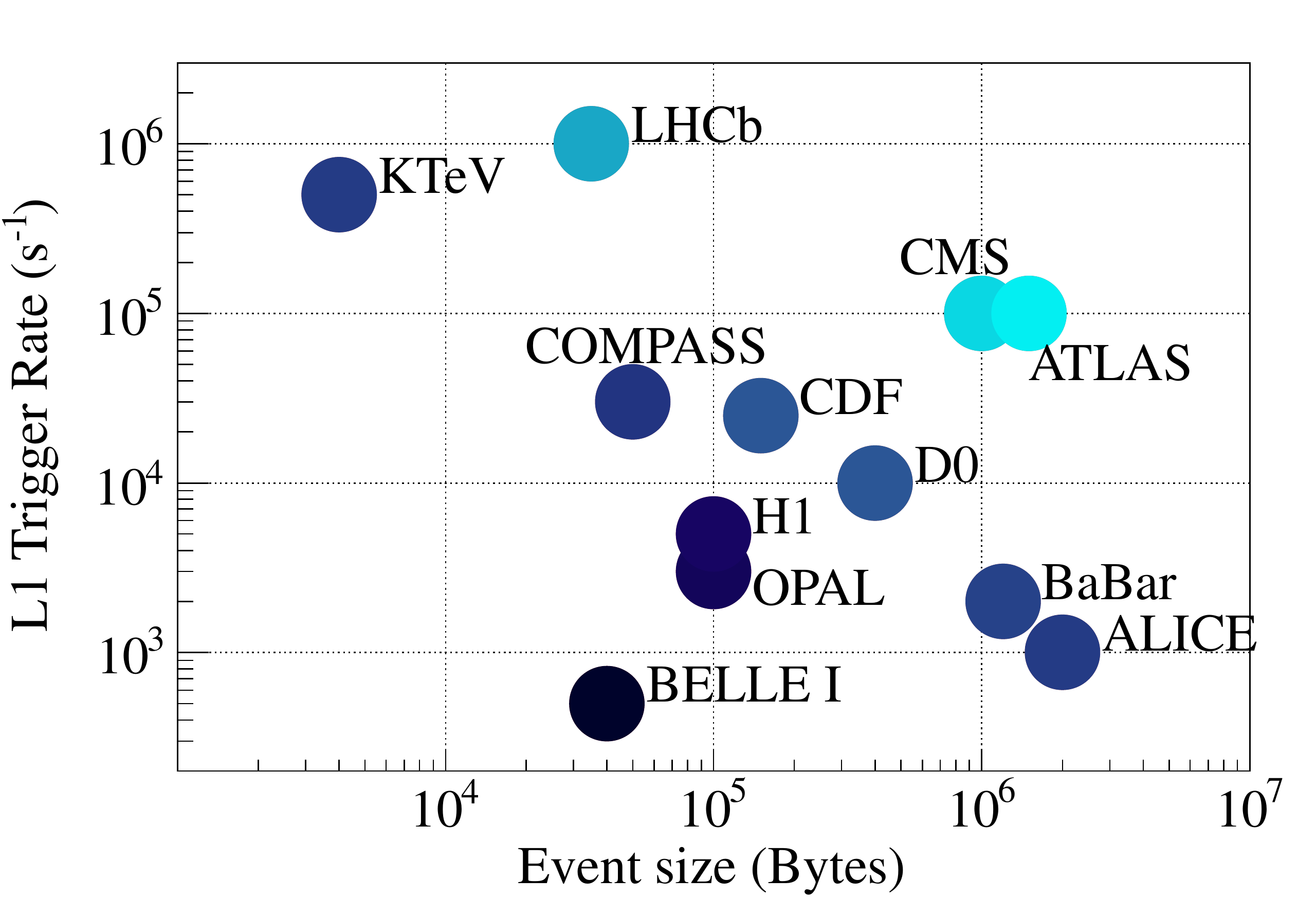}
  \end{center}
  \caption{\small Trigger rate versus event size. The COMPASS DAQ system is
    compared to several large-scale experiments. The comparison is done for
    first-level (L1) triggers or their equivalent.}
  \label{fig:daq-ratevssize}
\end{figure}

For typical hadron beam intensities of up to $5\cdot 10^7$ particles per spill,
the various COMPASS triggers combine to a total trigger rate of more than
$30~$kHz. The overall number of electronic channels is larger than $\sim
250~000$, and the generated event size has a mean value of $40~$kB.
Accordingly, a data rate of up to 1.2 GB/s is acquired during the $9.6\,\s$ long
SPS spill.  When averaged over a typical SPS accelerator `supercycle" of
$42\,\s$ (duty factor of 23\%), these data generate a flow of about
$270\,$MB/s. Note that as a function of the actual SPS accelerator usage, the
length of the supercycle may change by several seconds.  As a consequence, the
instantaneous data flow of the COMPASS readout system must be able to reach peak
throughput values of up to $350\,$MB/s.  In comparison with other large-scale
experiments, the COMPASS data acquisition system operates with relatively
moderate event sizes, but features high trigger rates
(Fig.~\ref{fig:daq-ratevssize}).

\begin{figure}[tbp]
  \begin{center}
    \includegraphics[width=\columnwidth]{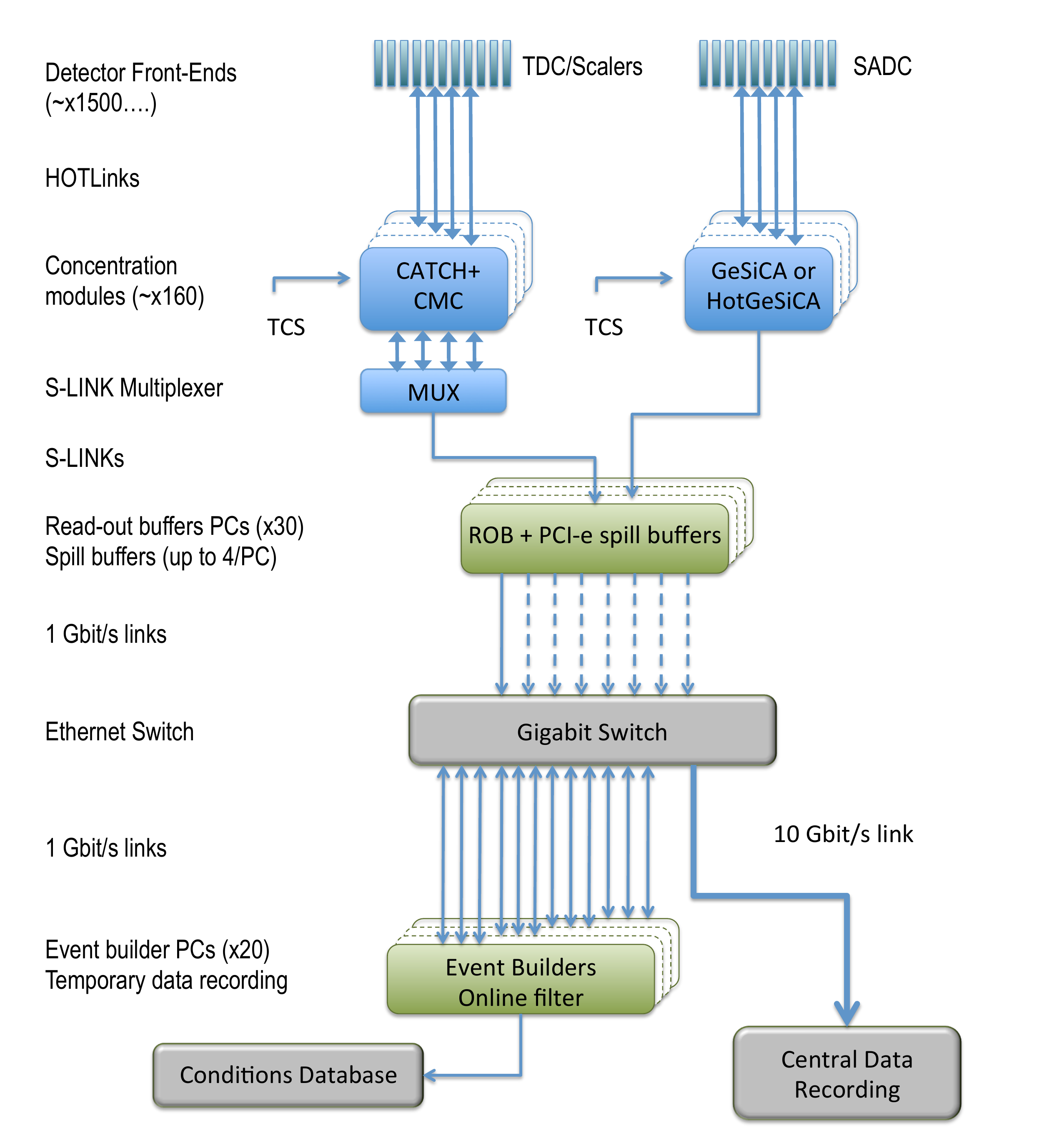}
  \end{center}
  \caption{\small Overview of the COMPASS DAQ system. Data coming from the
    detectors are first digitised in the front-end cards and then merged in the
    concentrator modules, either CATCH or GeSiCA(HotGeSiCA).  The data from the
    concentrator modules are first sent to the Readout Buffers and then
    transmitted to the Event Builders. The data are temporarily saved on disk,
    before being migrated to the Central Data Recording facility.}
  \label{fig:daq-overview}
\end{figure}

The DAQ system is controlled by the Trigger Control System (TCS). The TCS
synchronises the elements of the front-end electronics by providing a common
low-jitter system clock of 38.88 MHz, an encoded trigger signal, and event
identification information. More than 150 data concentrator modules (called
readout-driver modules in Ref.\,\cite{Abbon:2007pq}) receive this information
and distribute it further to the front-end cards.  The system has a star-like
network topology implemented using optical network elements and passive
splitters. The system clock is used for front-end pipeline data processing and
time measurements.

The overall structure of the COMPASS DAQ system is shown in
Fig.~\ref{fig:daq-overview}. The detector signals collected by the front-end
electronics are digitised in either TDC or Sampling ADC (SADC) cards. The data
coming from the digitisation cards are then transmitted to the concentrator
modules through a HOTLink interface \cite{Cypress}. Two types of concentrator
modules were specifically developed for the COMPASS detectors: CATCH and GeSiCA,
including a more recent version, called HotGeSiCA.  All modules are housed in
VME crates.  Spill and event numbers, reference clock and synchronisation
signals are provided to the concentrator modules by an optical link coming from
the TCS.  The HOTLink interface is also used to transmit this information to the
front-end cards.

Each 9U CATCH module \cite{Fischer:02a} houses four CATCH mezzanine cards (CMC)
that receive detector signals coming from Micromegas, scintillating fibres, wire
chambers, and hodoscope detectors.  The CATCH firmware merges the data from the
mezzanine cards and transmits them to the central readout buffer computers
through a S-LINK optical link \cite{Bij:96}.

For detectors with low occupancies the data of up to 4 CATCH modules are
multiplexed by an S-LINK MUX card before being transmitted to the readout
computers. The S-LINK MUX card houses one S-LINK source card and is mounted on
P3 connectors on the backside of a VME crate.

The 9U GeSiCA modules can read up to four 12-channel GEM or Silicon SADC cards.
The more recent 6U version named HotGeSiCA is also able to read SADC, MSADC, and
APV data from the RICH and PixelGEM detectors. Although smaller in size, the
HotGeSiCA module has eight RJ45 or optical HOTLink ports, instead of four for
GeSiCA.  In addition, the HotGeSiCA module can be equipped with 500 MB of memory
and a HOTLink output interface for cascading HotGeSiCA modules and concentrating
the data from up to 64 front-end cards.  This readout scheme is used for the
Rich Wall detector (see Fig.~\ref{rw_readout}).  Similar to the CATCH modules,
the GeSiCA and HotGeSiCA modules send data through a S-LINK interface to the
readout buffers, while the information coming from the TCS receiver is
transmitted to the front-end electronics through a HOTLink connection.
 
The DAQ system is composed of two main types of computers, called Readout
Buffers (ROB) and Event Builders (EB). All computers run Linux operating system
(see Table~\ref{tab:daq.hardware}).  Each ROB is equipped with up to four
spill-buffer PCI cards. The PCI cards collect the information from the
corresponding concentrator modules via S-LINK optical fibres. The data are
temporarily stored in a daughter SDRAM card of 512~MB or 1~GB memory during the
spill, before being fully transmitted to the computer.  All ROBs are connected
to all EBs through a Gigabit Ethernet interface. The role of each EB is to build
a complete event using the information from all the ROBs, to split the data into
files of 1~GB each (chunks), and to store these files on its internal disks. In
addition, the EBs run an on-line filtering software used for both data filtering
and data quality monitoring. The software used for the data acquisition is the
DATE acquisition framework \cite{DATE}, developed for the CERN experiment ALICE.

From the EBs disks, the data files are transferred to the CERN computer centre
into the CASTOR hierarchical storage system \cite{castor}.  Files are copied to
CASTOR disk pools by multiple TCP/IP streams through a $10\,$GB/s optical link
and then stored to tapes. Up to $20\,$TB of data per day can be stored on tape
when the experiment is running.

\subsection{Trigger rate and dead time settings}
In order to achieve high trigger rates with reasonable dead times, several
improvements of the data acquisition chain were necessary.  The dead time, as
defined by the Trigger Control System, depends on three minimum time
intervals. These are: the time interval between two consecutive triggers, the
time interval for three successive triggers, and the time interval for ten
successive triggers. During data taking with a muon beam these values were set
to $5$, $75$ and $250\,\mus$, respectively.

For data taking with hadron beams, the minimum time between two consecutive
triggers was decreased to $3\,\mus$. Smaller values were prevented due to a
noise correlated with the previous trigger and appearing on the front-end cards
of the MWPC detectors. The minimal time interval for no more than three triggers
was set at $30\,\mus$. Smaller values could lead to an overflow in the TDC
multi-event buffer.  A third minimal time interval is required by the internal
buffer of the APV chip, which can store up to 10 events. This time is set
according to the speed of the analogue output signal sent by the APV chip to the
SADC cards. Analogue values are sent by the APV chip at a frequency of
$20$\,MHz, corresponding to a digitisation time (including overheads) of
$21\,\mus$ per event. More than $200\,\mus$ are then necessary to read ten
events, justifying the $250\,\mus$ limit.  Dedicated tests have shown that in
the future the ten events interval can be reduced to $125\,\mus$ if the APV
read-out frequency is increased to $40$\,MHz and the SADC card firmware is
modified accordingly.

\begin{figure}[tbp]
  \begin{center}
    \includegraphics[width=0.8\columnwidth]{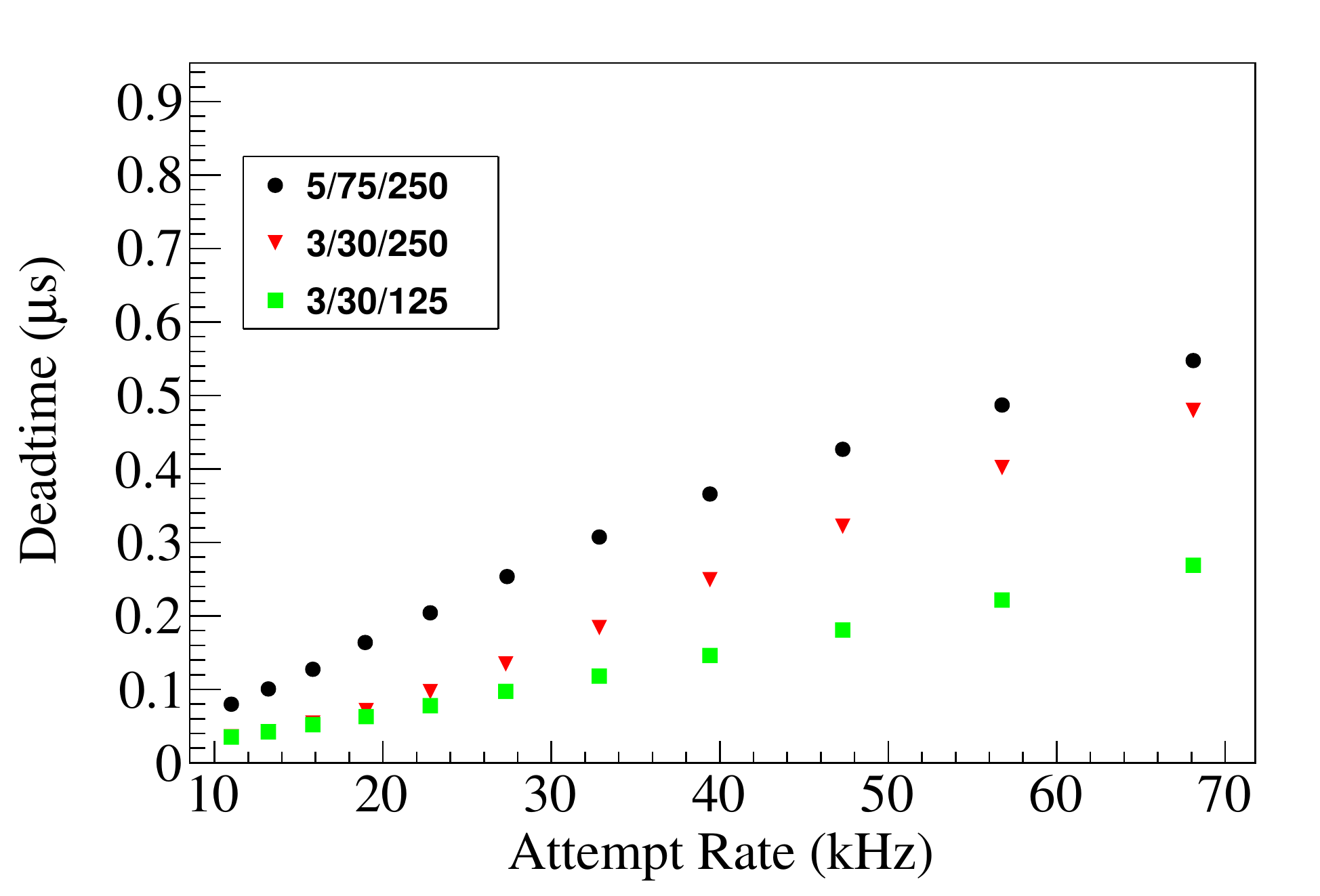}
  \end{center}
  \caption{\small Data acquisition dead time for three different TCS settings,
    as measured as a function of the attempted trigger rate. The settings used
    in 2008/2009 are shown in red triangles.}
\label{fig:daq-deadtime}
\end{figure}

Figure~\ref{fig:daq-deadtime} shows a comparison between the dead times
resulting from the different trigger settings. The improvement of the time
interval for three triggers results in a significant decrease of the dead
time. For the nominal hadron data taking trigger rates of $30$~kHz the new
settings generate a dead time of 15\% instead of 28\%.

\subsection{DAQ hardware and database architecture}
\label{sec:daq.db}

\begin{table*}[tbp]
  \centering
  \caption{Summary of the COMPASS data acquisition hardware, as used in 2009.
    Dual and quad-core processors are indicated in the parenthesis near the
    processor name. Both memory and disk sizes are given in GB. }
  \begin{tabular*}{\textwidth}{@{\extracolsep{\fill}}lrlrr} \hline
    {\bf Service}      & {\bf Nb} & {\bf Processor}              & {\bf Memory size} & {\bf Disk size} \\ 
\hline
    Event builders     & 12       & Xeon, 3~GHz ($\times$2)      & 4                 & 1100 \\ 
                       & 8        & Xeon, 2.5~GHz ($\times$4)    & 4                 & 5600 \\ 
    Read-out buffers   & 5        & Xeon, 2~GHz ($\times$4)      & 4                 &  500 \\ 
                       & 8        & Xeon, 3~GHz ($\times$2)      & 4                 &  250 \\ 
                       & 16       & Pentium-3, 866~MHz           & 1                 &   18 \\ 
    File servers       & 2        & Xeon, 3~GHz ($\times$2)      & 4                 & 1100 \\ 
    Gateways computers & 2        & Pentium-4, 3~GHz ($\times$2) & 2                 &   82 \\ 
    Database servers   & 3        & Xeon, 3~GHz ($\times$2)      & 4                 & 1100 \\ 
    Run control        & 4        & Pentium 4, 3~GHz ($\times$2) & 2                 &   82 \\
    Front-end CPUs     & 26       & Celeron, 336~MHz             & 0.256             &      \\
\hline
\end{tabular*}
\label{tab:daq.hardware}
\end{table*}

The COMPASS DAQ system operates with a large number of hardware
(Table~\ref{tab:daq.hardware}) and software components, which are controlled
through various parameters. These parameters are produced by on-line processes,
operator entries, slow-control of the detectors, or result from specific run
conditions. Together with other monitored quantities, such as trigger rates, run
and spill information, they are stored in several MySQL \cite{MySQL:05a} and
Oracle \cite{Oracle} databases.

The \textit{front-end configuration database} incorporates all information
relevant to the front-end modules and processors. The \textit{logbook database}
collects a large number of experimental parameters and operator comments
relevant for a specific data-taking period, usually defined as a run. Parameters
that are likely to change more frequently, such as beam information, beam line
settings, scalers and some monitoring values, are stored in a \textit{spill
  database}.  The \textit{DATE configuration database} is used for the
description of the on-line computers (ROBs, EBs, FSs,..) as well as for
parameters relevant for the data acquisition. This database also includes the
configuration of the on-line filter software. Another database, called
\textit{DATE message log database}, collects all process logs and messages.

The $130\,$GB of data from all MySQL databases are hosted on two physical
servers, synchronised through a master--master replication. Clients connect to
the database through a virtual address pointing to a third server. The third
server runs a MySQL Proxy software\,\cite{Jary:10a} that monitors the
communication between the client and the database. Besides the proxy, the third
server also hosts a web server Apache \cite{Apache:10a} and a monitoring service
(Nagios) \cite{Nagios:10a}. The web service provides interfaces to run logbook,
database administration programs, and diagnostic tools. Nagios monitors the
availability of database servers and the state of replication.

A specific database table is used for the Detector Control System (DCS),
e.g.\ for monitoring of ECAL1 and ECAL2 modules. On the other hand, read-out
values of parameters obtained by the DCS independently are copied to a dedicated
database table.

Full MySQL database backup is being executed regularly, whereby the binary log
that is created during the replication is regarded as incremental
backup. Furthermore, the databases are periodically replicated into the CERN
computer centre.

\subsection{Detector Control System}
\label{sec:daq.dcs}
The Detector Control System (DCS)~\cite{Bordalo:12} collects data from the
various detectors, hardware devices, and data acquisition elements with
programmable reading cycles.  For the COMPASS experiment the actual cycle times
range between 2 seconds and 30 minutes.  It provides a user-friendly interface,
which is used to set remotely most parameters relevant for operating the
experimental setup. When predefined conditions are met, namely if monitored
values go beyond predefined thresholds or settings, it displays on-line warnings
and alerts in the user interface, sounds acoustic alarms in the control room,
notifies predefined recipients by SMS and email and, when necessary, switches
off sensitive detector channels.  All values and alerts are stored in a
centralised Oracle database with a frequency of typically few minutes per
monitored parameter.  Queries on the database are executed regularly for storage
of data, or on demand.

The DCS architecture consists of three layers: the supervisory layer, the
front-ends layer and the devices layer. The supervisory layer of the DCS is
based on a commercial SCADA system (Supervisory Control and Data Acquisition),
PVSS-II~\cite{PVSS:05}, adopted by CERN. On top of PVSS-II, a package of
software tools called Joint COntrol Project (JCOP) Framework~\cite{JCOP} is also
used.  Developed at CERN, this package is specific for high-energy physics
applications.  The front-ends layer includes the drivers necessary for the
hardware devices and provides the communication protocol between the supervisory
layer and the devices layer. The devices layer comprises all hardware elements
and sensors.

The system is flexible enough to easily incorporate new detectors and monitored
parameters.  For the data taking with hadron beams in 2008/2009, a number of new
detectors were included in the system: RPD, Sandwich veto, Multiplicity Counter,
Beam Killers, Beam Counter, PixelGEMs, liquid hydrogen target, and the new
Silicon detectors with their cryogenic devices. In addition, the monitoring of
the CEDARs and the two electromagnetic calorimeters was considerably improved.

New high voltage and low voltage channels and VME crates were used and
integrated in the DCS for the hadron data taking. They are monitored and
controlled by OPC or DIM servers~\cite{OPC,DIM}, with which they communicate by
use of the CAN\footnote{Controller Area Network. ISO standard 11898, see
  e.g.\ www.iso.org.} and CAENet field buses~\cite{CAEN}.

The monitoring of CEDAR parameters (pressures, temperatures, HVs and motors) is
done via a DIP server maintained by CERN~\cite{DIP}. The ratio of the pressure
to the temperature is calculated for every spill. If the value is found outside
the appropriate range, a warning signal requesting a correction of the CEDAR gas
pressure is generated.

The monitoring of the two electromagnetic calorimeters required a substantial
extension in the number of monitored channels, namely 1500 for ECAL1 and 3068
for ECAL2. Between spills, the calorimeter modules of ECAL1 and ECAL2 are
flashed by a laser and LED light pulses, respectively (see
\Secref{subsec:pid.ecal.ecal1.calib} and \Secref{sec:pid.ecal.ecal2.calib}). The
DAQ on-line filtering software collects the responses from all modules,
calculates the average amplitudes for each spill, and stores them in the
conditions database. The DCS reads them, compares them to the reference values,
and defines its state of alert.  The voltage and the current of the powering
system of ECAL1 and ECAL2, the power supply of the LED monitoring system and the
status of the laser of ECAL1 were added to the list of controlled parameters.

\begin{figure*}[tbp]
  \begin{center}
    \includegraphics[width=\textwidth]{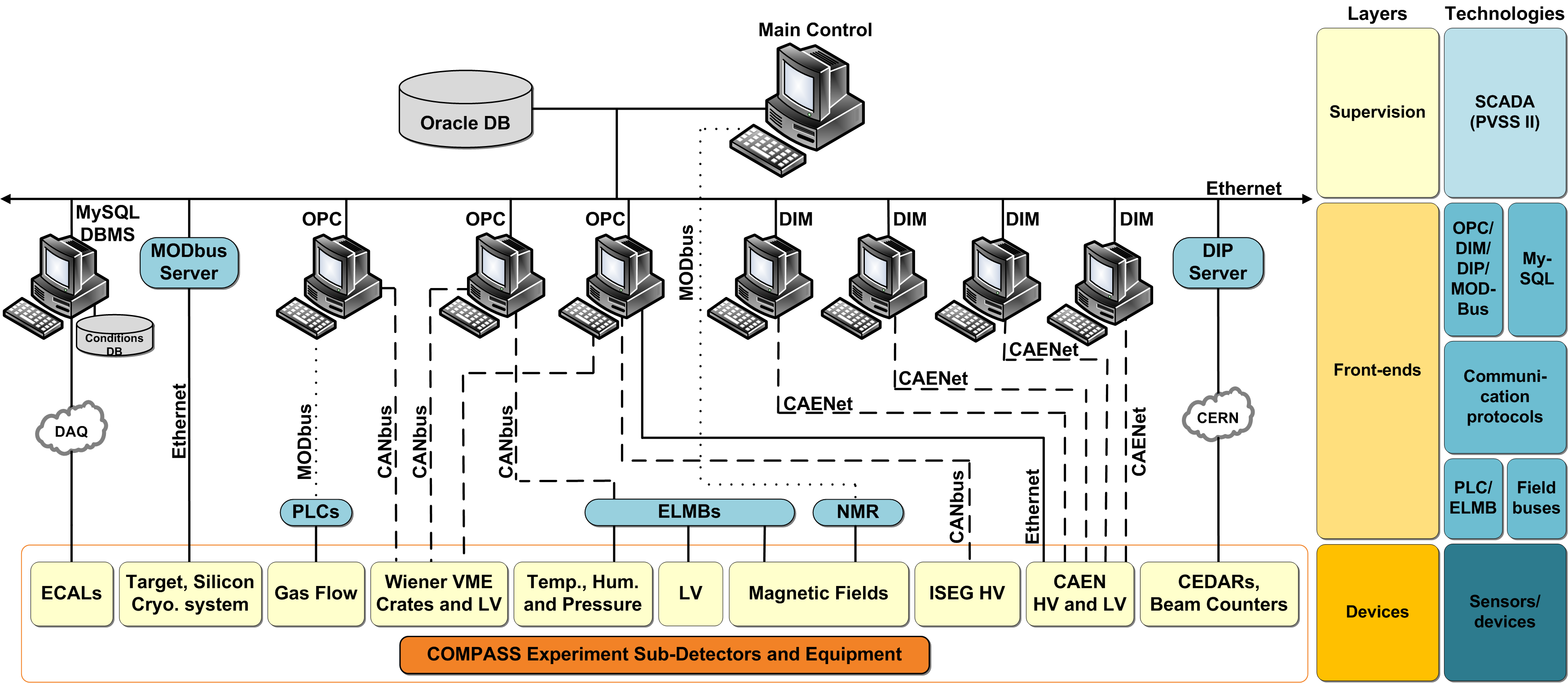}
  \end{center}
  \caption{Implementation of supervision, front-end and device layers of the
    Detector Control System.}
  \label{dcs.details}
\end{figure*}

The monitoring of the liquid hydrogen target and of the cryogenic system of the
Silicon detectors is done using a dedicated Programmable Logic Controller (PLC)
and a control system. Pressures, temperatures, vacuum gauge values and liquid
levels are transmitted to the DCS using a MODbus~\cite{MODbus} server.

The data generated by the DCS are temporarily saved on a local disk whose
contents are transmitted to a centralised CERN Oracle database with a cycle of a
few seconds. Parameters that are relevant for the physics analysis are regularly
copied to a COMPASS MySQL database or are provided to users as ROOT trees or
ASCII files.

\section{Event Reconstruction and Detector Performance}
\label{sec:reconstruction}
The main reconstruction software is called CORAL (COmpass Reconstruction and
AnaLysis); it is detailed in Ref.~\cite{Abbon:2007pq}.  CORAL outputs mDST
(``mini Data Summary Trees'') files that contain reconstructed events and
information related to the detectors.  The mDST files are organized in tree-like
structures based on the ROOT~\cite{Brun:05a} package.  The information stored in
the mDST is analysed using a dedicated software called PHAST (PHysics Analysis
Software Tool).  Besides accessing the mDST data, PHAST provides an environment
for physics analysis and includes tools for mDST creation, further data
processing and filtering.  The event reconstruction in CORAL comprises all
detectors except the RPD and the CEDARs, which are included at the PHAST level,
i.e. during the mDST processing stage.

This section describes the tracking method, the detector alignment procedure,
and the vertex reconstruction technique used in the analysis. It gives also
details on the analysis flow for detectors introduced into or upgraded for the
hadron setup, namely for RPD, RICH-1, CEDARs, and ECALs.

\subsection{Track Reconstruction}
\label{sec:reconstruction.tracking}
 
 The track reconstruction software reconstructs trajectories of charged
 particles, thereby determining such properties as their momentum and total
 radiation length traversed.  It uses the measurements from the tracking
 detectors and combines them with the description of the magnetic fields and
 material distribution in the setup.  For the material distribution the ROOT
 geometry package~\cite{Brun:2003xr} is used.  The included detectors are all
 tracking detectors, trigger hodoscopes and the beam telescope (SciFi and
 Silicon detectors before the target in Fig.~\ref{fig:layout_setup}).  Prior to
 the reconstruction process, a time cut relative to the trigger time is applied
 to all hits.

The track finding algorithm is subdivided into two steps. First, it searches for
straight track segments in the zones that are free of strong magnetic field or
of large material thickness. In these zones the particle trajectories can be
approximated by straight lines.  In a second step, called ``bridging", the
straight track segments from different zones are combined over dipole magnets
and hadron absorbers.  In order to account for the deviation from a perfect
straight line, as caused by fringe fields and multiple scattering, an iterative
approach is used with progressively wider search roads. This approach is
motivated by the idea of first solving the case of the straighter tracks, which
have higher momenta, and turning to the more difficult case of the
lower-momentum tracks only after the hits used in the first iterations have been
removed from the search procedure. This scheme still yields many ghost tracks,
particularly in the later iterations that have wider ``roads". To filter these
out, candidate tracks are checked against a lookup table of all tracks within
the accessible phase space, which was produced in a dedicated MC simulation.

 The track fitting procedure is based on a Kalman
 filter~\cite{Fruhwirth:1987fm}.  It comprises the treatment of multiple
 scattering, which is based on a prior estimate of the track momentum.  A
 ``forward" fit, which starts from the most upstream tracking plane, gives the
 best estimate of the track parameters in the plane of the detector with the
 last hit. A ``backward" fit, which starts from the most downstream tracking
 plane using the same hits, gives the best estimate at the first hit.  A process
 combining the results of the two fits, which is known in the Kalman formalism
 as ``smoothing", is used to determine the local best estimate at any position
 along the track. For each of these estimates, the procedure also determines the
 uncertainty in terms of the covariance matrix of the parameter vector.  Outlier
 detection and elimination as well as the resolution of left/right ambiguities
 in drift detectors are also done within this framework.

For the data taken with hadron beams, the sequential three-step structure of
straight zone track finding, bridging and fitting was adapted to the high rate.
A search for straight track projections in the vertical plane spanning several
zones of the spectrometer is attempted at an early stage. A re-evaluation of the
hit patterns is undertaken after bridging, once the momenta are fairly well
known and the number of competing candidate tracks is reduced.  These
modifications improve the tracking through the drift chambers that are located
in high fringe fields around the SM1 magnet and in the Silicon vertex detectors
that are placed in the close vicinity of the target and are densely packed with
hits.
 
The event reconstruction was further improved by allowing some update of the
track information after vertex reconstruction (see
\secref{sec:reconstruction.vertex}).  If the weighted mean time of the hits
associated to the incident particle of a primary vertex deviates from the
trigger time more than expected by statistical fluctuations, in the order of a
few ns, the latter is re-evaluated. After correcting it and updating the hits in
the drift detectors, the tracks are either refitted or the search is restarted
from scratch.

 For the evaluation of the performances of the tracking package, a GEANT3-based
 simulation of the COMPASS setup is used, as described in
 \secref{sec:performance.acceptance}. The evaluation is based on criteria of
 association, reconstruction, and reconstructibility. A track is associated to a
 MC particle if more than a fraction $f$ of the hits originate from this
 particle. Here, pions decaying into muons are considered the same single
 particle. A MC particle is declared reconstructed if a track fulfilling some
 {\it ad hoc} requirements can be associated to it. Not reconstructible (or not
 worth to be reconstructed) are those particles that fall outside the acceptance
 or are not relevant for the physics process being under study.  In the examples
 presented below, the fraction $f$ is set at 75\%, only primary particles are
 considered, the reconstructed tracks must be bridged over one or both magnets
 and must originate from the primary vertex.

\begin{figure}[h!]
    \centering
    \includegraphics[width=.8\textwidth]{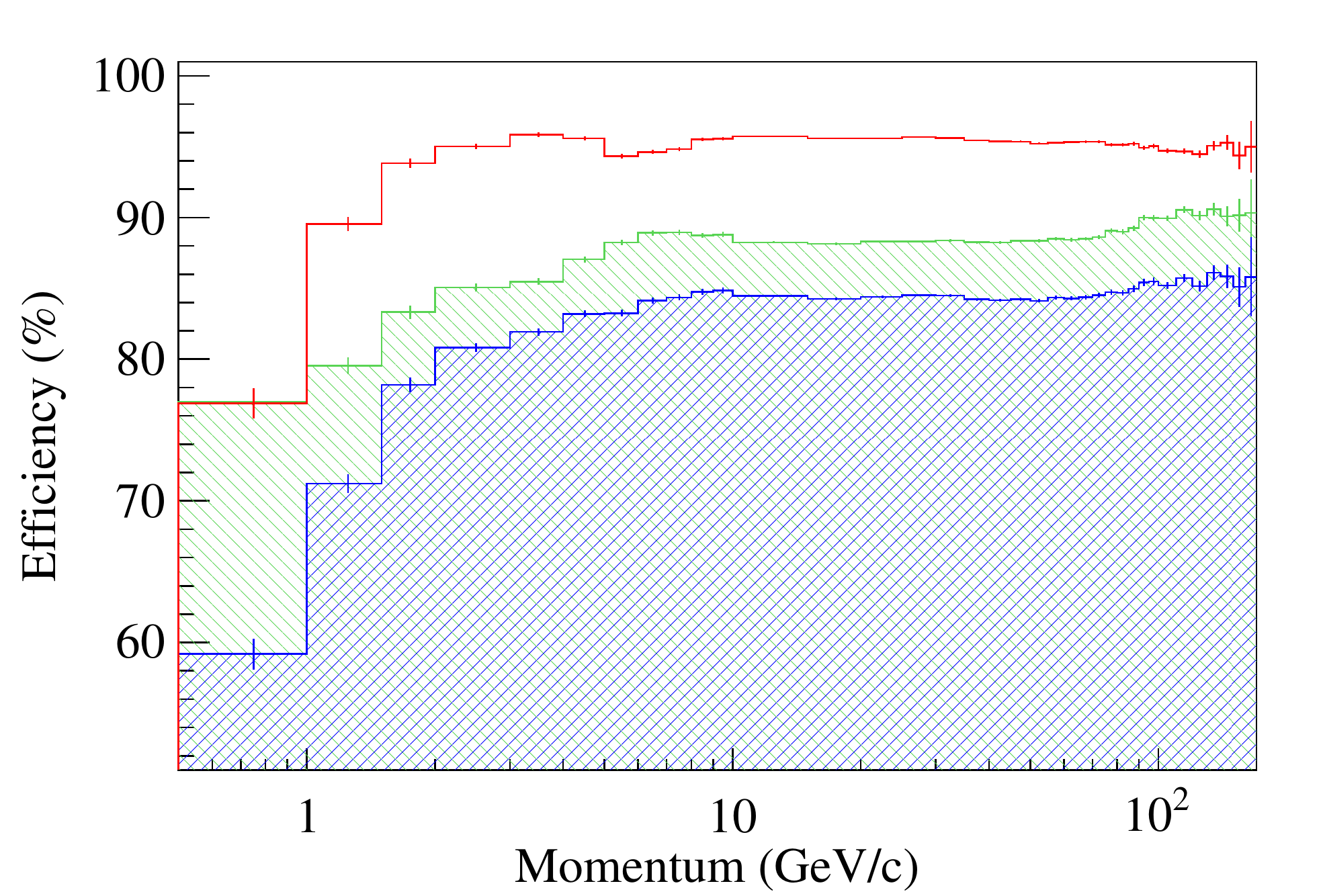}
    \caption{Efficiency of tracking and vertexing as a function of momentum with
      the efficiency of tracking software (red, solid line), setup efficiency
      (hatched, green area), and overall efficiency (crossed, blue area).  }
    \label{fig:track.efficiency}	  
\end{figure}

\begin{figure}[h!]
    \centering
    \includegraphics[width=.8\textwidth]{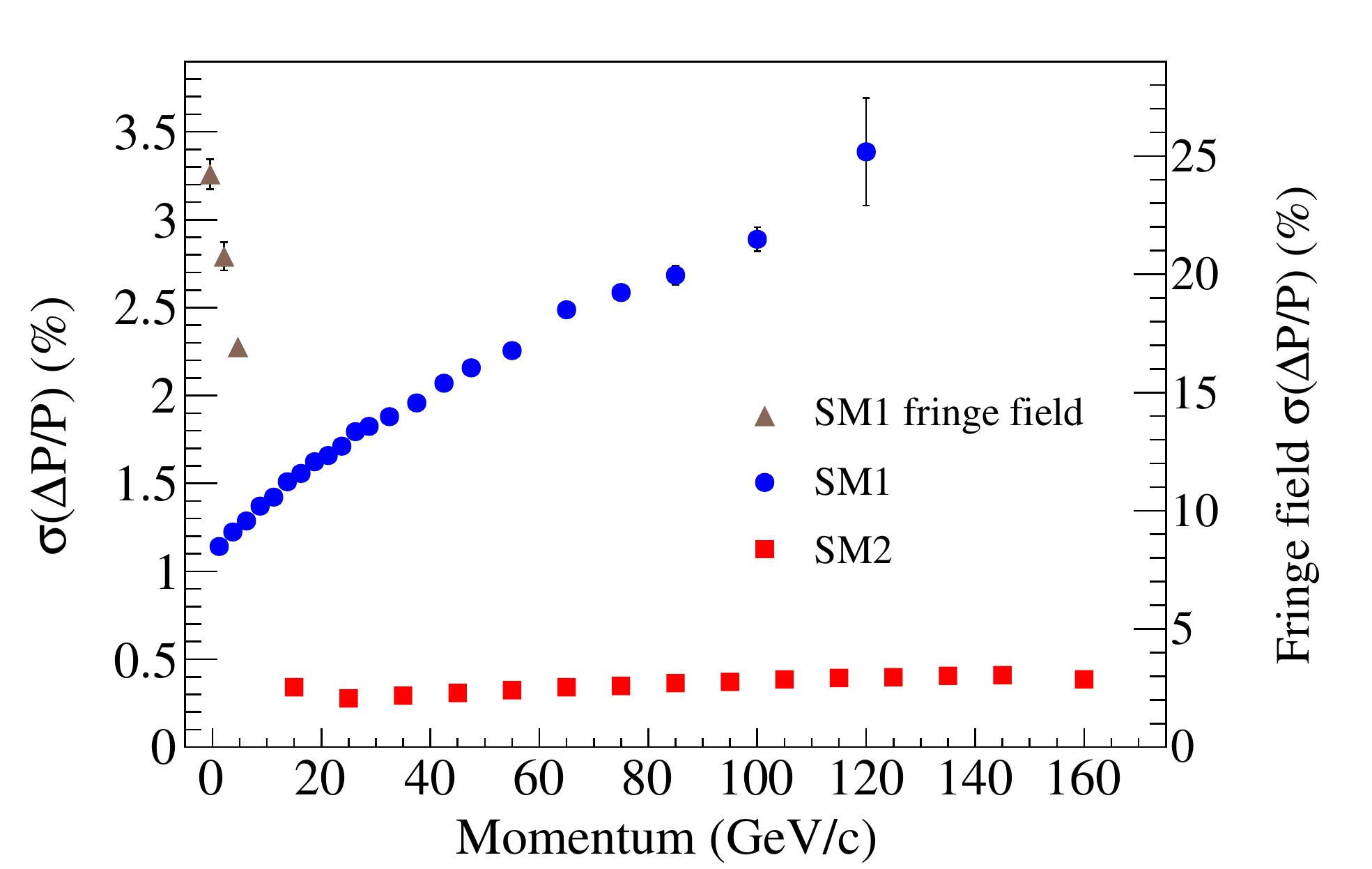}
    \caption{Relative momentum resolution as a function of track momentum.  The
      standard deviation of the reconstruction error is shown for tracks
      deflected by the SM2 magnet alone or by both SM1 and SM2 (squares), by the
      SM1 magnet alone (circles) and for those deflected by the fringe field of
      SM1 only (triangles, right scale).}
    \label{fig:track.resolution}	  
\end{figure}

Several MC samples were evaluated, corresponding to different final states of
the COMPASS hadron programme. They give similar performance values.  As an
example, Figs.~\ref{fig:track.efficiency},~\ref{fig:track.resolution} show the
results for the case of the dissociation of $190\,\GeV/c$ pions on a hydrogen
target into five charged pions with an invariant mass in the range $1\,\GeV/c^2
\leq M(5\pi) \leq 4\,\GeV/c^2$. This process was selected, because it has a
larger angular coverage compared to other processes.

The efficiency is defined as the ratio of number of reconstructed
reconstructible particles over the number of reconstructible ones. With the
criteria defined above, this corresponds to the fraction of primary particles
reconstructed with momentum and connected to a vertex.  The efficiency
represents the combined performance of all involved detectors of the
reconstruction software: a particle may fail to be reconstructed because it
decays, re-interacts, re-scatters, because of inefficiencies of the trackers or
because of deficiencies of the algorithm.  In order to isolate the contributions
of the software, the efficiency that an ideally performing algorithm would reach
is computed in a special mode by exceptionally making use of the Monte Carlo
truth information.  The usual track finding steps are bypassed and the track hit
patterns are determined instead by accumulating hits along the known
trajectories of the generated particles up to a point where multiple scattering
or reinteractions become dominant.  The overall efficiency is then factorised
into this ideal efficiency characterising the setup and a software contribution.
The two factors, as well as their product, are shown as a function of momentum
in Fig.~\ref{fig:track.efficiency}.  While beyond $10\,\GeV/c$ the overall
efficiency is nearly flat, it starts to \ decrease below that value. The
software contribution is stable at 95\% down to about $1.5\,\GeV/c$.

 The momentum resolution is obtained from the statistical distribution of the
 momentum residual for the reconstructed sample. The distribution is first
 binned as a function of the momentum, and in each bin it is fitted with a
 double Gaussian and the average standard deviation is taken as the
 resolution. This is done for particles bridged over SM2 (and possibly also over
 SM1), those bridged over SM1 only, and those only tracked in the fringe field
 of SM1. The latter have a very poor resolution, but can nonetheless be useful
 to reject unwanted final states. The simultaneously obtained angular resolution
 is dominated by the contribution from multiple Coulomb scattering in the target
 material (5\%~$X_0$ in case of the liquid hydrogen target).

\subsection{Alignment procedure}
\label{sec:reconstruction.alignment}
In order to achieve the optimal reconstruction performance, a precise knowledge
of the position and orientation in space of the more than 200 tracking detector
planes of the COMPASS spectrometer is mandatory. In many cases, the geometrical
survey of the experimental setup does not reach a precision that is comparable
to the spatial resolution of the detectors. It is used as the starting point for
an alignment procedure, which uses a sample of reconstructed tracks.  The whole
procedure is done in three steps with different sets of data. Each step is
repeated until the corrections become negligible compared to the detector
resolution.

The first step uses data recorded with a muon beam with the spectrometer magnets
switched off. Therefore, straight trajectories can safely be assumed and all
spectrometer arms including the beam telescope contribute to the reconstruction
of a particle track.  In order to reach a broad illumination of all spectrometer
parts, these data are recorded with a widely defocused muon beam and by using
both beam (see \Secref{sec:trigger.beam}) and veto counters (see
\Secref{sec:trigger.veto}) as trigger.  The alignment is performed by minimising
the total $\chi^2$ of all tracks in the sample, keeping four detector planes
(GM04XY, GM10XY) fixed. For these pivotal points, the positions determined by an
optical survey of the experimental setup has to be used in order to keep the
coordinate system fixed in space. For all other planes, corrections for a
translation along the measured coordinate, a rotation around the beam axis, and
the effective pitch are introduced. The effective pitch takes into account a
possible inclination of the detector plane with respect to the beam axis.  The
position along the beam axis is normally fixed to the position determined by the
geometrical survey; a fit is only attempted if the residual distribution with
respect to the beam axis of a given plane indicates a possible problem.  The
minimisation is done by the Millipede program~\cite{Blobel:02}, which
analytically inverts a large but sparsely populated matrix.

For the second step of the procedure, the detector planes downstream of the
target are aligned with the spectrometer magnets switched on. The magnetic field
not only shifts the positions of the mechanical support of some detectors, but
also influences the internal processes of charge propagation in gaseous
detectors. The effect is strongest for some of the small-area trackers in the
fringe field of SM1 (MM03 and GM01), where the Lorentz-force acting on drifting
and amplified charges results in an apparent translation of the detector planes
of up to $400\,\mum$. For these detectors, a correction in form of an effective
shift is applied, since the distortion is uniform over the active area of the
respective detector within the spatial resolution.

\begin{figure}[h!]
  \centering
  \includegraphics[width=.8\textwidth]{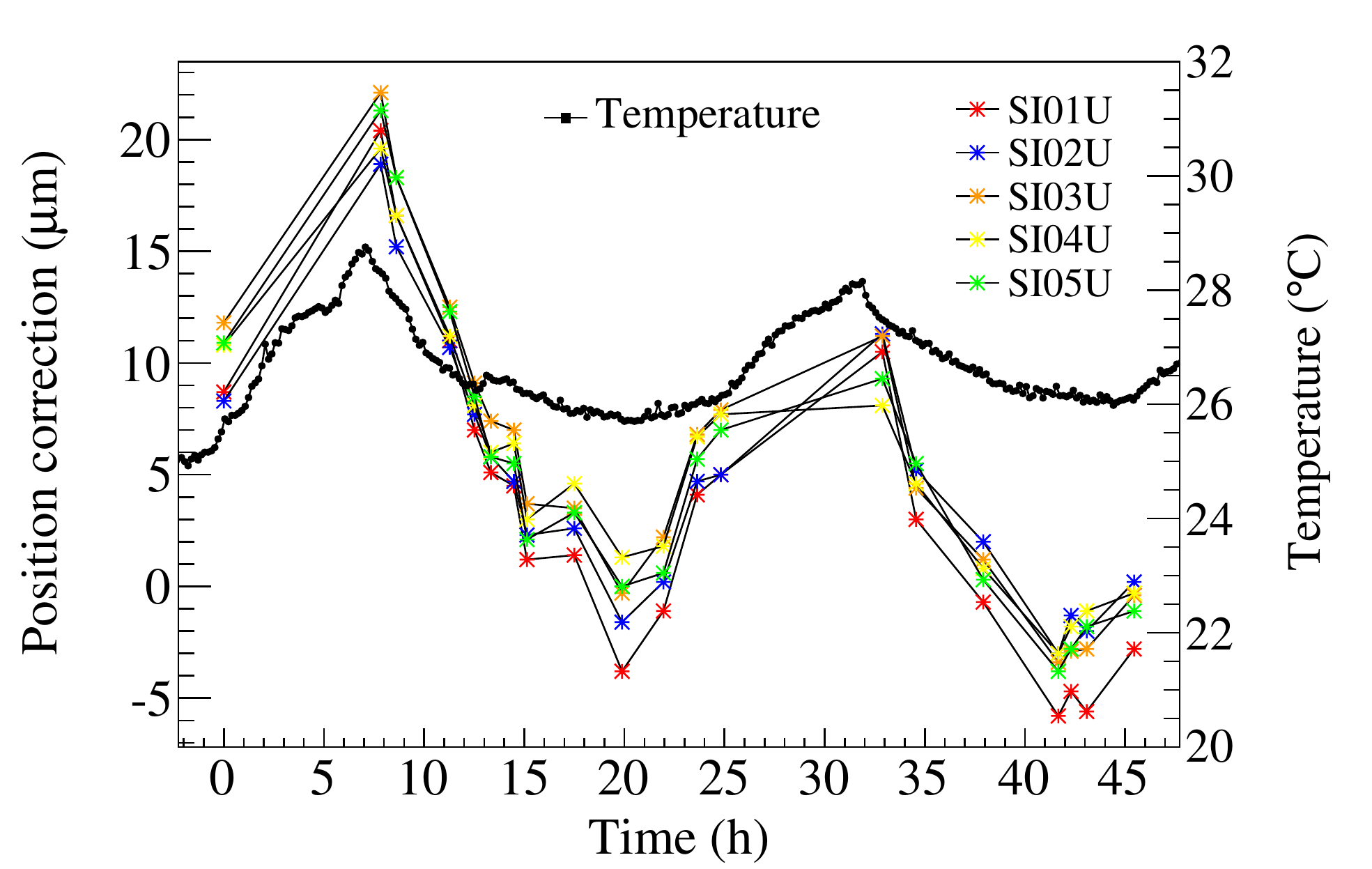}
  \caption{Run-by-run alignment correction applied to the silicon detector
    positions and correlation with ambient temperature.}
  \label{fig:SI_shift}
\end{figure}

As the third step, the beam telescope upstream of the target is aligned with
respect to the spectrometer that is kept fixed in space. This step is essential
to optimise the reconstruction of vertices in the target. As a result, a primary
vertex is reconstructed for up to $90\%$ of the triggered events.

\begin{figure}[h!]
  \begin{minipage}{.48\textwidth}
    \includegraphics[width=\textwidth]{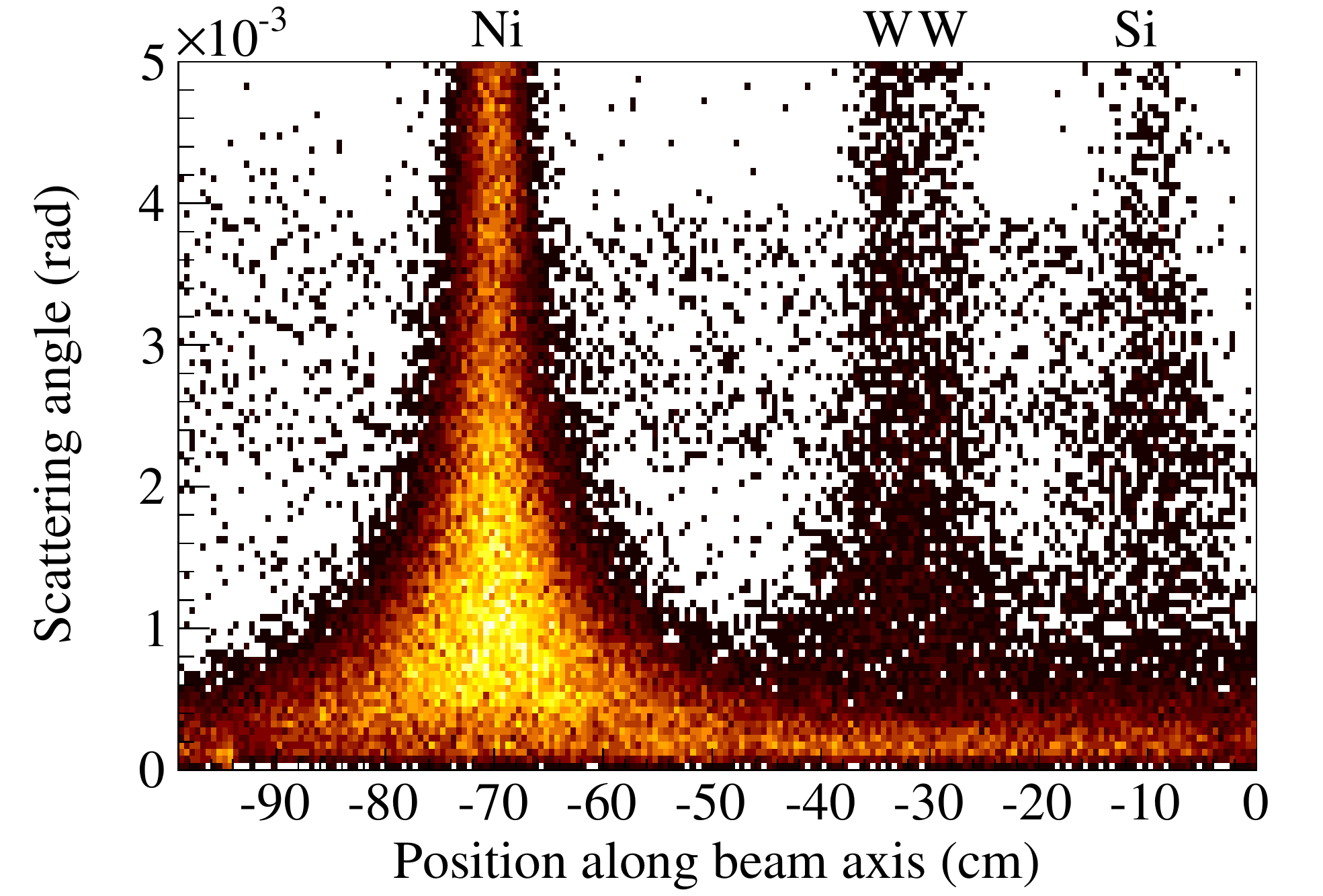}
  \end{minipage}
  \begin{minipage}{.48\textwidth}
    \includegraphics[width=\textwidth]{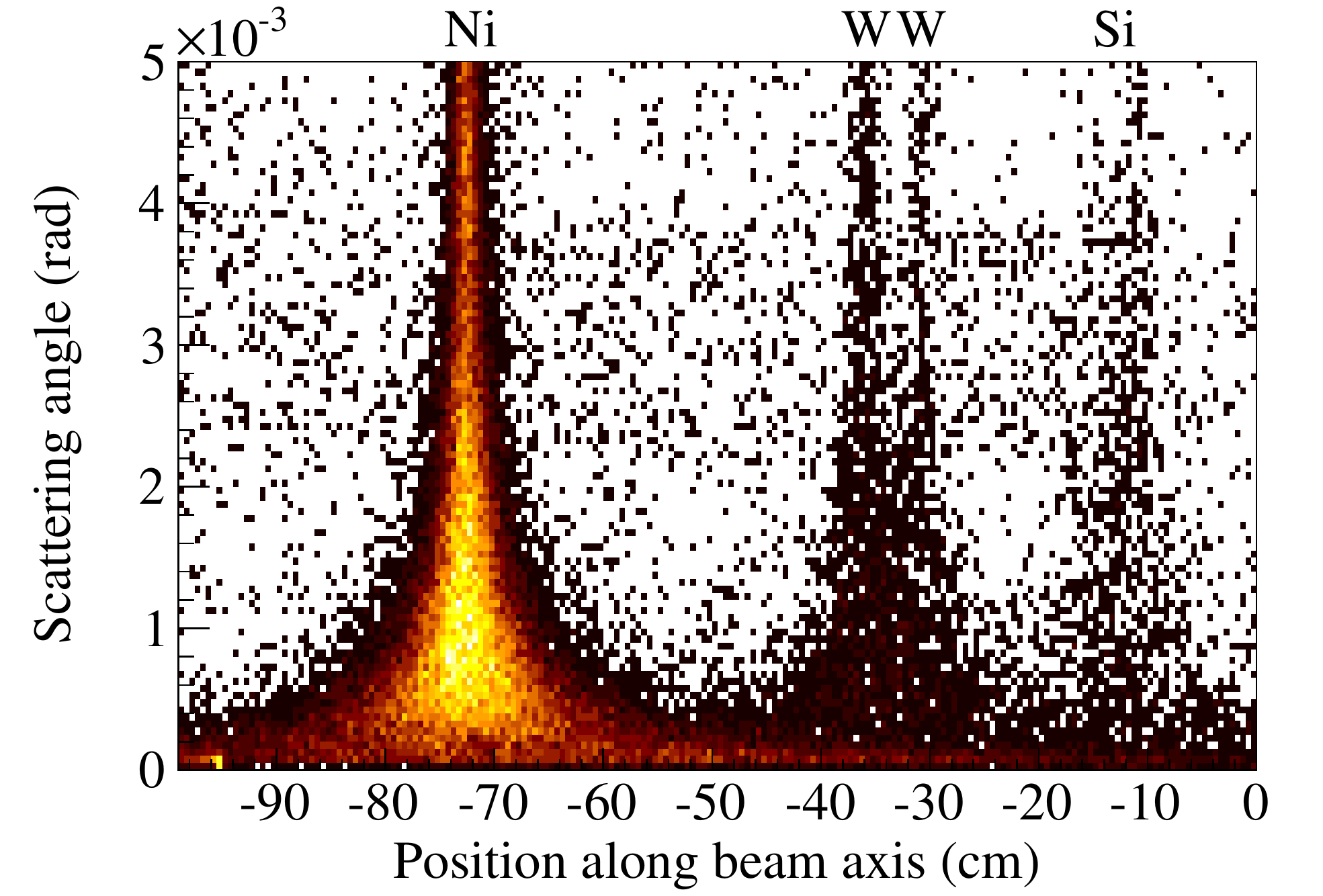}
  \end{minipage}
  \caption{Distribution of scattering angle of the outgoing pion vs the position
    of primary vertex along the beam axis from Primakoff data, illustrating the
    improvement of the vertex resolution between (left) standard alignment and
    (right) run-by-run alignment. The structures correspond to interactions in
    the different targets used in the measurement (see
    Table~\ref{tab:target.overview}) and in the first Silicon station downstream
    of the targets.}
  \label{fig:Improvement}
\end{figure}

The alignment of the tracking stations is completed by an separate procedure for
the silicon stations to fully exploit their high resolution.  For these
stations, displacements of up to $50\,\mum$ were observed, which are caused by
variations in the temperature of the support structure. Therefore, a separate
alignment for the silicon micro-strip telescope was produced for each run to
account for these variations. As an example, \figref{fig:SI_shift} shows the
corrections applied to the five silicon tracker stations in the horizontal plane
as a function of time. The effect of this time-dependent alignment on the
resolution is illustrated in \figref{fig:Improvement} for the distribution of
the scattering angle vs. the vertex position.  For the events with Primakoff
kinematics, in which the one outgoing track has a a very small scattering angle,
the improvement is substantial. The background can therefore be reduced by a
considerable fraction. As a result the distribution in
\figref{fig:Improvement}~(right) matches the simulation, in which a perfect
alignment is assumed (see \figref{fig:vertex_mc} in
\secref{sec:performance.acceptance}).
		    
The calorimeters are aligned with respect to the tracking detectors by using a
separate procedure that associates charged particle tracks with signals in the
calorimeters. Residuals are computed between the expected impact point of the
track and the reconstructed shower position in the calorimeter. The calorimeter
positions are adjusted accordingly in the plane transverse to the beam.

\subsection{Vertex Reconstruction}
\label{sec:reconstruction.vertex}
The vertex reconstruction uses as input the charged tracks reconstructed in the
spectrometer and in the beam telescope (see
\secref{sec:reconstruction.tracking}).  Only two kinds of vertex topologies are
considered: primary vertex and secondary vertex. The former designates the
association of one beam track with any finite number of spectrometer tracks,
whereas the latter corresponds to a combination between two oppositely charged
tracks with a common origin.

Tracks are fed into the vertexing procedure as vectors of parameter estimates
and their corresponding covariance matrices. Only tracks with momentum are
accepted.  Also, a cut is applied on the difference of the track times for the
incoming and outgoing tracks, except for those reconstructed only in drift
detectors.\footnote{For each reconstructed vertex, the output comprises its
  Cartesian coordinates and the list of its associated particles together with
  their reduced track parameters at the common origin, which encode only
  directional and momentum information.}

The contruction of a primary vertx is achieved in an iterative procedure that
starts with the set of all tracks compatible with a given beam track and
progressively removes outliers using an inverse Kalman filter. This procedure is
prone to failing if the initial set of tracks contains a large number of fakes,
because the preliminary estimate of the vertex position may then be too far from
the truth. A recovery mechanism is therefore applied in order to reconnect one
by one tracks that were unduly discarded. The overall procedure provides a good
vertex-finding efficiency (see \figref{fig:track.efficiency}).

\begin{figure}[htbp]
  \centering
  \includegraphics[width=.7\textwidth]{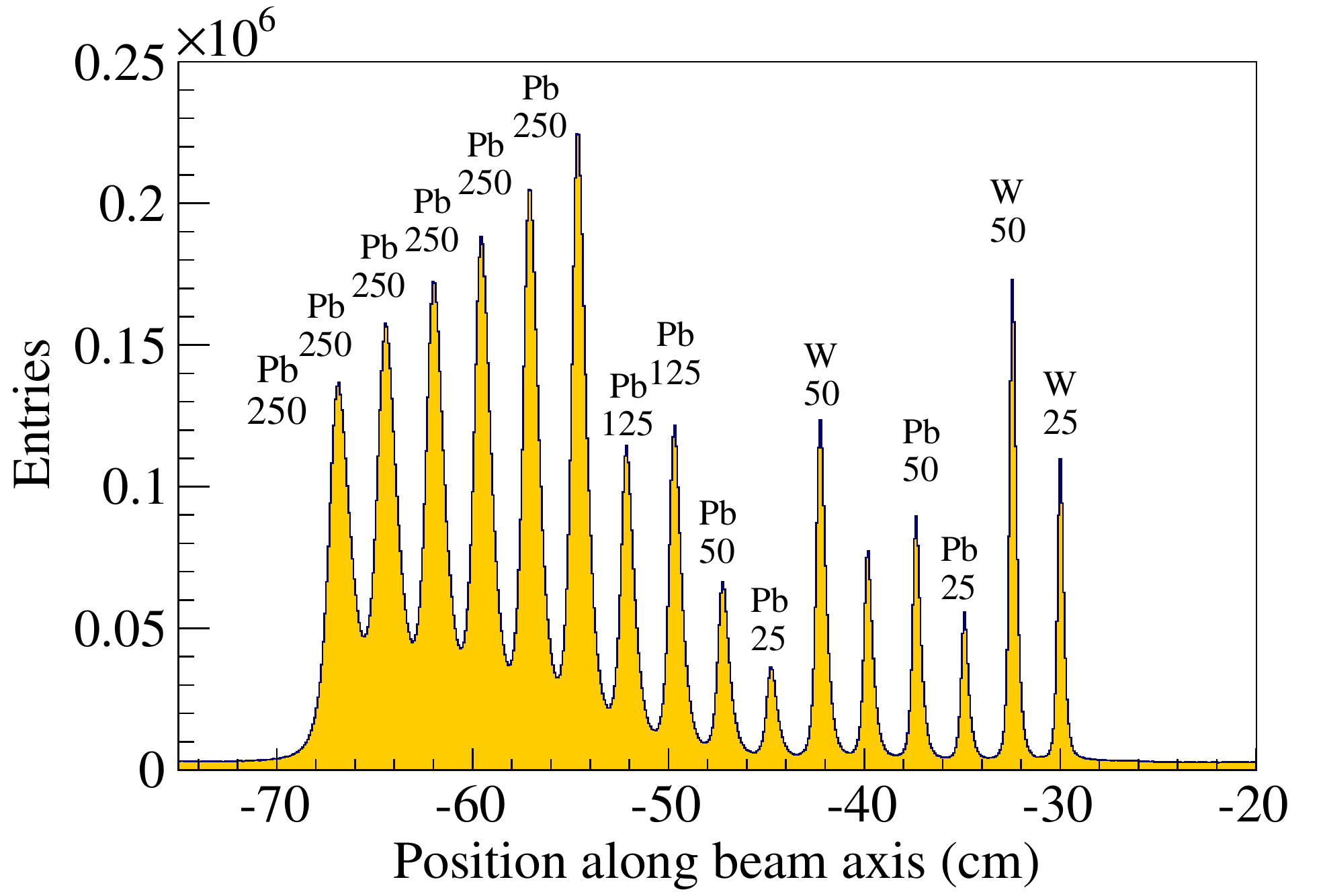}
  \caption{Distribution of reconstructed interaction vertices with three
    outgoing charged particles along the beam direction for exclusive
    events. For each solid state target the thickness is indicated (in $\mum$).}
\label{fig:targets.PbW.vertices}
\end{figure}

\begin{figure}[htbp]
  \centering
  \includegraphics[width=0.6\textwidth]{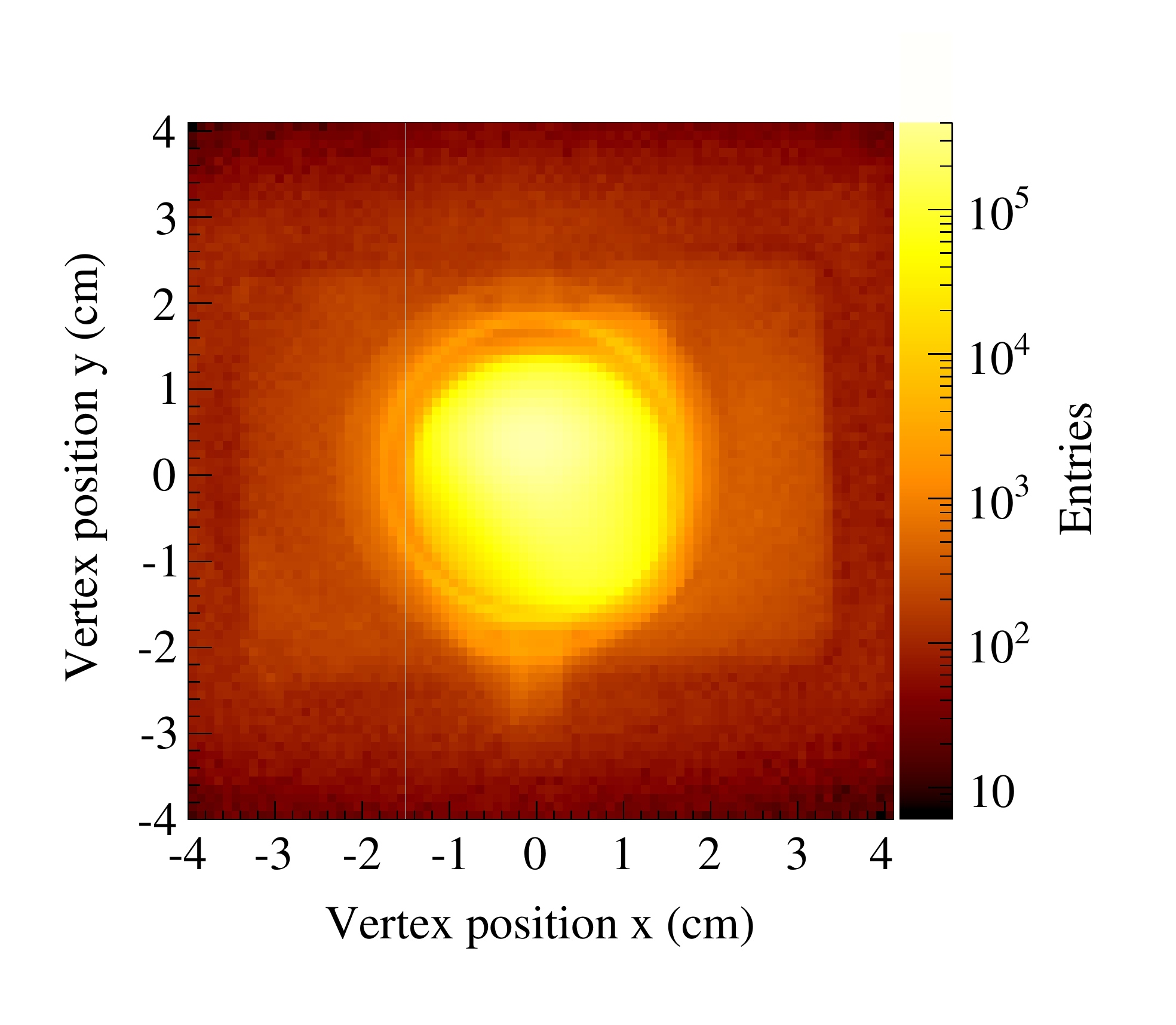}
  \caption{Vertex distributions for the liquid hydrogen target ($xy$ projection)
    for events with three charged tracks.}
  \label{fig:vertexdis_xy}
\end{figure}

\begin{figure}[htpb]
  \centering
  \includegraphics[width=.90\textwidth]{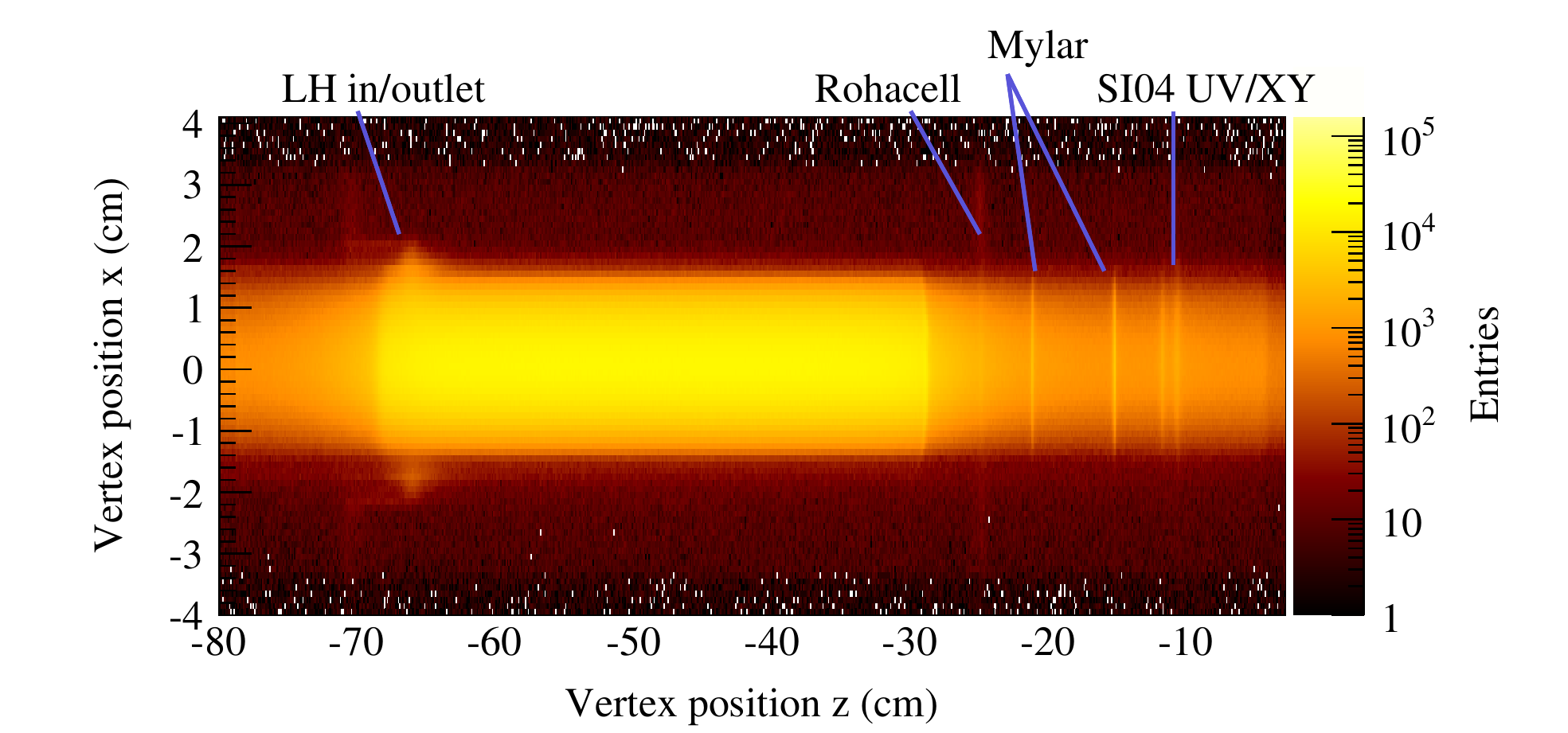}
  \caption{Vertex distributions for the liquid hydrogen target ($xz$ projection)
    for events with three charged tracks. For the explanation of the structures,
    see also \Figref{fig:targets.lh2.pic2}. }
  \label{fig:vertexdis_xz}
\end{figure}

The vertex resolution for the hadron setup is is found to be significantly
better than that of the muon setup described in~\cite{Abbon:2007pq}. The
improvement is due to the reduced multiple scattering in the thinner targets
employed with the hadron beam, as well as to the use of precise silicon
microstrip detectors at both ends of the targets. For example, on the 5$\pi$
sample already used to evaluate the tracking performances and for fully
reconstructed 5$\pi$ final states, the resolution along the beam axis varies
from $0.75$ to $4.7\,\mm$, depending on the 5$\pi$ invariant mass, while the
resolution across the beam axis lies in the $13$ to $16~\mum$ range.  The vertex
resolution achieved is illustrated in \figref{fig:Improvement} for a single
charged particle final state.  Similarly, \figref{fig:targets.PbW.vertices}
shows the system of nuclear targets described in \secref{sec:target.nuclear} as
reconstructed for a three-particle final state. The sixteen lead and tungsten
targets are all clearly separated. Thanks to the good resolution, the various
details of the liquid hydrogen target are distinctly visible in the
two-dimensional $xy$ and $xz$ distributions shown in \figref{fig:vertexdis_xy}
and \figref{fig:vertexdis_xz}.

Secondary vertices are reconstructed again using the Kalman filter for any
pair of oppositely charged tracks that satisfy a cut on the minimum distance
of approach.  Any track can thus be associated with several secondary vertices
and the primary vertex.  Reconstructed neutral particles can then be tested
against different particle hypotheses and the neutral particles can again be
combined with other charged tracks to study heavier hadrons (see
\secref{sec:performance.acceptance}).

\subsection{Recoil Particle Detection}
\label{sec:reconstruction.rpd}
The 12 PMTs of the inner ring and the 24 PMTs of the outer ring of the RPD
(described in \secref{sec:target.rpd}) provide information about the integrated
charge and a set of time hits. Each possible combination of upstream and
downstream PMTs is used to determine a coordinate along the longitudinal
direction of the scintillator and the time at which the particle crossed it.
Hits are discarded if their reconstructed position is outside the fiducial
dimensions of the scintillators with a safety margin of 20\,cm. Reconstructed
hits for the inner ring elements are associated to hits in the three
corresponding outer ring elements to form tracks. For each track, the momentum
is determined from the time of flight using the proton mass hypothesis and the
calculated position of the hits. The track is extrapolated backwards to the
vertex that is reconstructed using the beam track and the tracks of the
scattered particles. A correction on the momentum is determined by accounting
for the amount of the material crossed by the recoil particle. After
reconstruction, a set of RPD tracks is available for event selection and physics
analysis.

\begin{figure}[htbp]
  \centering
  \includegraphics[width=0.7\columnwidth]{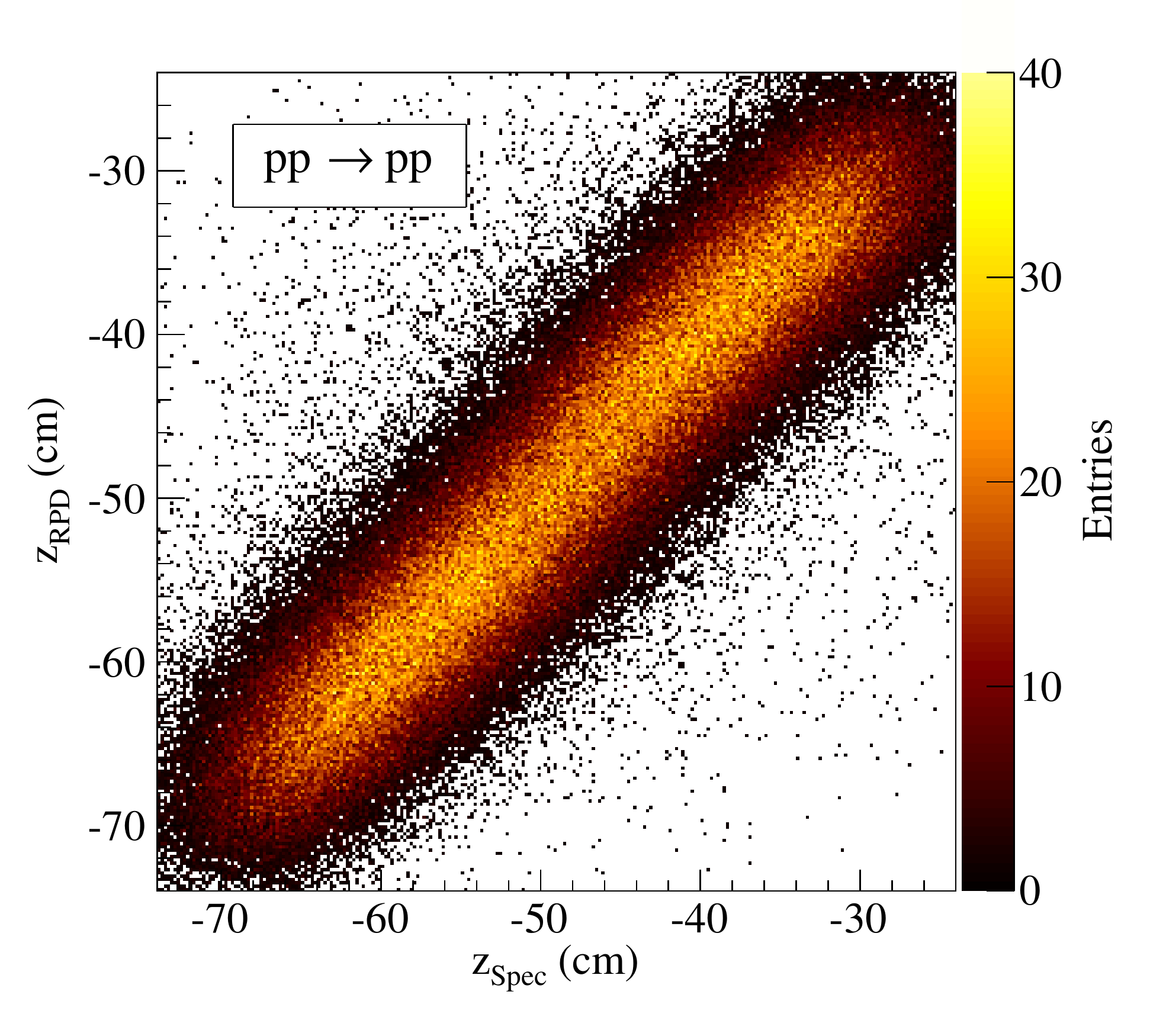}
  \caption{Correlation between the longitudinal vertex position $z$ determined
    with the RPD and the one determined with the spectrometer.}
  \label{fig:performance.rpd.RPD_Vertex_Correlation}
\end{figure}

The calibration of the RPD is done using proton-proton elastic scattering
events. The impact point of the scattered proton in the scintillator and its
momentum can be predicted from the kinematics of this reaction. Matching of
measurement and prediction allows for tuning the position offsets on each
individual counter and the global offset of the RPD position in the COMPASS
reference system. The correlation of the predicted longitudinal vertex position
and the one determined using the information from the RPD is shown in
Fig.~\ref{fig:performance.rpd.RPD_Vertex_Correlation}.  The momentum calibration
is done by adjusting time offsets between each possible pair of scintillators in
ring A and ring B.  The energy loss is calibrated using the features of the
energy loss distribution as a function of the velocity of the proton.  The
maximum in the energy loss distribution is adjusted to agree with a Monte Carlo
simulation (see Fig.~\ref{fig:target.rpd.rpd-deb-beta}).  For the inner ring,
the corresponding distribution does not show the rising part seen in
Fig.~\ref{fig:target.rpd.rpd-deb-beta}, hence the maximum energy loss $\Delta E$
is used for calibration.

\begin{figure}[htbp]
  \centering
  \includegraphics[width=0.5\columnwidth]{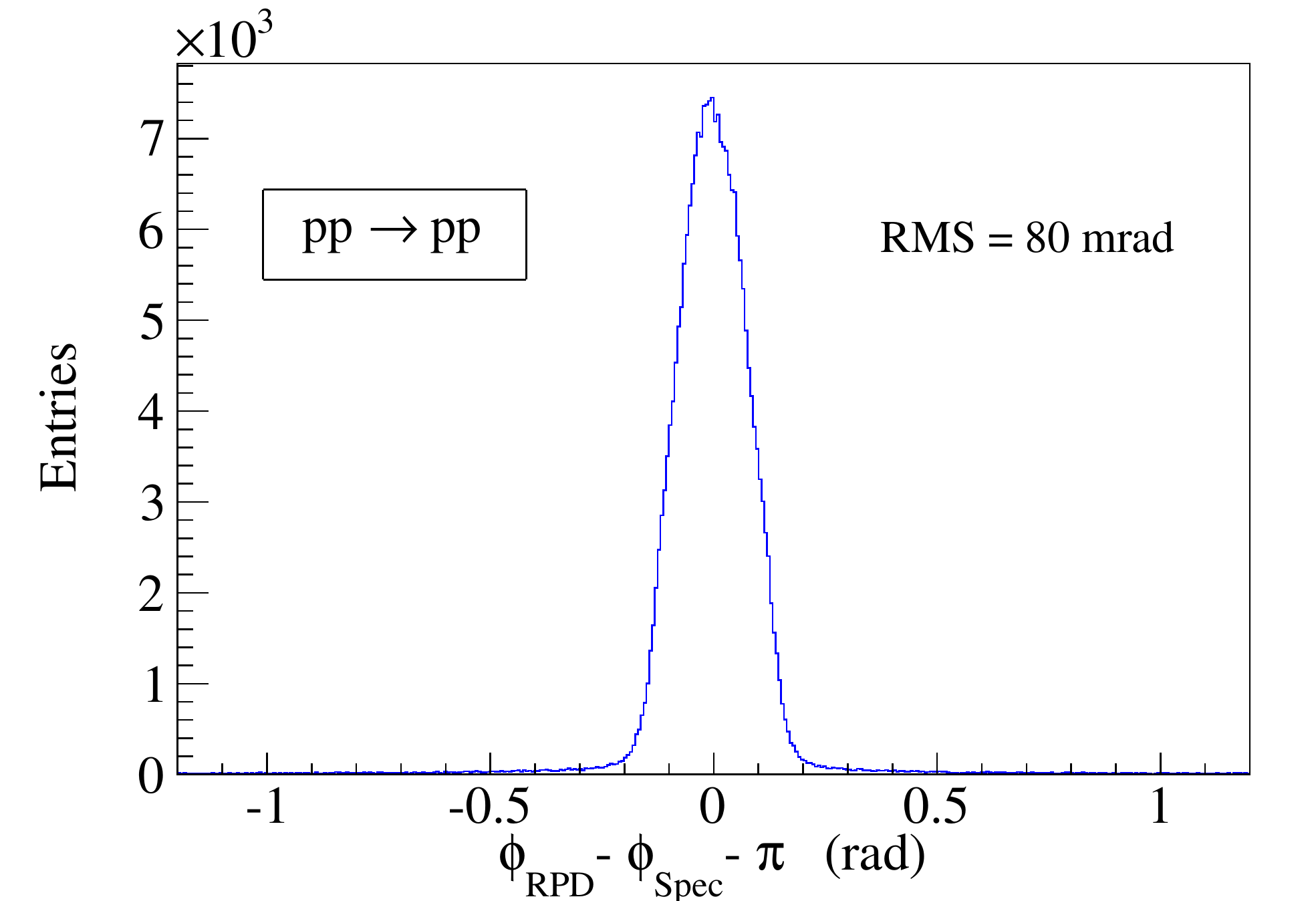}
    \caption{Correlation between the azimuthal angles of the recoil proton
      detected in the RPD and the scattered proton detected in the
      spectrometer.}
   \label{fig:performance.rpd.angle}
\end{figure}

\begin{figure}[htbp]
  \centering
  \includegraphics[width=0.5\columnwidth]{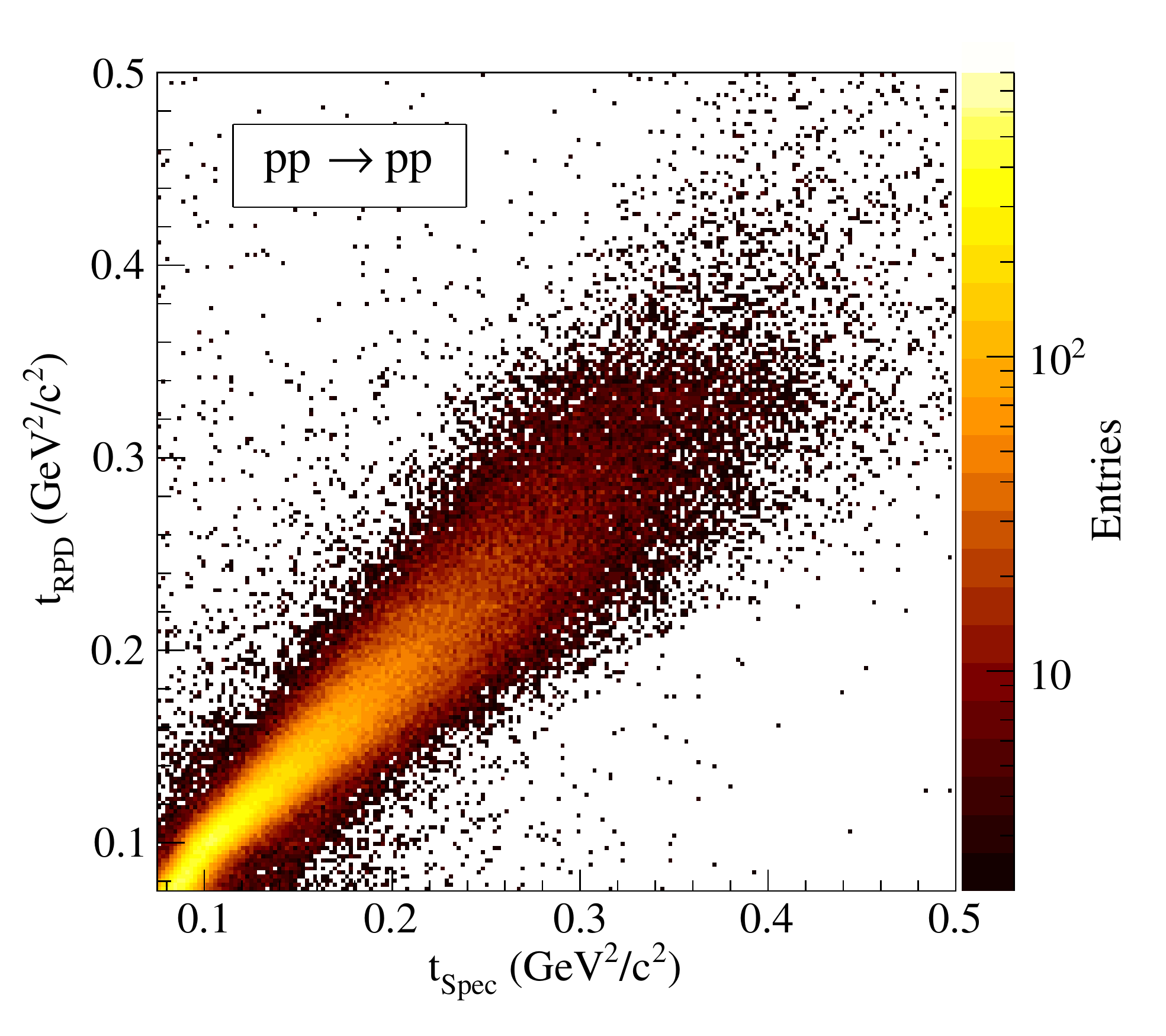}
   \caption{Momentum transfer correlation between the recoil proton detected in
     the RPD and the scattered proton detected in the spectrometer.}
   \label{fig:performance.rpd.mom}
\end{figure}

For elastic $pp$ scattering, the correlation (difference by $180^\circ$) between
the azimuthal angles $\phi_{Spec}$ determined from tracking the scattered proton
in the spectrometer and $\phi_{RPD}$ measured on the recoiling proton with the
RPD is shown in Fig.~\ref{fig:performance.rpd.angle}.  The value of the
corresponding resolution, of about $80$\,mrad, is a consequence of the 24-fold
segmentation of the outer ring barrel and the multiple scattering encountered by
the recoiling particle in the target.  The measured momentum transfer $|t|$ as
determined from forward and from RPD tracks is presented in
Fig.~\ref{fig:performance.rpd.mom}. A clear correlation between the two
measurements over the covered range of momentum transfer is observed.

\subsection{RICH-1}
\label{sec:reconstruction.rich}
\begin{figure}[htpb]
\begin{center}
\includegraphics[width=0.5\textwidth]{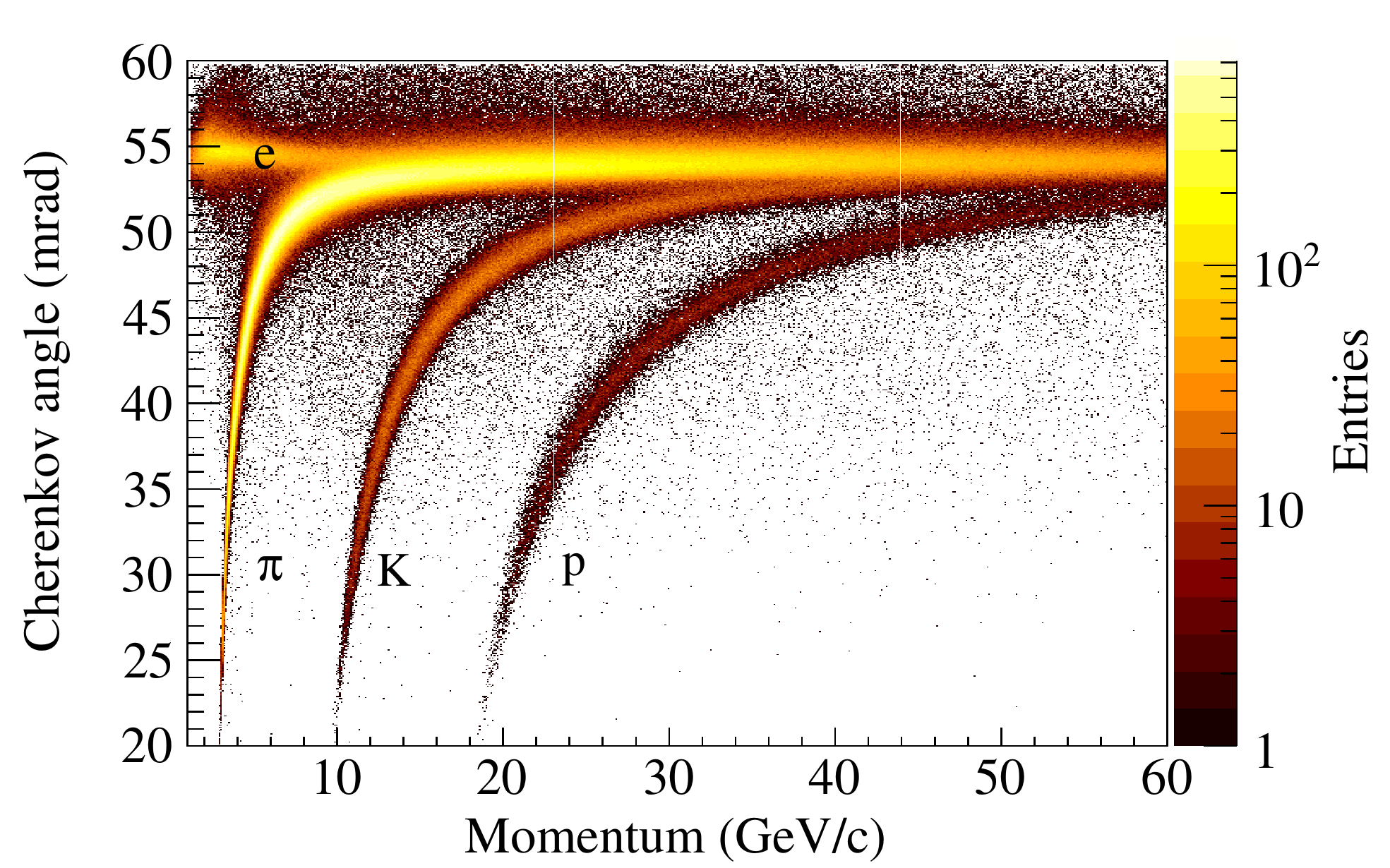}
\caption{Cherenkov angle for reconstructed rings as a function of the particle
  momentum for the C$_4$F$_{10}$ radiator.}
\label{fig:pid.rich.themom}
\end{center}
\end{figure}

\begin{figure}[htpb]
\begin{center}
\includegraphics[width=0.5\textwidth]{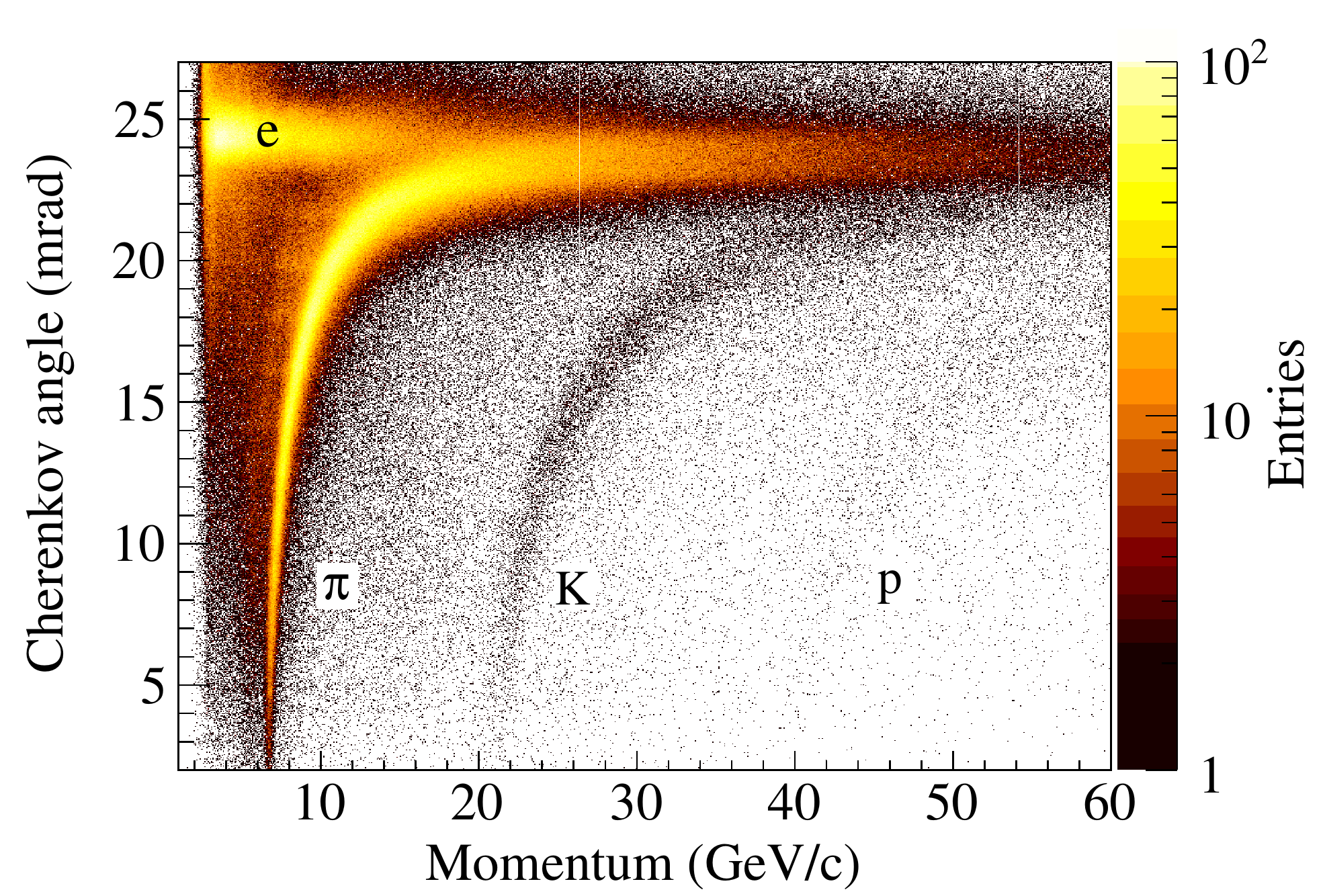}
\caption{Cherenkov angle for reconstructed rings as a function of the particle
  momentum for the N$_2$ radiator.}
\label{fig:pid.rich.n2}
\end{center}
\end{figure}

The separation between the different hadron types in RICH-1 is illustrated in
Fig.~\ref{fig:pid.rich.themom}, where the Cherenkov angles for reconstructed
rings are shown as a function of the particle momenta.  The four clearly visible
bands correspond to electrons, pions, kaons and protons.  For comparison, the
same picture for the N$_2$ radiator (see \Secref{sec:pid.rich1.char}) is shown
in Fig.~\ref{fig:pid.rich.n2}.  In this case, the Cherenkov angle at saturation
reaches only $24.5\,\mrad$, a value that is about a factor of two smaller than
the corresponding value for the C$_4$F$_{10}$ radiator.


\begin{figure}[htbp]
\begin{center}
\includegraphics[width=0.45\textwidth]{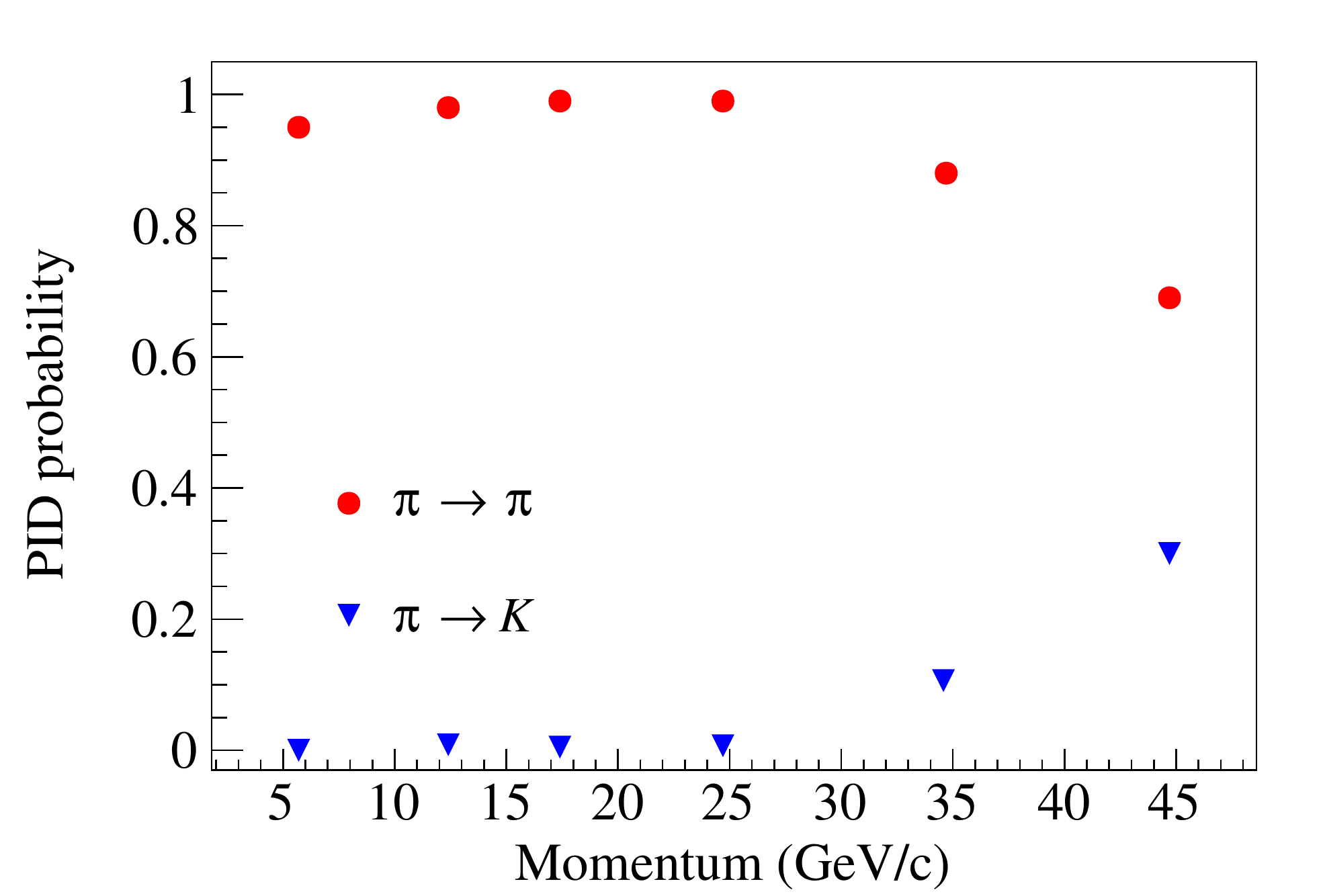}
\includegraphics[width=0.45\textwidth]{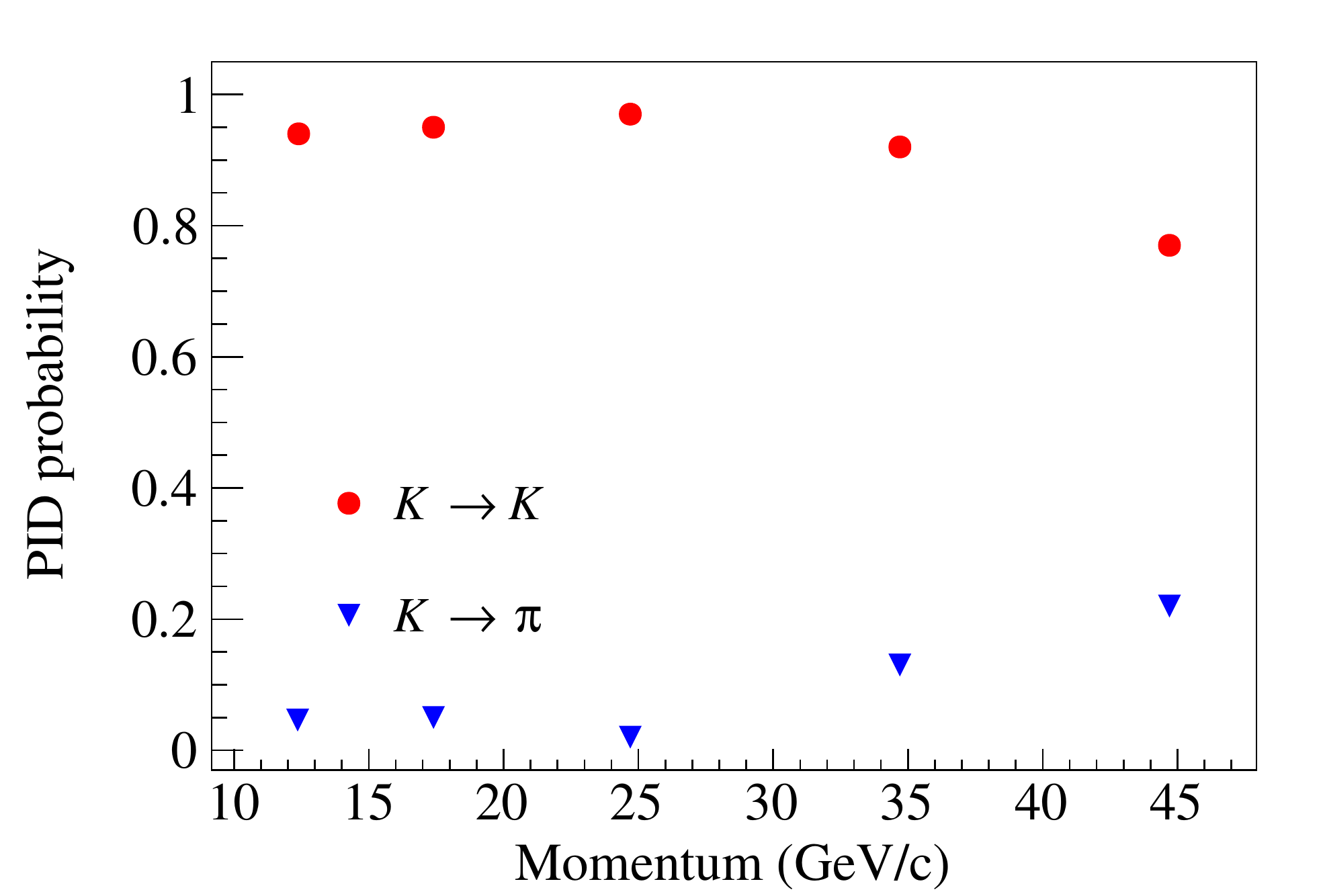}
\caption{Identification efficiency and mis-identification probabilities as a
  function of the particle momentum for (left) a pion sample and (right) a kaon
  sample.}
\label{fig:pid.rich.effmom}
\end{center}
\end{figure}

The particle identification (PID) efficiency was evaluated on samples of pions
and kaons from the decay of $\phi$ and $K^0_S$ mesons, respectively.  The PID
relies on an extended maximum-likelihood method.  For each particle, different
likelihood functions corresponding to the relevant mass hypotheses are computed
and then compared.  The likelihood function parametrises the photon distribution
taking into account both the photons emitted by the considered particle (the
Cherenkov signal) and the photons emitted by other particles in the event (the
background).  For the background parametrisation, the map of the integrated hits
(see \secref{sec:pid.rich1}) in the photon detector is used.  The PID
probabilities (efficiency and mis-identification probabilities) are shown in
Fig.~\ref{fig:pid.rich.effmom} as a function of the particle momentum.  The
efficiency is larger than 90\% in the region below $30\,\GeV/c$, where the
Cherenkov angles for different mass hypotheses are well
separated. Correspondingly, the mis-identification probabilities are close to
zero.  Above $30\,\GeV/c$, the Cherenkov angle starts to saturate, and as a
consequence the efficiency decreases and the mis-identification probability
increases.  Moreover, the high momentum region corresponds to small polar angle
values, and thus to a region with larger hadron multiplicity in the events.

\subsection{CEDARs}
\label{sec:reconstruction.cedar}
As detailed in \secref{sec:beam.cedar}, two CEDARs are used to select the
particle type in the hadron beam. Beam particles can be identified by requiring
a minimum number of hits in the eight PMTs attached to each of the two
detectors.  For protons, this method achieves high efficiency and purity due to
the good separation of proton and pion rings as discussed in
\Secref{sec:beam.cedar}.  On the other hand, an online efficiency for kaon
identification of only 35\% is obtained using hit multiplicities, which is due
to the large beam divergence and the small difference between kaon and pion ring
radii. In the offline analysis this efficiency increases to 48\% for physics
events with a vertex in the target.

For measurements with negative hadron beams both CEDARs are set on kaon
identification, requiring good kaon efficiency.  In order to further improve the
kaon identification, a different method was developed for offline analysis. It
is based on beam particles reconstructed in the beam telescope before the target
and makes use of the response of each PMT individually to improve PID for
particles that do not travel parallel to the CEDAR optical axis. In a first
step, the response of the PMTs for kaons and pions is determined as a function
of the horizontal and vertical angles between track and CEDAR optical axis,
$\theta_x$ and $\theta_y$.  These angles are obtained from tracks measured in
the beam telescope, which are traced back to the CEDAR position using the known
beam optics.

A clean kaon sample is obtained from data taken with the CEDAR kaon trigger plus
a beam trigger by selecting decays of beam kaon into three charged pions $K^-
\rightarrow \pi^- \pi^+ \pi^-$ outside the target region.  The probability for a
kaon to produce a signal in one of the photomultipliers is:
\begin{equation}
  P_{(\theta_x,\theta_y)}(\text{signal}|K) = \frac{\text{Number of beam
      particles with signals in PMT}} {\text{Number of beam particles in kaon
      sample}}. \label{P_signal}
\end{equation}
In order to identify a particle, the probability is needed that a signal is
produced by a kaon.  This probability can be calculated using Bayes' theorem:
\begin{equation}
	P_{(\theta_x,\theta_y)}(K|\text{signal}) = \frac{P_{(\theta_x,\theta_y)}
          (\text{signal}|K) \cdot
          P_{(\theta_x,\theta_y)}(K)}{P_{(\theta_x,\theta_y)}
          (\text{signal})}\,. \label{P_kaon}
\end{equation}
Here, $P(K)$ and $P(\text{signal})$ are the probabilities to have a kaon with
$(\theta_x,\theta_y)$ in the beam and to get a signal from any beam particle
with $(\theta_x,\theta_y)$, respectively.  Similar equations hold for a pion
sample, which is obtained using diffractive production of three charged pions on
liquid hydrogen, $\pi^- p \rightarrow \pi^- \pi^+ \pi^- p$.

\begin{figure}[!ht]
	\centering \includegraphics[trim = 0cm 8cm 0cm 0cm, clip, width=\textwidth]{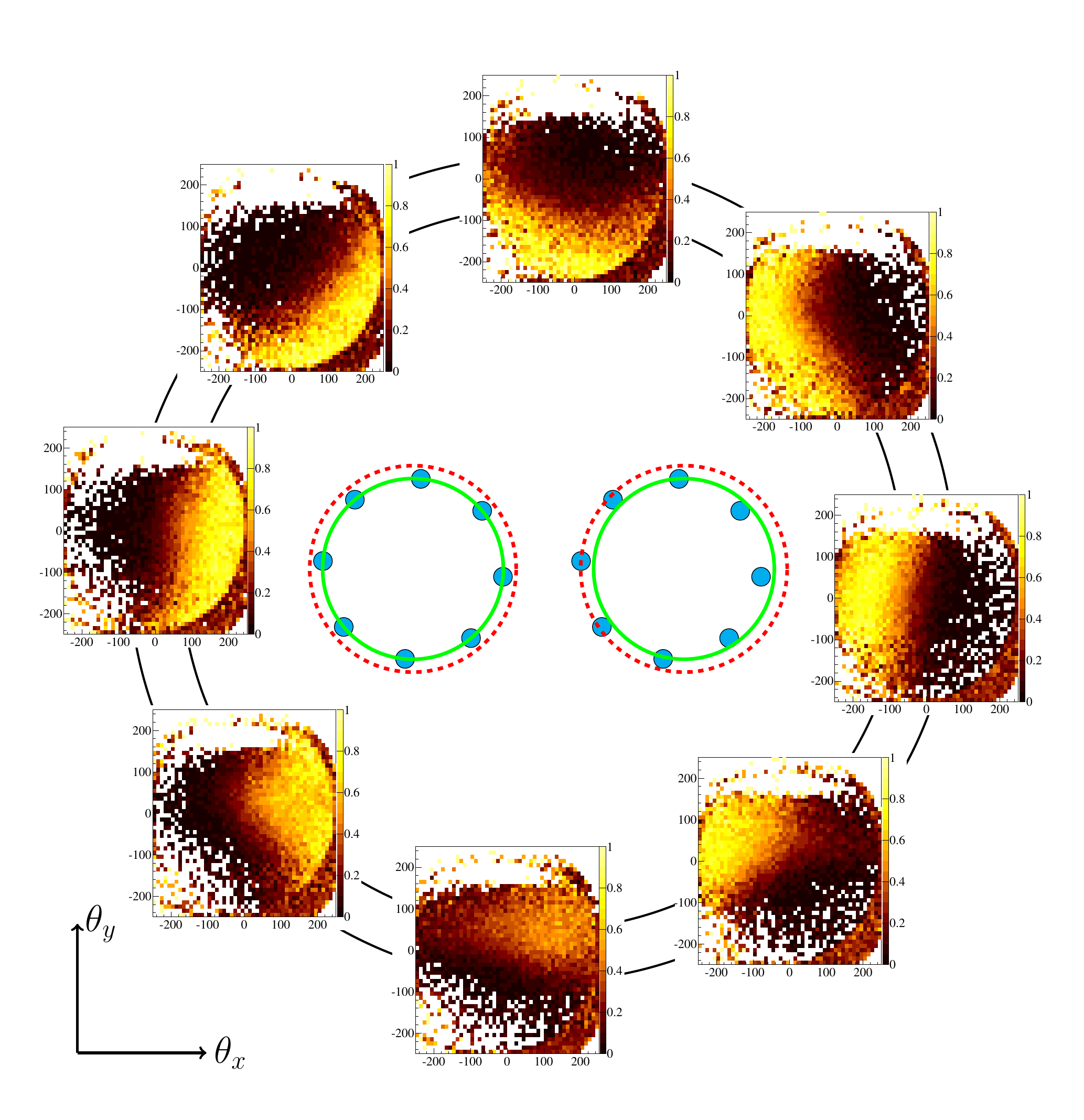}
\caption{Dependence of $P(\text{signal}|\pi)$ on $\theta_x$ (horizontal) and
  $\theta_y$ (vertical) for the eight PMTs of CEDAR 2 (arranged according to the
  CEDAR geometry). The range for both angles is from $-250\,\murad$ to
  $250\,\murad$.  The insets in the centre illustrate the position of a pion
  (dashed, red) and a kaon (green) ring relative to the PMT positions for
  $\theta_x=0$ and $\theta_y=0$ (left inset) and for $\theta_x>0$ and
  $\theta_y=0$ (right inset).  }
\label{fig:cedar.prob}
\end{figure}

Since the beam divergences for pions and kaons are the same, the probabilities
$P_{(\theta_x,\theta_y)}(\text{K})$ and $P_{(\theta_x,\theta_y)}(\pi)$ can be
dropped together with the common $P_{(\theta_x,\theta_y)}(\text{signal})$.  The
only quantities needed are the probabilities (Eq.~\eqref{P_signal}) for kaons
and pions to produce a signal in a PMT.  In order to avoid regions with low
statistics, a cut $(\theta_x^2 +\theta_y^2)^{1/2}<200\,\mrad$ is applied to the
data before further analysis.  As an example, the probability distributions
$P(\text{signal}|\pi)$ for all eight PMTs of CEDAR 2 are shown in
Fig.~\ref{fig:cedar.prob}.  The insets in the centre of the figure illustrate
the position of a pion and a kaon ring relative to the PMT positions for
$\theta_x=0$ and $\theta_y=0$ (left) and for $\theta_x>0$ and $\theta_y=0$
(right).  No pions are expected to be detected for $\theta_x=0$ and $\theta_y=0$
as the detector was set on kaons for the negative beam (left inset in
Fig.~\ref{fig:cedar.prob}) yielding photon rings for pions larger than the
diaphragm. However, PMT hits are expected when the photon ring intersects with
the diaphragm due to beam divergence. As an example, pion and kaon photon rings
with positive $\theta_x$ are shown in the right inset. The photon ring from the
pion illuminates the PMTs on the left side, thus reducing the kaon
identification efficiency.

\begin{figure}[h]
  \centering 
  \subfloat[kaon sample]{\includegraphics[width=0.47\textwidth]{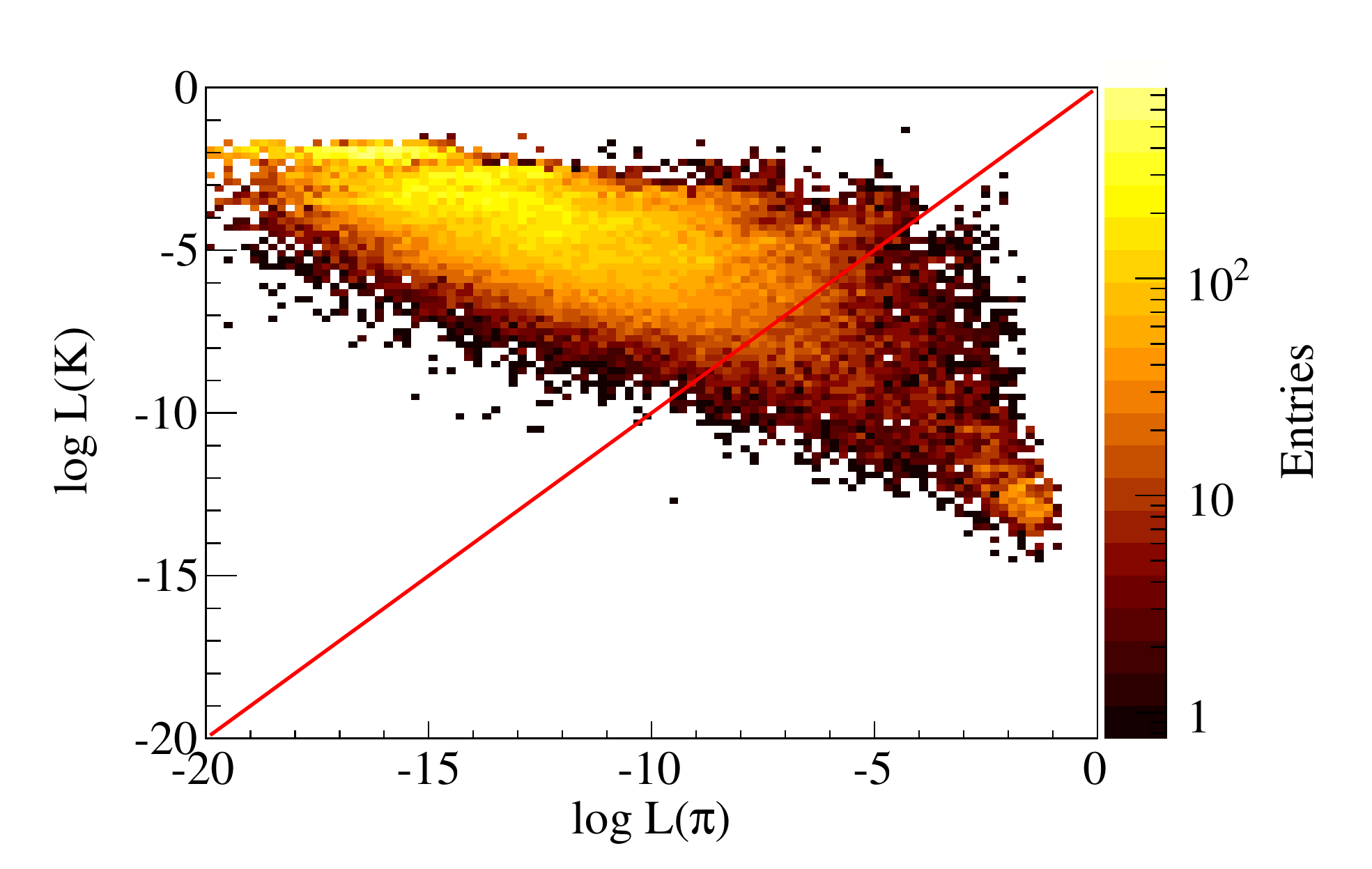}}
  \subfloat[pion sample]{\includegraphics[width=0.47\textwidth]{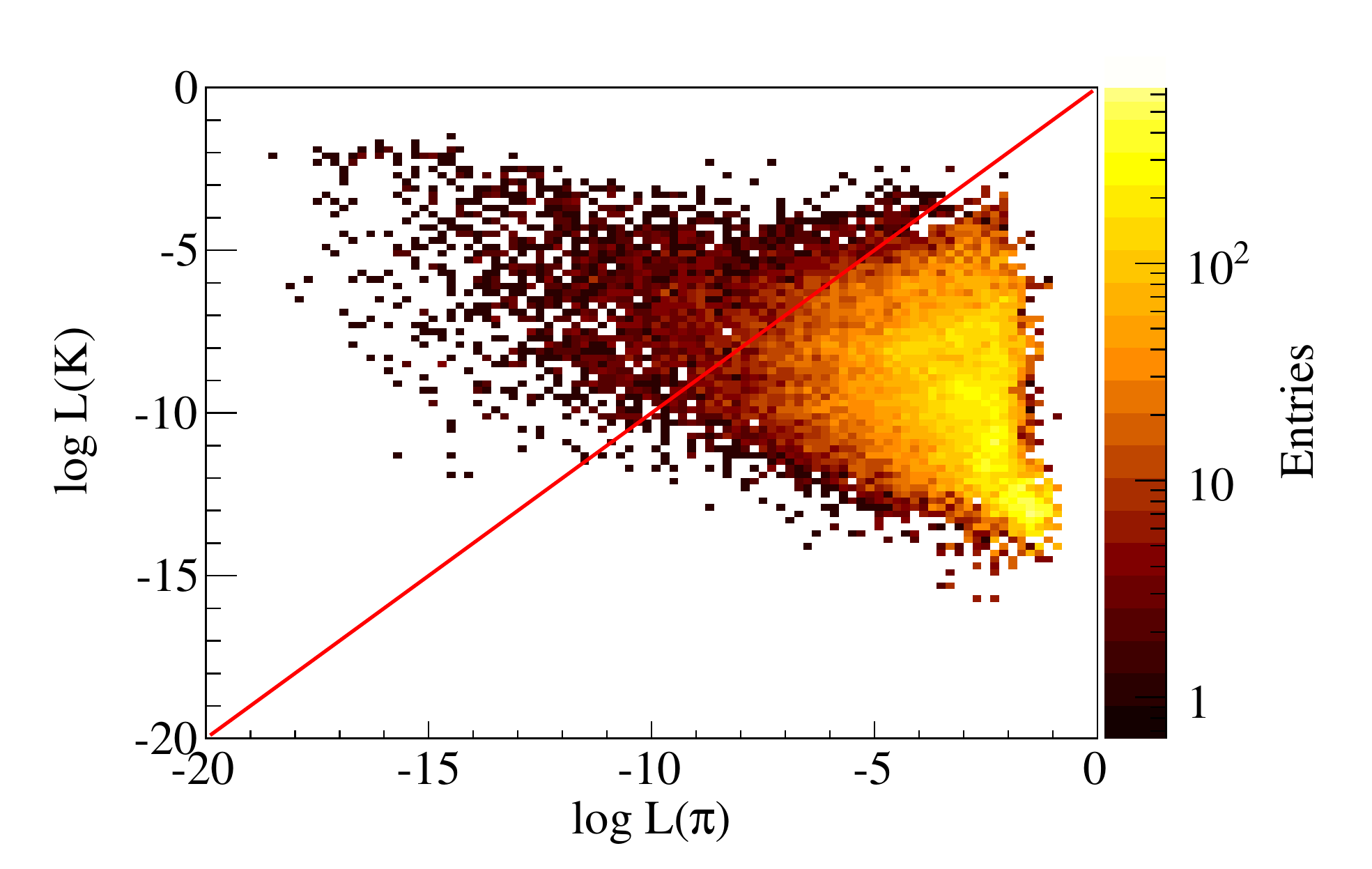}}\\
  \subfloat[beam sample]{\includegraphics[width=0.47\textwidth]{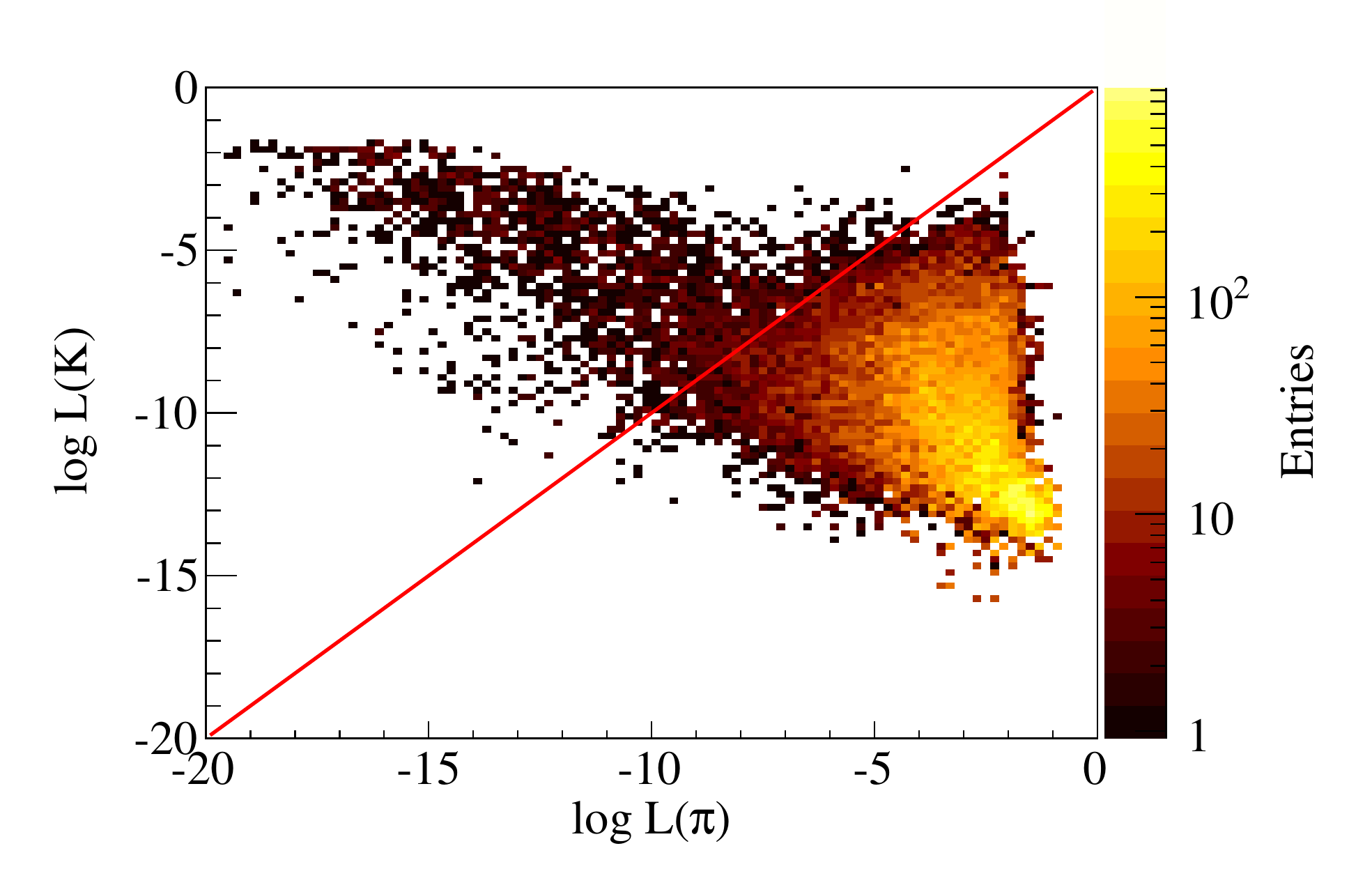}}\hfill
  \caption{Values for the log-likelihoods function for different samples
    obtained from CEDAR 2 calculated for (a) the kaon sample, (b) the pion
    sample and (c) an unbiased beam sample.  The red line indicates $\log L(\pi)
    = \log L(K)$.}
\label{fig:likes}
\end{figure}

Using the probabilities for all PMTs, the log-likelihood for a beam particle
being a kaon is calculated according to:
\begin{equation}
\log L(K) = \sum\limits_{\text{signal}}\log
P_{(\theta_x,\theta_y)}(\text{signal}|K) + \sum\limits_{\text{no signal}}
\log\left[1-P_{(\theta_x,\theta_y)}(\text{signal}|K)\right]\,,
\end{equation}
where the first sum only counts photomultipliers with a signal and the second
sum only those without a signal. A corresponding equation holds for the
log-likelihood for a beam particle to be a pion. Figure \ref{fig:likes} shows
the distribution of $\log L(K)$ vs.  $\log L(\pi)$ for (a) the kaon and (b) the
pion sample, while (c) shows the results for an unbiased beam sample. The
intensity in (c) reflects the beam composition, namely that the kaon component
is nearly two orders of magnitude smaller than the pion component.
   
Kaons and pions are identified requiring a certain difference between $\log
L(K)$ and $\log L(\pi)$.  The particle is identified as kaon If $\log L(K)> \log
L({\pi}) + A$, and as a pion if $\log L({\pi}) > \log L({K}) +B$. In all other
cases no PID is given.  The likelihood differences $A$ and $B$ are chosen by
maximising purity and efficiency simultaneously.  A good balance between high
efficiency and high purity is achieved for the choice of $A=B=1$.

In order to determine the purity of the CEDAR identification, the reactions
$\pi^- p \rightarrow K^- K^0_{S}\,p\,$ and $K^- p \rightarrow \pi^-
\bar{K}^0_{S}\,p\,$ are used. Due to conservation of strangeness, the incoming
hadron is tagged by the outgoing hadron. The $K^0_{S}$ and $\bar{K}^0_{S}$ are
reconstructed using the two-pion invariant mass distribution. The negatively
charged outgoing particle is identified using RICH-1 information.  After
selecting incoming kaons with the CEDAR, its purity is determined by the ratio
of identified pions in RICH-1 divided by the total number of identified
particles. Thus the kaon purity $p(K)$ is given by
\begin{equation}
  p(K) = \frac{N_{\text{RICH}}(\pi)}{N_{\text{RICH}}(K)+N_{\text{RICH}}(\pi)}~.
\end{equation}
The purity for pions is obtained in the same way.

In order to determine the efficiency for pions and kaons, their numbers as
obtained from the CEDARs are divided by the respective numbers of pions and
kaons assuming the known beam decomposition (see \Secref{sec:beam.hadron}).  The
values for kaon efficiency and purity are given in Table~\ref{tab:cedar.result}.
\par
\begin{table}[tbp]
\centering
\caption{Efficiencies and purities for the likelihood method ($A=B=1$) in
  comparison with the multiplicity method. Only statistical errors are given.}
\label{tab:cedar.result}
\begin{tabular}{lrr} \hline \hline
		& Kaon efficiency & Kaon purity \\ \hline Multiplicity method &
  $(48.4\pm0.2)\%$ & $(86.9\pm0.9)\%$ \\ Likelihood method & $(80.3\pm0.4)\%$ &
  $(85.4\pm0.9)\%$ \\ \hline \hline
\end{tabular}
\end{table}
The kaon identification efficiency is improved by almost a factor of two for the
likelihood method in comparison with the multiplicity method when applied
offline, while the corresponding purities are nearly identical.  With the
multiplicity method pion identification is not possible as the pressure was
adjusted for kaon identification for the data taken with the negative hadron
beam.  The likelihood method allows for pion identification as well. The values
obtained for pions are similar to those obtained for kaons.

In the analysis of Primakoff data, the CEDAR information is needed for an
efficient kaon rejection.  Using an optimisation of the likelihood method method
described above, pions are identified with an efficiency higher than 95\%, while
the kaon component is suppressed by more than a factor of 20.

\subsection{Electromagnetic calorimeters}
\label{sec:reconstruction.ecal}
Event reconstruction in ECAL1 and ECAL2 is performed by using time and signal
amplitude information as directly extracted from the SADC samples. The signal
amplitude for each module is converted into energy applying conversion
coefficients that were derived from the electron beam calibration.  The
variation of the amplitudes over the data taking period is accounted for by
using the information provided by the Laser and LED monitoring systems. Details
about signal extraction, electron calibration, and data monitoring are given in
\secref{sec:pid.ecal.ecal1}.

The energy calibration of each module is further improved by using the data
derived from an analysis of the $\pi^0\rightarrow \gamma \gamma$ decay
process. The $\pi^0$ calibration is performed prior to the final data
reconstruction on a fraction of the collected events. The two decay photons are
singled out after having defined clusters of deposited energy and performed fits
based on the definition of a shower profile.  During the final data analysis,
additional corrections are applied according to the specific data set, namely
diffractive dissociation or Primakoff scattering.

\subsubsection{Clusters and showers}
\label{sec:reconstruction.ecal.clusters}
For both ECALs, the event reconstruction consists of associating an energy
deposit in one or several adjacent modules to a single incident particle.  A set
of energy deposits that is assumed to originate from a single particle is called
in the following a \textit{shower}; the full energy deposit and hit position of
the particle are calculated from it.  In many cases, two or more showers overlap
and form a \textit{cluster}.  Thus precise knowledge of the shower profile
facilitates the separation of overlapping showers.  In addition, it improves
spatial and energy resolutions and limits the impact of inefficient or noisy
cells.  Clusters of two or more particles can result from electromagnetic
showers initiated in the material upstream of the ECAL or from decay photons
that hit the ECAL at a distance smaller than the lateral shower size.

The data analysis procedure starts by defining a cluster of neighbouring
modules, in which the deposited charges are larger than a pre-defined threshold
(see \secref{sec:reconstruction.ecal.reconstruction}).  The cluster is then
split into showers employing a parametrisation for the lateral spread of the
shower profile~\cite{Lednev:95}.  The shower parameters for the lead glass and
Shashlik modules are determined using electrons from a dedicated calibration
beam (see \secref{sec:beam.electron}).  For Primakoff data taking in which
mainly high-energy photons are detected, no electron beam with the corresponding
energy is available.  The shower parameters are therefore derived using single
photons from real data events.

\subsubsection{Shower profile}
\label{sec:reconstruction.ecal.profile}
The shower profiles used in the reconstruction are based on an empirical
cumulative function, as defined in \cite{Lednev:95}. If the energy deposited by
a shower is projected onto a transverse axis with the shower center at $0$, the
fraction of the total shower energy accumulated between $-\infty$ and a position
$x$ on this axis can be described by:

\begin{equation}
  F\left(x\right)=\frac{1}{2} +
  \frac{1}{\pi}\sum_{i}a_{i}\cdot\arctan\frac{x}{b_{i}}\,.
  \label{eq:reconstruction.ecal.showerprofile1D}
\end{equation}

\begin{figure*}[h]
  \begin{center}
      \includegraphics[width=.45\textwidth]{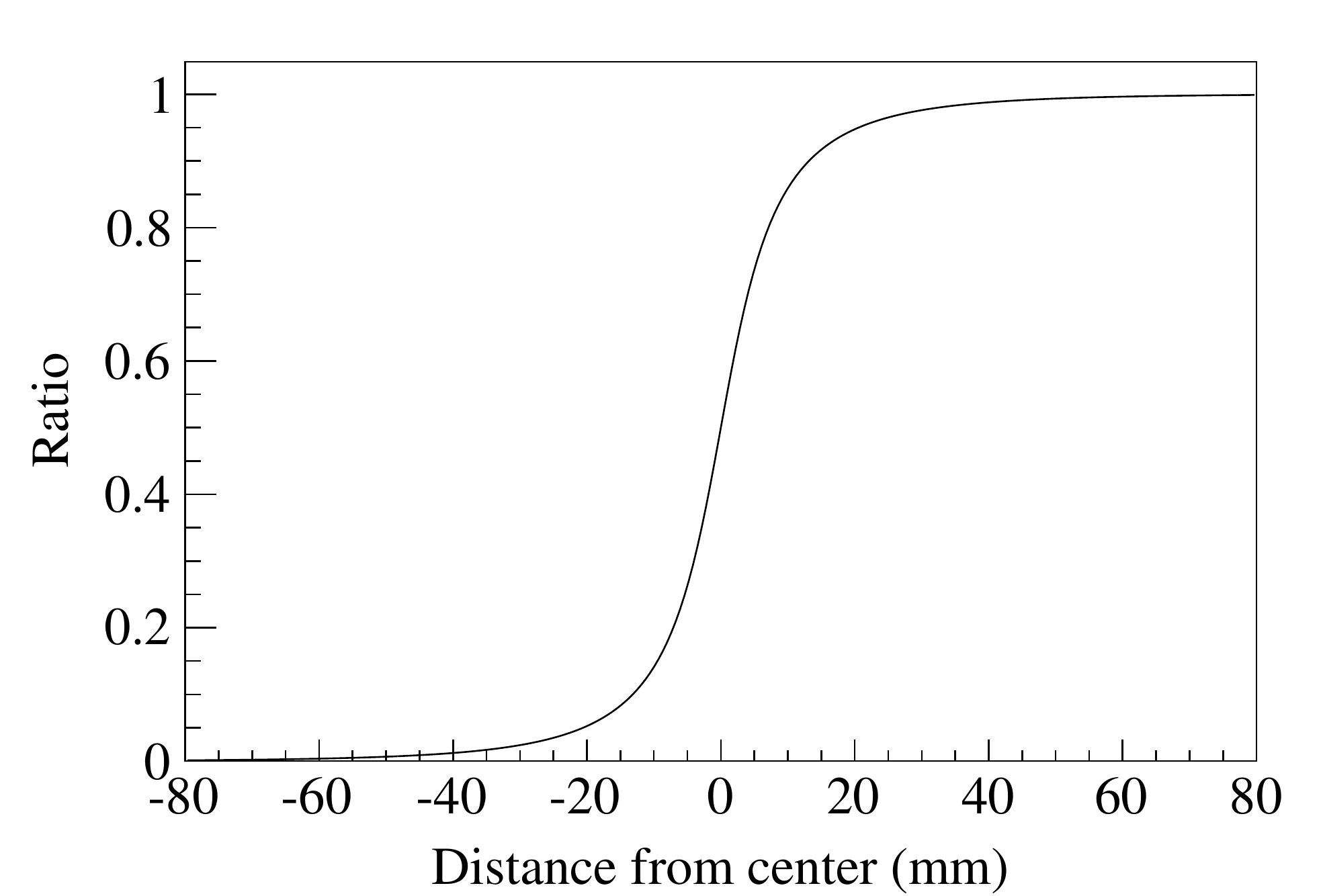}
      \includegraphics[width=.45\textwidth]{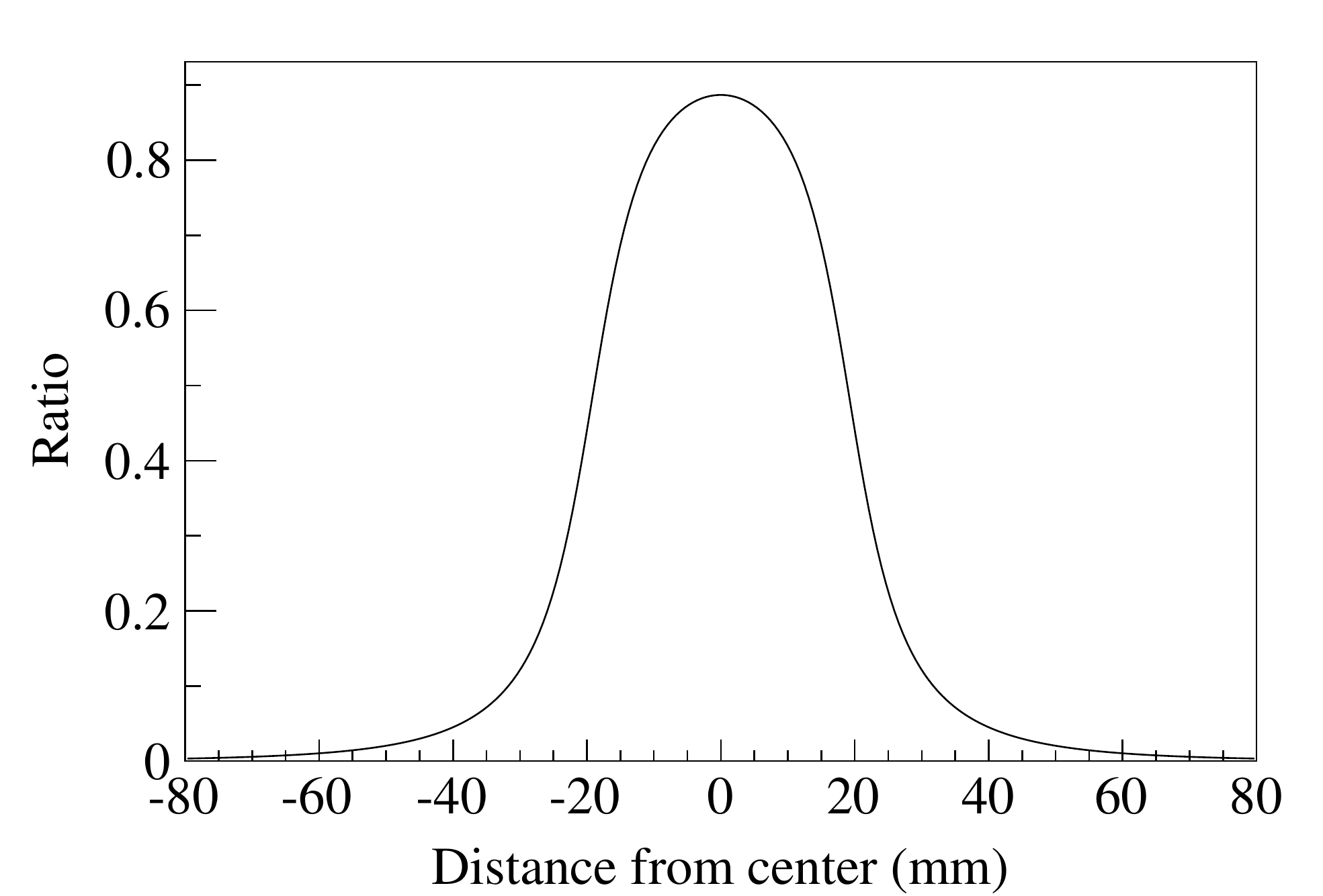}
    \caption{Shower reconstruction: (left) fraction of total shower energy
      collected from $-\infty$ up to a particular distance from the shower
      center (Eq.~\ref{eq:reconstruction.ecal.showerprofile1D}); (right)
      fraction of the total energy deposited in a column as a function of its
      distance from the shower center.}
    \label{fig:performance.ecal.showerprofile1D}
    \end{center}
 \end{figure*}

In addition to providing a good description of this ratio,
Eq.~\ref{eq:reconstruction.ecal.showerprofile1D} is conveniently related to the
energy deposited in each module~\cite{Lednev:95}.  Up to three contributing
shower components (denoted by the index $i$) are summed up, with parameters
$a_i$ and $b_i$ describing the relative weight and width of each component,
respectively.  \\ In order to obtain the shower profile parameters, the
following procedure is applied.  A column-wise calculation of the ratio of the
energy accumulated so far over the total energy of clusters, taking the simple
centre-of-gravity as the central position, yields a distribution that is fitted
with the cumulative function Eq.~\ref{eq:reconstruction.ecal.showerprofile1D}
describing the shower.  This is illustrated in
Fig.~\ref{fig:performance.ecal.showerprofile1D}~(left), which represents the
fraction of the total energy deposited up to a particular column at a given
distance from the shower center.  The fraction of the total energy deposited in
a column as a function of its distance from the shower center is shown in
Fig.~\ref{fig:performance.ecal.showerprofile1D}~(right).

This concept can be extended to two dimensions. In this case, the ratio between
the accumulated energy up to a point ($x$,$y$) and the total energy of the
shower is given by:
\begin{equation}
  F\left(x,y\right)=\frac{1}{4} +
  \frac{1}{2\pi}\sum_{i}a_{i}\cdot\left(\arctan\frac{x}{b_{i}}+
  \arctan\frac{y}{b_{i}}+\arctan\frac{x\cdot
    y}{b_{i}\sqrt{b_{i}^2+x^2+y^2}}\right)\,.
  \label{eq:reconstruction.ecal.showerprofile}
\end{equation}
The first two terms account for the ratio along $x$ and $y$ projections, while
the third term adds an asymmetry along the diagonal.

Different sets of parameters are used for each module type. In ECAL1 the same
profile with three contributions is used for all modules. The parameters were
obtained for the GAMS-2000 spectrometer~\cite{Lednev:95}. Since COMPASS makes
use of the same lead glass modules, the parameters are unchanged.

In ECAL2, the profiles of both lead glass and radiation-hardened lead glass
modules are described by two contributions, with parameters derived from the
electron beam calibration. For the Shashlik modules, the profile consists of
three contributions. The corresponding parameters are obtained from Primakoff
data events that contain a single high-energy cluster. In both cases it has to
be assumed that a cluster contains only one shower.

The two-dimensional ratio defined in
Eq.~\ref{eq:reconstruction.ecal.showerprofile} is used to calculate the relative
energy deposited by a shower at the position $\left(u_j,v_j\right)$ in each
module:

\begin{eqnarray}
  G_j\left(x,y\right) & = & F\left(u_{j}+\Delta,v_{j}+\Delta\right) -
  F\left(u_{j}+\Delta,v_{j}-\Delta\right) \nonumber\\ & - &
  F\left(u_{j}-\Delta,v_{j}+\Delta\right) +
  F\left(u_{j}-\Delta,v_{j}-\Delta\right)\,.
  \label{eq:reconstruction.ecal.showerdist}
\end{eqnarray}

Here $u_j = X_j - x$ and $v_j = Y_j - y$ are local coordinates relative to the
centre ($X_{j}$,$Y_{j}$) of each module, $j$ denotes an index over all modules
and $\Delta$ is half the transverse size of a module.

\subsubsection{Clustering and shower reconstruction}
\label{sec:reconstruction.ecal.reconstruction}
In a first step, signals from adjacent modules are combined to form a cluster.
Starting from the first module not yet used in the cluster, each of the
(vertical, horizontal, and diagonal) neighbours is checked for a measured energy
above a threshold of $100\,\MeV$ for ECAL1, and of $200\,\MeV$ for ECAL2.  For
each new module added to the cluster, its neighbours are treated the same way.

In a second step, a fit of the shower profiles to the cluster data is
performed. The fit improves the spatial resolution of the calorimeter and
separates overlapping showers. The fit is first done with a single
shower. Further showers are added one by one with a new fit being performed
after each added shower.  The parameters of the first shower are initialised to
those of the module with the highest energy in the cluster and its neighbours.
The energy is set to the sum of the energies of those modules, the position and
time are set to the mean of the respective information weighted with the energy
of each of those modules.  When adding more showers, the module searched is the
one with the largest relative discrepancy between its measured energy and the
energy predicted to be deposited by all showers fitted to the data so far.  The
centre of the new shower candidate must be located at a distance larger than
$\sqrt{2}/2$ times the module width from the centre of the nearest shower.  A
difference in energies is also required: the central module of the new shower
must contain at least $20\%$ of the energy deposited by all other showers in
this module.  If such a module is found, the information it provides together
with the information from the neighbouring modules is used to initialise the
parameters of a new shower.

According to Eq.~\ref{eq:reconstruction.ecal.showerdist}, the energy deposited
by a shower of energy $e_{i}$ at impact point ($x_{i}$,$y_{i}$) in a module at
position ($X_{j}$,$Y_{j}$) can be calculated as:

\begin{equation}
  E_{j,i}^{\textnormal{pred}} = e_i\cdot G_{j}\left(x_{i},y_{i}\right)\,.
\end{equation}

As several showers might be fitted into the same cluster, the total energy of
all showers in the module is given by:

\begin{equation}
  E_{j}^{\textnormal{pred}} = \sum_i E_{j,i}^{\textnormal{pred}} = \sum_i e_i\cdot G_{j}\left(x_{i},y_{i}\right)\,.
\end{equation}

In addition to the energy, the time information is also used. The time $t_{i}$
of a shower is defined as the mean value of the times of all modules
contributing to the shower, weighted with the energy deposited in each module.
Similarly to the predicted energy deposit in a block, the predicted time is
calculated as:

\begin{equation}
T_{j}^{\textnormal{pred}}
=
{\sum_i e_i\cdot G_{j}\left(x_{i},y_{i}\right)}
\frac{\sum_i e_i\cdot G_{j}\left(x_{i},y_{i}\right)\cdot t_i}{\sum_i e_i\cdot G_{j}\left(x_{i},y_{i}\right)}\,.
\end{equation}

The predicted energy and time are compared to the measured energy
$E_{j}^{\textnormal{meas}}$ and time $T_{j}^{\textnormal{meas}}$ in each module.
The Minuit fitter from the ROOT package \cite{Brun:05a} is used to optimise the
shower parameters to maximise the likelihood:

\begin{equation}
  -\log L = \frac{1}{2}\sum_{j}\left(\frac{\left(E_{j}^{\textnormal{meas}}-
        E_{j}^{\textnormal{pred}}\right)^{2}}{\sigma_{j,E}^{2}} + 
    \frac{\left(T_{j}^{\textnormal{meas}}-T_{j}^{\textnormal{pred}}\right)^{2}}{\sigma_{j,T}^{2}}\right)\,.
\end{equation}

The errors on the measured energies $\sigma_{j,E}^{2}$ and times
$\sigma_{j,T}^{2}$ (Fig.~\ref{fig:pid.ecal2.time-resolution}) are calculated
from an energy dependent parametrisation that has been determined from data
beforehand. With this procedure the fit distinguishes between in-time showers
and pile-up events.

The procedure of trying to add a new shower is stopped if the fit describes the
data well or if the maximum number of showers in a cluster has been reached.
The decision, whether the last shower added improves the fit, is based on a
comparison of the log-likelihood normalised to the number of degrees of freedom.
It is also checked that all showers have energies above the energy threshold and
that they fulfill the requirements on the distance between two showers described
above.

\begin{figure*}[h]
  \begin{center}
      \includegraphics[width=.45\textwidth]{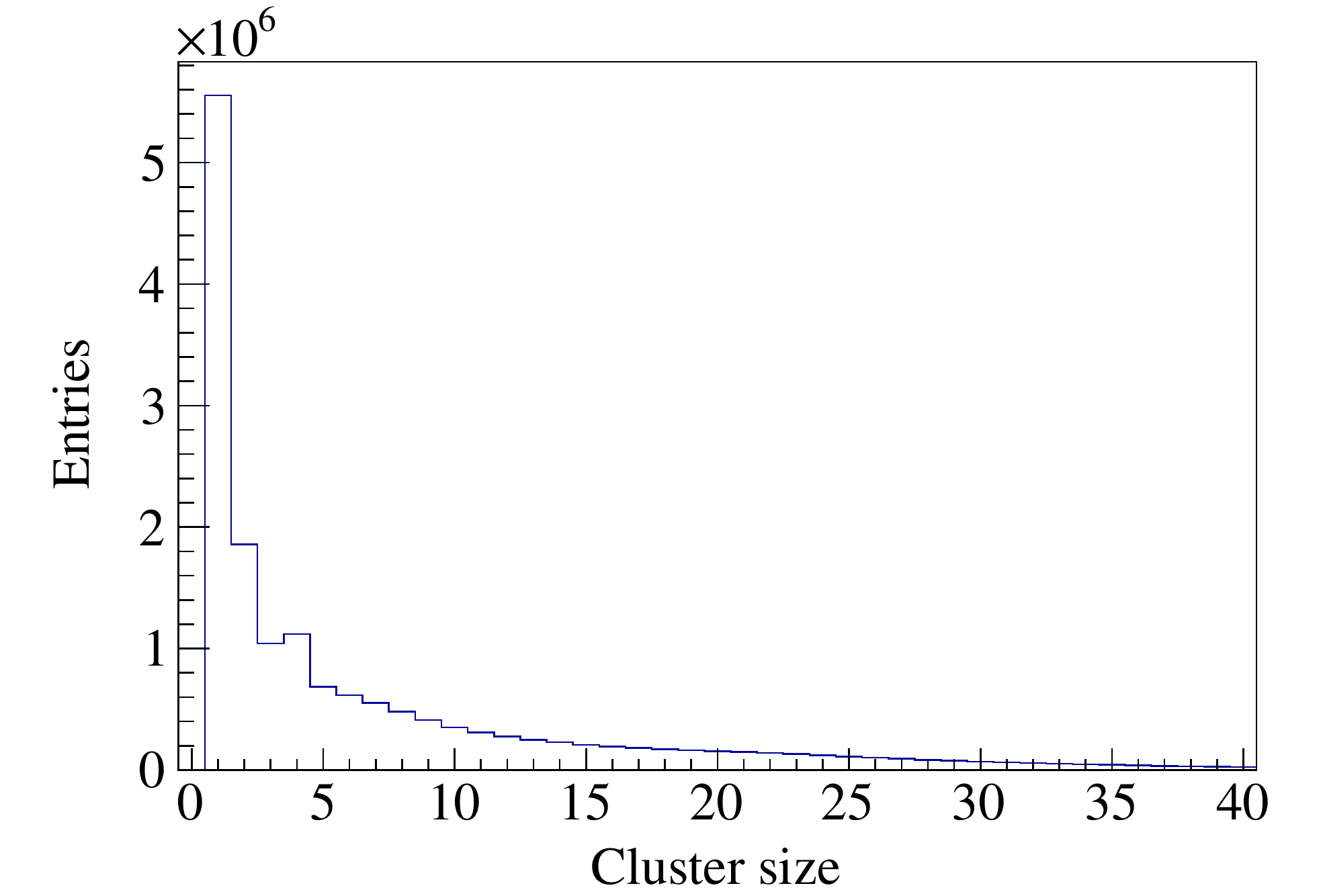}
      \includegraphics[width=.45\textwidth]{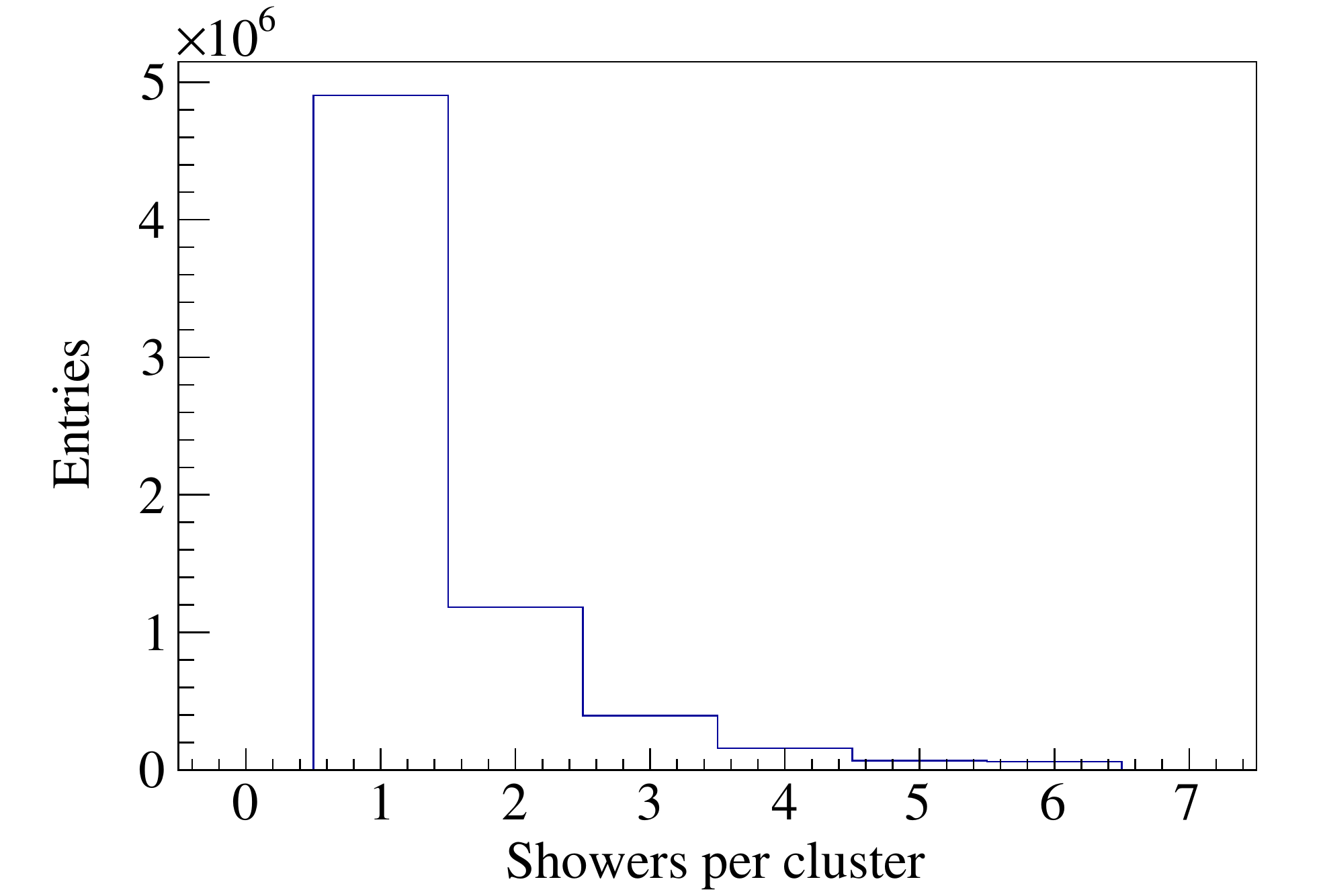}
   \caption{ECAL2 fit results for (left) number of modules per cluster and
     (right) number of fitted showers per cluster.}
   \label{fig:performance.ecal.clustersize}
   \end{center}
\end{figure*}

This fitting procedure returns the energy $e_i$, the position
$\left(x_i,y_i\right)$ and the time $t_i$ of each shower fitted into a
cluster.\\ The number of modules contributing to the total shower energy reaches
$5\times5$ for the highest energy photons. A cluster may contain from one to six
showers, the distribution varying as a function of the beam intensity and the
trigger conditions. Figure~\ref{fig:performance.ecal.clustersize}~(left) shows
the distribution of the number of modules in one cluster (cluster size) in ECAL2
for the diffractive dissociation data. The number of reconstructed showers per
cluster is shown in Fig.~\ref{fig:performance.ecal.clustersize}~(right).

For ECAL1 a simplified version of the procedure described above is employed.
For the larger MAINZ and OLGA modules, an improvement of the performance by
using shower profiles was not found.  Each module with a deposit larger than
that of any of its neighbours is used as a starting point for a new
shower. Energy and position of this module are used to initialise the parameters
of the shower.  If larger than the energy threshold, the information contained
in the neighbouring modules is then used to improve the position of the shower
by calculating its centre of gravity.

\subsubsection{Calibration with $\pi^0\rightarrow \gamma \gamma$ decays}
\label{sec:reconstruction.ecal.pi0calib}
The $\pi^0$ calibration procedure is performed, prior to the final analysis,
using a fraction (equivalent to 1 to 2 days of data taking) of the physics
events.  The reconstruction of the incident and outgoing particle tracks is
required, with a definition of a primary interaction vertex.  Only showers with
energies $E_{\gamma}$ larger than $1\,\GeV$ for ECAL1 and $3\,\GeV$ for ECAL2
are taken into account.  Showers associated with charged tracks are
discarded. In order to minimise combinatorial background, only events with less
than 5 showers are used.

\begin{figure*}[h]
  \begin{center}
      \includegraphics[width=.49\textwidth]{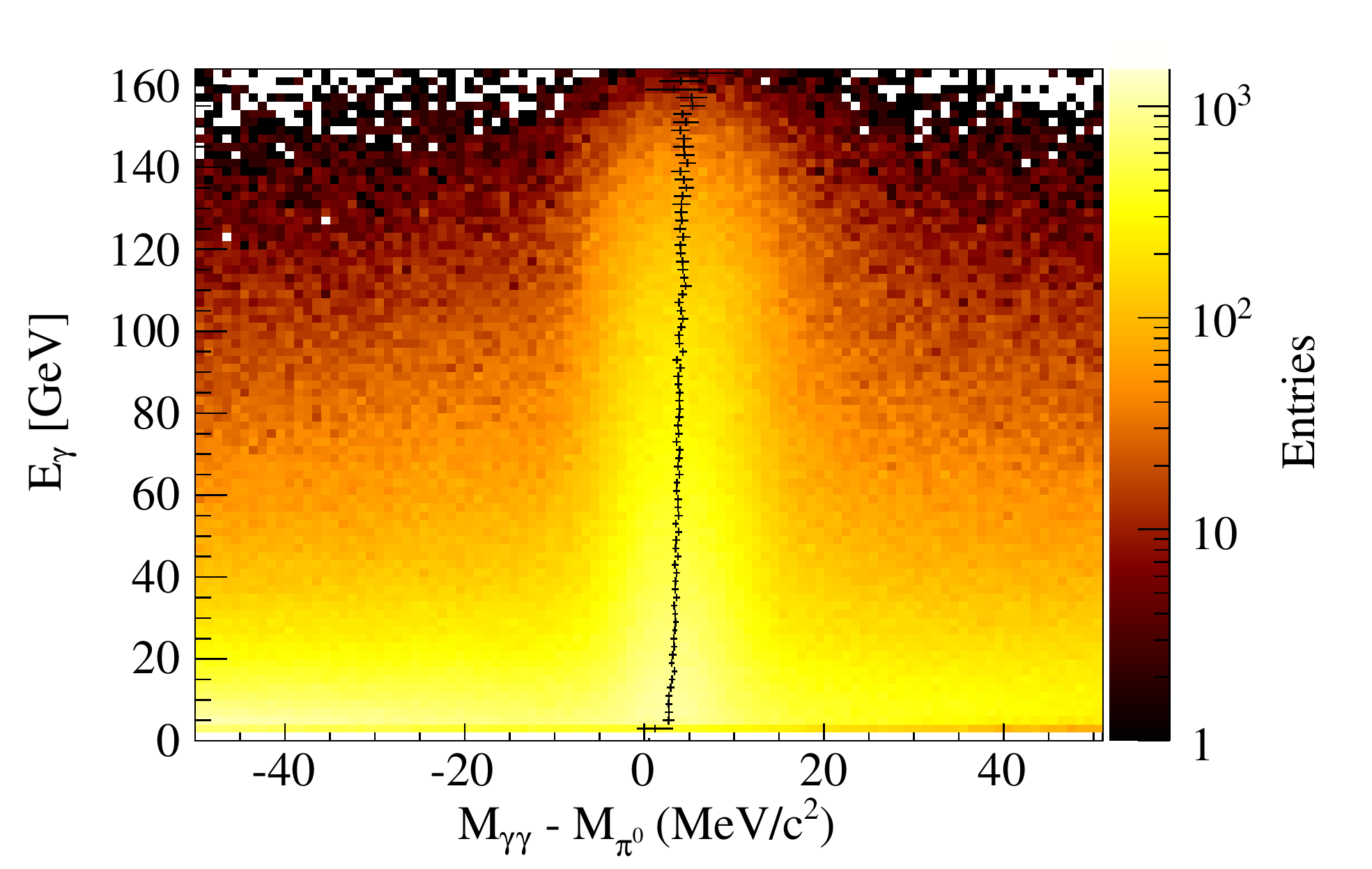}
      \includegraphics[width=.49\textwidth]{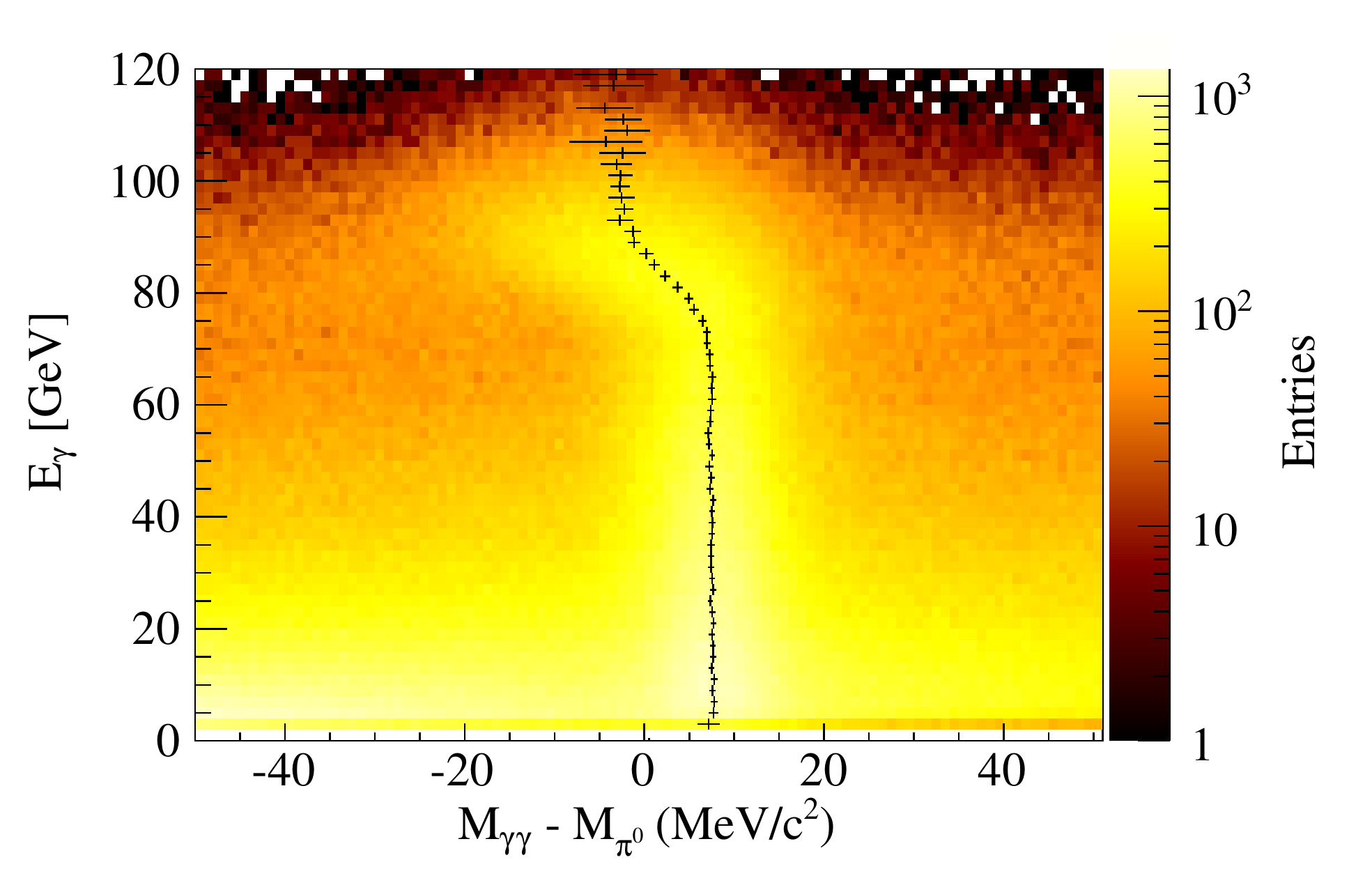}
    \caption{Energy deposition in two ECAL2 modules as a function of the
      difference between reconstructed and nominal $\pi^0$ mass for (left) a
      module with typical behaviour and (right) a module with an unusual
      behaviour.}
    \label{fig:performance.ecal.ribfit}
    \end{center}
 \end{figure*}

The two-photon invariant mass $M_{\gamma \gamma}$ is calculated for every pair
of showers, assuming that both photons originate from the interaction
vertex. Only pairs with invariant masses within $\pm 50\,\MeV/c^2$ around the
nominal $\pi^0$ mass, $M_{\pi^0}$, are considered as valid $\pi^0$ candidates.
For each of two showers, a two-dimensional histogram $E_{\gamma}$ vs $(M_{\gamma
  \gamma}-M_{\pi^0})$ is filled, which is associated with the shower's central
module.  The central module is defined as the module which contains the highest
fraction of the deposited energy.  The values stored in the two-dimensional
histograms are used as a starting point for the calibration procedure.

\begin{figure*}[t]
  \begin{center}
      \includegraphics[width=.49\textwidth]{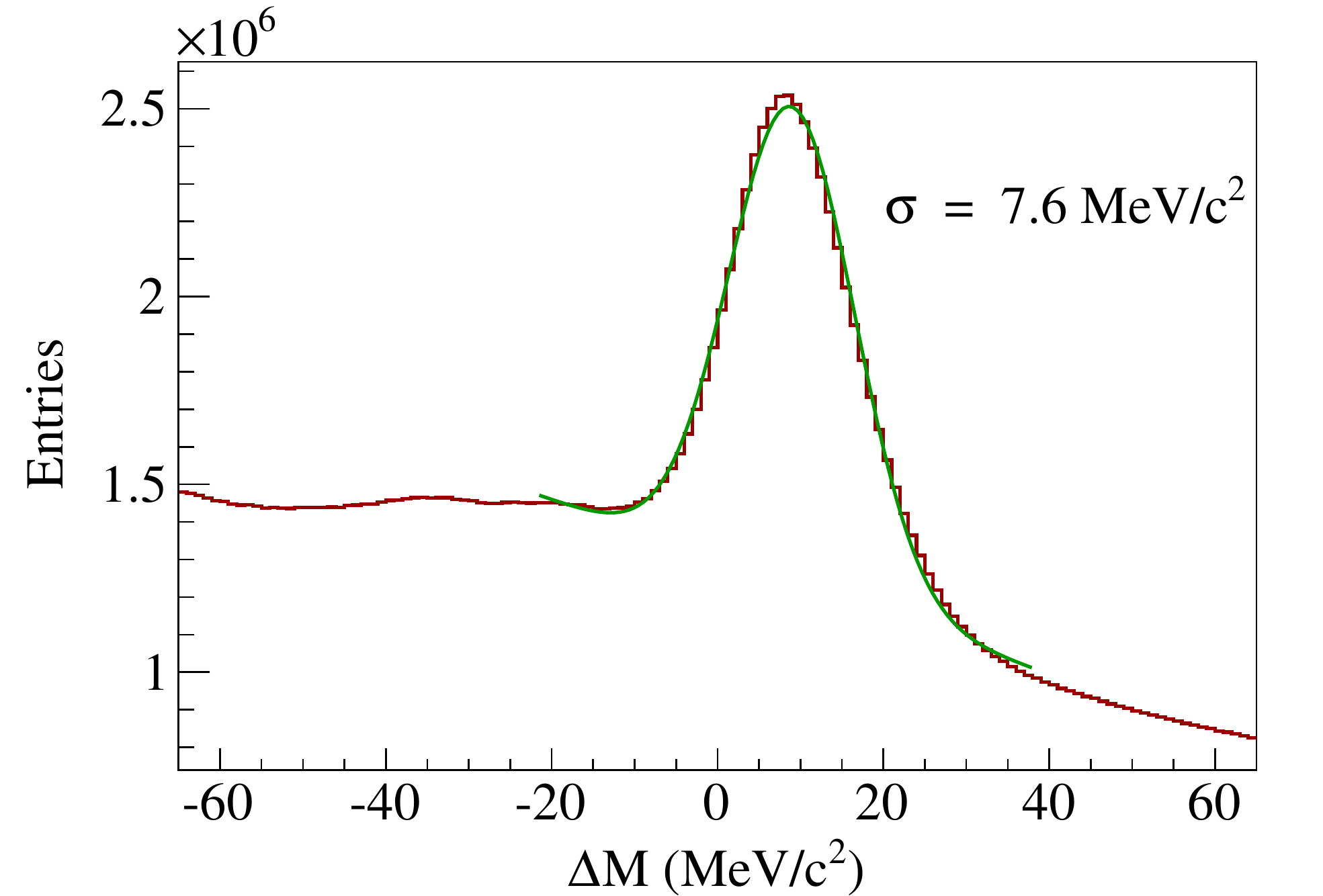}
      \includegraphics[width=.49\textwidth]{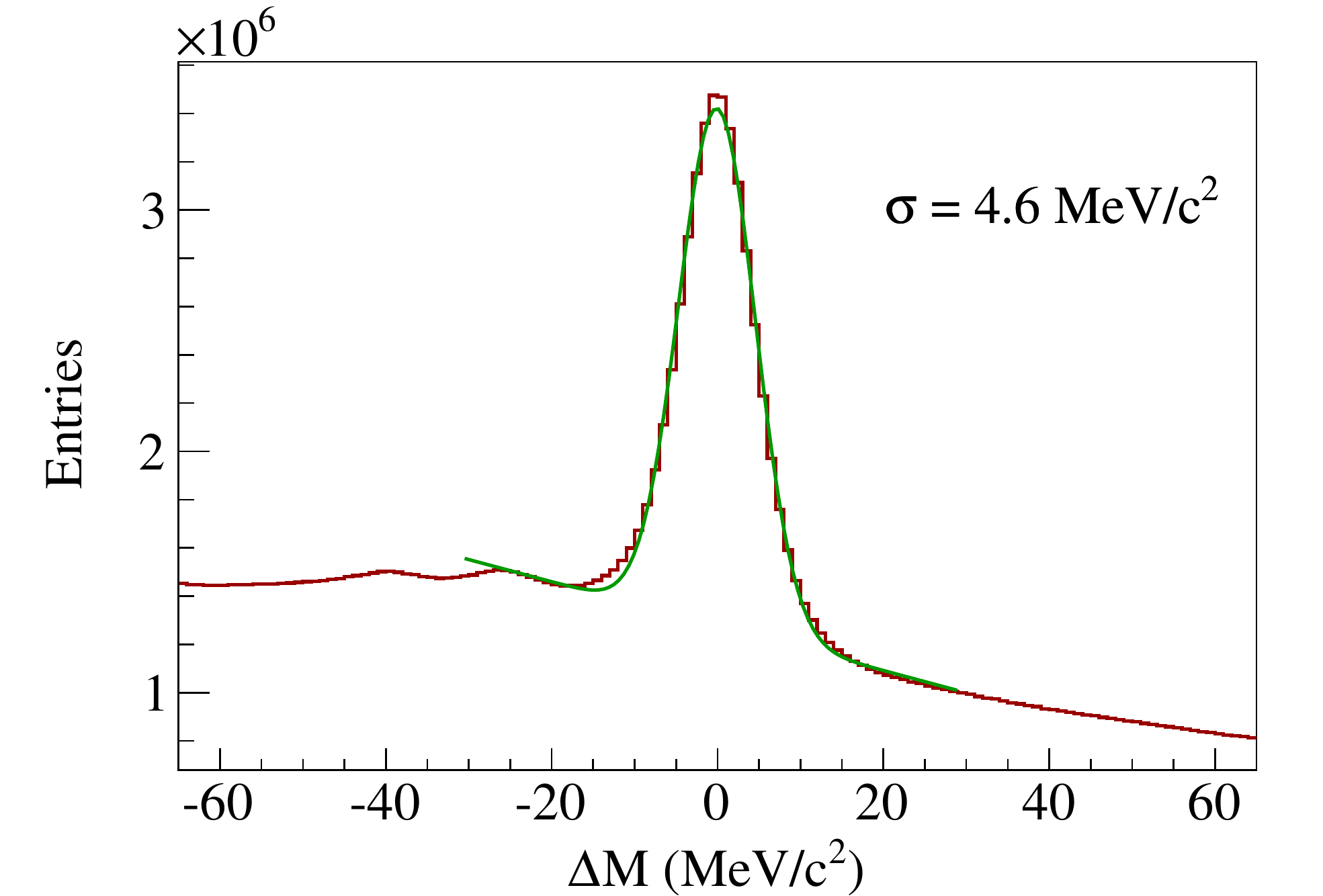}
    \caption{Difference $\Delta M$ between reconstructed and nominal $\pi^0$
      masses in ECAL2 for (left) before calibration and (right) after
      calibration.}
    \label{fig:performance.ecal.pi0s}
    \end{center}
 \end{figure*}

For most ECAL modules a slight energy dependence is observed, as illustrated in
Fig.~\ref{fig:performance.ecal.ribfit}~(left). However, various types of unusual
behaviours may also be present, e.g. as the one shown in
Fig.~\ref{fig:performance.ecal.ribfit}~(right), which resuls from a saturated
photomultiplier tube.  The variations are accounted for by introducing
correction factors that depend on the photon energy.  The correction factors are
calculated in energy slices of 2\,GeV.  In each slice the spectrum is fitted
with a Gaussian for the $\pi^0$ peak and with a first order polynomial for the
background, in an interval of $\pm 20\,\MeV/c^2$ around the $\pi^0$ peak. The
results of the fits are displayed in Fig.~\ref{fig:performance.ecal.ribfit} as
black crosses; its horizontal and vertical tick marks represent the $3\sigma$
fit error and the bin size, respectively. The fitted mass differences are then
used to calculate the correction factor $\alpha_{i}$ for each energy slice $i$,
\begin{equation}
  \alpha_{i} = \frac{1}{(1+\frac{\Delta M_i}{M_{\pi^0}})^{2}},
\end{equation}
where $\Delta M_i$ is the fitted mass offset. The correction factor for each
module is calculated assuming that the energy of the second decay photon is
measured precisely.  Since this is not the case, the $\pi^0$ calibration is done
iteratively, each iteration adding corrections to the result from the previous
iteration. Typically, after 8 to 10 iterations the procedure converges.  The
result is a significant improvement of the $\pi^0$ mass resolution and of the
$\pi^0$ mass offset, as shown in Fig.\,\ref{fig:performance.ecal.pi0s}.  After
calibration, the mean value of the peak position shifts from $9.0\,\MeV/c^2$ to
$0.0\,\MeV/c^2$.  The mass resolution improves from $7.6\,\MeV/c^2$ to
$4.6\,\MeV/c^2$.
 
\begin{figure*}[b]
  \begin{center}
      \includegraphics[width=.49\textwidth]{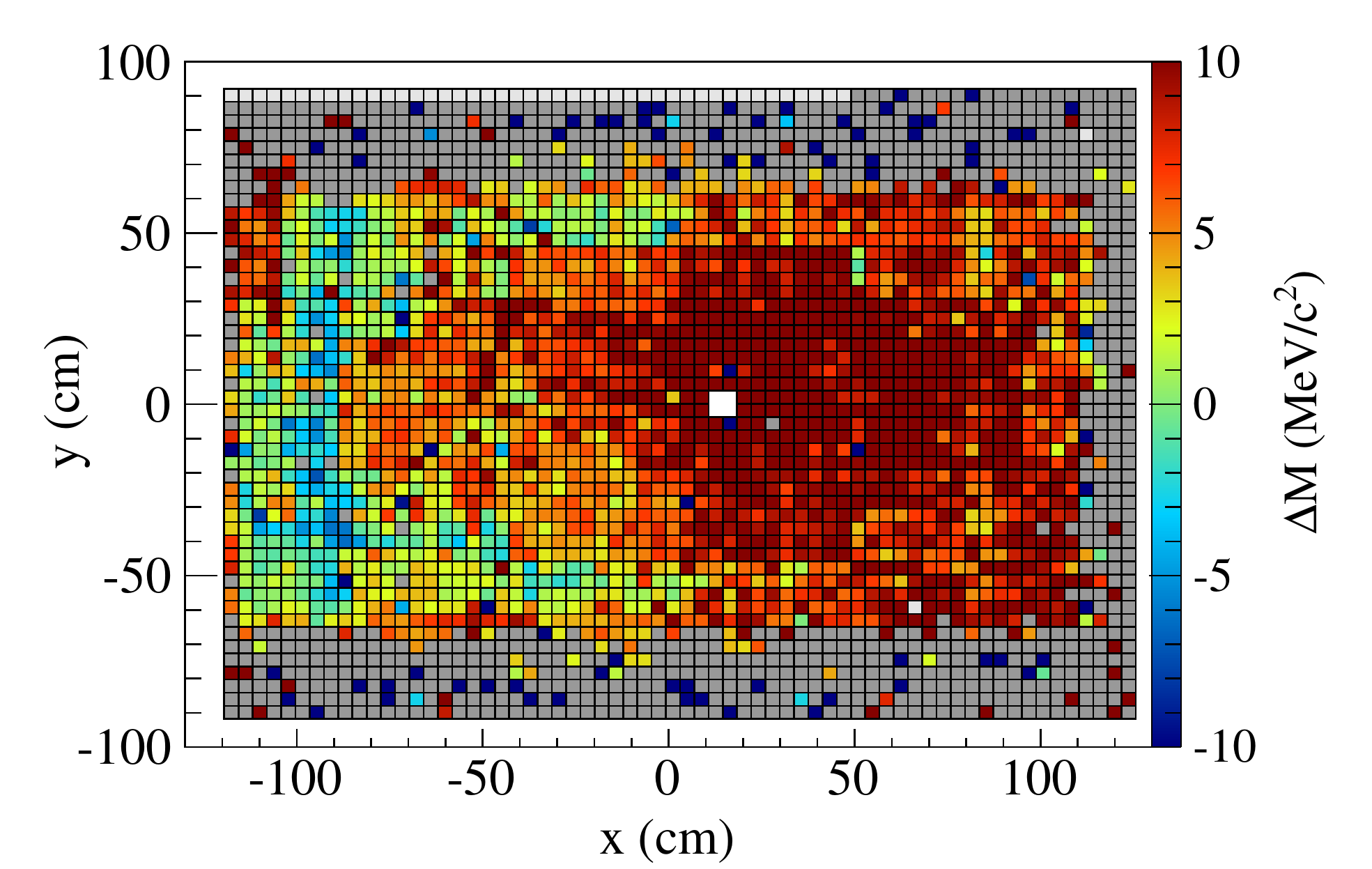}
      \includegraphics[width=.49\textwidth]{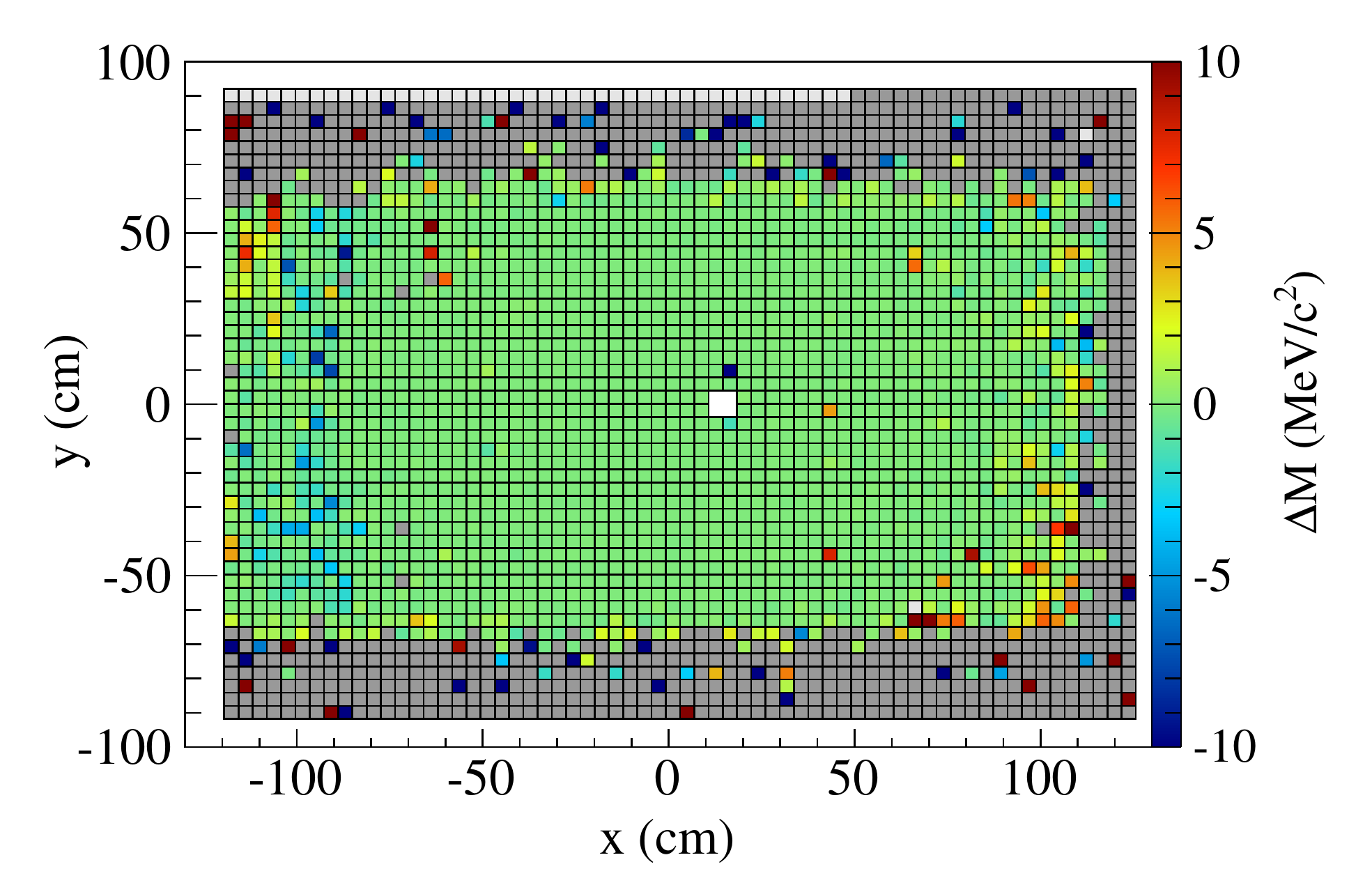}
    \caption{Difference between reconstructed and nominal $\pi^0$ masses as a
      function of the impact position for the ECAL2 modules for (left) before
      $\pi^0$ calibration and (right) after $\pi^0$ calibration.  The difference
      is calculated using the mean value of the fitted (with a Gaussian)
      X-projection of $E_{\gamma}$ vs $(M_{\gamma \gamma}-M_{\pi^0})$
      histograms.  The grey rows at the top and bottom ends and on the right
      side of ECAL2 are located beyond the angular acceptance for photons coming
      from the target (see \secref{sec:pid.ecal.ecal2}.)}
    \label{fig:pid.ecal2.pi0calib_2d}
    \end{center}
 \end{figure*}

The calibration significantly improves the response of the individual 
ECAL2 modules, as illustrated in Fig.~\ref{fig:pid.ecal2.pi0calib_2d}. 
A similar improvement is observed for ECAL1. 
For most modules, the 
reconstructed pion mass after calibration agrees within less than 
$1\,\MeV/c^2$ with the nominal $\pi^0$ mass for $\pi^0$ energies up to  
$160\,\GeV$.

The resulting calibration is used to correct the individual module responses 
during the event reconstruction procedure. The precise value of the correction 
factor $\alpha$, which corresponds to the actual energy deposited in each module, 
is determined by interpolation.

\subsubsection{Additional corrections}
\label{sec:reconstruction.ecal.corr-rods}
\begin{figure}[th]	 
  \centering
  \includegraphics[width=0.5\textwidth]{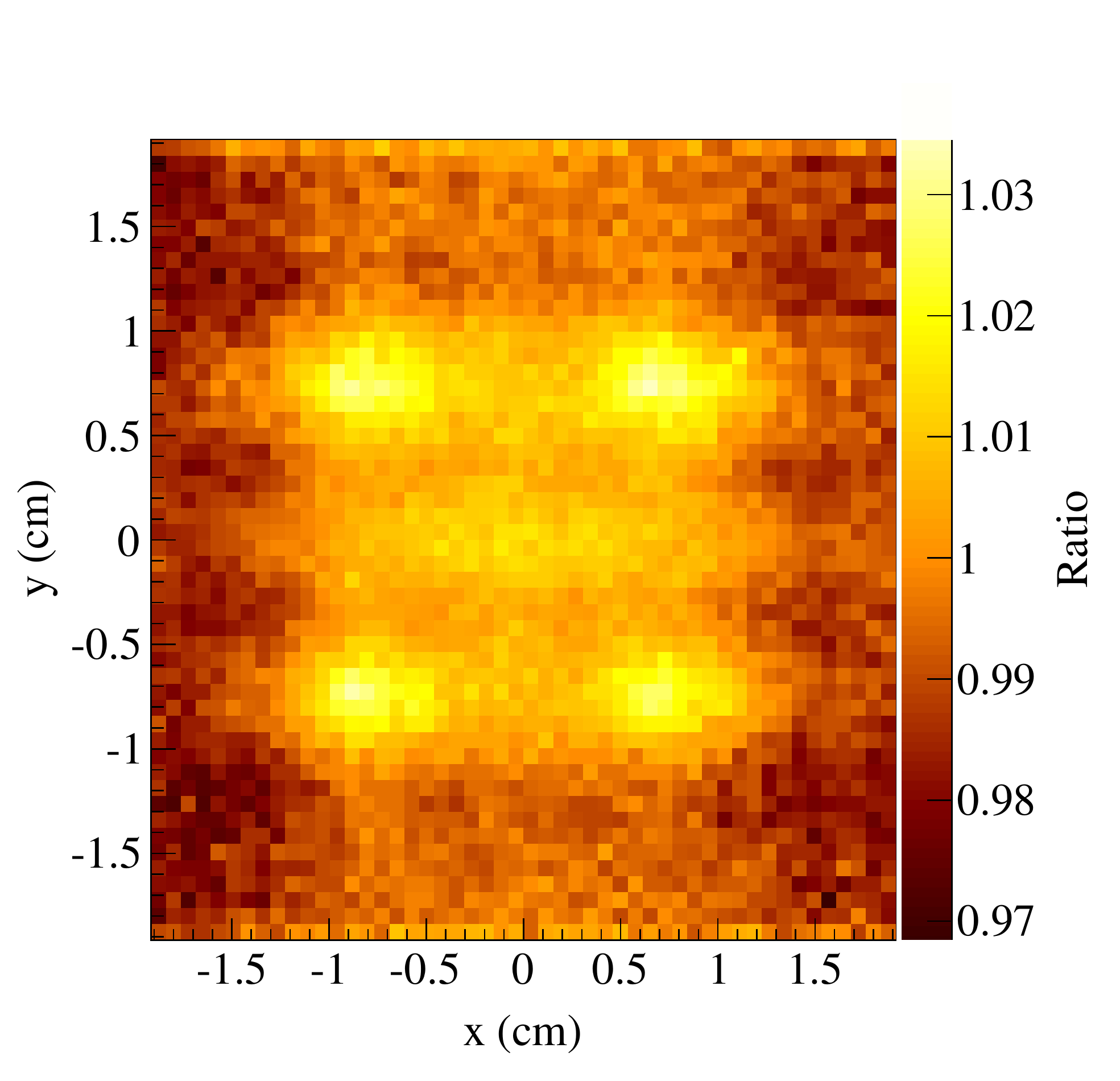}
    \caption{Ratio of track momentum over calorimeter energy as a function of
      the impact position in a Shashlik module relative to its centre. The four
      central spots with a ratio larger than one correspond to the four module
      rods.}
  \label{fig:performance.ecal.shashlik_ratio}
\end{figure}

Additional corrections, which are evaluated independently for the different data
sets, are applied on top of the shower fit result. For the diffractive
dissociation data, the reconstructed energy $E$ as measured in a Shashlik module
for an electron is compared with the corresponding charged track momentum $p$.
A position dependence of the ratio $p/E$ is then observed, as indicated in
Fig.~\ref{fig:performance.ecal.shashlik_ratio}.  This dependence reflects slight
inefficiencies in the vicinity of the four central rods. It is accounted for by
using the hit position as determined by the shower fit.
	 
The photons detected in the calorimeters cover an energy domain that extends
from less than 1~GeV for ECAL1 to more than 120~GeV for ECAL2. The measured
cluster times for both ECAL1 and ECAL2 show a slight energy dependence, mainly
for low photon energies.  This dependence, which is always smaller than 1~ns, is
fitted to the data and accounted for.
 
\subsubsection{High-energy photons}
\label{sec:reconstruction.ecal.high-gamma}
\begin{figure}[tbp]
    \centering
    \includegraphics[width=0.6\textwidth]{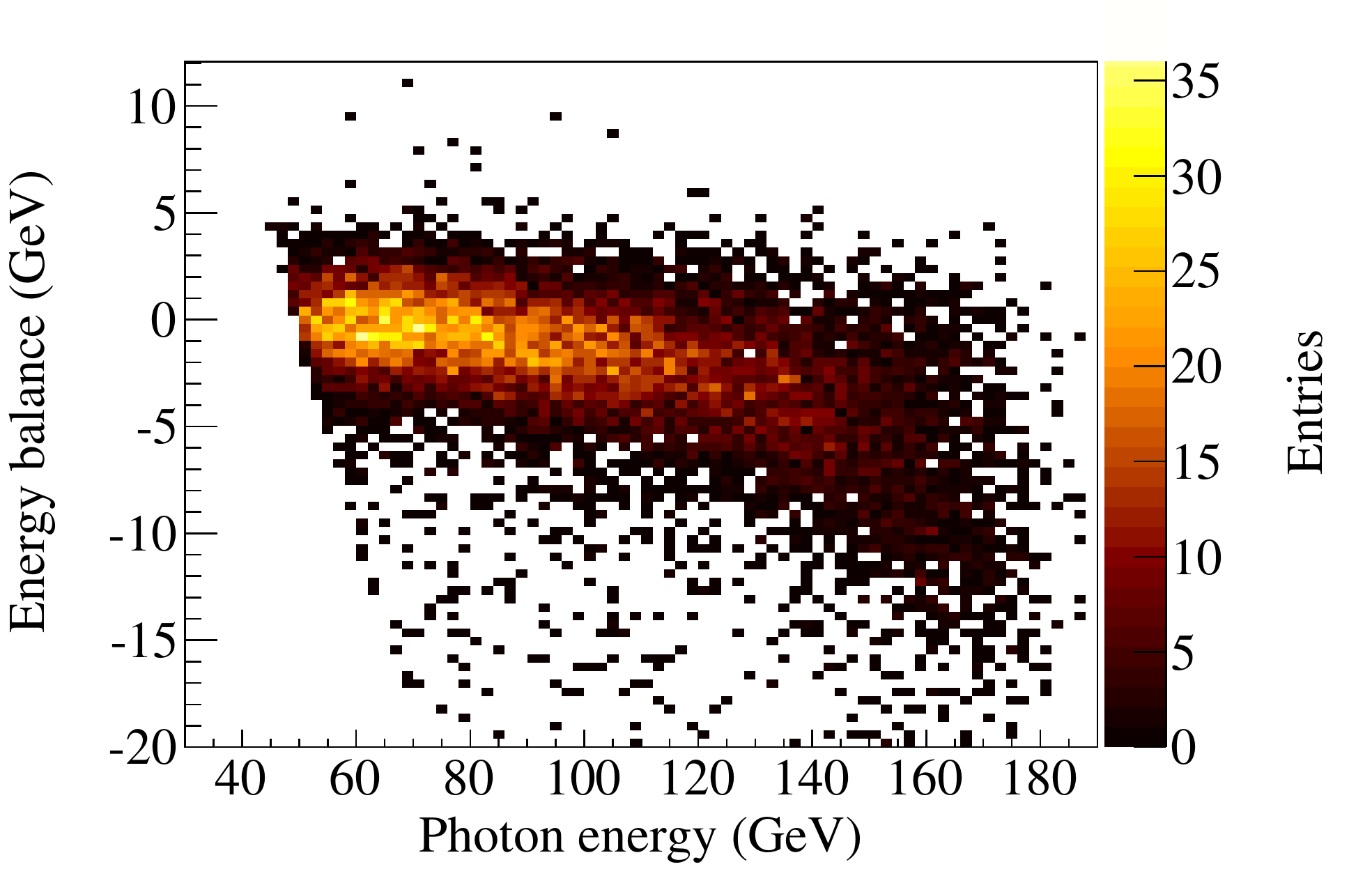}
    \caption{Difference between the beam energy and the total measured energy as
      a function of the photon energy.}
    \label{fig:performance.ecal.prim-corr.mu}
\end{figure}

For photon energies above 80\,GeV, the kinematics of the Primakoff-Compton
reaction, $\pi^{-}+(A,Z)\rightarrow\pi^{-}+\gamma+(A,Z)$, constrain the detected
photons to the central $4\times4$ ECAL2 modules. The energy deposited in these
modules must be accurately determined, even for values as large as 95\,\% of the
beam energy.  With the $\pi^0$ calibration alone this cannot be achieved. For
higher energies a different method is applied.  During the Primakoff data-taking
period, muon beam data for systematic studies are periodically collected. These
data also contain Primakoff-Compton events with photon energies nearly as high
as the beam energy. Moreover, since the Beam Momentum Station (BMS) is present
in the beam line, the muon incident momentum is known. Since the scattered muon
momentum is also measured, the energy conservation (exclusivity) in the process
provides an independent prediction of the energy of the emitted photon. The
comparison with the actual ECAL2 measurement based on the $\pi^0$ mass
calibration exhibits a slightly falling slope as a function of the photon
energy, as shown in Fig.~\ref{fig:performance.ecal.prim-corr.mu}.

\begin{figure}[tbp]
    \centering
    \includegraphics[width=0.6\textwidth]{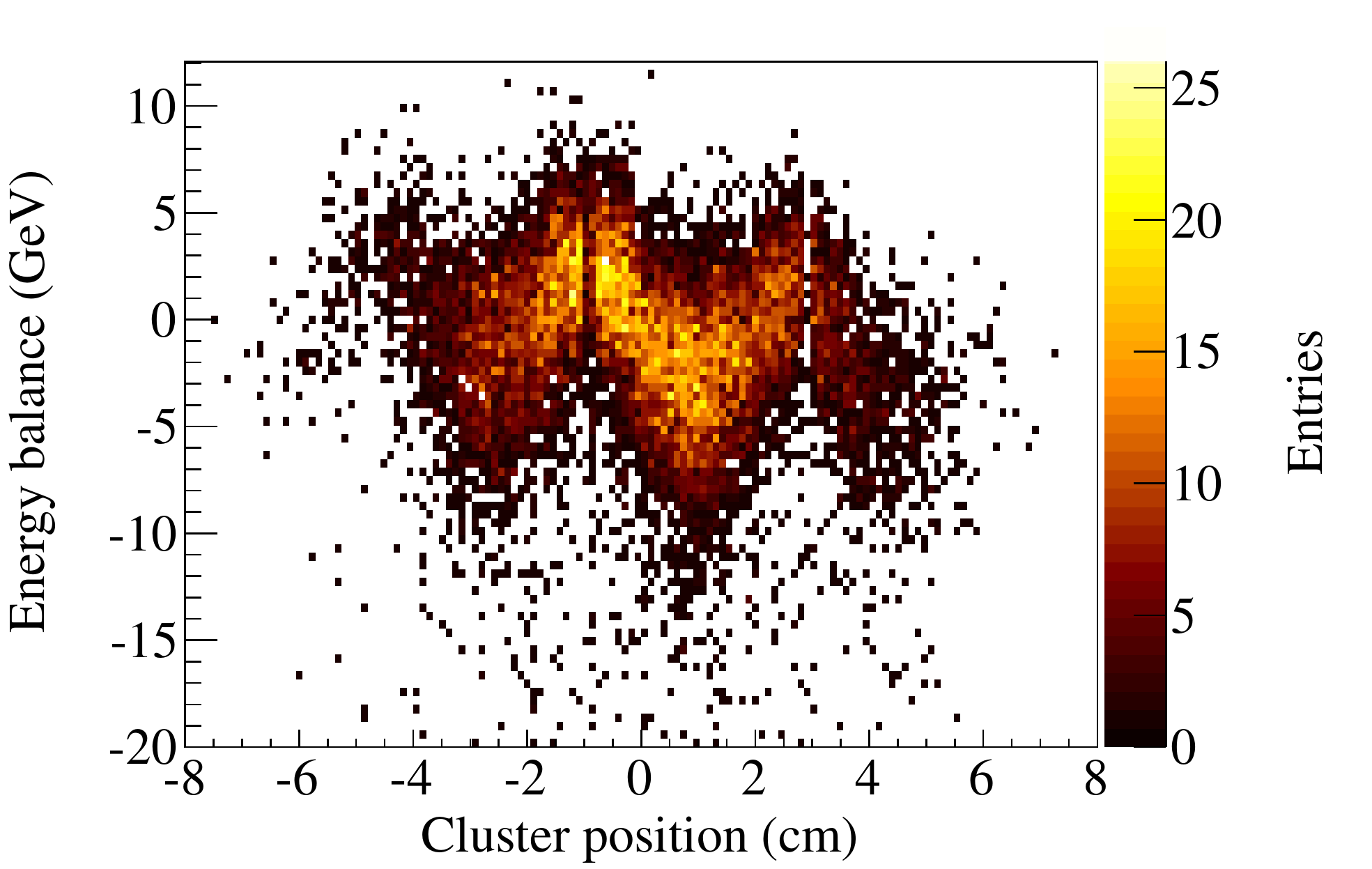}
    \caption{Intra-cell energy variation as a function of the distance to the
      cell centre.}  
  \label{fig:performance.ecal.prim-corr.clust}
\end{figure}

This trend does not only depend on the photon energy but also on the actual hit
position within the Shashlik modules. The observed dependence is fitted with a
three-dimensional function that includes both intra-cell coordinates and the
shower energy.  The correction reaches values of up to $+7\,\GeV$ and
$-12\,\GeV$, as shown in Fig.~\ref{fig:performance.ecal.prim-corr.clust}. The
major part of this correction is due to the steel rods that tie the Shashlik
stack together, as previously explained in
\secref{sec:reconstruction.ecal.corr-rods}.
\begin{figure}[ht]
  \begin{minipage}[ht]{0.45\textwidth} \centering
    \includegraphics[width=\textwidth]{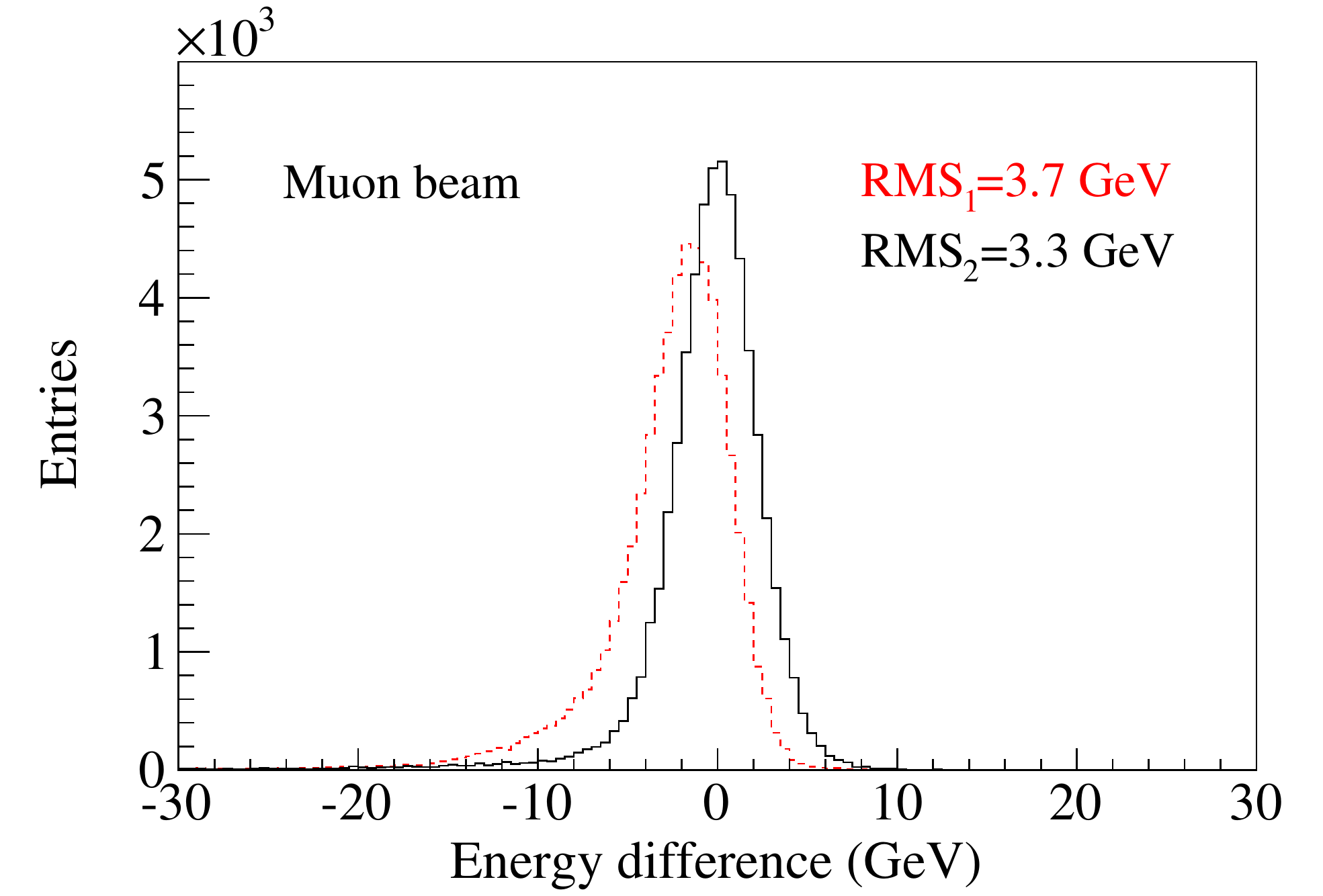}
  \end{minipage} \hfill
  \begin{minipage}[ht]{0.45\textwidth} \centering
    \includegraphics[width=\textwidth]{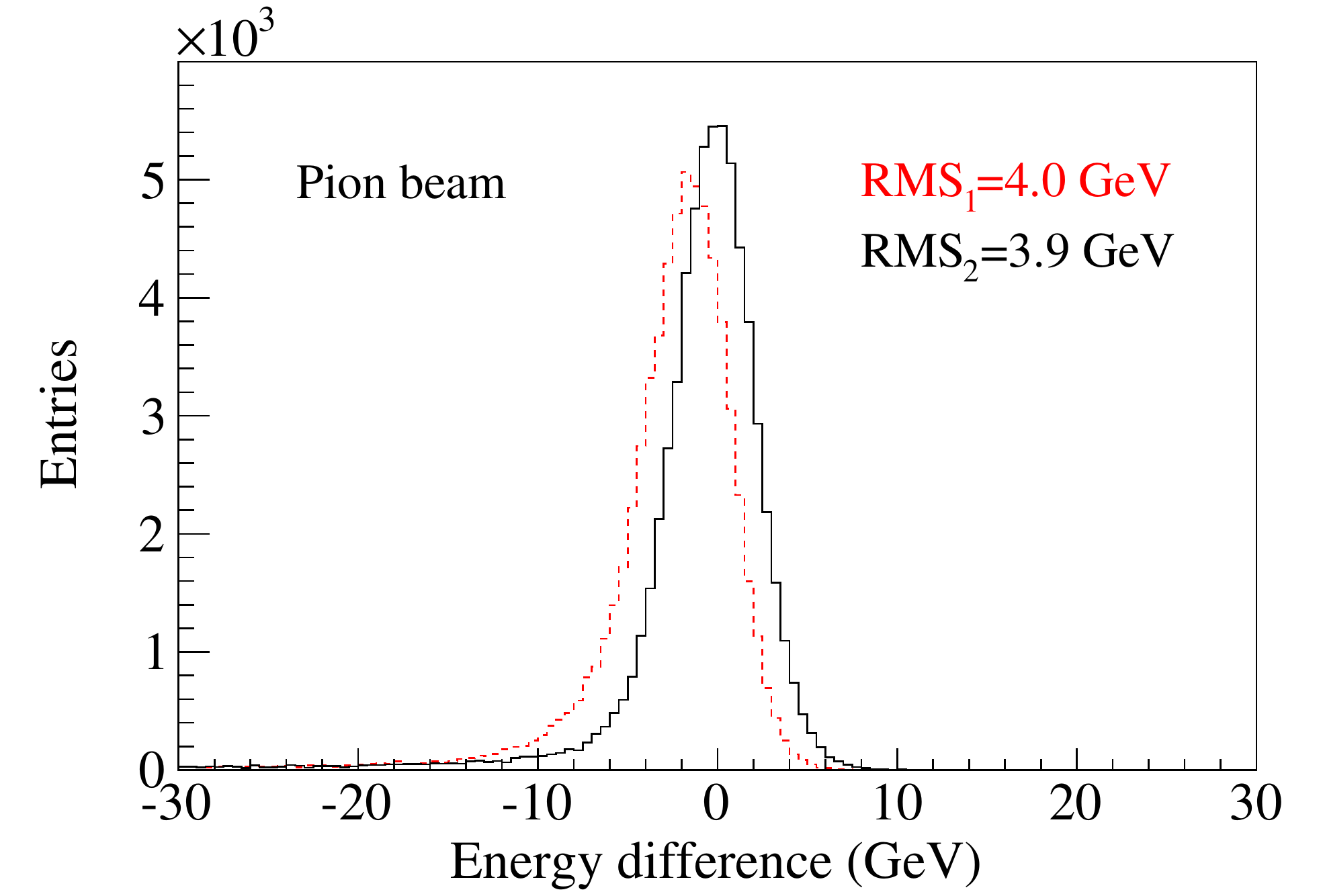}
  \end{minipage} 
  \caption{Difference between beam and measured energies (energy balance) for
    Primakoff-Compton scattering (left) with a muon beam and (right) with a pion
    beam. The distributions are displayed with the standard $\pi^0$ calibration
    only (dashed curve) and with linearity and intra-cell position corrections
    (solid curve); the corresponding RMS$_1$ and RMS$_2$ values are indicated.}
  \label{fig:performance.ecal.exclusivity}
\end{figure}
During data taking with a pion beam, the BMS is removed and no measurement of
the incident pion momentum can be performed. It is assumed that the corrections
to the ECAL2 calibration for muon and pion beams are identical. This assumption
is supported by the data itself; after applying the above intra-cell
corrections, both the position and the standard deviation of the exclusivity
peak with a pion beam improve, similarly to the improvement achieved with a muon
beam. Fig. \ref{fig:performance.ecal.exclusivity} illustrates the effect of
these corrections on the central modules of ECAL2.

\subsubsection{Efficiency of photon detection}
\label{sec:reconstruction.ecal.efficiency}
The efficiency for the reconstruction of single photons is defined as the
fraction of photons that originate from the target reconstructed in one of the
calorimeters. It thus includes effects of geometric acceptance for photons like
dead material, and the intrinsic performance of the calorimeters related to
thresholds, etc.  The reconstruction efficiency is evaluated by a Monte-Carlo
procedure using diffractive events for the $\pi^-\pi^0\pi^0$ channel (see
\secref{sec:performance.acceptance}) with $0.1<t'<1.0\,\GeV^2/c^2$. The
thresholds applied correspond to those in the analysis of physics data
($0.6\,\GeV$ for ECAL1, $1.2\,\GeV$ for ECAL2).

\begin{figure}[tbp]
  \begin{minipage}[t]{0.45\textwidth} \centering
    \includegraphics[width=\textwidth]{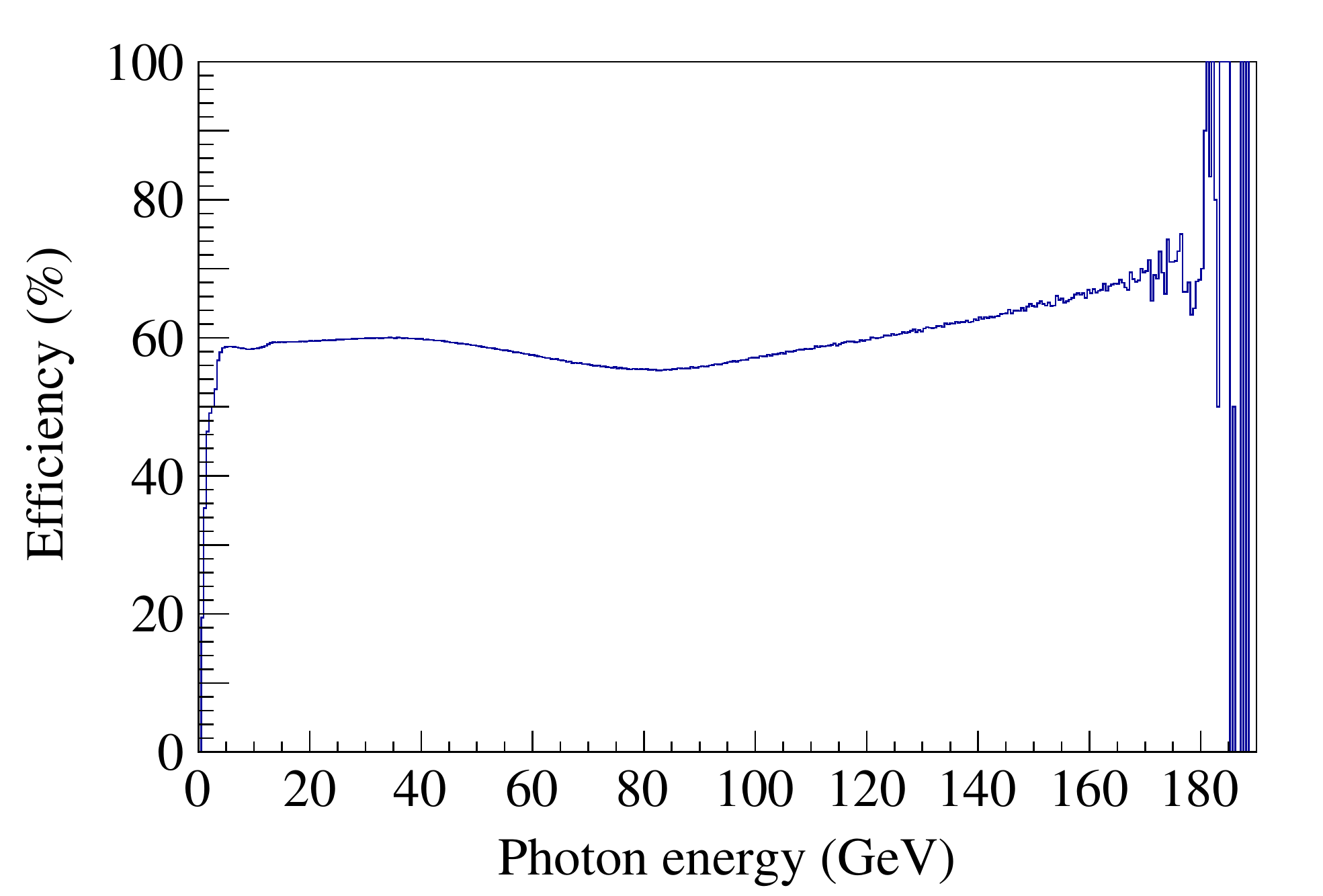}
  \end{minipage} \hfill
  \begin{minipage}[t]{0.45\textwidth} \centering
    \includegraphics[width=\textwidth]{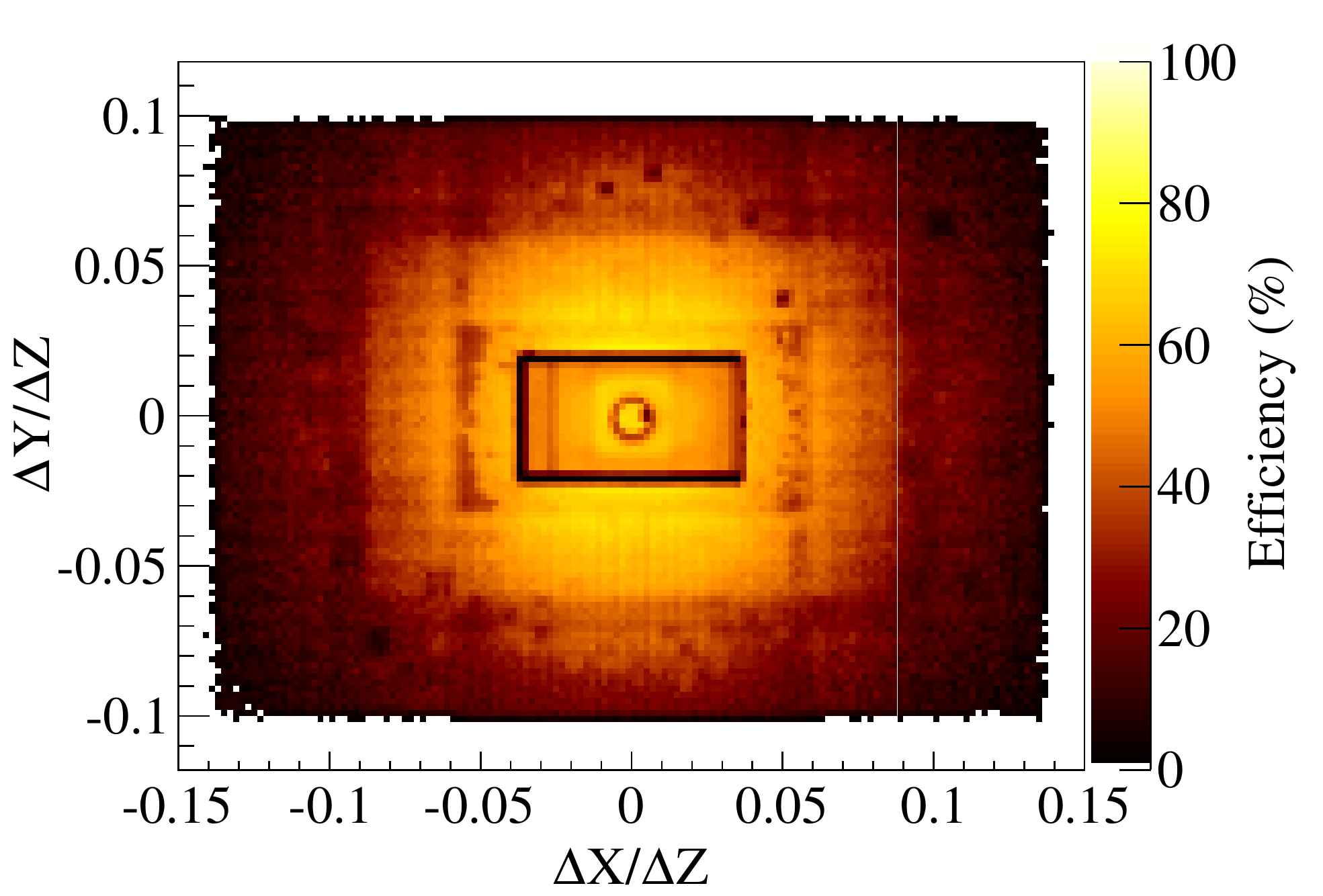}
  \end{minipage} \hfill
  \caption{Simulated photon efficiency (left) as a function of the photon energy
    and (right) as a function of the photon direction in the laboratory system.}
  \label{fig:performance.ecal.efficiency}
\end{figure}

\Figref{fig:performance.ecal.efficiency} shows the single-photon reconstruction
efficiency as a function of photon energy, and as a function of the photon
direction in the laboratory system.
The single-photon reconstruction efficiency is rather uniform at a level of
$60\%$ for photon energies above approximately $5\,\GeV$. However, the
acceptance strongly depends on the direction of the photon. In the centre a drop
due to the beam hole in ECAL2 can be seen at $\Delta y/\Delta z=0$ and $\Delta
x/\Delta z=0.005$. The circular shape overlapping with the hole is attributed to
photons lost in the beam pipe of RICH-1. Further outside, at $\Delta y/\Delta
z\approx\pm 0.02$, the shadow of HCAL1 and SM2 on ECAL2 is visible as a
horizontal line, while the vertical lines at $\Delta x/\Delta z\approx\pm 0.04$
stem from a non-perfect overlap of ECAL1 and ECAL2. A general trend of lower
acceptance towards outer regions is also visible, which is caused by the loss of
photons with an energy below the ECAL1 threshold.

\section{Monte Carlo simulation and performance of the setup}
\label{sec:performance}
The interpretation of physics processes that involve hadron beams and several
particles in the final-state particles requires a thorough understanding of the
experimental setup. This requirement can only be achieved through a realistic
simulation of the apparatus and a detailed knowledge of its acceptance as
function of any of the kinematic variables that are relevant in a particular
physics process. The Monte-Carlo code used to describe the setup and to
determine its acceptance is described below. It is followed by a selection of
characteristic experimental results, for each of the two beam polarities and for
various particles in the final state.  All results were obtained with the
nominal hadron beam momentum of $190\,\GeV/c$.

\subsection{Monte Carlo simulation and acceptance}
\label{sec:performance.acceptance}
The simulation of the COMPASS setup is performed using a dedicated Monte-Carlo
(MC) code called COMGEANT.  The code can be linked to external event generators
specific to the reaction mechanism that is dominant in a given
channel. Final-state particles are then propagated through the setup. The
digitisation of the MC data and the subsequent reconstruction are carried out
with the same software that is used for reconstructing the measured events.

Three different event generators are used to simulate diffractive reactions,
central production, and Primakoff reactions.  The partial-wave analysis method
employed for diffractively and centrally produced $n$-body final states takes
into account the acceptance of the apparatus using Monte-Carlo pseudo data,
where the final-state hadrons are distributed isotropically in the $n$-body
phase space.  In addition to the decay phase space, the generators also simulate
the production kinematics.  Diffractive events are generated with a $t'$
distribution that is tuned to the data. The central-production
generator~\cite{uman_qnp09} simulates exponential $t'$ distributions for both
beam and target vertices.  For Primakoff-Compton scattering, the generator
calculates differential cross sections with contributions due to polarisability,
first-order Compton vertex corrections, and soft photon
emission~\cite{Friedrich:12}.

For all generators, the beam phase space spanned by the positions and angles of
the incoming particles is generated using parametrizations extracted from real
data.  Since the incident energy is not measured, it is reconstructed from the
kinematics of fully exclusive events.  Interactions in the target volume are
distributed in the target material according to the target positions.  Primary
interactions of the beam particles in materials and detectors surrounding the
target are not generated.

Scattered and secondary particles are propagated through the spectrometer by the
simulation code COMGEANT, based on GEANT 3.21~\cite{geant_man}.  Multiple
scattering, energy loss, shower development, and secondary interactions are
taken into account.  This includes interactions of electrons and photons with
detector material and creation of electromagnetic showers by these particles.
Additional physics processes like hadron interactions and in-flight decays are
also taken into account.  Furthermore, pile-up events due to two or more
particles occurring in the same time window can be generated.

The digitisation of the simulated events is performed in CORAL (see
\secref{sec:reconstruction.tracking}).  Dead and active materials along the
tracks are accounted for with the ROOT geometry package~\cite{Brun:2003xr}.
Charged-particle tracks are reconstructed from the simulated hits in the
tracking detectors using the same procedure as for the real data (see
\secref{sec:reconstruction.tracking}).  Detector properties such as efficiencies
and resolutions are implemented in the reconstruction software using information
from the experimental data.  For RICH-1, the purity and efficiency of the
detector are determined from the measured events and separately unfolded from
the simulated data. For the electromagnetic calorimeters, shower profiles are
extracted from the ECAL MC data, and a $\pi^0$ calibration is performed as for
real data (see \secref{sec:reconstruction.ecal}).  The RPD information is passed
directly to the PHAST physics analysis software.

\begin{figure*}[h]
  \begin{center}
    \includegraphics[width=.50\textwidth]{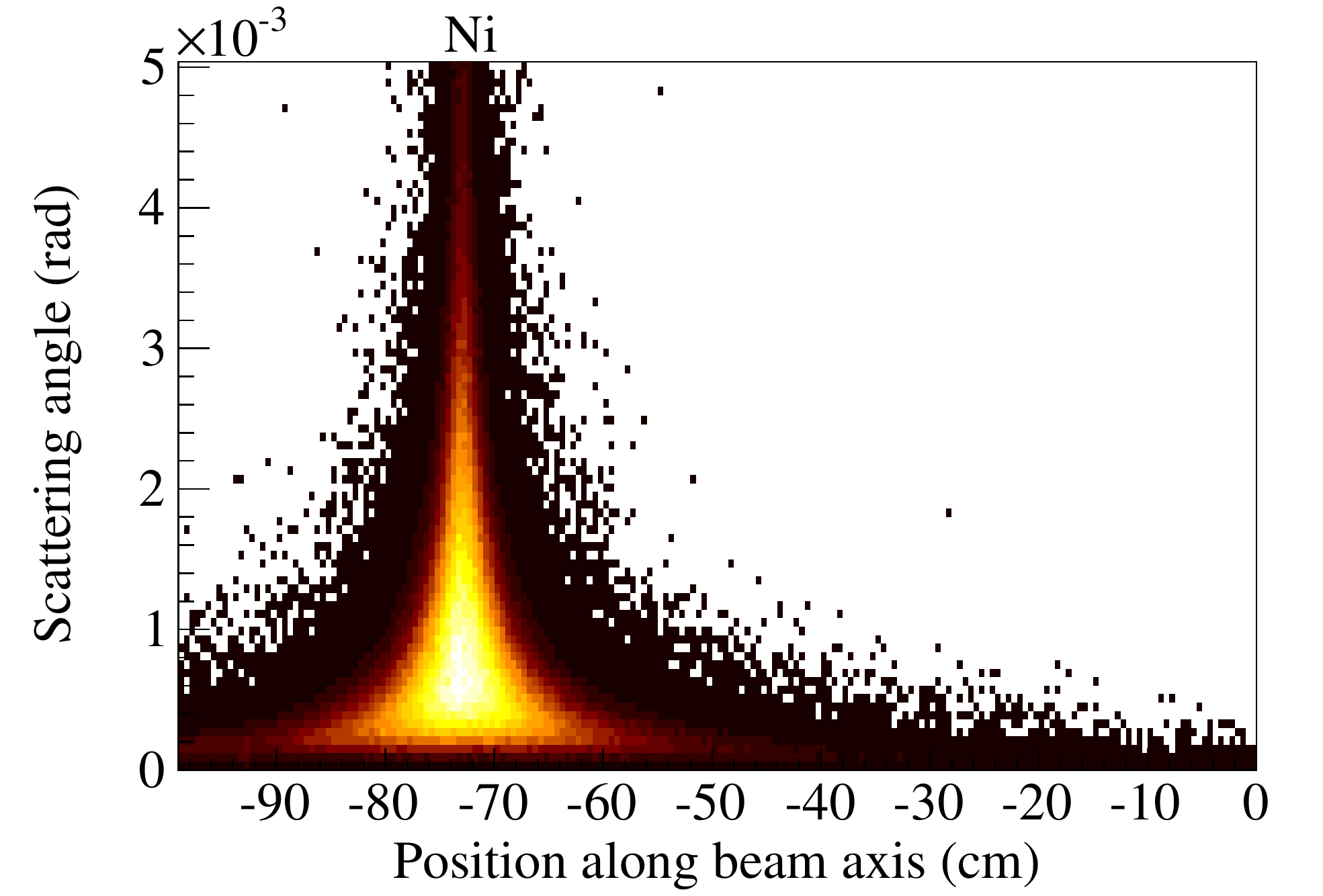}
    \caption{Monte Carlo simulation of the Primakoff-Compton reaction, showing
      the reconstructed position of the primary vertex along the beam direction
      as a function of the scattering angle of the outgoing pion. Note that
      interactions outside the target material are not simulated.}
    \label{fig:vertex_mc}
  \end{center}
\end{figure*}

An example of the good MC description achieved for the Primakoff-Compton
reaction is shown in \figref{fig:vertex_mc} for the reconstructed primary vertex
as a function of the pion scattering angle. A qualitative comparison with
\figref{fig:Improvement} (see \secref{sec:reconstruction.vertex}) shows that the
agreement between MC and data is good. Background effects, e.g.  from
interactions with the detectors downstream of the target, are minimized by
applying selection cuts identical to those used for the specific physics
process.

Other observables relevant for the Primakoff-Compton reaction are discussed in
\Secref{sec:performance.setup}. The momentum distributions of the
electromagnetic component for pion and muon interactions with a solid target are
well reproduced by the MC simulations (see
\figref{fig:performance.setup.primakoff-q}).  The $\pi^-\pi^+\pi^-$ decay of the
$K^-$ mesons in the beam, a process used for flux normalisation, is accurately
simulated as illustrated by the momentum transfer distributions in
\figref{fig:performance.setup.t.kaons}.

The MC simulation of the photon reconstruction efficiency for channels with
final-state photons is validated by comparing the acceptance-corrected particle
decay yields for different decay channels.  For example, the resulting branching
ratios of $\omega \to \pi^-\pi^+\pi^0$ and $\omega \to \pi^0\gamma$ agree within
5\% with the PDG values.

\begin{figure*}[t]
  \begin{center}
    \includegraphics[width=.49\columnwidth]{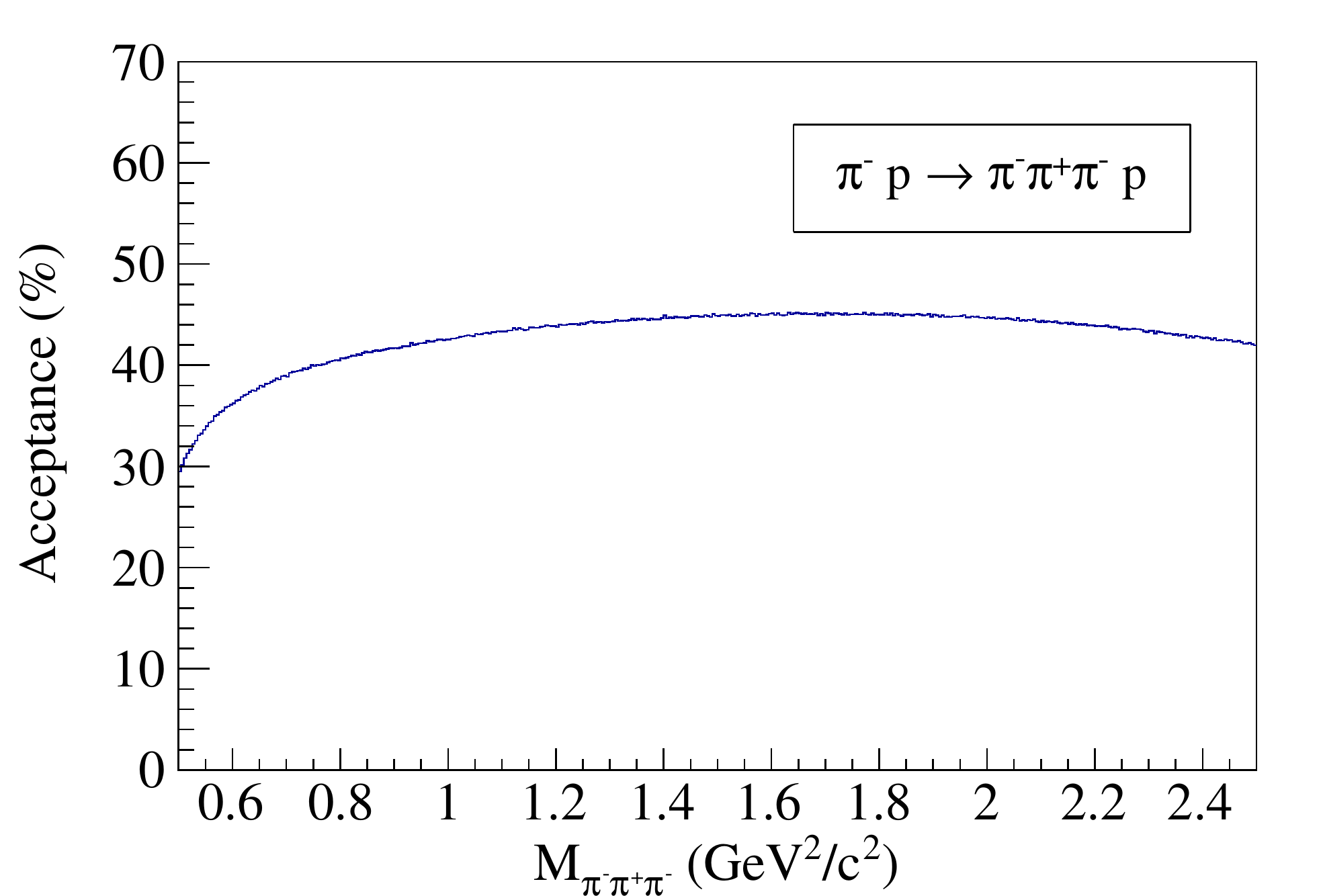}
    \includegraphics[width=.49\columnwidth]{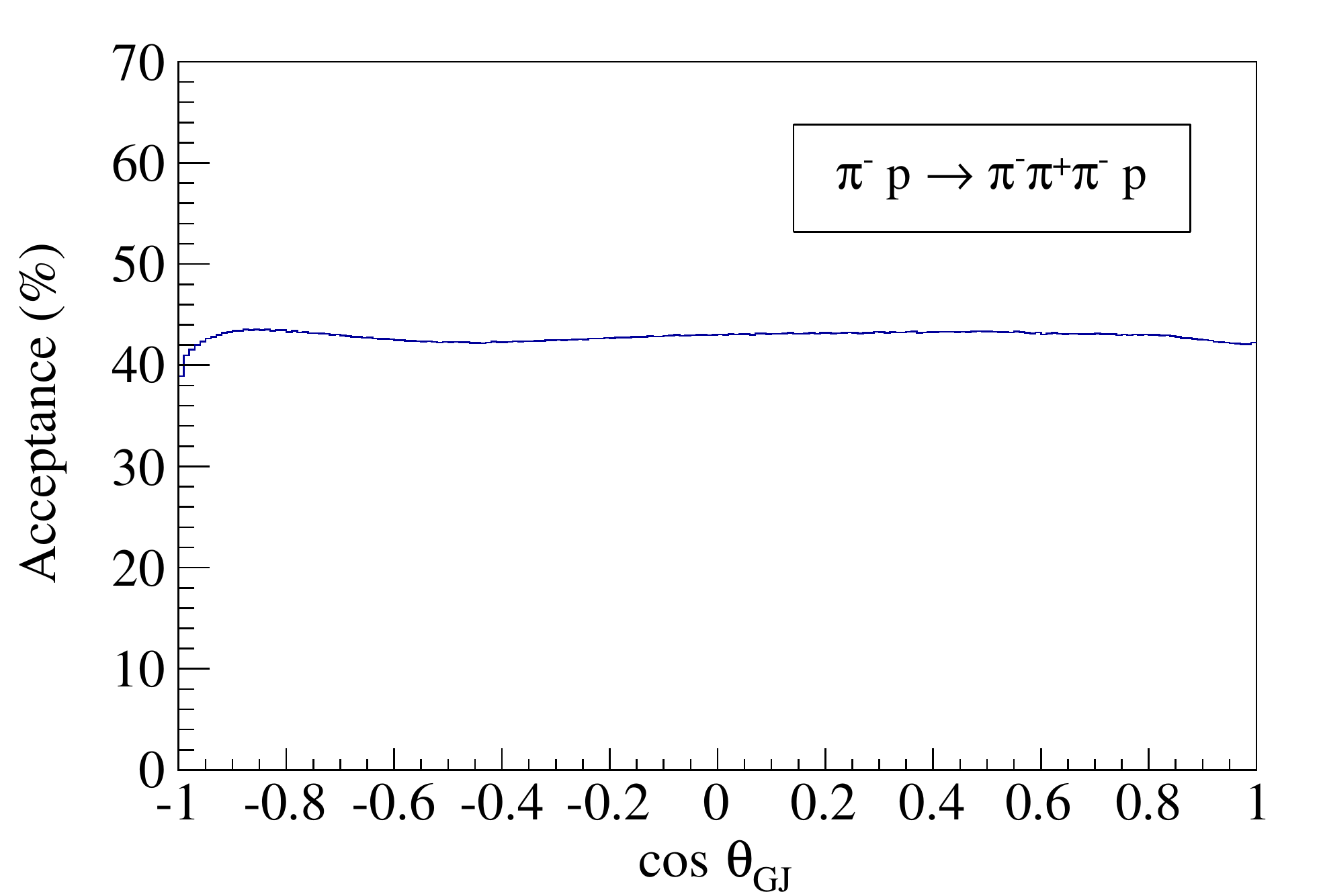}
    \caption{Acceptance for the diffractively produced $\pi^-\pi^+\pi^-$ final
      state (left)~as a function of the $3\pi$ invariant mass and (right)~as a
      function of the polar angle of the $\pi^+\pi^-$ isobar in the
      Gottfried-Jackson frame.}
    \label{fig:3pi_acceptances}
  \end{center}
\end{figure*}

The acceptance of the apparatus is determined by comparing the reconstructed and
generated MC events. In the Partial Wave Analysis (PWA) formalism employed for
diffractive scattering, the full multidimensional acceptance for a given final
state is used. For example, in three-body analyses (like $\pi^-\pi^+\pi^-$) for
fixed four-momentum transfer and three-body mass the acceptance depends on five
kinematic variables.  The acceptance determined by the simulated phase-space
events is then used as an input to the fits performed in the PWA formalism.

\begin{figure*}[b]
  \begin{center}
     \includegraphics[width=.49\textwidth]{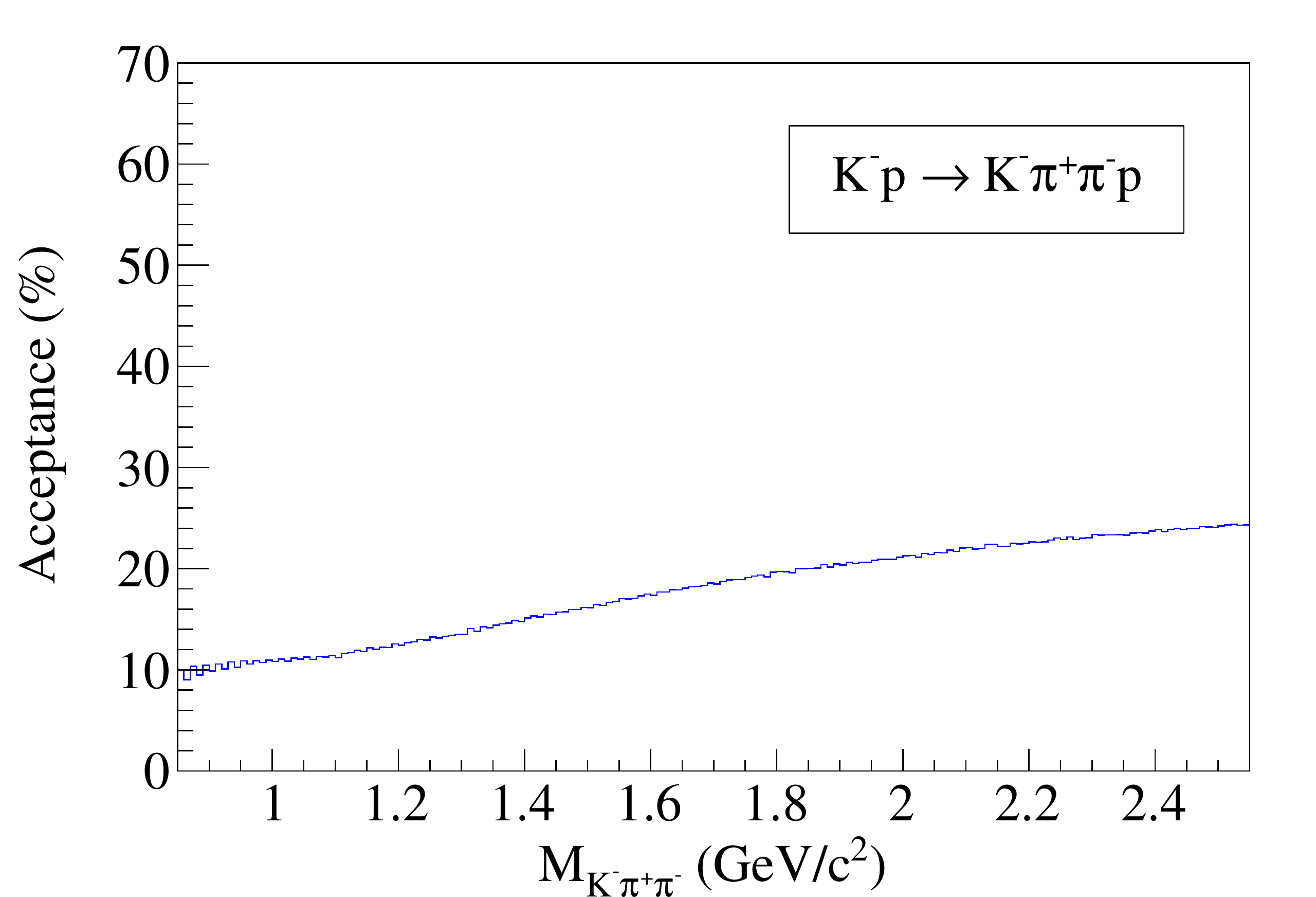}
     \includegraphics[width=.49\textwidth]{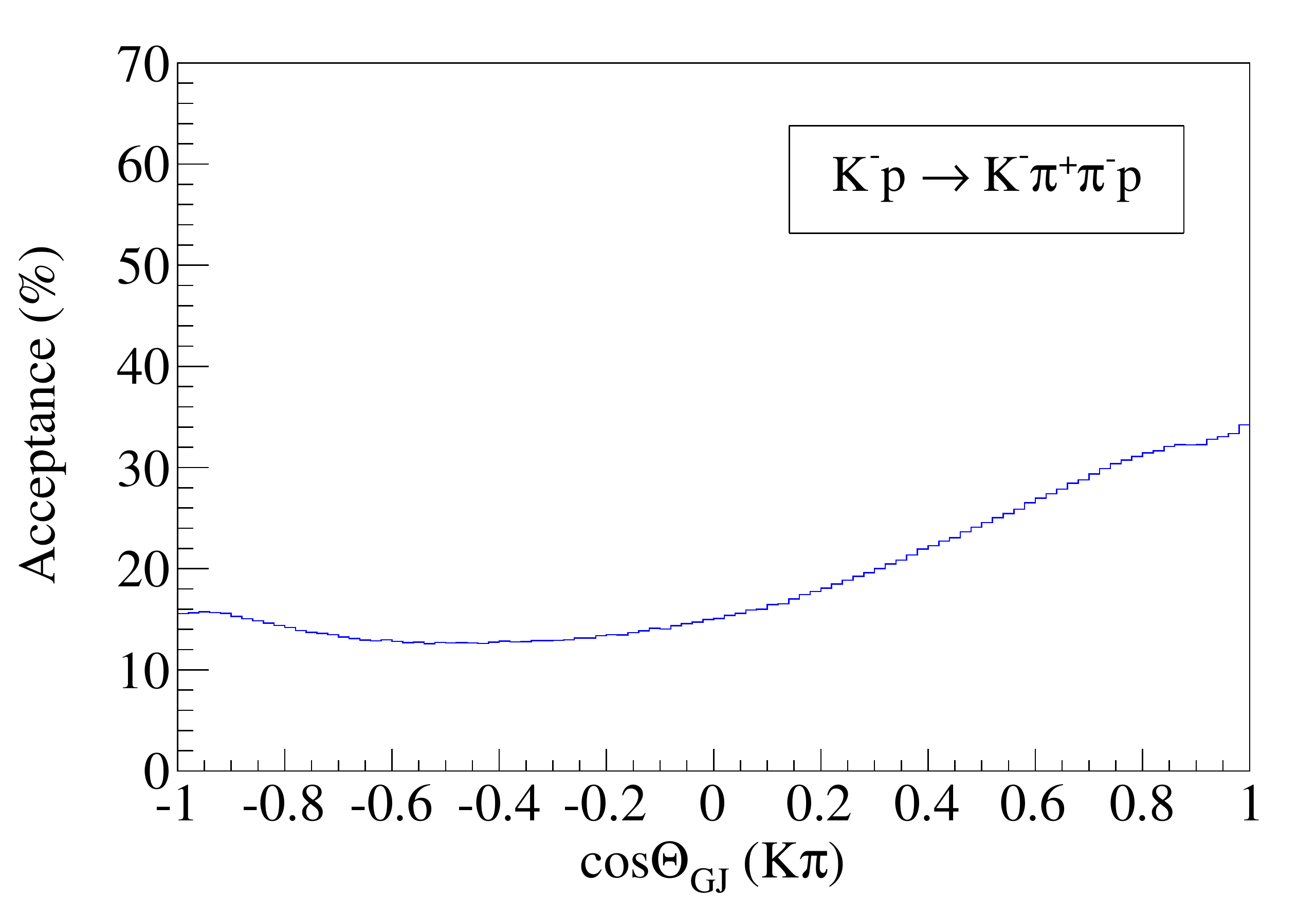}
     \caption{Acceptance for the diffractively produced $K^-\pi^+\pi^-$ final
       state (left)~as a function of the $K\pi\pi$ invariant mass, and
       (right)~as a function of the polar angle of the $\pi^+K^-$ isobar in the
       Gottfried-Jackson frame.}
    \label{fig:Kpipi_acceptances}
  \end{center}
\end{figure*}

For the purpose of illustration, the acceptance for a particular variable can be
determined by projecting the multi-dimensional acceptance onto this variable.
In \figref{fig:3pi_acceptances}, the acceptance for the $\pi^{-}\pi^{+}\pi^{-}$
final state of diffractive dissociation is evaluated in the $t'$ range between
0.1 and $1\,\GeV^2/c^2$. It is a fairly flat function of the invariant
three-pion mass from near threshold up to $2.5\,\GeV/c^2$ and of the polar angle
of the $\pi^+\pi^-$ isobar in the Gottfried-Jackson frame (see definition in
\secref{sec:intro}).

More pronounced modulations of the acceptance are observed for channels where
one or several final-state particles are identified by the RICH-1 detector.  The
impact of the particle identification on the acceptance in the
$K^{-}\pi^{+}\pi^{-}$ channel is depicted in \figref{fig:Kpipi_acceptances} for
both the $K\pi\pi$ invariant mass distribution and the Gottfried-Jackson angle
of the $\pi^+K^-$ isobar.  The reduction of the acceptance is mainly due to the
limited momentum range available for kaon identification (see
\secref{sec:reconstruction.rich}).

\begin{figure*}[h]
  \begin{center}
    \includegraphics[width=.49\columnwidth]{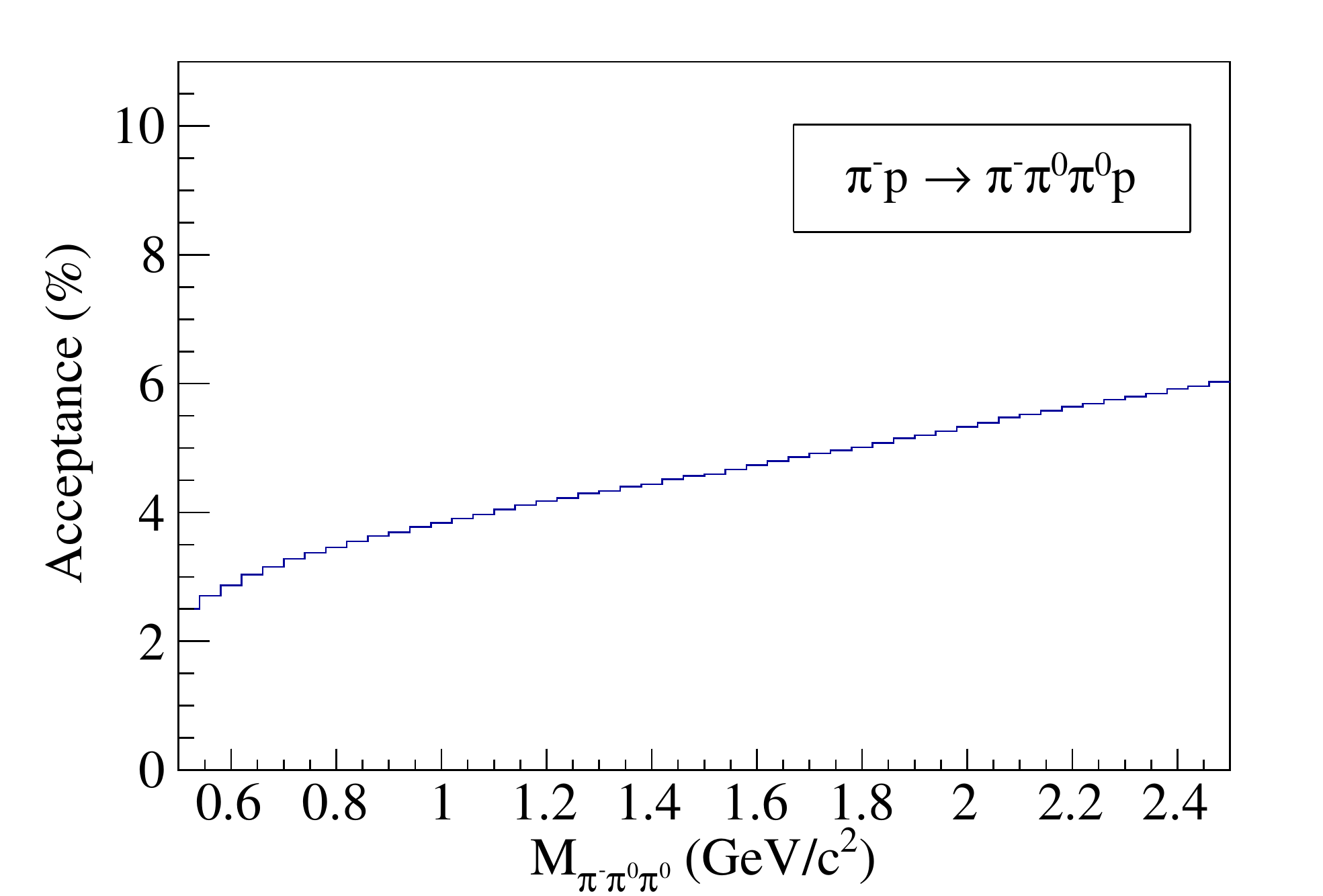}
    \includegraphics[width=.49\columnwidth]{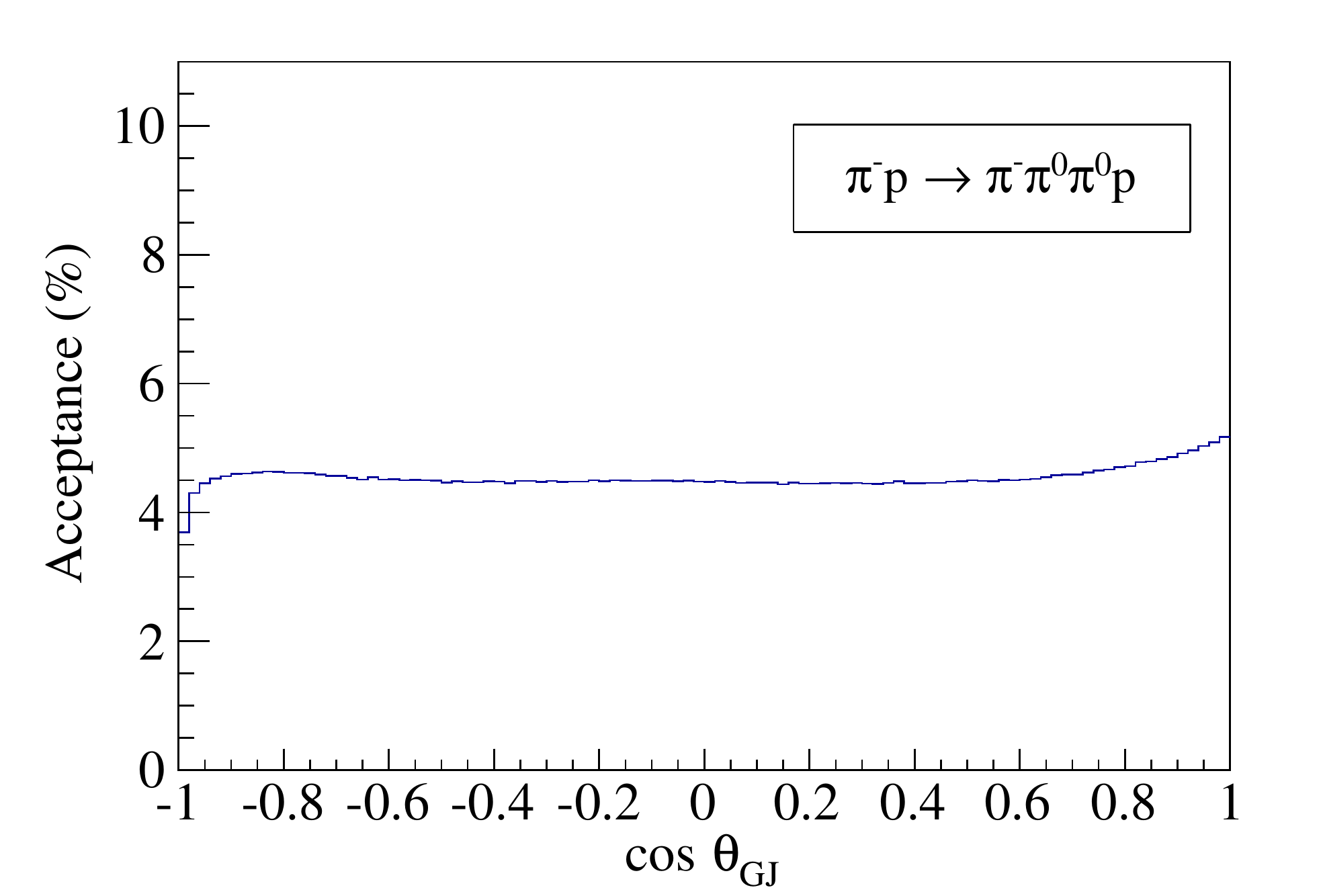}
    \caption{Acceptance for the diffractively produced $\pi^-\pi^0\pi^0$ final
      state (left)~as a function of the $3\pi$ invariant mass and (right)~as a
      function of the polar angle of the $\pi^0\pi^0$ isobar in the
      Gottfried-Jackson frame.}
    \label{fig:3pi0_acceptances}
  \end{center}
\end{figure*}

\Figref{fig:3pi0_acceptances} shows the corresponding acceptance plots for the
$\pi^-\pi^0\pi^0$ final state. Compared to the charged-pion channel, the
acceptance for the channel containing neutral pions is smaller and its
dependence on the three-pion mass is more pronounced. The decrease of the
acceptance is mainly due to the photon detection efficiency, which is lower than
that for charged particles as photons may get absorbed in passive materials
before reaching the calorimeters.  The largest absorption is caused by the beam
pipe of the RICH-1 detector (see \secref{sec:pid.rich1}), an effect which is
mainly important for forward-going photons. Nevertheless, the detection of four
photons smears the effect, so that the angular modulation of the acceptance
remains weak and similar to that of the charged-pion case.

\subsection{Performance of the hadron setup}
\label{sec:performance.setup}
The performances of the individual detectors and of the reconstruction software
were presented in the previous sections. Here, the main characteristics of the
setup are presented with examples from various physics processes.

\subsubsection{Kinematic distributions}
\label{sec:performance.setup.kinematics}
\begin{figure}[tbp] \centering
  \includegraphics[width=0.45\textwidth]{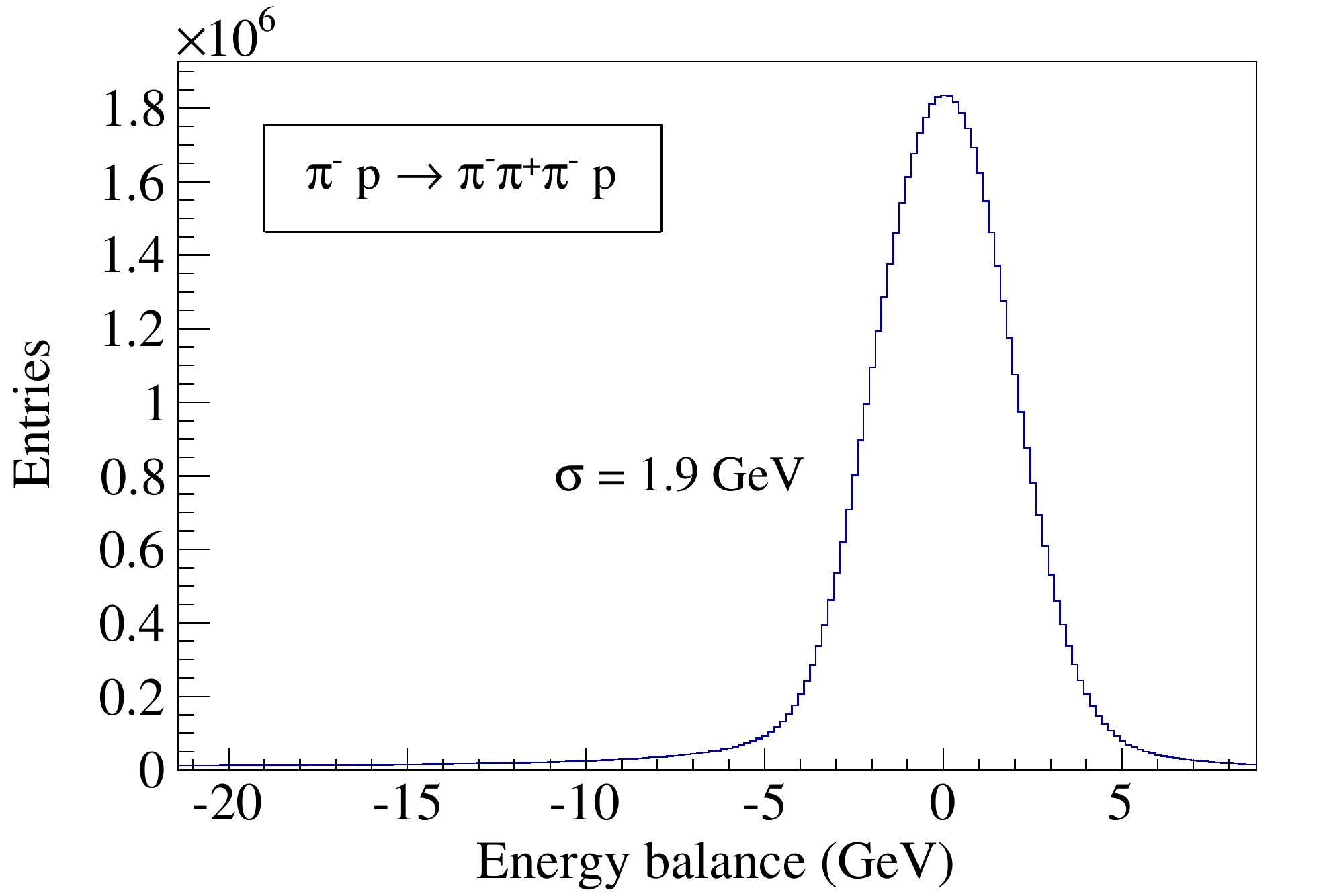}
  \includegraphics[width=0.45\textwidth]{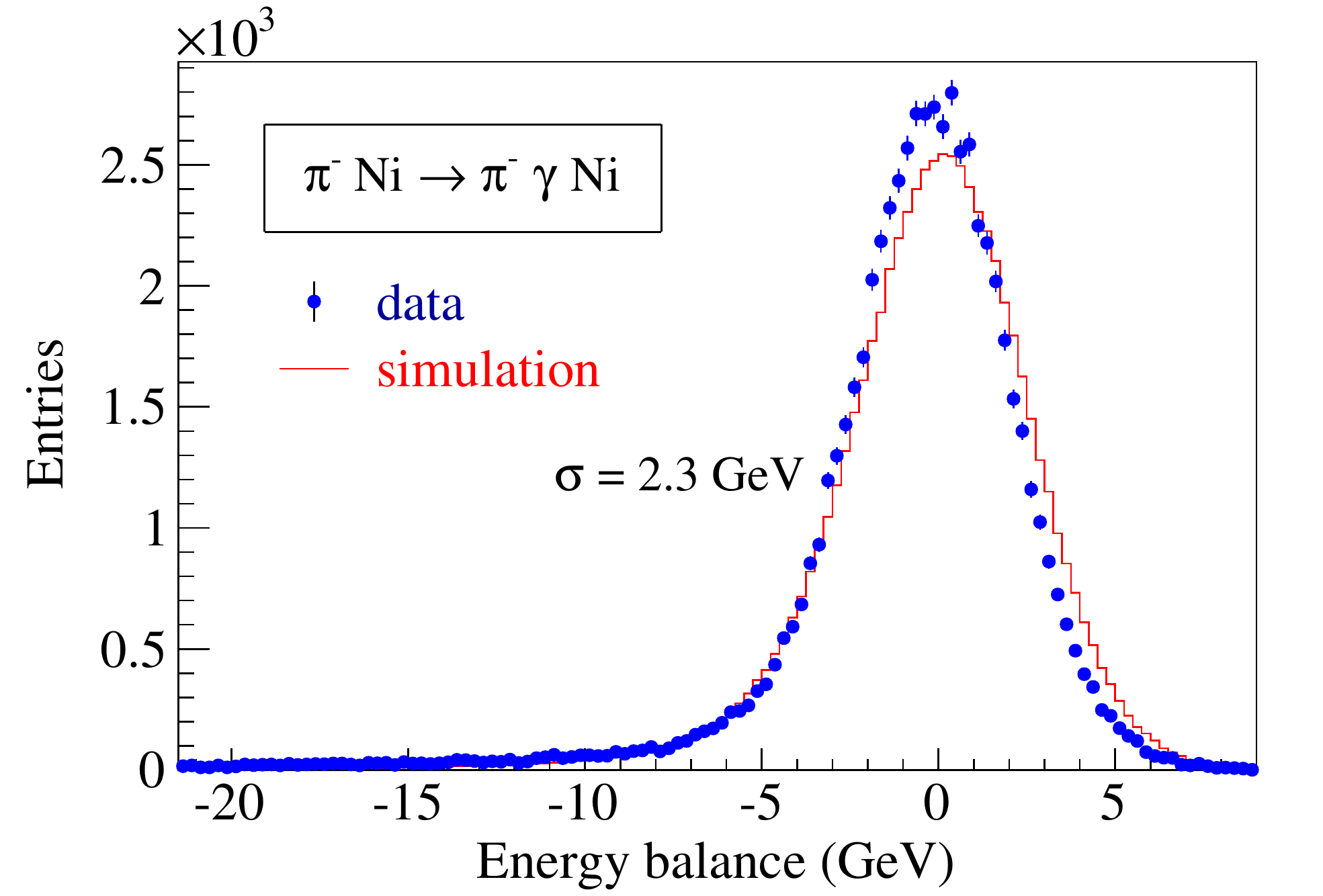}
  \caption{Energy balance between outgoing and incoming particles for (left)
    diffractive dissociation with three charged pions in the final state and
    (right) for Primakoff scattering.}
  \label{fig:performance.setup.exclusivity}
\end{figure}

The selection of exclusive events with a primary vertex inside the target is a
prerequisite for most analyses performed on the data with hadron
beams. Exclusive events are selected by requiring energy conservation and
transverse momentum balance between incoming and outgoing particles.
\Figref{fig:performance.setup.exclusivity} shows the distributions of the
difference between the energy of the outgoing particles and of the incoming beam
particle, for the diffractive dissociation into three charged pions and for the
Primakoff pion Compton scattering.  The beam momentum station, which is used to
determine the momentum of each incoming beam particle when operating the muon
beam, is removed for hadron beams in order to reduce the amount of material in
the beam.  Therefore, the beam energy is set to the value determined by the beam
line settings.  In the diffractive process, the energy of the outgoing pions is
combined with the recoil proton energy measured by the RPD.  In the Primakoff
reaction, the incident pion energy is shared between the scattered pion and the
emitted photon, whereby the contribution of the target recoil remains
negligible.  The widths of the energy balance distributions shown in
\figref{fig:performance.setup.exclusivity} are dominated by the momentum spread
of the beam particles (see \secref{sec:beam.hadron}), with a smaller
contribution from the finite momentum resolution for charged particles and also
from the finite energy resolution for photons in the case of the Primakoff
reaction.

The physics processes studied in COMPASS can be identified via their
characteristic dependence on the reduced squared four-momentum transfer $t'$,
which is calculated from the four-momenta of the incoming beam particle and the
outgoing particles according to the equations given in \secref{sec:intro}. While
Primakoff reactions proceeding through the exchange of quasi-real photons
dominate the cross section at $t'<0.001\,\GeV^2/c^2$, diffractive and central
production reactions prevail at larger values of $t'$. Here we show measured
$t'$ distributions for the different physics triggers mentioned in
\secref{sec:trigger.physics}.

\begin{figure}[tbp] 
  \begin{minipage}[t]{.48\textwidth}
    \centering \includegraphics[width=0.95\textwidth]{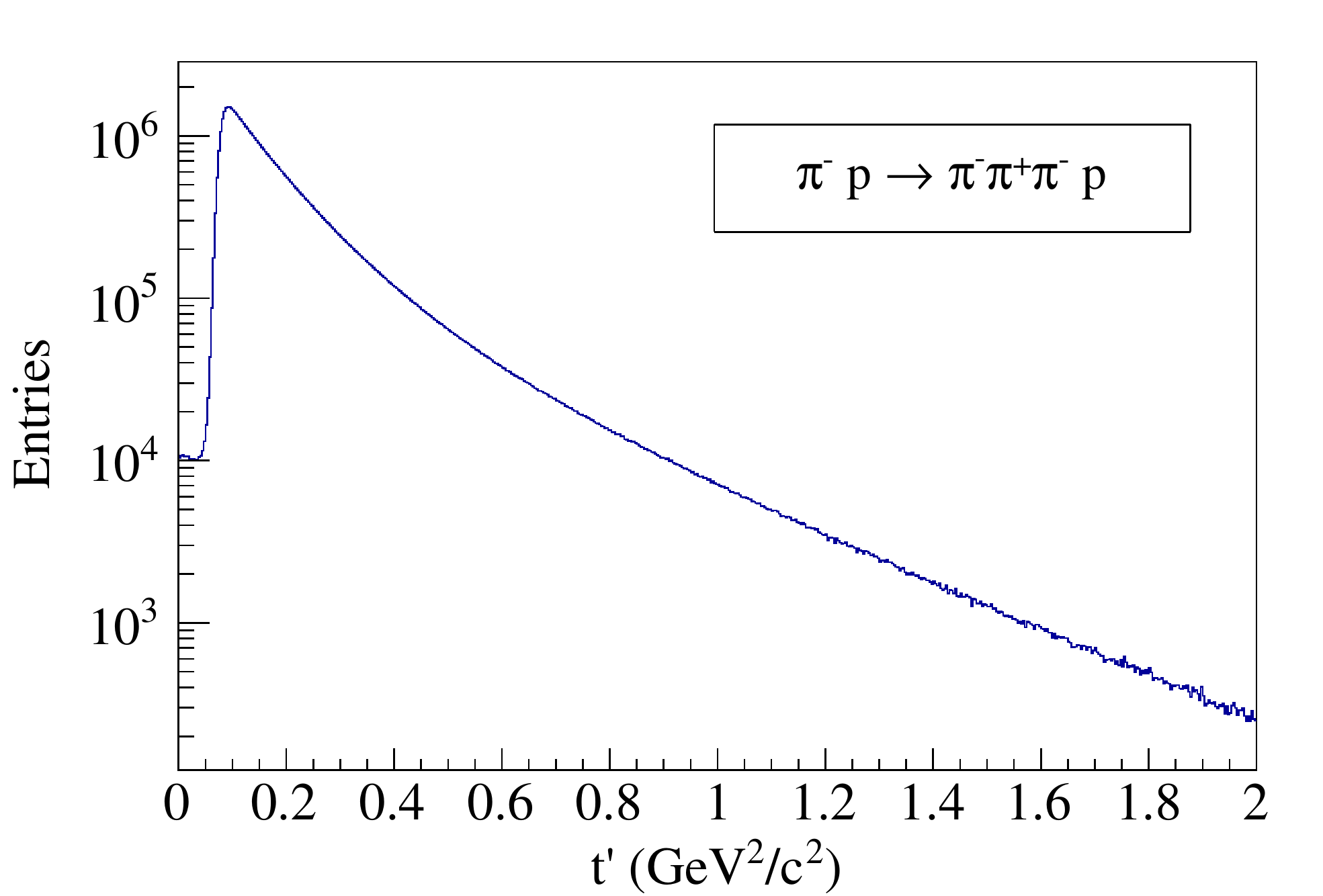}
    \caption{Squared four-momentum transfer for $\pi^-\pi^+\pi^-$ events
      produced by a pion beam impinging on a liquid hydrogen target, and
      selected by the DT0 trigger.}
    \label{fig:performance.setup.t.DT0}
\end{minipage}
\hfill
 \begin{minipage}[t]{.48\textwidth}
   \centering \includegraphics[width=0.95\textwidth]{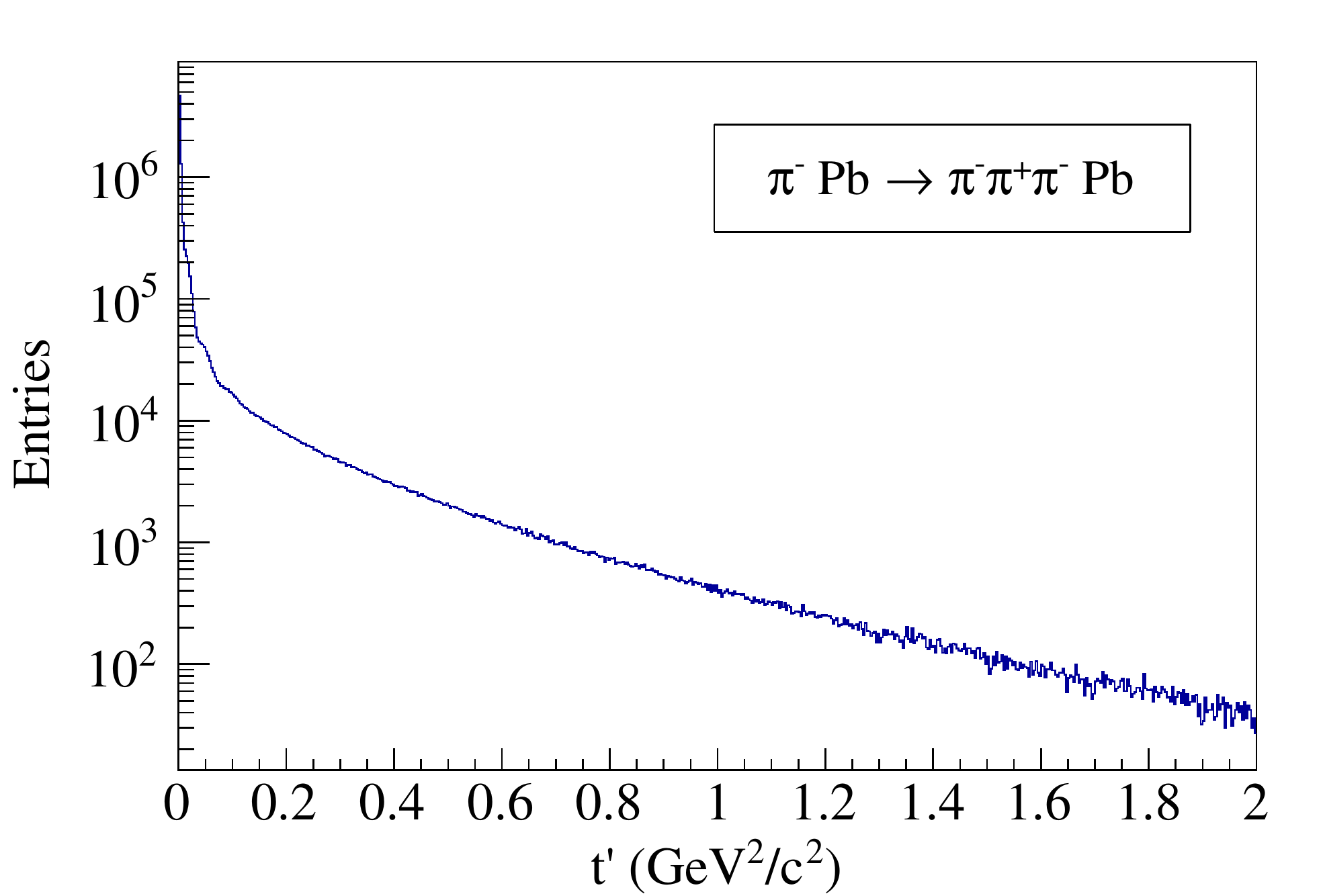}
   \caption{Squared four-momentum transfer for $\pi^-\pi^+\pi^-$ events produced
     by pions hitting a lead target, and selected by the multiplicity trigger.}
   \label{fig:performance.setup.t.LT}
 \end{minipage}
\end{figure}

\Figref{fig:performance.setup.t.DT0} shows the $t'$ distribution recorded with
the diffractive trigger DT0, determined from events with three charged pions in
the final state.
The cut at $t'\approx 0.07\,\GeV^2/c^2$ is due to the requirement of a signal in
the RPD. The DT0 trigger thus enhances events with high $t'$. The small leakage
of events with $t'<0.07\,\GeV^2/c^2$ presumably originates from $\delta$
electrons or pions accidentally firing the RPD.

The multiplicity triggers LT1 and LT2 are used to also include events with lower
values of $t'$, since no recoil proton is required.  The corresponding $t'$
distribution measured with a pion beam and solid nuclear targets
(\secref{sec:target.nuclear}), shown in \figref{fig:performance.setup.t.LT},
exhibits an exponential increase of the number of events towards low values of
$t'$. In the first few bins also Primakoff events contribute, in addition to
those generated by strong interaction.

\begin{figure}[tbp] 
  \centering \includegraphics[width=0.5\textwidth]{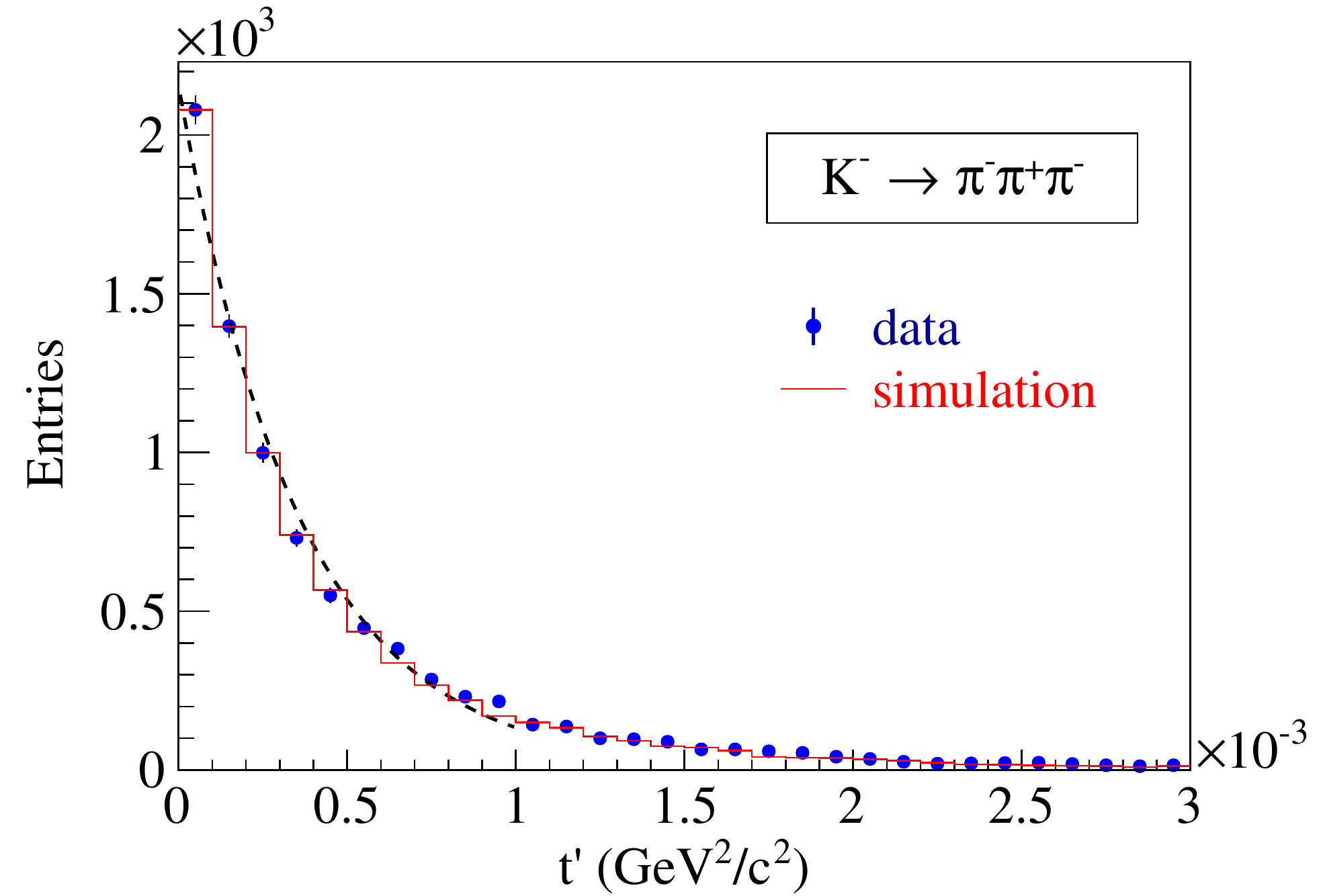}
  \caption{Squared four-momentum transfer of reconstructed beam kaons (data
    points) compared to the Monte Carlo simulation of purely electromagnetic
    interaction (solid lines). The dashed line is an exponential fit, used to
    determine the resolution.}
  \label{fig:performance.setup.t.kaons}
\end{figure}

A good resolution on the measurement of $t'$ is important in order to
distinguish between Primakoff and diffractive scattering. The resolution at very
small values of $t'$ is determined using the decay of beam kaons into two or
three charged pions.  For free-particle decays, $t'$ is by definition zero, and
the measured width of the $t'$ distribution, shown in
\figref{fig:performance.setup.t.kaons}, gives a direct estimate of the
resolution. A width of $3.6\EE{-4}\,\GeV^2/c^2$ is obtained from the data, in
good agreement with the resolution from Monte Carlo simulations.  At higher
values of $t'$, the resolution can only be determined from Monte Carlo
simulations.  A value of $7\cdot10^{-3}\,\GeV^2/c^2$ is obtained for
$0.1\,\GeV^2/c^2<t'<1.0\,\GeV^2/c^2$, from the simulation of diffractive
production of three charged pions.
 
\begin{figure}[tbp] 
  \begin{minipage}[t]{.48\textwidth}
    \centering
    \includegraphics[width=0.95\textwidth]{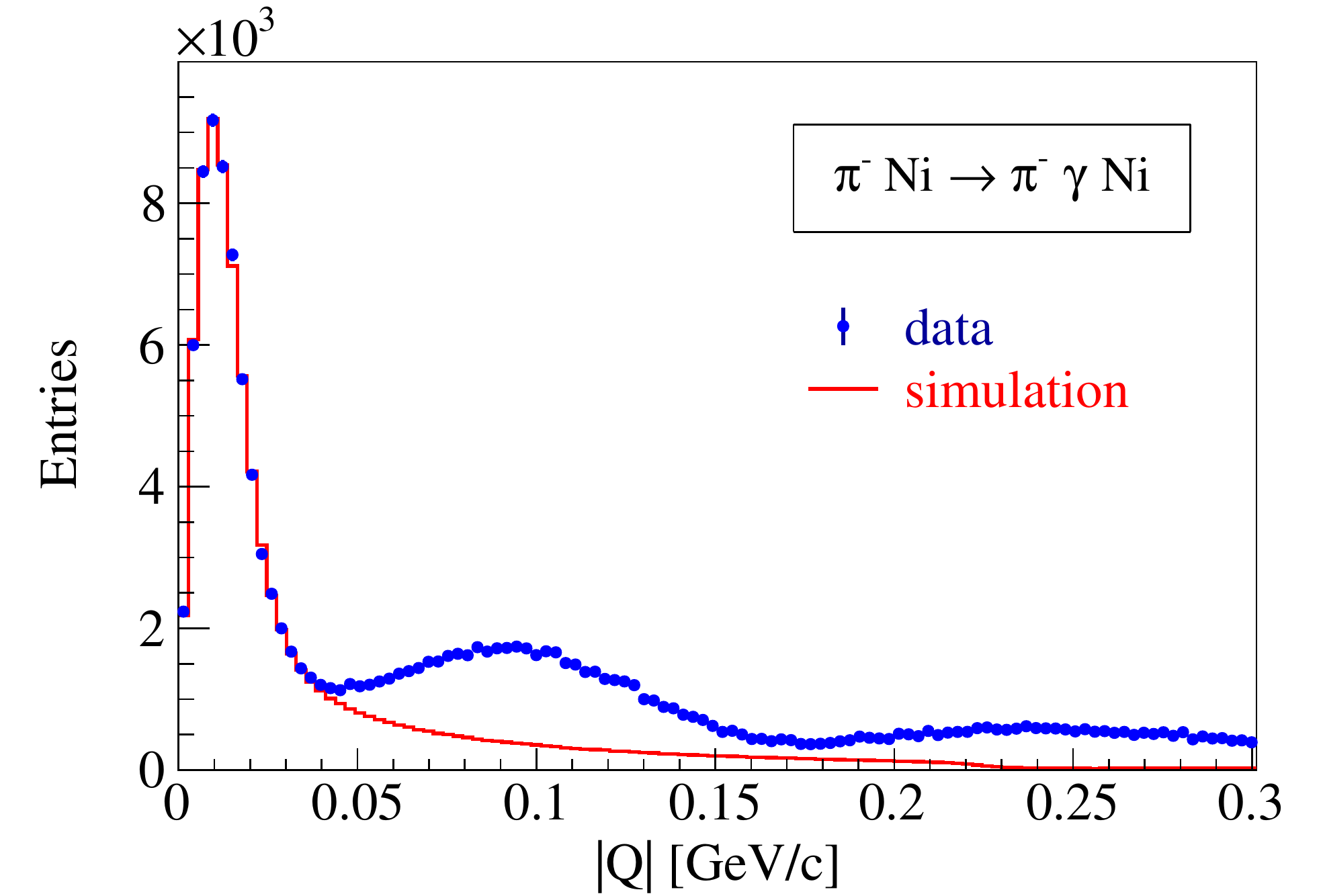}
  \end{minipage}
  \hfill
  \begin{minipage}[t]{.48\textwidth}
    \centering
    \includegraphics[width=0.95\textwidth]{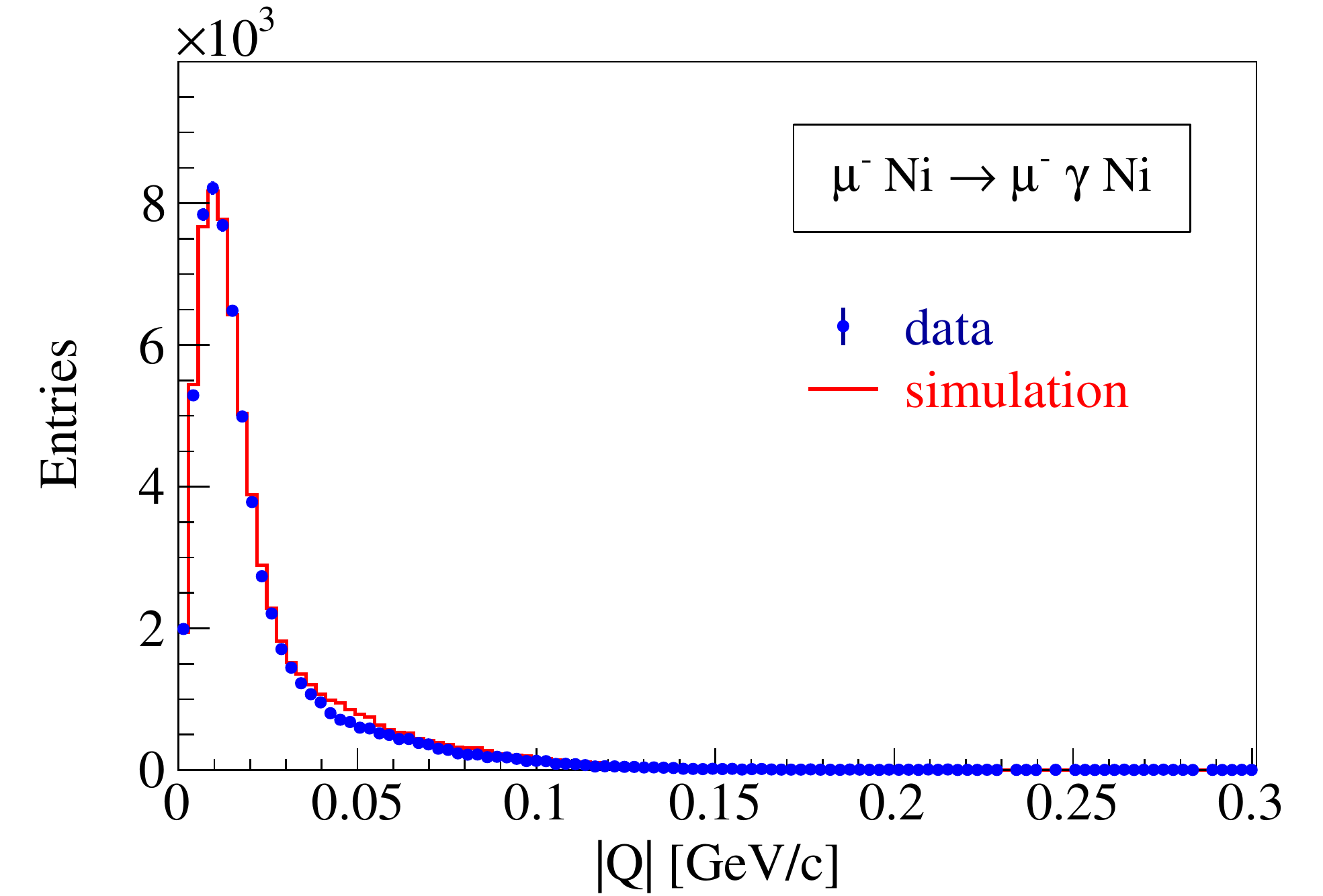}
    \end{minipage}
  \caption{Momentum transfer distributions for exclusive (left) $\pi^-\gamma$
    and (right) $\mu^-\gamma$ events. The data (dotted lines) are compared to
    the MC simulation (solid lines).}
  \label{fig:performance.setup.primakoff-q}
\end{figure}

For the measurement of the pion polarisability, exclusive $\pi^-\gamma$ events
are selected from the data sample collected with the calorimeter trigger (see
\secref{sec:trigger.ecal}).  The left panel of
\Figref{fig:performance.setup.primakoff-q} shows the distribution of the
four-momentum transfer $\left|Q\right|=\sqrt{t'}$, chosen here to emphasize its
shape at small values.  The peak at $\left|Q\right|\approx 0.02\,\GeV/c$ mainly
contains quasi-real photoproduction events. The fact that the interaction is
purely electromagnetic at very low values of $t'$, which correspond to large
impact parameters, becomes clear when comparing it to the right panel of
\Figref{fig:performance.setup.primakoff-q} that shows the corresponding
distribution taken under the same conditions, but with a $\mu^-$ beam instead of
a $\pi^-$ beam.  For the pion beam, the strong interaction dominates at
$\left|Q\right|$ values above $0.05\,\GeV^2/c^2$, resulting in typical
diffractive structures. The Monte Carlo simulation, superimposed as solid line
in both panels, describes both cases very well.

\subsubsection{Mass resolution}
\label{sec:performance.setup.mass_resolution}
The mass resolution of the spectrometer is determined using known narrow states
that are reconstructed in the spectrometer via their decay into neutral and/or
charged particles. Here we show distributions for two-photon decays ($\pi^0$,
$\eta$), for decays into final states with charged particles only ($K_S^0$,
$\phi$, $\Lambda$, $\Xi$), and for decays into final states containing both
charged and neutral particles ($\eta$, $\omega$, $\eta'$, $f_1$).

\begin{figure}[tbp]
  \centering
  \includegraphics[width=.45\textwidth]{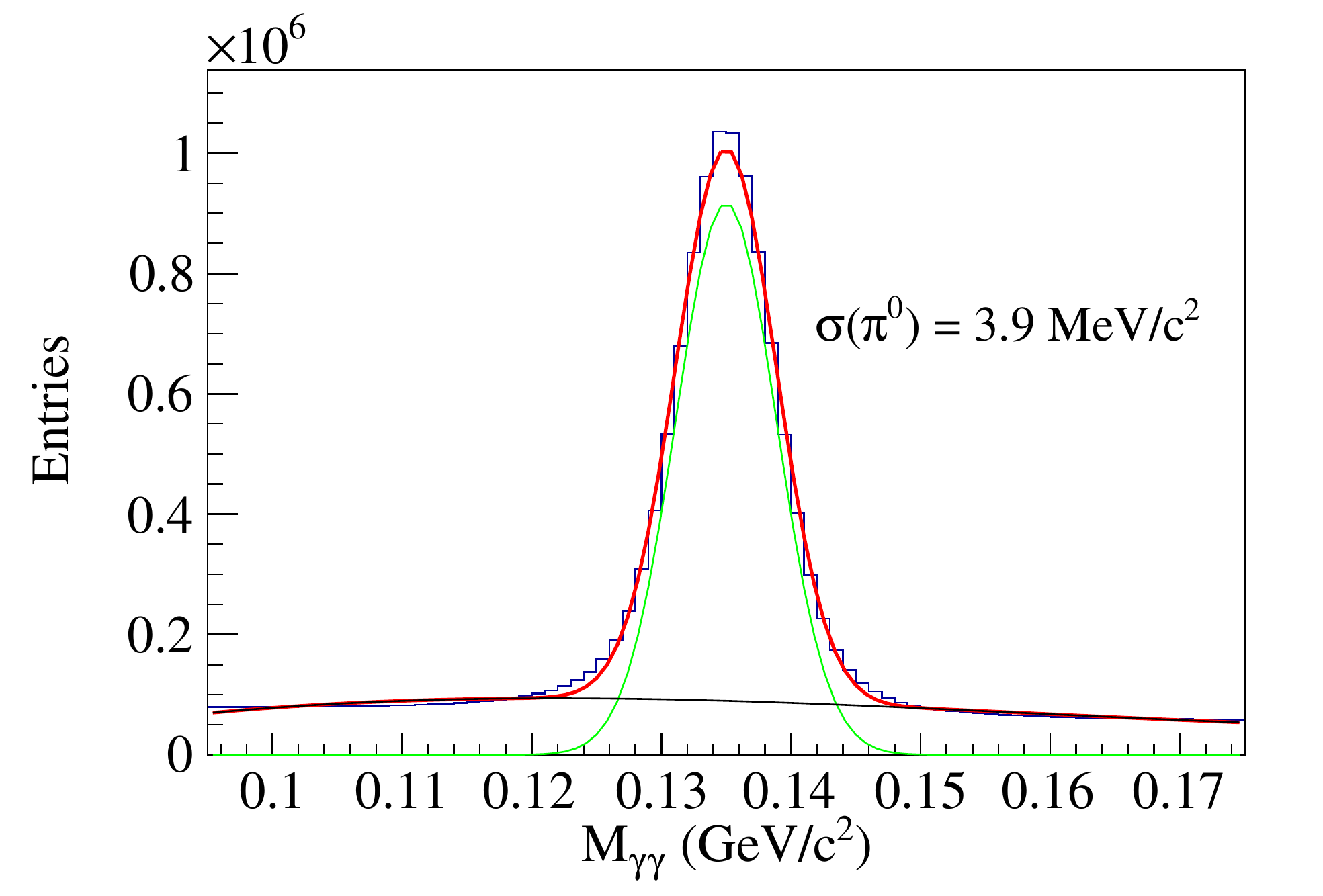}
  \includegraphics[width=.45\textwidth]{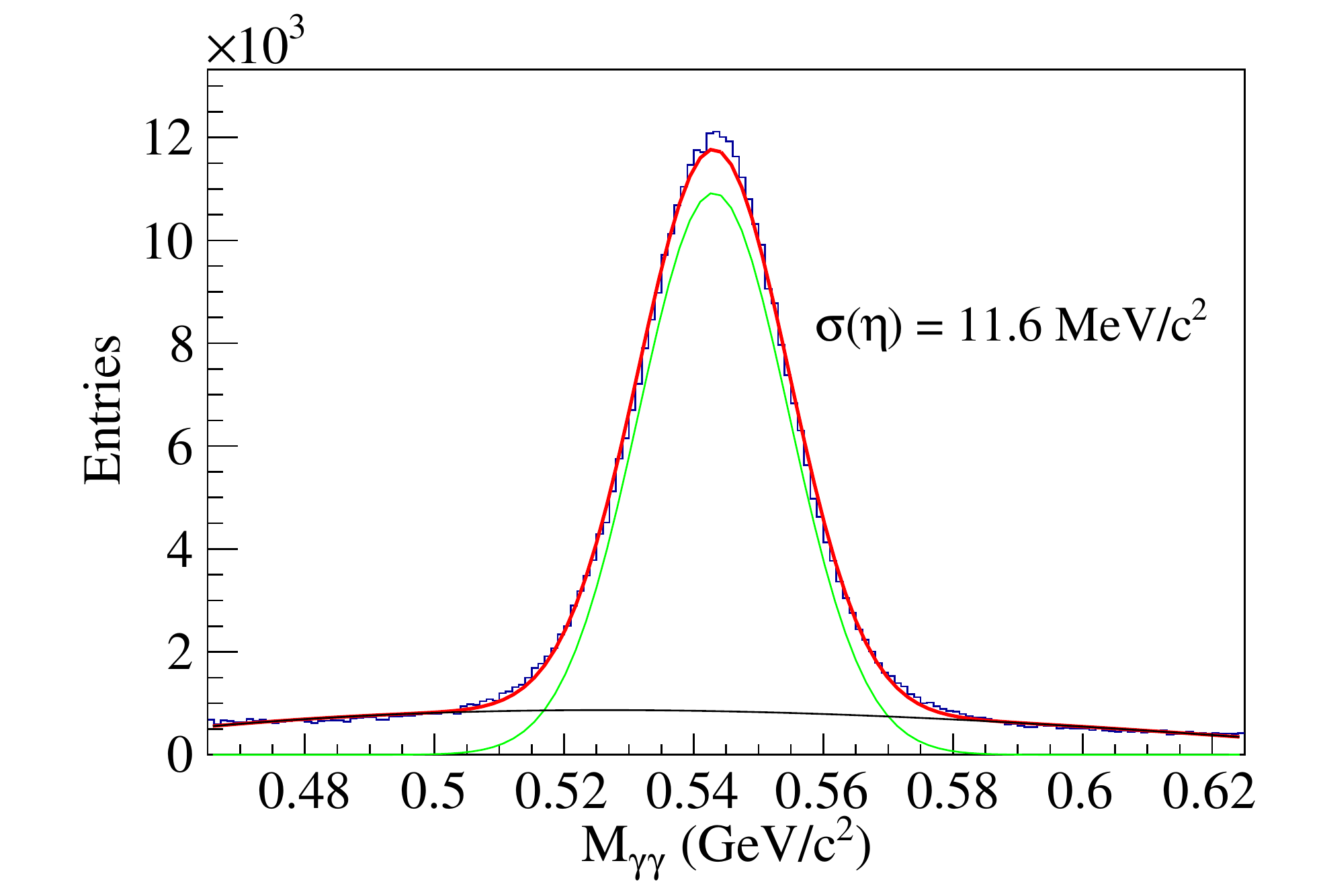}
  \caption{Two-photon invariant mass distribution as measured in ECAL2, in the
    (left) $\pi^0$ mass region and (right) $\eta$ mass region. The solid curves
    are fits to the signal and to the background. The values of the resolution
    achieved are indicated in each plot.}
  \label{fig:performance.ecal.pion-eta}
\end{figure}

The invariant mass distributions of photon pairs in the $\pi^0$ and $\eta$ mass
regions, as measured by ECAL1 and ECAL2, are shown in
Fig.~\ref{fig:performance.ecal.pion-eta}.
The distributions are obtained from diffractive interactions of a $190\,\GeV$
negative hadron beam impinging on a liquid hydrogen target.  Apart from the
standard event selection mentioned in \secref{sec:performance.setup.kinematics},
only clusters with energies larger than $0.6\,\GeV$ in ECAL1 and $1.2\,\GeV$ in
ECAL2 are selected.  Energy resolution values of $7.8\,\MeV/c^2$ and
$3.9\,\MeV/c^2$ are obtained for the two-photon $\pi^0$ peaks detected in ECAL1
and in ECAL2, respectively.  The corresponding values for the $\eta$ meson are
$19.5\,\MeV/c^2$ and $11.7\,\MeV/c^2$.

\begin{figure}[tbp]
  \centering
  \includegraphics[width=.49\textwidth]{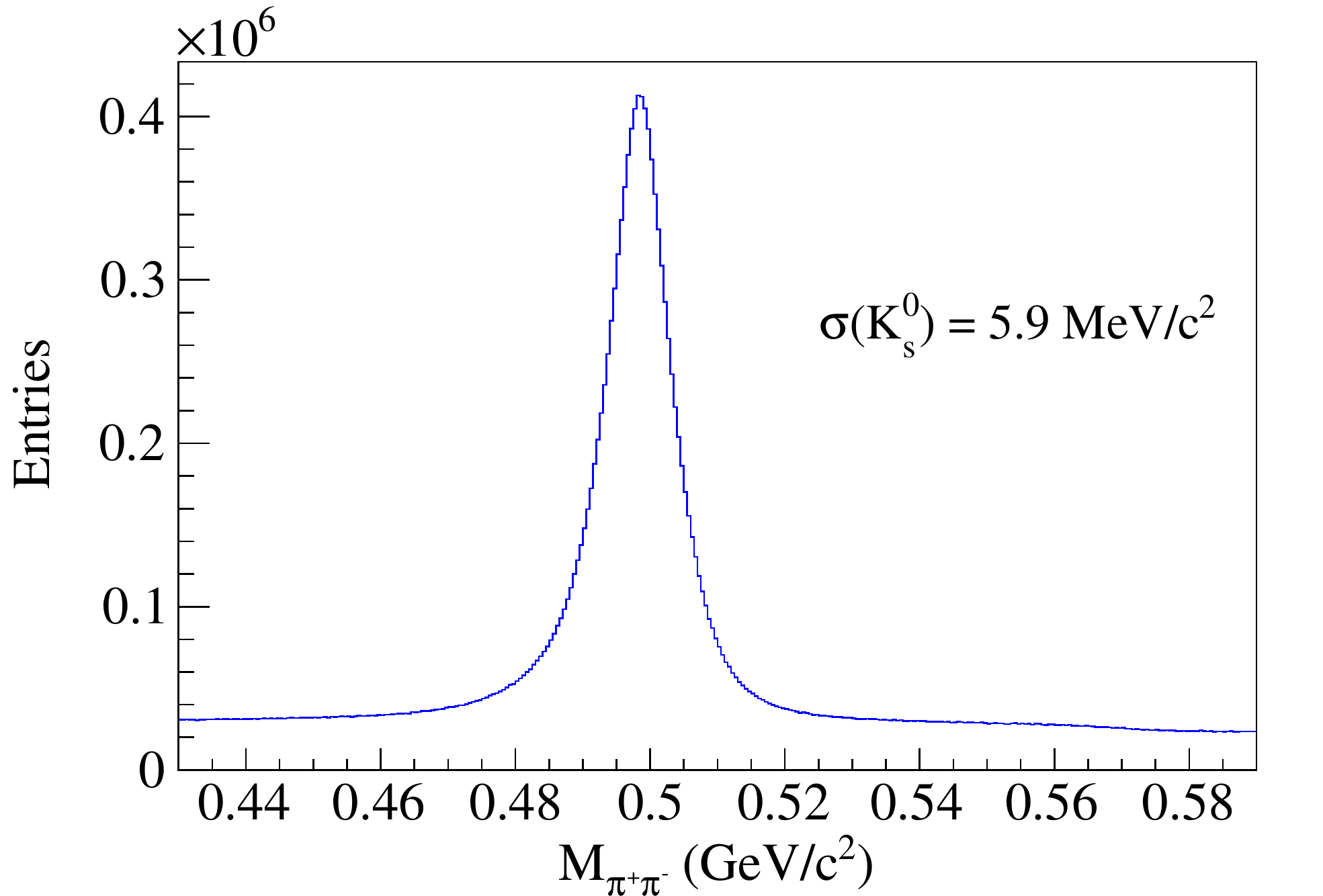}
  \includegraphics[width=.49\textwidth]{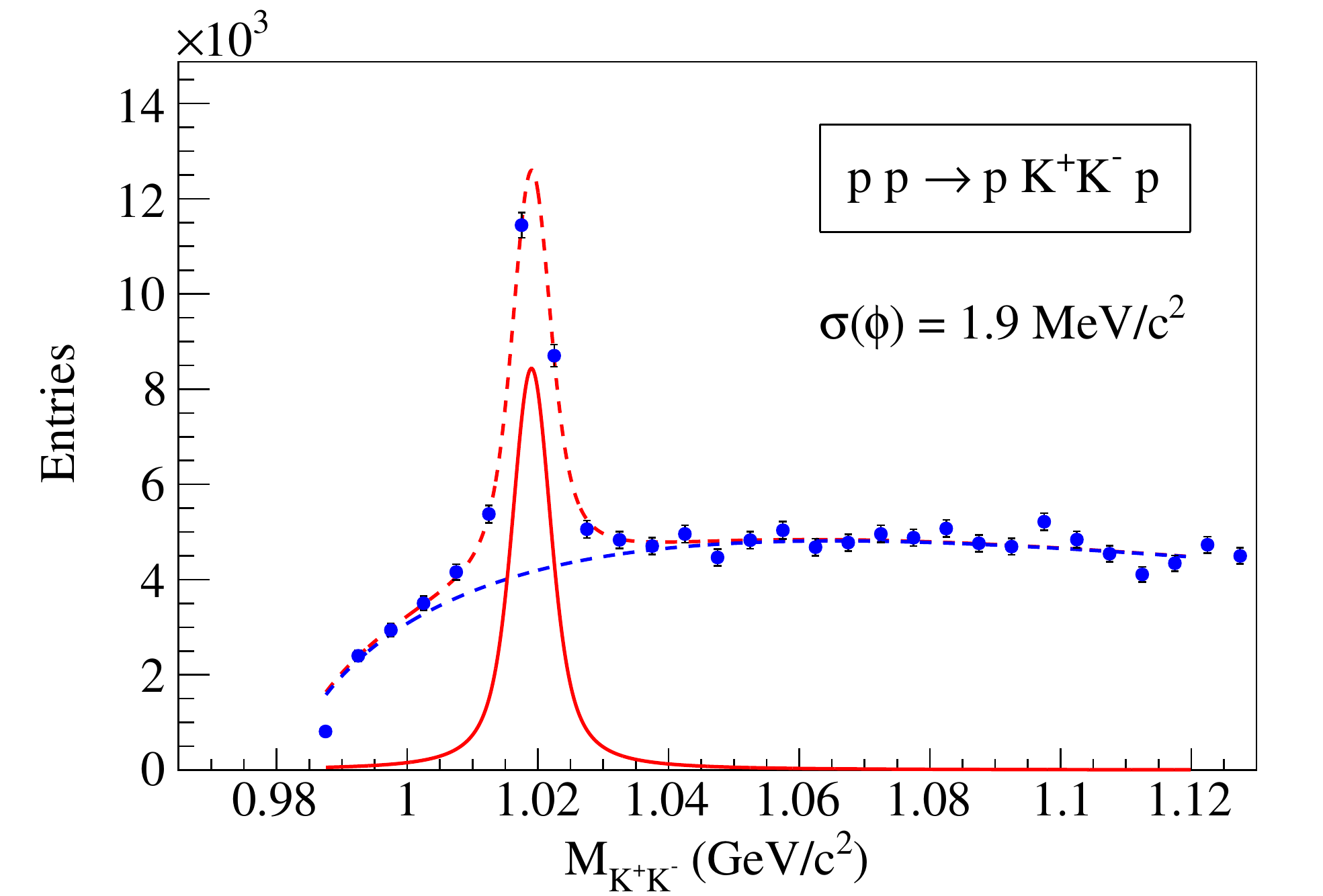}
  \includegraphics[width=.49\textwidth]{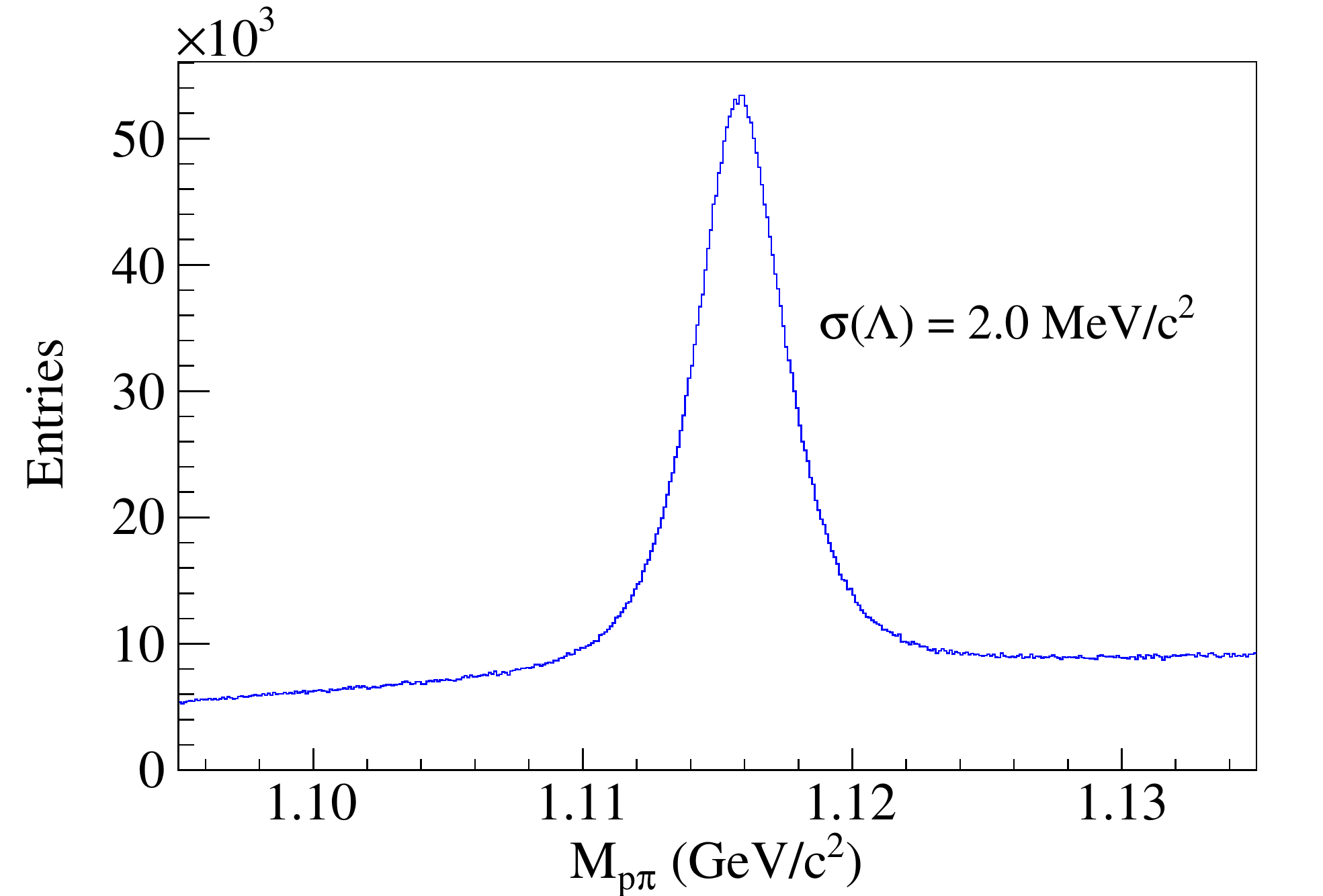}
  \includegraphics[width=.49\textwidth]{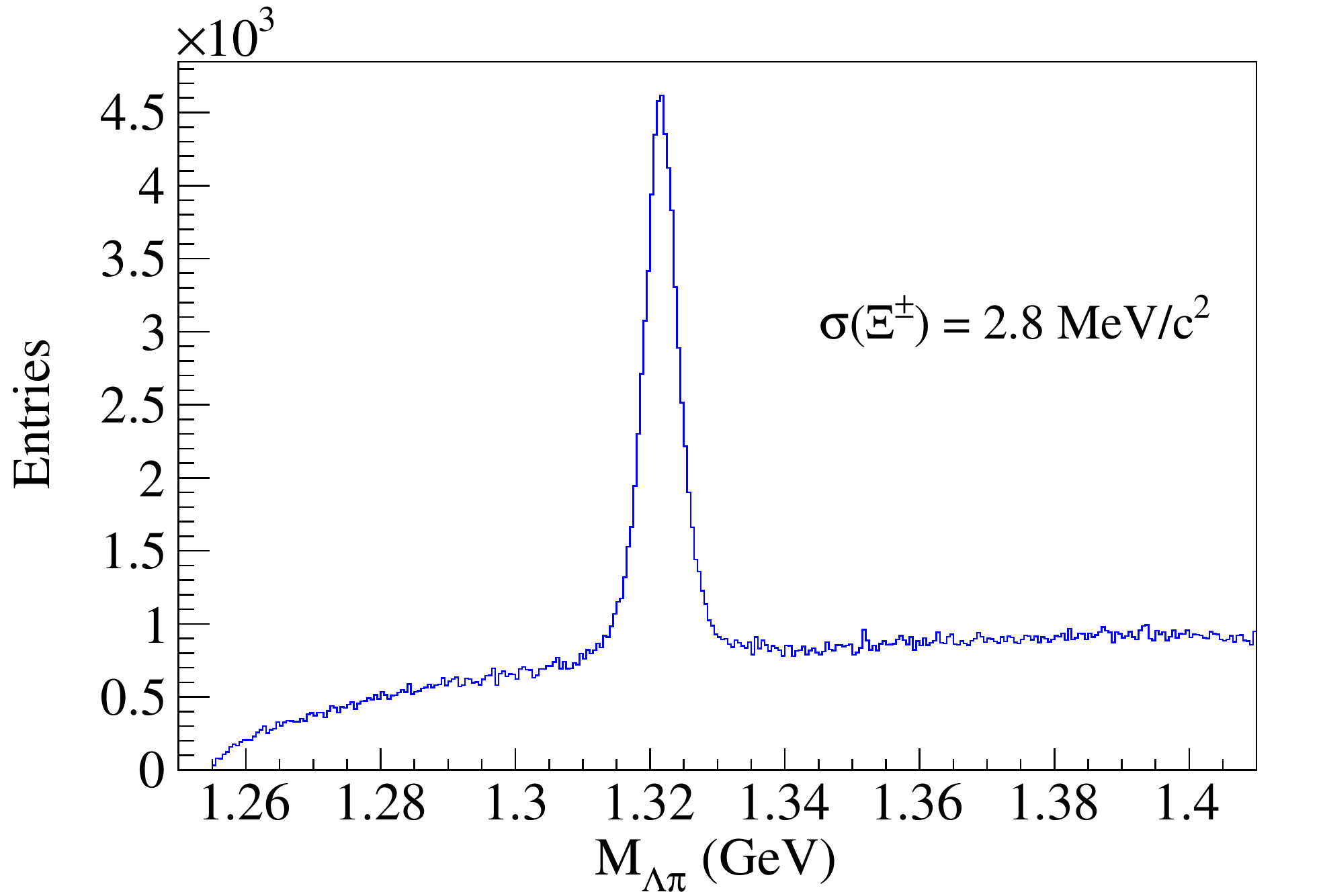}
  \caption{Reconstructed invariant masses for charged particles in the final
    state.  The peaks shown are for (top left) $K^0_S(498)$, (top right)
    $\phi(1020)$, (bottom left) $\Lambda(1115)$, and (bottom right)
    $\Xi^\pm$. The $K^0_S$, $\Lambda$, and $\Xi^\pm$ particles are produced in
    inclusive reactions. The dashed curve in the $\phi(1020)$ plot is a fit to
    the background.  }
\label{fig:secondary}
\end{figure}

Hadrons reconstructed from decay modes that contain only charged particles are
shown in \figref{fig:secondary}.  In an inclusive selection, the following
resolutions are obtained: $5.90\,\MeV/c^2$ for the $K^0_S$, $1.9\,\MeV/c^2$ for
the $\phi(1020)$, $1.99\,\MeV/c^2$ for the $\Lambda$ and $\bar\Lambda$ and
$2.80\,\MeV/c^2$ for the $\Xi^{\pm}$.
 
\begin{figure}[tbp] 
  \begin{minipage}[t]{.48\textwidth}
    \centering
    \includegraphics[width=0.95\textwidth]{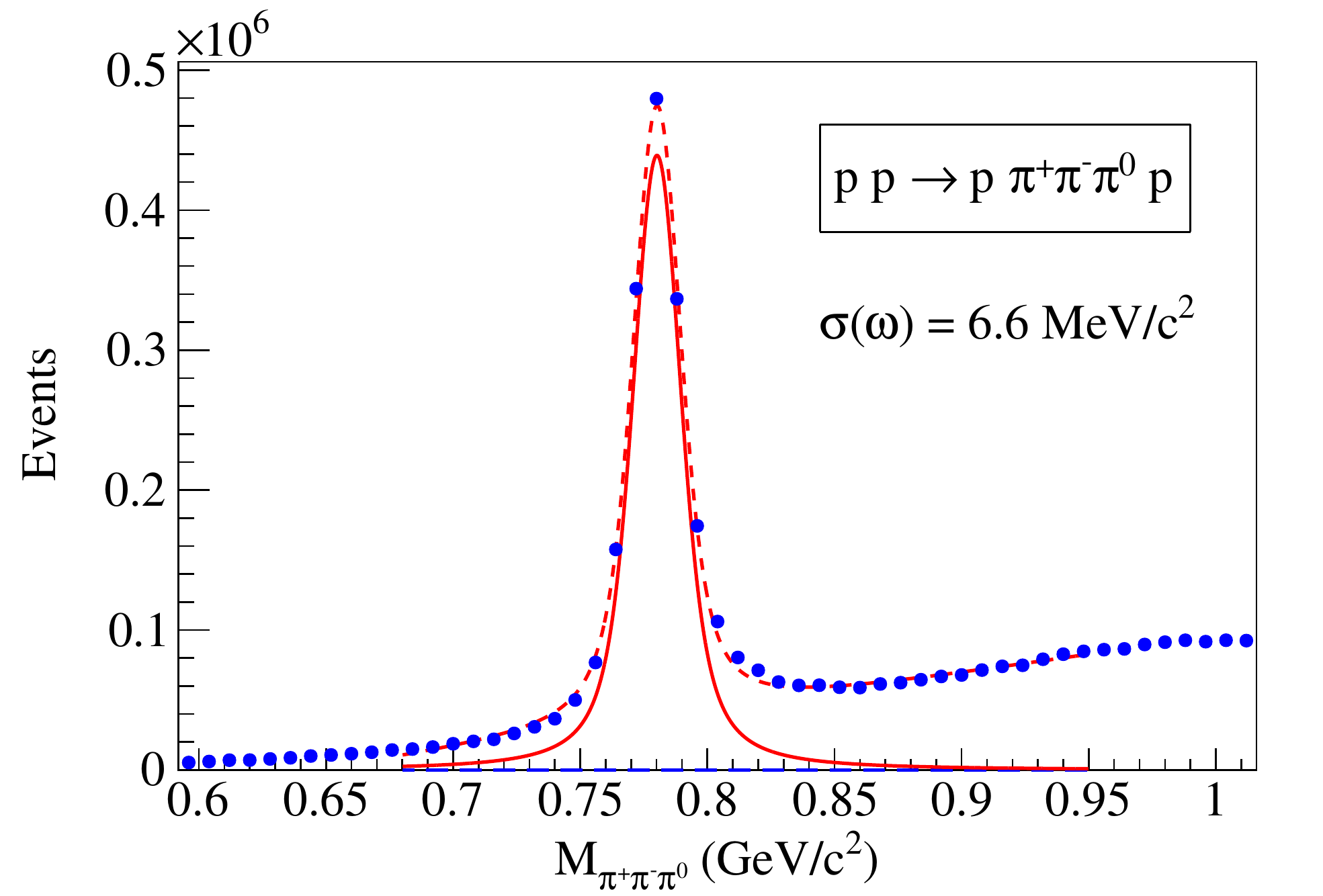}
  \end{minipage}
  \hfill
  \begin{minipage}[t]{.48\textwidth}
    \centering
    \includegraphics[width=0.95\textwidth]{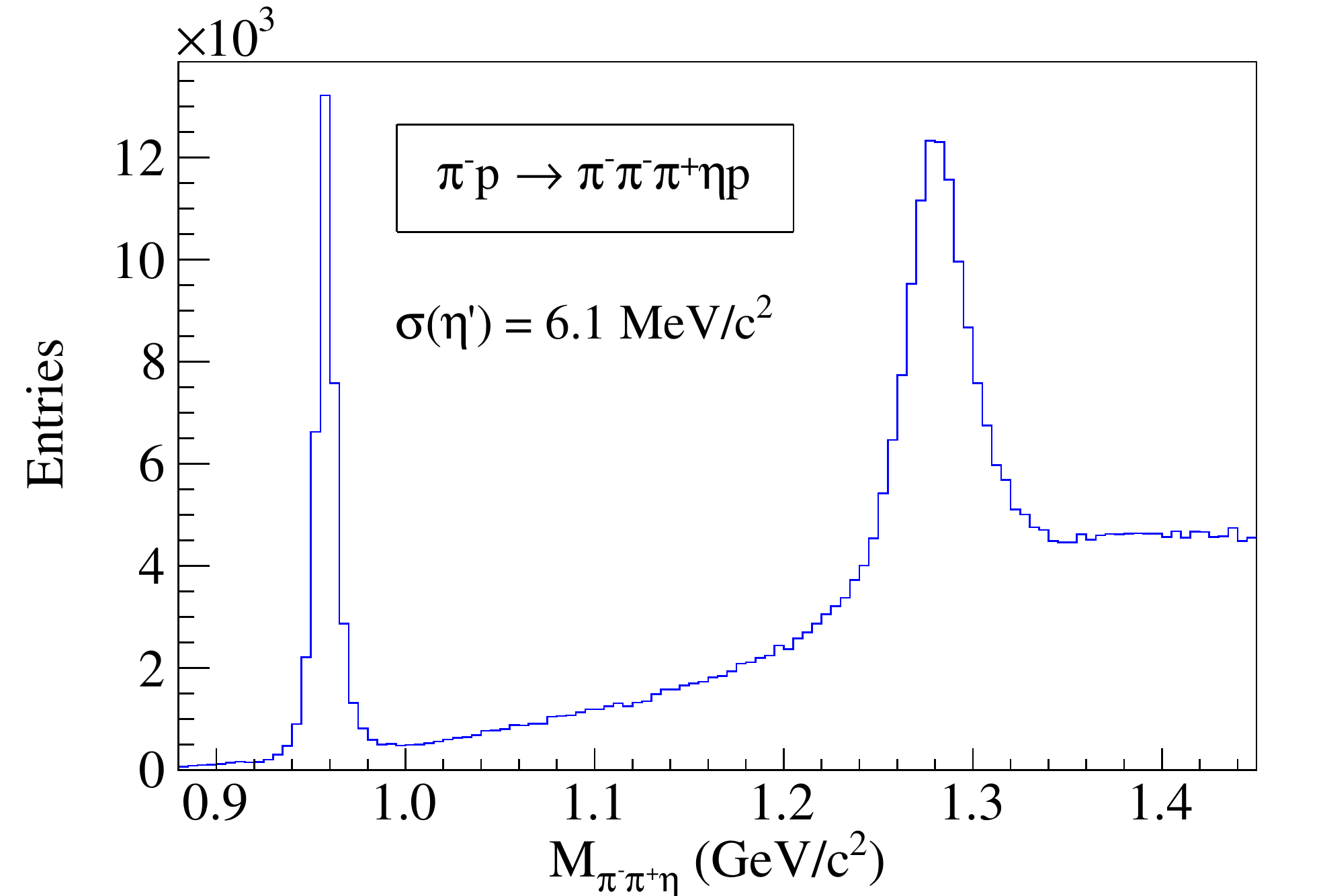}
  \end{minipage}
  \caption{Invariant mass spectra for (left) $\pi^-\pi^+\pi^0$ and (right)
    $\pi^-\pi^+\eta$ systems. The full line in the left panel is a fit to the
    $\omega$ peak only; the dashed line includes also the background.}
  \label{fig:performance.setup.mixed}
\end{figure}

Complex resonance decays with more than three particles in the final state are
reconstructed e.g.\ by combining a $\pi^0$ or $\eta$ in the $\gamma\gamma$
channel with a neutral pair of pions ($\pi^+\pi^-$) leaving the primary vertex.
The left panel of \figref{fig:performance.setup.mixed} shows the invariant mass
spectrum of the $\pi^-\pi^+\pi^0$ final state in the $\omega(782)$ mass region
from central production reactions of a proton beam with the liquid hydrogen
target.  As shown in the right panel of \figref{fig:performance.setup.mixed},
selecting the $\eta$ instead of the $\pi^0$ gives access to the decays
$\eta'(958)\to\pi^-\pi^+\eta$ and $f_1(1285)\to\pi^-\pi^+\eta$, which are
reconstructed from diffractive $\pi^-p\to\pi^-\pi^+\pi^-p\gamma\gamma$ events.
Deconvoluting the natural width of the $\omega$, a resolution of the
spectrometer of $6.6\,\MeV/c^2$ is obtained.  The natural width of the $\eta'$
is negligible, so the width of the peak directly gives a resolution of
$6.1\,\MeV/c^2$ in this mass range.

\begin{figure}[tbp] \centering
  \includegraphics[width=0.7\textwidth]{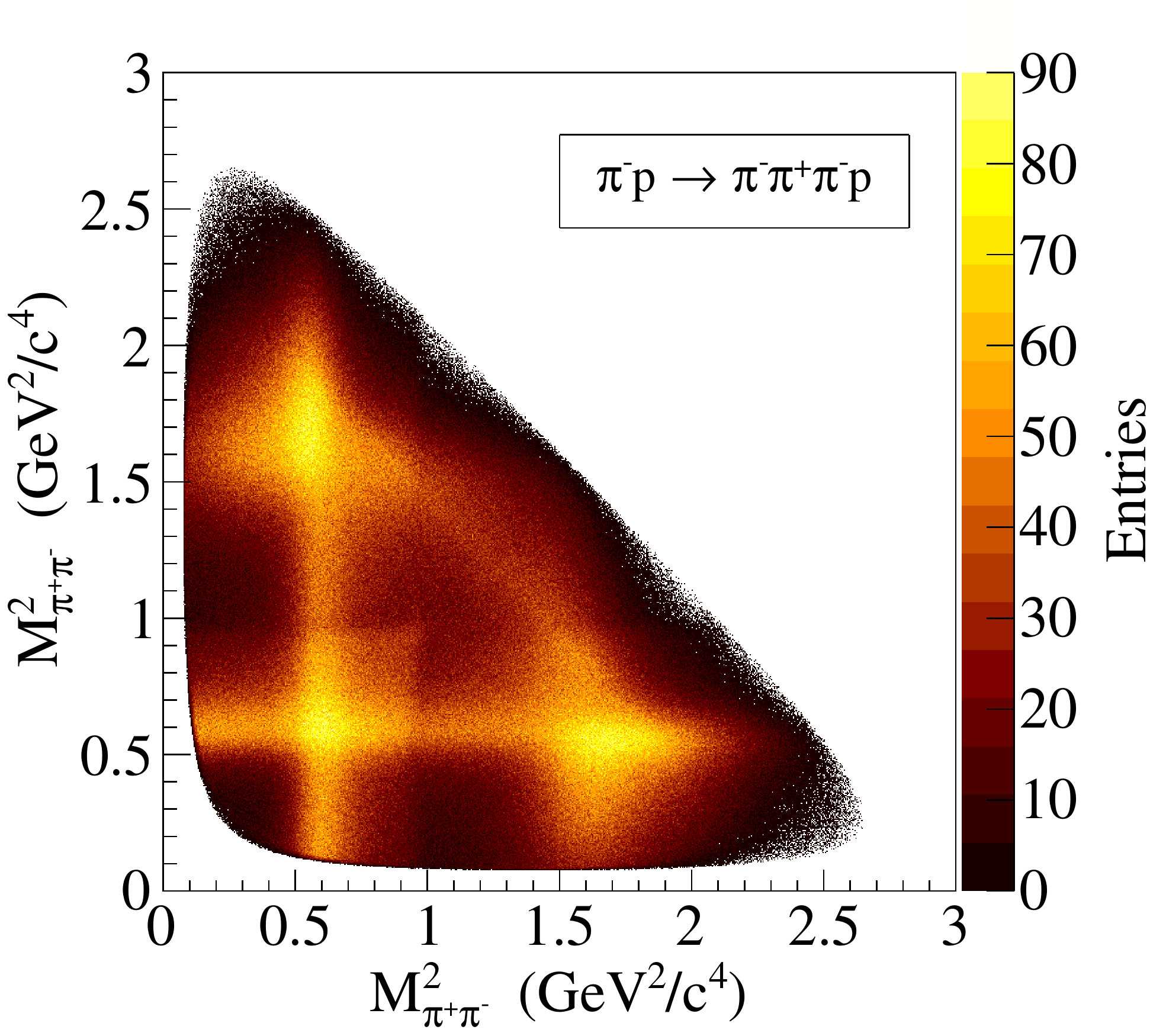}
  \caption{Dalitz plot for three diffractively produced charged pions after a
    cut of $\pm 130\,\MeV/c^2$ around the $\pi_2(1670)$ mass.}
  \label{fig:performance.setup.dalitz}
\end{figure}

Three-body decays of short-lived resonances with correspondingly larger widths
can be studied in Dalitz plots or by using the technique of PWA. A
high-statistics Dalitz plot for the $\pi^-\pi^+\pi^-$ final state ($5\EE{7}$
events) is depicted in \figref{fig:performance.setup.dalitz}, where the
invariant mass of the $3\pi$ system was required to be within $\pm
130\,\MeV/c^2$ around the nominal mass of the $\pi_2(1670)$ resonance. The bands
correspond to the decays $\pi_2(1670)\rightarrow\rho\pi$ and
$\pi_2(1670)\rightarrow f_2(1270)\pi$.

\section{Summary}
\label{sec:summary}
In this paper, a detailed description of the COMPASS experimental setup as used
for the physics programme with hadron beams is given.  Operational since 2002,
the setup was designed for both hadron structure and hadron spectroscopy
studies. It makes use of the various beams available at the CERN M2 beam line,
namely positive and negative hadrons, positive and negative muons, and
electrons.  The apparatus operates with beams in the energy range of 100 to
200~GeV and is able to detect charged and neutral hadrons in the final
state. Two large-aperture dipole magnets provide wide angular and momentum
acceptances. While the major part of the setup remains essentially unchanged,
its target region is reconfigurable as a function of the specific experimental
programme.

For several years, the COMPASS setup was successfully used with muon beams and
with a large-size polarised target ,or spin structure studies.  After a week of
exploratory data taking period in 2004, an important part of the COMPASS hadron
programme was conducted in 2008 and 2009.  Over the years, several new
components were added to the setup, according to the requirements of the hadron
physics programme and also to improve the overall performance of the apparatus.

Immediately upstream of the COMPASS setup, two CEDAR detectors were installed
into the M2 beam line.  Based on the Cherenkov effect, the CEDARs identify the
hadron beam particle, separating kaons, pions and protons. A new target system,
consisting of either a solid-state target holder or a liquid hydrogen target was
built. A Recoil Proton Detector, surrounding the target, provides access to
exclusive measurements. An accurate vertex resolution was achieved by adding
nitrogen-cooled Silicon microstrip detectors upstream and downstream of the
target.
 
Several new PixelGEM detectors were positioned along the setup for particle
tracking at very small angles. Modified Micromegas detectors were used for
tracking immediately downstream of the target in the presence of high hadron
fluxes.  Two additional large-size drift chamber detectors were also installed
in order to improve the detection at large polar angles.

Both charged and neutral particle identifications were considerably improved.
An important upgrade of the RICH-1 detector was carried out, resulting in higher
efficiency and increased rate capability.  The ECAL1 calorimeter was completed
and added to the setup, while ECAL2 was modified to withstand the high flux in
the case of hadron beams.

The main part of the trigger system was rebuilt for use with hadron
beams. Several new trigger and veto elements such as recoil proton detector,
multiplicity counter, and sandwich veto were added, thereby optimizing the
system for diffractive scattering. A new digital calorimeter trigger was
developed for selecting Primakoff reactions. The data acquisition system was
further tuned in order to stand high trigger rates with low dead time. The
detector control system was adapted to include the new detectors and upgraded
with new monitoring features.

All new detectors were successfully included in the full software analysis
chain.  The tracking, reconstruction, simulation, and analysis tools were
updated and adapted to the use with hadron beams.  The acceptance of the
apparatus covers large angular and momentum ranges and is nearly uniform for all
kinematical variables.  The overall characteristics of the setup illustrate its
important potential for hadron spectroscopy studies. Invariant masses of up to 3
GeV$/c^2$ are covered with statistical accuracies significantly better than in
previous experiments. The good energy resolutions achieved allow access to a
large number of meson and baryon resonances. In summary, the upgraded COMPASS
setup is fully operational for use with the various hadron beams available at
CERN.

\section*{Acknowledgements}
We gratefully acknowledge the CERN laboratory and the CERN BE, EN, IT, TE and PH
departments for providing constant and efficient support during the upgrade
phase of our experimental setup and during data taking. We express our gratitude
to the numerous engineers and technicians from our home institutions, who have
contributed to the construction and later to the maintenance of our detectors
and equipment. We are also grateful to A.~Alting\"un for help in the preparation
of the numerous figures.

We acknowledge support from MEYS Grants ME492 and LA242 (Czech Republic), CEA
(France), Bundesministerium f\"ur Bildung und Forschung, DFG cluster of
excellence ``Origin and Structure of the Universe'' and DFG Research Training
Group Programme 1102 ``Physics at Hadron Accelerators'' (Germany), CERN-RFBR
Grants 08-02-91009 and 12-02-91500, Israel Science Foundation, founded by the
Israel Academy of Sciences and Humanities (Israel), INFN and MIUR (Italy), MEXT
and JSPS Grants Nos. 18002006, 20540299 and 18540281, Daiko Foundation and
Yamada Foundation (Japan), SAIL (CSR) (Government of India), NCN Grant
DEC-2011/01/M/ST2/02350 (Poland), Funda\c{c}\={a}o para a Ci\^{e}ncia e
Tecnologia, COMPETE and QREN, Grants CERN/\-FP/109323/2009,
CERN/\-FP/\-116376/2010 and CERN/FP/123600/2011 (Portugal) and from European
Union FP7 (HadronPhysics3, Grant Agreement number 283286).

\newcommand{\SortNoop}[1]{} \newcommand{\OneLetter}[1]{#1}
\newcommand{\SwapArgs}[2]{#2#1}

\end{document}